\documentclass[a4paper, 10pt, oneside]{article}
\usepackage[a4paper,inner=2cm,outer=2cm,top=3cm,bottom=2.5cm]{geometry}

\usepackage[tbtags]{amsmath}
\usepackage{amssymb}
\usepackage{amsthm}
\usepackage{dsfont}
\usepackage{slashed}
\usepackage{mathrsfs}
\usepackage[tbtags]{mathtools}
\usepackage{color}

\usepackage[english]{babel}

\usepackage[utf8x]{inputenc}

\usepackage{graphicx,epsfig}

\usepackage{setspace}

\usepackage{booktabs}

\usepackage{float}

\usepackage{dcolumn}

\usepackage{nextpage}

\usepackage{newclude}

\usepackage[small,bf]{caption}
\usepackage{subcaption}

\usepackage{url}

\usepackage[colorlinks=true,
	linkcolor=black,
	citecolor=black,
	urlcolor=black,
	filecolor=black
]{hyperref}

\usepackage[T1]{fontenc}

\usepackage{cancel}

\usepackage{pifont}

\usepackage{etex}

\usepackage{appendix}

\usepackage[absolute]{textpos}

\usepackage[numbers,sort&compress]{natbib}

\newcommand{\minidiag}[2]{\begin{minipage}{2cm} \includegraphics[height=2cm]{images/#2} \end{minipage}}
\newcommand{\minidiagSize}[3]{\begin{minipage}{#3} \includegraphics[height=#3]{images/#2} \end{minipage}}

\renewcommand{\O}{\mathcal{O}}
\newcommand{\p}{\partial}

\renewcommand{\Im}{\mathrm{Im}}

\newcommand{\<}{\langle}
\renewcommand{\>}{\rangle}

\newcommand{\atan}{\mathrm{arctan}}
\newcommand{\dprime}{{\prime\prime}}

\newcommand{\eps}{\epsilon}

\newcommand{\remark}[1]{}

\newcommand{\hateq}{\mathrel{\widehat{=}}}

\newcommand{\mytag}{\\[-\baselineskip] \stepcounter{equation}\tag{\theequation}}

\newcolumntype{.}{D{.}{.}{2} } 
\newcolumntype{d}{D{.}{.}{2.2} }

\makeatletter
\let\oldvec\vec
\renewcommand{\vec}[1]{\oldvec{#1}\@ifnextchar{^}{\mkern3.5mu\vphantom{#1}}{\mkern2.5mu\vphantom{#1}}}
\makeatother

\makeatletter
\renewcommand{\maketag@@@}[1]{\hbox{\m@th\normalsize\normalfont#1}}%
\makeatother

\makeatletter
\renewcommand\paragraph{\@startsection{paragraph}{4}{\z@}%
  {-3.25ex \@plus -1ex \@minus -0.2ex}%
  {0.01pt}%
  {\bfseries}%
}
\makeatother

\usepackage{abstract}

\numberwithin{equation}{section}

\allowdisplaybreaks[1]

\begin{document}

\mbox{}

\bigskip

\begin{center}
{\LARGE{\bf Dispersion relation for hadronic light-by-light scattering:\\[2mm] theoretical foundations}}

\vspace{0.5cm}

Gilberto Colangelo${}^a$, Martin Hoferichter${}^{b,c,d,a}$, Massimiliano Procura${}^{e,a}$, Peter Stoffer${}^{f,a}$

\vspace{1em}

\begin{center}
\it
${}^a$Albert Einstein Center for Fundamental Physics, Institute for Theoretical Physics, \\
University of Bern, Sidlerstrasse~5, 3012 Bern, Switzerland \\
\mbox{} \\
${}^b$Institut f\"ur Kernphysik, Technische Universit\"at Darmstadt, 64289 Darmstadt, Germany \\
\mbox{} \\
${}^c$ExtreMe Matter Institute EMMI, GSI Helmholtzzentrum f\"ur Schwerionenforschung GmbH, \\
64291 Darmstadt, Germany \\
\mbox{} \\
${}^d$Institute for Nuclear Theory, University of Washington, Seattle, WA 98195-1550, USA \\
\mbox{} \\
${}^e$Fakult\"at f\"ur Physik, Universit\"at Wien, Boltzmanngasse 5, 1090 Wien, Austria \\
\mbox{} \\
${}^f$Helmholtz-Institut f\"ur Strahlen- und Kernphysik (Theory) and Bethe Center for Theoretical Physics, University of Bonn, 53115 Bonn, Germany
\end{center} 

\end{center}

\vspace{1em}

\hrule

\begin{abstract}
In this paper we make a further step towards a dispersive description of
the hadronic light-by-light (HLbL) tensor, which should ultimately lead
to a data-driven evaluation of its contribution to $(g-2)_\mu$. We first
provide a Lorentz decomposition of the HLbL tensor performed according to
the general recipe by Bardeen, Tung, and Tarrach, generalizing and extending our
previous approach, which was constructed in terms of a basis of helicity amplitudes.
Such a tensor decomposition has several advantages: 
the role of gauge invariance and crossing symmetry becomes fully transparent;
the scalar coefficient functions are free of kinematic singularities and zeros, and thus fulfill a Mandelstam double-dispersive representation; and the explicit relation for the HLbL contribution to $(g-2)_\mu$ in terms of the coefficient functions simplifies substantially.  
We demonstrate explicitly that the dispersive approach defines both the pion-pole and
the pion-loop contribution unambiguously and in a model-independent way. 
The pion loop, dispersively defined as pion-box topology, is proven to coincide exactly with the one-loop scalar QED amplitude, multiplied by the appropriate pion vector form factors.
\end{abstract}

\setlength{\TPHorizModule}{1cm}
\setlength{\TPVertModule}{1cm}
\textblockorigin{\paperwidth}{0cm}
\begin{textblock}{4}(-6,2)
\raggedleft
\footnotesize
INT-PUB-15-019 \\
UWTHPH-2015-10
\end{textblock}

\hrule

\vspace{1em}
\setcounter{tocdepth}{3}
\tableofcontents


\section{Introduction}

The anomalous magnetic moment of the muon, $a_\mu = (g-2)_\mu / 2$, is one
of the very rare quantities in particle physics where a significant
discrepancy between its experimental determination and the Standard-Model
evaluation still persists. 
A robust, reliable estimate of the theoretical uncertainties
is thus mandatory, especially in view of the projected accuracy of the
forthcoming $(g-2)_\mu$ experiments at FNAL~\cite{Grange:2015fou} and
J-PARC~\cite{Saito:2012zz}, to settle whether this discrepancy can be
attributed to New Physics.

The uncertainty on the theory side is dominated by hadronic contributions,
see e.g.~\cite{Jegerlehner2009, Prades2009,Benayoun:2014tra}, the largest
error arising from hadronic vacuum polarization (HVP). Since HVP is
straightforwardly related to the total $e^+e^-$ hadronic cross section via
a dispersion relation, the evaluation of this contribution is expected to
become more accurate within the next few years~\cite{Blum2013} thanks to
upcoming improvements in the experimental input,\footnote{For a recently
  suggested alternative approach to determine the leading hadronic
  correction to $a_\mu$, based on space-like data extracted from Bhabha
  scattering, see~\cite{Calame:2015fva}.}  although reducing further the
present sub-percent accuracy will be challenging.  This implies that
hadronic light-by-light (HLbL) scattering is likely to soon dominate the
theory uncertainty in $a_\mu$.\footnote{At this order in the fine-structure
  constant also two-loop diagrams with HVP insertion
  appear~\cite{Calmet:1976kd}. Higher-order hadronic contributions have
  been investigated in~\cite{Kurz:2014wya,Colangelo:2014qya}.} Using
previous approaches~\cite{Rafael1994, Bijnens1995, Bijnens1996,
  Bijnens2002, Hayakawa1995, Hayakawa1996, Hayakawa1998, Knecht2002a,
  Knecht2002, Ramsey-Musolf2002, Melnikov2004, Goecke2011}, systematic
errors are difficult, if not impossible, to quantify. A novel strategy is
required to go beyond the state of the art, to avoid or at least reduce
model dependence to a minimum, make a solid estimate of theoretical errors,
and possibly reduce them. Lattice QCD is a natural candidate to achieve
this goal, but it is not clear yet if or when this method will become
competitive~\cite{Hayakawa2006, Blum2012,Blum:2014oka}. Alternatively, as
we recently argued~\cite{Colangelo2014a}, it is possible to establish a
rigorous framework based on dispersion theory that directly links
$a_\mu^{\rm HLbL}$ to experimentally accessible on-shell form factors and
scattering amplitudes~\cite{Colangelo2014b}, contrary to what has been
previously conjectured~\cite{Kinoshita1985, Bijnens1995, Blum2012,
  Rafael2013}.\footnote{For a different approach aiming at a data-driven
  evaluation of $a_\mu^{\rm HLbL}$, based on a dispersive description of
  the Pauli form factor instead of the HLbL tensor,
  see~\cite{Pauk:2014rfa}.}

Our dispersive formalism is based on the fundamental principles of
unitarity, analyticity, crossing symmetry, and gauge invariance. The
derivation of such a systematic framework becomes challenging due to the
fact that HLbL scattering is described by a hadronic four-point function
whose properties are significantly more complicated than those of the
two-point function entering HVP. Besides the opportunity to achieve a
data-driven evaluation of $a_\mu^{\rm HLbL}$, this approach allows us to
unambiguously define and evaluate the various low-energy contributions to
HLbL scattering, most notably pion pole and pion loop.

The key result of~\cite{Colangelo2014a} was a master formula giving the
contribution to $a_\mu^{\rm HLbL}$ from the pion-pole term and intermediate
$\pi \pi$ states expressed in terms of the pion transition form factor and
helicity partial waves for $\gamma^* \gamma^* \to \pi \pi$. In the present
paper we reformulate this approach in a more general context, by employing
a different generating set for the Lorentz structures of the HLbL tensor.
This new set, which we constructed following the prescription by Bardeen,
Tung~\cite{Bardeen1968}, and Tarrach~\cite{Tarrach1975} (BTT), has the
property of being explicitly free of kinematic singularities and
zeros~\cite{Stoffer2014}. Therefore, the scalar coefficient functions
associated with each Lorentz structure of this generating set are also free
of kinematic singularities and zeros, and thus fulfill a Mandelstam
double-dispersive representation.  Further motivation for deriving such a
BTT set is provided by the fact that the role of gauge invariance and
crossing symmetry becomes fully transparent, with constraints from
soft-photon zeros incorporated automatically. Moreover, the absence of
kinematic singularities makes it much easier to perform the angular averages
necessary to calculate the contribution to $(g-2)_\mu$ --- indeed both the
derivation and the final form of the master formula are simplified
considerably in the present setting, and allow for a simpler inclusion of
partial-wave contributions beyond $S$-waves. In the derivation of the master
formula it is very useful to perform a Wick rotation of the integration
variables. This has been already used in several papers in the literature
without ever addressing an important subtlety which is relevant in the
present case: namely if the Wick rotation goes through without changes even
if the integrand is a loop function showing anomalous thresholds. We
discuss this point in detail and prove that this is indeed the case.
Finally, the BTT formulation facilitates the proof that both pion-pole and
pion-loop contributions are uniquely defined in the dispersive approach,
and confirms the relation to the pion transition form factor and the pion
electromagnetic form factor given in~\cite{Colangelo2014a}. We explicitly
proof that the scalar QED (sQED) pion loop dressed with pion form factors
(denoted by FsQED in~\cite{Colangelo2014a}) is identical to the
contribution from $\pi\pi$ intermediate states with a pion-pole left-hand
cut (LHC).

In summary, this paper contains three main results: (i) the explicit construction of
the BTT decomposition of the HLbL tensor; (ii) the derivation of the master
formula expressing the HLbL contribution in terms of the BTT scalar
functions; (iii) the proof that in the dispersive approach the FsQED
contribution is uniquely defined. Everything which goes beyond the FsQED
contribution is amenable to partial-wave expansion and will be discussed in
full detail in a forthcoming publication.

The present paper is structured as follows. In Sect.~\ref{sec:SubProcess}, we first
review some aspects of the process $\gamma^*\gamma^*\to\pi\pi$, mainly as
an illustration of the techniques that we apply afterwards to HLbL. In
Sect.~\ref{sec:LorentzStructureHLbLTensor}, we derive the decomposition of
the HLbL tensor into a set of scalar functions that are free of kinematics.
In Sect.~\ref{sec:HLbLContributionToGminus2}, this BTT decomposition is
then used to derive a master formula for $a_\mu^{\rm HLbL}$ that is
parametrized by the corresponding scalar functions.  We verify the validity
of the Wick rotation of the two-loop integral in the calculation of
$a_\mu^{\rm HLbL}$, even in the case of anomalous thresholds in the HLbL
tensor.  In Sect.~\ref{sec:MandelstamRepresentation}, we derive a
Mandelstam representation of the BTT scalar functions, and, by comparison
to the loop formulation of the sQED pion loop, demonstrate that it indeed
coincides with the contribution of $\pi\pi$ intermediate states with a
pion-pole LHC.  We conclude with an outlook in
Sect.~\ref{sec:HLbLDiscussionConclusion}, while some of the lengthier
expressions and derivations are collected in the appendices.


\section{The sub-process $\boldsymbol{\gamma^*\gamma^*\to\pi\pi}$}

\label{sec:SubProcess}

As a prelude, we discuss in this section the process
$\gamma^*\gamma^*\to\pi\pi$, which will become important as a sub-process
when we write down a dispersion relation for the HLbL tensor. Because the
Lorentz structure of $\gamma^*\gamma^*\to\pi\pi$ is much simpler than the
one of light-by-light scattering, it also allows us to illustrate a
technique for the construction of amplitudes that are free of kinematic
singularities and zeros, which we will apply afterwards to the more
complicated case of HLbL. In order to render this analogy manifest, we
slightly change conventions compared to~\cite{Colangelo2014a}: we identify
here $\gamma^*\gamma^*\to\pi\pi$ as the $s$-channel process instead of
$\gamma^*\pi\to\gamma^*\pi$. This assignment is slightly unnatural when it
comes to constructing dispersion relations for a process with these
crossing properties,
see~\cite{Hite:1973pm,Buettiker:2003pp,Hoferichter2011,Ditsche:2012fv}, but
makes the inclusion into the unitarity relation in HLbL more
straightforward.

\subsection{Kinematics and matrix element}

\begin{figure}[h]
	\centering
	\includegraphics[width=4cm]{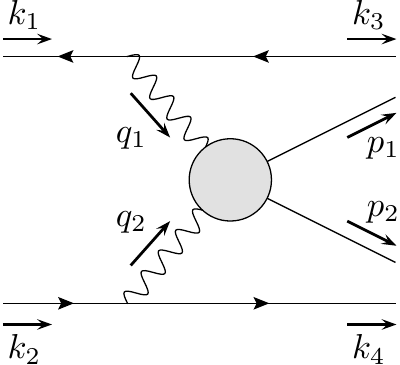}
	\caption{$\gamma^*\gamma^*\to\pi\pi$ as a sub-process of $e^+e^-\to e^+e^-\pi\pi$.}
	\label{img:eetopipi}
\end{figure}

Consider the process
\begin{align}
	e^+(k_1) e^-(k_2) \to e^+(k_3) e^-(k_4) \gamma^*(q_1) \gamma^*(q_2) \to e^+(k_3) e^-(k_4) \pi^a(p_1) \pi^b(p_2),
\end{align}
shown in Fig.~\ref{img:eetopipi}. At $\O(e^4)$, the amplitude for this process is given by
\begin{align}
	\begin{split}
		i \mathcal{T} &= \bar v(k_1) (-i e \gamma_\alpha) v(k_3) \bar u(k_4) (-i e \gamma_\beta) u(k_2) \\
			& \quad \times \frac{-i}{q_1^2} \left( g^{\alpha\mu} - (1-\xi) \frac{q_1^\alpha q_1^\mu}{q_1^2} \right) \frac{-i}{q_2^2} \left( g^{\beta\nu} - (1-\xi) \frac{q_2^\beta q_2^\nu}{q_2^2} \right) i e^2 W_{\mu\nu}^{ab}(p_1,p_2,q_1) ,
	\end{split}
\end{align}
where $\xi$ is an arbitrary gauge parameter for the photon propagators and the tensor $W_{\mu\nu}^{ab}$ is defined as the pure QCD matrix element
\begin{align}
	W^{\mu\nu}_{ab}(p_1,p_2,q_1) = i \int d^4x \, e^{-i q_1 \cdot x} \<\pi^a(p_1) \pi^b(p_2) | T \{ j_\mathrm{em}^\mu(x) j_\mathrm{em}^\nu(0) \} | 0 \> .
\end{align}
The contraction thereof with appropriate polarization vectors may be understood as an amplitude for the off-shell process
\begin{align}
	\gamma^*(q_1,\lambda_1) \gamma^*(q_2,\lambda_2) \to \pi^a(p_1) \pi^b(p_2) ,
\end{align}
where $\lambda_{1,2}$ denote the helicities of the off-shell photons. We define the connected part of such a matrix element by
\begin{align}
	\begin{split}
		\< \pi^a(p_1) & \pi^b(p_2) | \gamma^*(q_1,\lambda_1) \gamma^*(q_2,\lambda_2) \> \\
		:=&{} - e^2 \epsilon_\mu^{\lambda_1}(q_1) \epsilon_\nu^{\lambda_2}(q_2) \int d^4x \, d^4y \, e^{-i (q_1 \cdot x + q_2 \cdot y)} \<\pi^a(p_1) \pi^b(p_2)| T \{ j_\mathrm{em}^\mu(x) j_\mathrm{em}^\nu(y) \} | 0 \> \\
		=&{} - e^2 (2\pi)^4 \delta^{(4)}(p_1+p_2-q_1-q_2) \epsilon_\mu^{\lambda_1}(q_1) \epsilon_\nu^{\lambda_2}(q_2) \\
			&\quad \times \int d^4x \, e^{-i q_1 \cdot x} \<\pi^a(p_1) \pi^b(p_2)| T \{ j_\mathrm{em}^\mu(x) j_\mathrm{em}^\nu(0) \} | 0 \>  \\
		=&{} i e^2 (2\pi)^4 \delta^{(4)}(p_1+p_2-q_1-q_2) \epsilon_\mu^{\lambda_1}(q_1) \epsilon_\nu^{\lambda_2}(q_2) W^{\mu\nu}_{ab}(p_1,p_2,q_1) .
	\end{split}
\end{align}
The helicity amplitudes are given by the contraction with polarization vectors:
\begin{align}
	 \epsilon_\mu^{\lambda_1}(q_1) \epsilon_\nu^{\lambda_2}(q_2) W^{\mu\nu}_{ab}(p_1,p_2,q_1) = e^{i(\lambda_1-\lambda_2)\phi} H_{\lambda_1\lambda_2}^{ab} .
\end{align}
We introduce the following kinematic variables:\footnote{In this Sect.~\ref{sec:SubProcess} only, $s$, $t$, and $u$ refer to the Mandelstam variables of $\gamma^*\gamma^*\to\pi\pi$ and not to the ones of the HLbL tensor.}
\begin{align}
	\begin{split}
		s &:= (q_1+q_2)^2 = (p_1+p_2)^2 , \\
		t &:= (q_1-p_1)^2 = (q_2-p_2)^2 , \\
		u &:= (q_1-p_2)^2 = (q_2-p_1)^2 ,
	\end{split}
\end{align}
which satisfy $s+t+u = q_1^2 + q_2^2 + 2 M_\pi^2$.

\subsection{Tensor decomposition}

\label{sec:ggtopipiTensorDecomposition}

The tensor $W^{\mu\nu}_{ab}$ can be decomposed based on Lorentz covariance as (we drop isospin indices)
\begin{align}
	W^{\mu\nu} = g^{\mu\nu} W_1 + q_i^\mu q_j^\nu W_2^{ij},
\end{align}
where we abbreviate $q_i = \{ q_1, q_2, p_2-p_1 \}$ and where double indices are summed. The ten coefficient functions $\{W_1, W_2^{ij}\}$ cannot contain any kinematic but only dynamic singularities. However, they have to fulfill kinematic constraints that are required e.g.\ by gauge invariance, hence they contain kinematic zeros. Conservation of the electromagnetic current implies the Ward identities
\begin{align}
	q_1^\mu W_{\mu\nu} = q_2^\nu W_{\mu\nu} = 0 ,
\end{align}
hence a priori six relations between the scalar functions $\{W_1, W_2^{ij}\}$. Only five of them are linearly independent and thus reduce the set of scalar functions to five independent ones.

We now construct a set of scalar functions which are free of both kinematic singularities and zeros. We follow the recipe given by Bardeen, Tung~\cite{Bardeen1968}, and Tarrach~\cite{Tarrach1975}. As was shown in~\cite{Tarrach1975}, the basis (consisting of five functions) constructed according to the recipe of~\cite{Bardeen1968} is not free of kinematic singularities. However, a redundant set of six structures can be constructed, which fulfills the requirement.

In the case of $\gamma^*\gamma^*\to\pi\pi$, there exists in addition to gauge invariance and crossing symmetry of the photons also the crossing (Bose) symmetry of the pions, which is responsible for additional kinematic zeros. In fact, this can be used to circumvent the introduction of a sixth function, reducing the set again to five scalar functions free of both kinematic singularities and zeros. This was first derived in~\cite{Drechsel1998} in the context of doubly-virtual nucleon Compton scattering.\footnote{This does not mean that the Lorentz structures are independent in all kinematic limits, hence Tarrach's conclusion that no minimal basis exists is not invalidated. We thank J.~Gasser for bringing this to our attention.}

We define the projector
\begin{align}
	I^{\mu\nu} = g^{\mu\nu} - \frac{q_2^\mu q_1^\nu}{q_1 \cdot q_2} ,
\end{align}
which satisfies
\begin{align}
	\begin{split}
		q_1^\mu I_{\mu\nu} &= q_2^\nu I_{\mu\nu} = 0 , \\
		{I_\mu}^\lambda W_{\lambda\nu} &= W_{\mu\lambda} {I^\lambda}_\nu = W_{\mu\nu} ,
	\end{split}
\end{align}
i.e.~the tensor $W^{\mu\nu}$ is invariant under contraction with the projector, but contracting the projector with any Lorentz structure produces a gauge-invariant structure.

We apply this projector for both photons:
\begin{align}
	W_{\mu\nu} &= I_{\mu\mu^\prime} I_{\nu^\prime\nu} W^{\mu^\prime\nu^\prime} = \sum_{i=1}^5 \bar T_{\mu\nu}^i \bar B_i ,
\end{align}
where
\begin{align*}
	\bar T_1^{\mu\nu} &= g^{\mu\nu} - \frac{q_2^\mu q_1^\nu}{q_1 \cdot q_2} ,  & \bar B_1 &= W_1 , \\
	\bar T_2^{\mu\nu} &= q_1^\mu q_2^\nu - \frac{q_1^2 q_2^\mu q_2^\nu}{q_1\cdot q_2} - \frac{q_2^2 q_1^\mu q_1^\nu}{q_1 \cdot q_2} + \frac{q_1^2 q_2^2 q_2^\mu q_1^\nu}{(q_1 \cdot q_2)^2} ,  & \bar B_2 &= W_2^{12} , \\
	\bar T_3^{\mu\nu} &= q_1^\mu q_3^\nu - \frac{q_1^2 q_2^\mu q_3^\nu}{q_1 \cdot q_2} - \frac{q_2 \cdot q_3 q_1^\mu q_1^\nu}{q_1 \cdot q_2} + \frac{q_1^2 q_2 \cdot q_3 q_2^\mu q_1^\nu}{(q_1 \cdot q_2)^2} ,  & \bar B_3 &= W_2^{13} , \\
	\bar T_4^{\mu\nu} &= q_3^\mu q_2^\nu - \frac{q_2^2 q_3^\mu q_1^\nu}{q_1 \cdot q_2} - \frac{q_1 \cdot q_3 q_2^\mu q_2^\nu}{q_1 \cdot q_2} + \frac{q_2^2 q_1 \cdot q_3 q_2^\mu q_1^\nu}{(q_1 \cdot q_2)^2} ,  & \bar B_4 &= W_2^{32} , \\
	\bar T_5^{\mu\nu} &= q_3^\mu q_3^\nu - \frac{q_1 \cdot q_3 q_2^\mu q_3^\nu}{q_1 \cdot q_2} - \frac{q_2 \cdot q_3 q_3^\mu q_1^\nu}{q_1 \cdot q_2} + \frac{q_1 \cdot q_3 q_2 \cdot q_3 q_2^\mu q_1^\nu}{(q_1 \cdot q_2)^2} ,  & \bar B_5 &= W_2^{33} . \mytag
\end{align*}
As the functions $\bar B_i$ are a subset of the original scalar functions, they are still free of kinematic singularities, but contain zeros, because the Lorentz structures contain singularities. We have to remove now the single and double poles in $q_1 \cdot q_2$ from the Lorentz structures $\bar T_i^{\mu\nu}$. This is achieved as follows \cite{Bardeen1968}:
\begin{itemize}
	\item remove as many double poles as possible by adding to the structures linear combinations of other structures with non-singular coefficients,
	\item if no more double poles can be removed in this way, multiply the structures that still contain double poles by $q_1 \cdot q_2$,
	\item proceed in the same way with single poles.
\end{itemize}

It turns out that no double poles in $q_1 \cdot q_2$ can be removed by adding to the structures multiples of the other structures, hence $\bar T_2^{\mu\nu}$, \ldots, $\bar T_5^{\mu\nu}$ have to be multiplied by $q_1 \cdot q_2$. The resulting simple poles can be removed by adding multiples of $\bar T_1^{\mu\nu}$. In the end, we have to multiply $\bar T_1^{\mu\nu}$ by $q_1\cdot q_2$ in order to remove the last pole. We then arrive at the following representation:
\begin{align}
	W_{\mu\nu} &= \sum_{i=1}^5 \tilde T_{\mu\nu}^i \tilde B_i ,
\end{align}
where
\begin{align}
\label{BT_basis}
	\begin{split}
		\tilde T_1^{\mu\nu} &= q_1 \cdot q_2 g^{\mu\nu} - q_2^\mu q_1^\nu , \\
		\tilde T_2^{\mu\nu} &= q_1^2 q_2^2 g^{\mu\nu} + q_1 \cdot q_2 q_1^\mu q_2^\nu - q_1^2 q_2^\mu q_2^\nu - q_2^2 q_1^\mu q_1^\nu , \\
		\tilde T_3^{\mu\nu} &= q_1^2 q_2 \cdot q_3 g^{\mu\nu} + q_1 \cdot q_2 q_1^\mu q_3^\nu - q_1^2 q_2^\mu q_3^\nu - q_2 \cdot q_3 q_1^\mu q_1^\nu , \\
		\tilde T_4^{\mu\nu} &= q_2^2 q_1 \cdot q_3 g^{\mu\nu} + q_1 \cdot q_2 q_3^\mu q_2^\nu - q_2^2 q_3^\mu q_1^\nu - q_1 \cdot q_3 q_2^\mu q_2^\nu , \\
		\tilde T_5^{\mu\nu} &= q_1 \cdot q_3 q_2 \cdot q_3  g^{\mu\nu} + q_1 \cdot q_2 q_3^\mu q_3^\nu - q_1 \cdot q_3 q_2^\mu q_3^\nu - q_2 \cdot q_3 q_3^\mu q_1^\nu ,
	\end{split}
\end{align}
and
\begin{align*}
	\tilde B_1 &= \frac{1}{q_1 \cdot q_2} W_1 - \frac{q_1^2 q_2^2}{(q_1 \cdot q_2)^2} W_2^{12} - \frac{q_1^2 q_2 \cdot q_3}{(q_1\cdot q_2)^2} W_2^{13} - \frac{q_2^2 q_1\cdot q_3}{(q_1\cdot q_2)^2} W_2^{32} - \frac{q_1 \cdot q_3 q_2 \cdot q_3}{(q_1 \cdot q_2)^2} W_2^{33} , \\
	\tilde B_2 &= \frac{1}{q_1 \cdot q_2} W_2^{12} , \\
	\tilde B_3 &= \frac{1}{q_1 \cdot q_2} W_2^{13} , \\
	\tilde B_4 &= \frac{1}{q_1 \cdot q_2} W_2^{32} , \\
	\tilde B_5 &= \frac{1}{q_1 \cdot q_2} W_2^{33} . \mytag
\end{align*}
The Lorentz structures~\eqref{BT_basis} agree with the basis used in~\cite{Colangelo2014a}, upon $q_2\to -q_2$, $p_1\to -p_1$ due to crossing and the identification
\begin{equation}
 \tilde T_{1,\mu\nu}=T_{1,\mu\nu}^\text{\cite{Colangelo2014a}},\qquad 
  \tilde T_{2,\mu\nu}=T_{3,\mu\nu}^\text{\cite{Colangelo2014a}}, \qquad
   \tilde T_{3,\mu\nu}=\frac{1}{2}T_{4,\mu\nu}^\text{\cite{Colangelo2014a}}, \qquad
    \tilde T_{4,\mu\nu}=-\frac{1}{2}T_{5,\mu\nu}^\text{\cite{Colangelo2014a}}, \qquad
     \tilde T_{5,\mu\nu}=\frac{1}{4}T_{2,\mu\nu}^\text{\cite{Colangelo2014a}}.
\end{equation}

As shown by Tarrach, this basis is not free of kinematic singularities and
zeros \cite{Tarrach1975}. The structures $\tilde T_i^{\mu\nu}$ form a basis
for $q_1 \cdot q_2 \neq 0$, but are degenerate for $q_1 \cdot q_2 =
0$. This degeneracy implies that there is a linear combination of the
structures $\tilde T_i^{\mu\nu}$ that is proportional to $q_1 \cdot
q_2$:
\begin{align}
	q_1 \cdot q_3 q_2\cdot q_3 \tilde T_2^{\mu\nu} - q_2^2 q_1\cdot q_3 \tilde T_3^{\mu\nu} - q_1^2 q_2\cdot q_3 \tilde T_4^{\mu\nu} + q_1^2 q_2^2 \tilde T_5^{\mu\nu} = q_1 \cdot q_2 \tilde T_6^{\mu\nu} ,
\end{align}
where
\begin{align}
	\tilde T_6^{\mu\nu} = \left(q_1^2 q_3^\mu - q_1 \cdot q_3 q_1^\mu \right) \left(q_2^2 q_3^\nu - q_2 \cdot q_3 q_2^\nu \right) .
\end{align}
Hence, in order to have a generating set even at $q_1 \cdot q_2=0$ this sixth
structure $\tilde T_6^{\mu\nu}$
has to be added by hand. Projected on the basis, it gives coefficients with poles in $q_1 \cdot q_2$. Although the basis is not free of kinematic singularities and zeros, we have found the exact form of the singularities:
\begin{align}
	W_{\mu\nu} &= \sum_{i=1}^5 \tilde T_{\mu\nu}^i \tilde B_i =  \sum_{i=1}^6 \tilde T_{\mu\nu}^i B_i ,
\end{align}
where
\begin{align}
	\begin{split}
		\tilde B_1 &= B_1 , \\
		\tilde B_2 &= B_2 + \frac{q_1 \cdot q_3 q_2 \cdot q_3}{q_1 \cdot q_2} B_6 , \\
		\tilde B_3 &= B_3 - \frac{q_2^2 q_1 \cdot q_3}{q_1 \cdot q_2} B_6 , \\
		\tilde B_4 &= B_4 - \frac{q_1^2 q_2 \cdot q_3}{q_1 \cdot q_2} B_6 , \\
		\tilde B_5 &= B_5 + \frac{q_1^2 q_2^2}{q_1 \cdot q_2} B_6 .
	\end{split}
\end{align}
The functions $B_i$ are free of kinematic singularities. The tensor structures can be written in terms of the Mandelstam variables $t$ and $u$ as
\begin{align}
	\begin{split}
		\tilde T_1^{\mu\nu} &= q_1 \cdot q_2 g^{\mu\nu} - q_2^\mu q_1^\nu , \\
		\tilde T_2^{\mu\nu} &= q_1^2 q_2^2 g^{\mu\nu} + q_1 \cdot q_2 q_1^\mu q_2^\nu - q_1^2 q_2^\mu q_2^\nu - q_2^2 q_1^\mu q_1^\nu , \\
		\tilde T_3^{\mu\nu} &= q_1 \cdot q_2 q_1^\mu q_3^\nu - q_1^2 q_2^\mu q_3^\nu - \frac{1}{2} (t-u) q_1^2 g^{\mu\nu} + \frac{1}{2} (t-u) q_1^\mu q_1^\nu , \\
		\tilde T_4^{\mu\nu} &= q_1 \cdot q_2 q_3^\mu q_2^\nu - q_2^2 q_3^\mu q_1^\nu + \frac{1}{2} (t-u) q_2^2 g^{\mu\nu} - \frac{1}{2} (t-u) q_2^\mu q_2^\nu , \\
		\tilde T_5^{\mu\nu} &= q_1 \cdot q_2 q_3^\mu q_3^\nu - \frac{1}{4} (t-u)^2  g^{\mu\nu} + \frac{1}{2} (t-u) \left( q_3^\mu q_1^\nu - q_2^\mu q_3^\nu \right) , \\
		\tilde T_6^{\mu\nu} &= q_1^2 q_2^2 q_3^\mu q_3^\nu + \frac{1}{2}(t-u) \left(q_1^2 q_3^\mu q_2^\nu - q_2^2 q_1^\mu q_3^\nu \right) - \frac{1}{4} (t-u)^2 q_1^\mu q_2^\nu .
	\end{split}
\end{align}

The derivation of the Lorentz decomposition for $\gamma^*\gamma^*\to\pi\pi$
up to this point should illustrate the techniques that will be applied to
the much more complicated case of HLbL in
Sect.~\ref{sec:LorentzStructureHLbLTensor}. In the case of
$\gamma^*\gamma^*\to\pi\pi$, there is in fact a solution to get rid of the
additional redundant structure without introducing kinematic
singularities~\cite{Drechsel1998}. The necessary steps are discussed in the
next subsection. Note that it is an open question whether an analogous procedure to get rid of all redundancies in HLbL exists, see the discussion below~\eqref{eq:AmbiguityFixingCondition}.

\subsection{Kinematic zeros due to crossing antisymmetry}

We have now constructed a redundant set of six Lorentz structures. All kinematic constraints from gauge invariance are implemented and the scalar coefficient functions $B_i$ are free of kinematic singularities. However, due to the crossing symmetry of the pions, additional kinematic zeros are present in the functions $B_i$. As noted in~\cite{Drechsel1998}, this can be exploited to eliminate the redundancy and to work again with basis coefficient functions free of kinematic singularities and zeros.

Since the two-photon state is even under charge conjugation, so must be the two-pion state. Therefore, the isospin $I=1$ amplitude vanishes. Bose symmetry implies that the amplitude and hence the tensor $W^{\mu\nu}$ is invariant under $p_1 \leftrightarrow p_2$ or equivalently $q_3 \leftrightarrow -q_3$. Under this transformation, the tensor structures $\tilde T_i^{\mu\nu}$ are even for $i=1,2,5,6$ and odd for $i=3,4$. Hence,
$B_{1,2,5,6}$ are even, while $B_{3,4}$ must be odd. They contain a kinematic zero of the form
\begin{align}
	B_{3,4} = (t-u) \hat B_{3,4} ,
\end{align}
where $\hat B_{3,4}$ are free of kinematic singularities.

Crossing symmetry of the photons requires the invariance of $W^{\mu\nu}$ under $q_1 \leftrightarrow q_2$, $\mu \leftrightarrow \nu$. While $\tilde T_{1,2,5,6}^{\mu\nu}$ are invariant under this transformation, we observe the crossing relation $\tilde T_3^{\mu\nu} \leftrightarrow \tilde T_4^{\mu\nu}$. Hence, for fixed Mandelstam variables, $B_{1,2,5,6}$ and $\hat B_3 - \hat B_4$ are even under $q_1^2 \leftrightarrow q_2^2$, while $\hat B_3 + \hat B_4$ is odd and contains a kinematic zero $q_1^2-q_2^2$.

We make a basis change
\begin{align}
	W_{\mu\nu} = \sum_{i=1}^5 T^i_{\mu\nu} A_i = \sum_{i=1}^6 \tilde T^i_{\mu\nu} B_i ,
\end{align}
where
\begin{align}
	\begin{split}
		T_1^{\mu\nu} &:= \tilde T_1^{\mu\nu} , \\
		T_2^{\mu\nu} &:= \tilde T_2^{\mu\nu} , \\
		T_3^{\mu\nu} &:= (t-u)(\tilde T_3^{\mu\nu} - \tilde T_4^{\mu\nu}) , \\
		T_4^{\mu\nu} &:= \tilde T_5^{\mu\nu} , \\
		T_5^{\mu\nu} &:= \tilde T_6^{\mu\nu} ,
	\end{split}
\end{align}
i.e.\ we trade off the combination $\tilde T_3^{\mu\nu} + \tilde T_4^{\mu\nu}$, which is even under photon crossing, against the Tarrach structure $\tilde T_6^{\mu\nu}$ and absorb the kinematic zero in $t-u$ into $T_3^{\mu\nu}$. The projected basis functions are then:
\begin{align}
	\begin{split}
		\label{eq:ggpipiFinalBasisFunctions}
		A_1 &= B_1 , \\
		A_2 &= B_2 + \frac{(t-u)^2}{2}\frac{\hat B_3 + \hat B_4}{q_1^2 - q_2^2} , \\
		A_3 &= \frac{\hat B_3 - \hat B_4}{2} + \frac{q_1^2+q_2^2}{2} \frac{\hat B_3 + \hat B_4}{q_1^2 - q_2^2} , \\
		A_4 &= B_5 - 2 q_1^2 q_2^2 \frac{\hat B_3 + \hat B_4}{q_1^2 - q_2^2} , \\
		A_5 &= B_6 + (s-q_1^2-q_2^2)\frac{\hat B_3 + \hat B_4}{q_1^2 - q_2^2} .
	\end{split}
\end{align}
The apparent kinematic singularities in $q_1^2 - q_2^2$ are canceled by the corresponding zero in $\hat B_3 + \hat B_4$. Hence, the basis functions $A_i$ are free of both kinematic singularities and zeros.

\subsection{Helicity amplitudes and soft-photon zeros}

In the following, we construct the helicity amplitudes with the momenta and polarization vectors in the $s$-channel center-of-mass frame. We define the particle momenta as
\begin{align}
	\begin{split}
		q_1 &= ( E_{q_1}, 0, 0, |\vec q| ) , \quad q_2 = ( E_{q_2}, 0, 0, -|\vec q| ) , \\
		p_1 &= ( E_p, |\vec p| \sin \theta \cos\phi, |\vec p| \sin \theta \sin\phi, |\vec p| \cos\theta ) , \\
		p_2 &= ( E_p, -|\vec p| \sin \theta \cos\phi, -|\vec p| \sin \theta \sin\phi, -|\vec p| \cos\theta ) ,
	\end{split}
\end{align}
where
\begin{align}
	\begin{split}
		E_{q_1} &= \sqrt{ q_1^2 + \vec q^2} = \frac{s + q_1^2 - q_2^2}{2\sqrt{s}}, \quad E_{q_2} = \sqrt{ q_2^2 + \vec q^2} = \frac{s - q_1^2 + q_2^2}{2\sqrt{s}}, \quad |\vec q| = \frac{\lambda^{1/2}(s,q_1^2,q_2^2)}{2\sqrt{s}} , \\
		E_p &= \sqrt{ M_\pi^2 + \vec p^2} = \frac{\sqrt{s}}{2} , \quad |\vec p| = \sqrt{\frac{s}{4} - M_\pi^2} = \frac{\sqrt{s}}{2} \sigma_\pi(s) .
	\end{split}
\end{align}
We use the notation
\begin{align}
	\sigma_\pi(s) &:= \sqrt{ 1 - \frac{4M_\pi^2}{s}}  , \quad \lambda(a,b,c) := a^2 + b^2 + c^2 - 2(a b + b c + c a ) .
\end{align}
The scattering angle is then given by
\begin{align}
	z := \cos\theta = \frac{t-u}{4 |\vec q| |\vec p|} = \frac{t-u}{\sigma_\pi(s) \lambda^{1/2}(s,q_1^2,q_2^2)} .
\end{align}
We define the polarization vectors by
\begin{align}
	\begin{split}
		\epsilon_\pm(q_1) &= \mp \frac{1}{\sqrt{2}}( 0, 1, \pm i, 0 ) , \\
		\epsilon_0(q_1) &= \frac{1}{\xi_1}( |\vec q|, 0, 0, E_{q_1} ) , \\
		\epsilon_\pm(q_2) &= \mp \frac{1}{\sqrt{2}}( 0, 1, \mp i, 0 ) , \\
		\epsilon_0(q_2) &= \frac{1}{\xi_2}( -|\vec q|, 0, 0, E_{q_2} ) .
	\end{split}
\end{align}
The longitudinal polarization vectors are normalized to one for $\xi_i = \sqrt{q_i^2}$. However, as the off-shell photons are not physical states, the choice of $\xi_i$ has no influence on any physical observable. The fact that the dependence on the normalization has to drop out in the end provides a useful check on the calculation, so that we will keep $\xi_i$ general in the following. 

The helicity amplitudes are then given by
\begin{align*}
	\label{eq:ggpipiHelicityAmplitudes}
	H_{++} = H_{--} &= -\frac{1}{2}(s-q_1^2-q_2^2) A_1 - q_1^2 q_2^2 A_2 + \frac{1}{2s}(s-4M_\pi^2)\lambda_{12}(s)z^2( q_1^2 + q_2^2 ) A_3 \\
		& \quad + \frac{1}{4} (s-4M_\pi^2)\left( (s - q_1^2 - q_2^2) + \left( \frac{(q_1^2-q_2^2)^2}{s} - (q_1^2 + q_2^2)\right) z^2 \right) A_4 \\
		& \quad + \frac{1}{2} q_1^2 q_2^2 (s-4M_\pi^2) (1-z^2) A_5 , \\
	H_{+-} = H_{-+} &= - \frac{1}{4} (s-4M_\pi^2) (1-z^2) \bigg( (s-q_1^2-q_2^2) A_4 + 2 q_1^2 q_2^2 A_5 \bigg) , \\
	H_{+0} = - H_{-0} &= \frac{q_2^2}{4 \xi_2} \sqrt{\frac{2}{s}} (s-4M_\pi^2) z \sqrt{1-z^2} \bigg( \lambda_{12}(s) A_3 - (s+q_1^2-q_2^2) A_4  - q_1^2 (s-q_1^2 + q_2^2) A_5 \bigg) , \\
	H_{0+} = - H_{0-} &= \frac{q_1^2}{4\xi_1} \sqrt{\frac{2}{s}} (s-4M_\pi^2) z \sqrt{1-z^2} \bigg( \lambda_{12}(s) A_3  - (s-q_1^2+q_2^2) A_4  - q_2^2 (s + q_1^2 - q_2^2) A_5 \bigg) , \\
	H_{00} &= \frac{q_1^2 q_2^2}{\xi_1 \xi_2} \begin{aligned}[t]
		&\bigg( - A_1 - \frac{1}{2}(s - q_1^2 - q_2^2) A_2 - \frac{1}{s}(s-4M_\pi^2)\lambda_{12}(s)z^2 A_3 \\
		& + (s-4M_\pi^2) z^2 A_4 + \frac{1}{4s} (s - 4M_\pi^2) \left(s^2 - (q_1^2 - q_2^2)^2\right) z^2 A_5 \bigg) , \end{aligned} \mytag
\end{align*}
where $\lambda_{12}(s) := \lambda(s,q_1^2,q_2^2)$. Since the functions $A_i$ are free of kinematic singularities and zeros, we can read off from these equations the soft-photon zeros \cite{Low1958, Moussallam2013}. In the limit $q_1\to0$, the Mandelstam variables become
\begin{align}
	s = q_2^2, \quad t = u = M_\pi^2 .
\end{align}
We conclude that the helicity amplitudes must vanish at this point apart from terms containing a dynamic singularity, which will be discussed in Sect.~\ref{sec:PionPoleggpipi}. The second soft-photon limit, $q_2\to0$, leads to
\begin{align}
	s = q_1^2, \quad t = u = M_\pi^2 ,
\end{align}
and the same arguments apply for the helicity amplitudes.

Crossing symmetry of the photons implies that under the transformation $q_1^2\leftrightarrow q_2^2$ (and fixed Mandelstam variables), $H_{++}$, $H_{+-}$, and $H_{00}$ remain invariant, the other two helicity amplitudes transform as $H_{+0}\leftrightarrow H_{0+}$.

\subsection{Partial-wave expansion}

To make the connection to~\cite{Colangelo2014a}, we briefly discuss the consequences of the decomposition~\eqref{eq:ggpipiHelicityAmplitudes} for the kernel functions that appear in partial-wave dispersion relations for $\gamma^*\gamma^*\to\pi\pi$. Partial waves are most conveniently defined for helicity amplitudes.
Using the formalism of~\cite{Jacob1959}, we write the partial-wave expansions as
\begin{align}
	H_{\lambda_1\lambda_2}(s,t,u) = \sum_J (2J + 1) d_{m0}^J(z) h_{J,\lambda_1\lambda_2}(s) ,
\end{align}
where $d_{m0}^J$ is the Wigner $d$-function, $m=|\lambda_1-\lambda_2|$, and the helicity partial waves $h_{J,\lambda_1\lambda_2}$ depend implicitly on the photon virtualities $q_1^2$ and $q_2^2$. Since the isospin of the two-pion system is $I=0,2$, only even partial waves are allowed. For $m=0$, i.e.~$H_{++}$ and $H_{00}$, the partial-wave expansion starts at $J=0$, otherwise at $J=2$. Note also that $d_{00}^J(z) = P_J(z)$ are the Legendre polynomials.

The partial waves can be obtained by projection:
\begin{align}
	h_{J,\lambda_1\lambda_2}(s) = \frac{1}{2} \int_{-1}^1 dz \, d_{m0}^J(z) H_{\lambda_1\lambda_2}(s,z) .
\end{align}

Since the functions $A_i$ are free of kinematic singularities and zeros, they are appropriate for a dispersive description. To this end, the Roy--Steiner treatment of~\cite{Hoferichter2011} can be generalized to the doubly-virtual case. One starts by writing down hyperbolic dispersion relations
\begin{align}
	\begin{split}
		A_i(s,t,u) &= A_i^{\text{Born}}(s,t,u) + \frac{1}{\pi} \int_{4M_\pi^2}^\infty ds^\prime \frac{\Im A_i(s^\prime, z^\prime)}{s^\prime - s} \\
			&\quad + \frac{1}{\pi} \int_{t_0}^\infty dt^\prime \Im A_i(t^\prime,u^\prime) \left( \frac{1}{t^\prime-t} + \frac{1}{t^\prime-u} - \frac{1}{t^\prime-a} \right) ,
	\end{split}
\end{align}
with hyperbola parameter $a$~\cite{Hite:1973pm}.
If we invert~\eqref{eq:ggpipiHelicityAmplitudes} to express the $A_i$ in terms of the helicity amplitudes and insert the partial-wave expansion both on the left- and right-hand side of the dispersion relation, we obtain a set of Roy--Steiner equations
\begin{align}
	\label{eq:RoySteinerSystem}
	h_{J,i}(s) = \sum_{J^\prime} \sum_{j} \frac{1}{\pi} \int_{4M_\pi^2}^\infty ds^\prime K_{J J^\prime}^{ij}(s,s^\prime) \Im h_{J^\prime,j}(s^\prime) + \ldots ,
\end{align}
where $i,j \in \{1,2,3,4,5\}$,
\begin{align}
	\begin{split}
		\label{eq:RescaledHelicityPartialWaves}
		h_{J,1} &:= h_{J,++} , \quad h_{J,2} := h_{J,+-}, \\
		h_{J,3} &:= \frac{\xi_2}{q_2^2} h_{J,+0} + \frac{\xi_1}{q_1^2} h_{J,0+} , \quad h_{J,4} := \frac{\xi_2}{q_2^2} h_{J,+0} - \frac{\xi_1}{q_1^2} h_{J,0+}, \quad h_{J,5} := \frac{\xi_1 \xi_2}{q_1^2 q_2^2} h_{J,00},
	\end{split}
\end{align}
and $K_{JJ^\prime}^{ij}$ are integral kernels. The ellipses in~\eqref{eq:RoySteinerSystem} stand for the contribution of the partial waves of the crossed channels. The normalization of the helicity amplitudes is reabsorbed in~\eqref{eq:RescaledHelicityPartialWaves}, so that the partial waves $h_{J,i}$ remain finite in the limit $q_i^2\to0$.

If only $S$-waves are taken into account, the relation between the scalar functions and the helicity partial waves in the $s$-channel is given by
\begin{align}
	\begin{split}
		\label{eq:SWavesSubProcessScalarFunctions}
		A_1 &= \frac{2}{\lambda_{12}(s)} \left( 2 q_1^2 q_2^2 h_{0,5}(s) - (s - q_1^2 - q_2^2) h_{0,1}(s) \right) , \\
		A_2 &= \frac{2}{\lambda_{12}(s)} \left( 2 h_{0,1}(s) - (s - q_1^2 - q_2^2) h_{0,5}(s) \right) , \\
		A_3 &= A_4 = A_5 = 0.
	\end{split}
\end{align}
In this case, the scalar functions depend only on $s$. If we take into account $D$-waves as well, the scalar functions $A_3$, $A_4$, and $A_5$ no longer vanish, but they depend only on $s$, while the scalar functions $A_1$ and $A_2$ are now second order polynomials in $t$ and $u$. The explicit expressions are given in App.~\ref{sec:AppendixPartialWavesggpipi}.

For the diagonal kernel functions in~\eqref{eq:RoySteinerSystem}, we find:
\begin{align*}
	K_{00}^{11}(s,s^\prime) &= K_{00}^{55}(s,s^\prime) = \frac{1}{s^\prime - s} - \frac{s^\prime - q_1^2 - q_2^2}{\lambda_{12}(s^\prime)} , \\
	K_{22}^{11}(s,s^\prime) &= K_{22}^{55}(s,s^\prime) = \frac{s^\prime}{s} \frac{s - 4M_\pi^2}{s^\prime - 4M_\pi^2} \frac{\lambda_{12}(s)}{\lambda_{12}(s^\prime)} \left( \frac{1}{s^\prime - s} - \frac{s^\prime - q_1^2 - q_2^2}{\lambda_{12}(s^\prime)} \right) , \\
	K_{22}^{22}(s,s^\prime) &= \frac{s-4M_\pi^2}{s^\prime - 4M_\pi^2} \left( \frac{1}{s^\prime - s} - \frac{s^\prime - q_1^2 - q_2^2}{\lambda_{12}(s^\prime)} \right) , \\
	K_{22}^{33}(s,s^\prime) &= \sqrt{\frac{s^\prime}{s}} \frac{s - 4M_\pi^2}{s^\prime - 4M_\pi^2} \frac{\lambda_{12}(s)}{\lambda_{12}(s^\prime)} \frac{1}{s^\prime - s} ,\\
	K_{22}^{44}(s,s^\prime) &= \sqrt{\frac{s^\prime}{s}} \frac{s - 4M_\pi^2}{s^\prime - 4M_\pi^2} \left( \frac{1}{s^\prime - s} - \frac{s^\prime - q_1^2 - q_2^2}{\lambda_{12}(s^\prime)} \right) .\mytag
\end{align*}
Because we use here a different basis, the kernel functions are slightly different from the ones given in~\cite{Colangelo2014a}: in order to avoid fictitious poles in $q_1^2-q_2^2$, one has to work with the linear combinations $h_{2,+0} \pm h_{2,0+}$, which are symmetric/antisymmetric under $q_1^2\leftrightarrow q_2^2$. The remaining difference to~\cite{Colangelo2014a} is a polynomial, which can be reabsorbed into subtraction constants.

\subsection{Pion-pole contribution}

\label{sec:PionPoleggpipi}

In order to define the contribution to $\gamma^*\gamma^*\to\pi^+\pi^-$ due to the exchange of a single pion, we employ a dispersive picture. In this subsection, we demonstrate in detail that the pure pion-pole contribution in the sense of unitarity is exactly the Born contribution in scalar QED (sQED), multiplied by pion vector form factors. The dispersive picture cleanly defines this as the correct dependence on the photon virtualities $q_i^2$. We also discuss the different meaning of Feynman and unitarity diagrams in this context.

To keep the presentation as transparent as possible, we omit subtleties due to isospin conventions here and simply present the plain sQED expressions. 
Compared to~\cite{Colangelo2014a}, this results in an overall sign for the amplitude of $\gamma^*\gamma^*\to\pi^+\pi^-$. For HLbL, this convention has no effect, but the choice in~\cite{Colangelo2014a} is advantageous when transforming to the isospin basis and reconstructing partial waves in accordance with Watson's theorem~\cite{Watson:1954uc}.

We assume that the asymptotic behavior of $\gamma^*\gamma^*\to\pi^+\pi^-$ in the crossed Mandelstam variables $t$ and $u$ permits an unsubtracted fixed-$s$ dispersion relation\footnote{Whether a fixed-$s$ dispersion relation for $\gamma^*\gamma^*\to\pi\pi$ requires subtractions or not has no influence on the pion pole. Therefore, also the Mandelstam representation for the pion-box contribution to the HLbL tensor, which we discuss in this article, remains unaffected by possible subtractions in the sub-process.} for the scalar functions:
\begin{align}
	\label{eq:FixedSDispRelggpipi}
	A_i^s(s,t,u) &= \frac{\hat\rho_{i;t}^s(s)}{t-M_\pi^2} + \frac{\hat\rho_{i;u}^s(s)}{u-M_\pi^2} + \frac{1}{\pi} \int_{4M_\pi^2}^\infty dt^\prime \frac{\hat D_{i;t}^s(t^\prime;s)}{t^\prime - t} + \frac{1}{\pi} \int_{4M_\pi^2}^\infty du^\prime \frac{\hat D_{i;u}^s(u^\prime;s)}{u^\prime - u} ,
\end{align}
where $\hat\rho_{i;t,u}^s$ denote the pole residues and $\hat D_{i;t,u}^s$ the discontinuities along the $t$- and $u$-channel cuts. Both are determined by unitarity. Consider the $t$-channel unitarity relation:
\begin{align}
	\begin{split}
		\Im^t &\left( e^2 (2\pi)^4 \delta^{(4)}( p_1 + p_2 - q_1 - q_2) \epsilon_\mu^{\lambda_1}(q_1) {\epsilon_\nu^{\lambda_2}}^*(-q_2) W^{\mu\nu}(p_1,p_2,q_1) \right) \\
			&= \sum_n \frac{1}{2S_n} \left( \prod_{i=1}^n \int \widetilde{dk_i} \right) \< n;\{k_i\} | \pi^-(p_2) \gamma^*(-q_2,\lambda_2) \>^* \< n;\{k_i\} | \pi^-(-p_1) \gamma^*(q_1,\lambda_1) \> ,
	\end{split}
\end{align}
where $S_n$ is the symmetry factor for the intermediate state $|n\>$ and we denote the Lorentz-invariant measure by
$\widetilde{dk} := \frac{d^3 k}{(2\pi)^3 2 k^0}$.
If we single out the one-pion state from the sum over intermediate states, we find
\begin{align}
	\Im_\pi^t W^{\mu\nu} &= \frac{1}{2} \int \widetilde{dk} \, (2\pi)^4 \delta^{(4)}(q_1 - p_1 - k) (k-p_1)^\mu (k+p_2)^\nu F_\pi^V(q_1^2) F_\pi^V(q_2^2) ,
\end{align}
where $F_\pi^V$ is the electromagnetic form factor of the pion, defined by
\begin{align}
	\< \pi^+(k) | j^\mu_\mathrm{em}(0) | \pi^+(p) \> = (k + p)^\mu \; F_\pi^V\big((k-p)^2\big) .
\end{align}
After performing the trivial integral, we obtain
\begin{align}
	\label{eq:ggpipiPionPoleUnitarityTChannel}
	\Im_\pi^t W^{\mu\nu} &= F_\pi^V(q_1^2) F_\pi^V(q_2^2) \, \pi \delta( t - M_\pi^2 ) \left( q_3^\mu q_1^\nu - q_2^\mu q_3^\nu - q_2^\mu q_1^\nu + q_3^\mu q_3^\nu \right) .
\end{align}
$t$- and $u$-channel are related by $p_1 \leftrightarrow p_2$:
\begin{align}
	\label{eq:ggpipiPionPoleUnitarityUChannel}
	\Im_\pi^u W^{\mu\nu} &= F_\pi^V(q_1^2) F_\pi^V(q_2^2) \, \pi \delta( u - M_\pi^2 ) \left( q_2^\mu q_3^\nu - q_3^\mu q_1^\nu - q_2^\mu q_1^\nu + q_3^\mu q_3^\nu \right) .
\end{align}
Note that these expressions are only gauge-invariant due to the presence of the delta function: the contraction of the bracket in~\eqref{eq:ggpipiPionPoleUnitarityTChannel} with $q_1^\mu$ or $q_2^\nu$ is proportional to $t-M_\pi^2$.

If we project the imaginary parts of $W^{\mu\nu}$ onto the scalar functions, making use of the delta function, we obtain
\begin{align}
	\begin{split}
		\Im_\pi^t A_1 &= F_\pi^V(q_1^2) F_\pi^V(q_2^2) \, \pi \delta( t - M_\pi^2 ) , \\
		\Im_\pi^t A_4 &= \frac{2}{s-q_1^2-q_2^2}F_\pi^V(q_1^2) F_\pi^V(q_2^2) \, \pi \delta( t - M_\pi^2 ) , \\
		\Im_\pi^u A_1 &= F_\pi^V(q_1^2) F_\pi^V(q_2^2) \, \pi \delta( u - M_\pi^2 ) , \\
		\Im_\pi^u A_4 &= \frac{2}{s-q_1^2-q_2^2}F_\pi^V(q_1^2) F_\pi^V(q_2^2) \, \pi \delta( u - M_\pi^2 ) ,
	\end{split}
\end{align}
while the one-pion contributions to the imaginary parts of the remaining scalar functions vanish.

\begin{figure}[t]
	\centering
	\begin{gather*}
		\minidiagSize{HLbL}{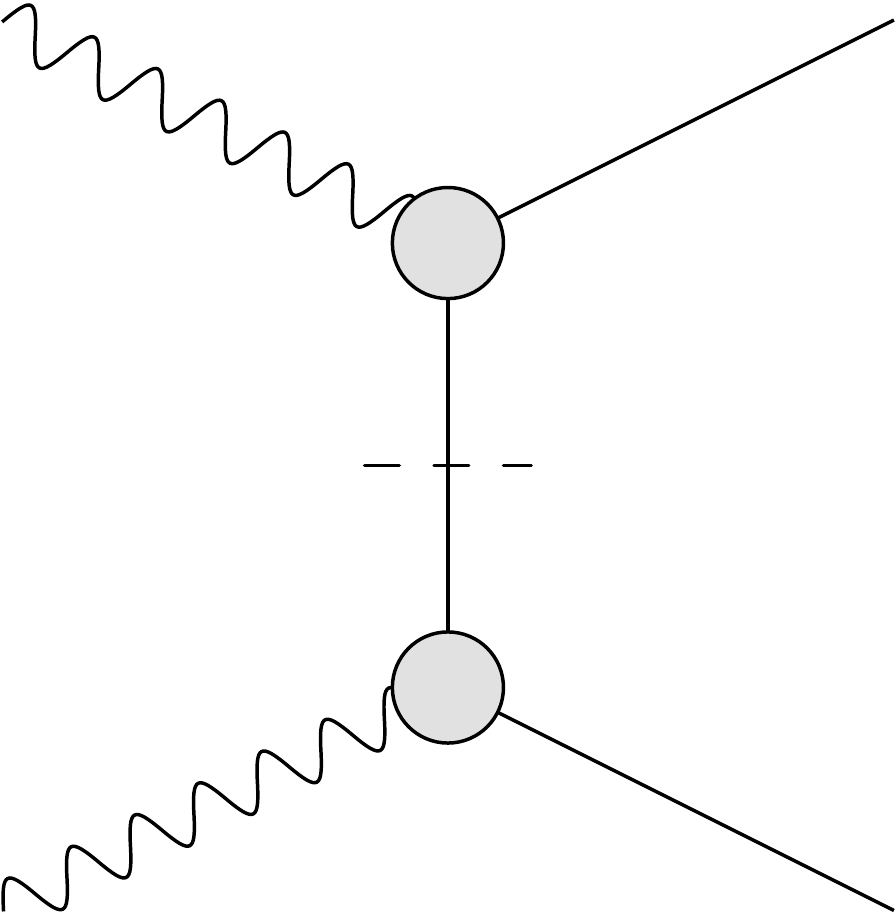}{2cm} + \minidiagSize{HLbL}{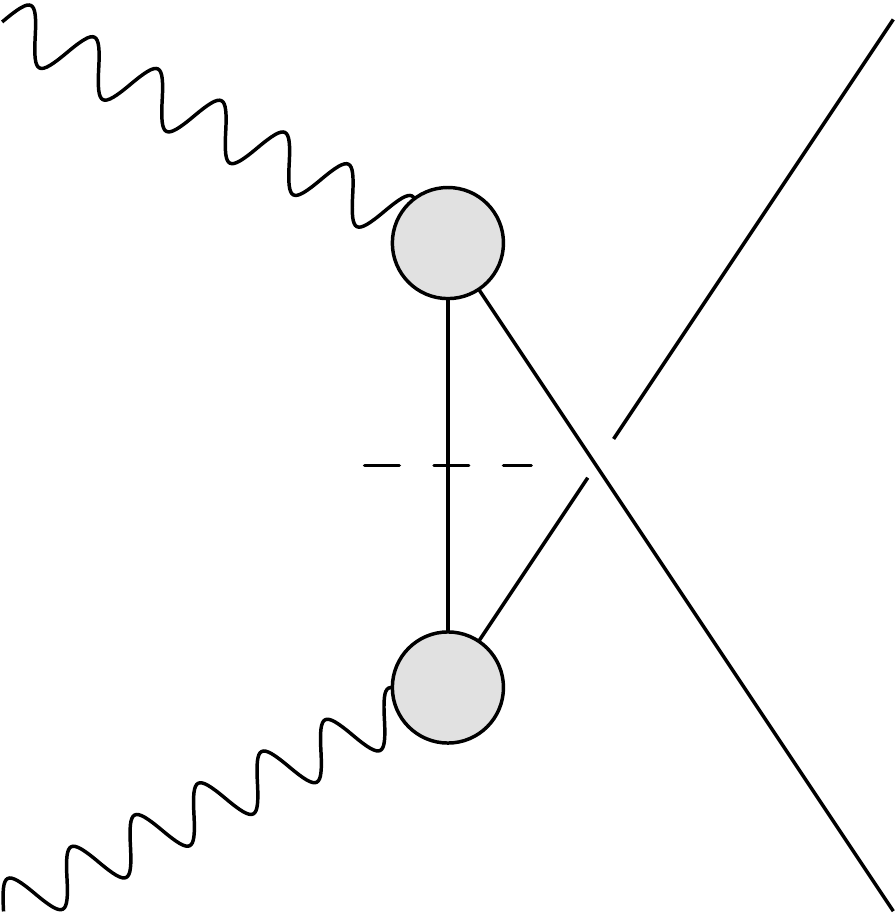}{2cm} \quad \hateq \quad F_\pi^V(q_1^2) F_\pi^V(q_2^2) \times \left( \; \minidiagSize{HLbL}{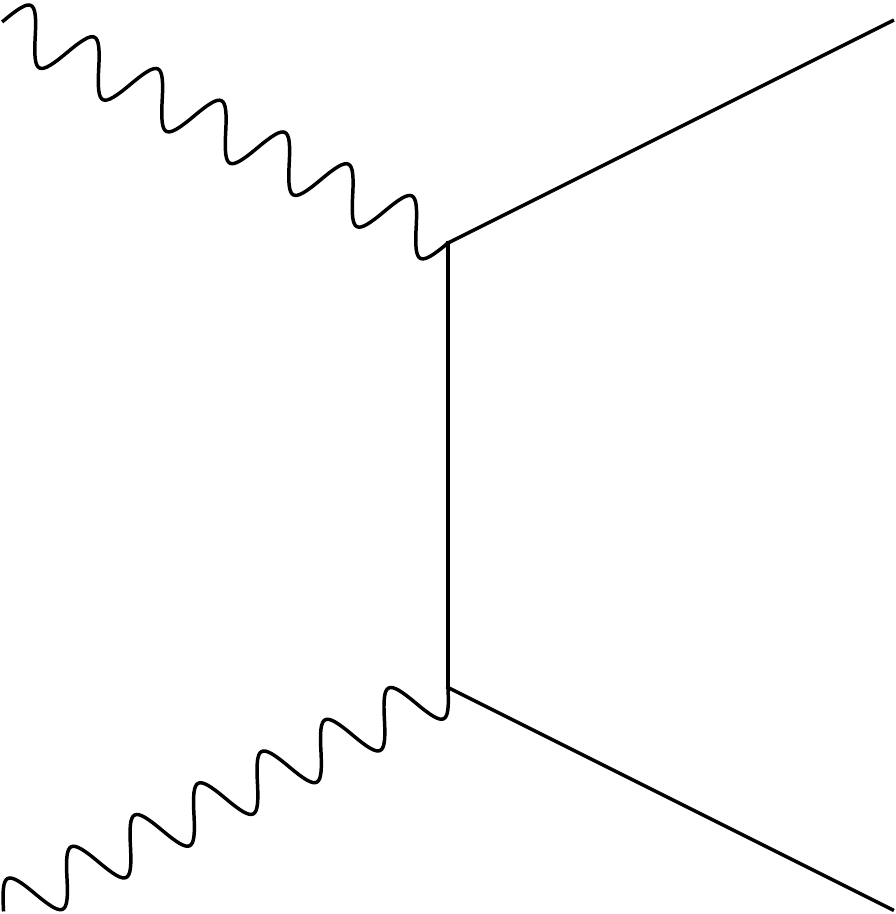}{2cm} +  \minidiagSize{HLbL}{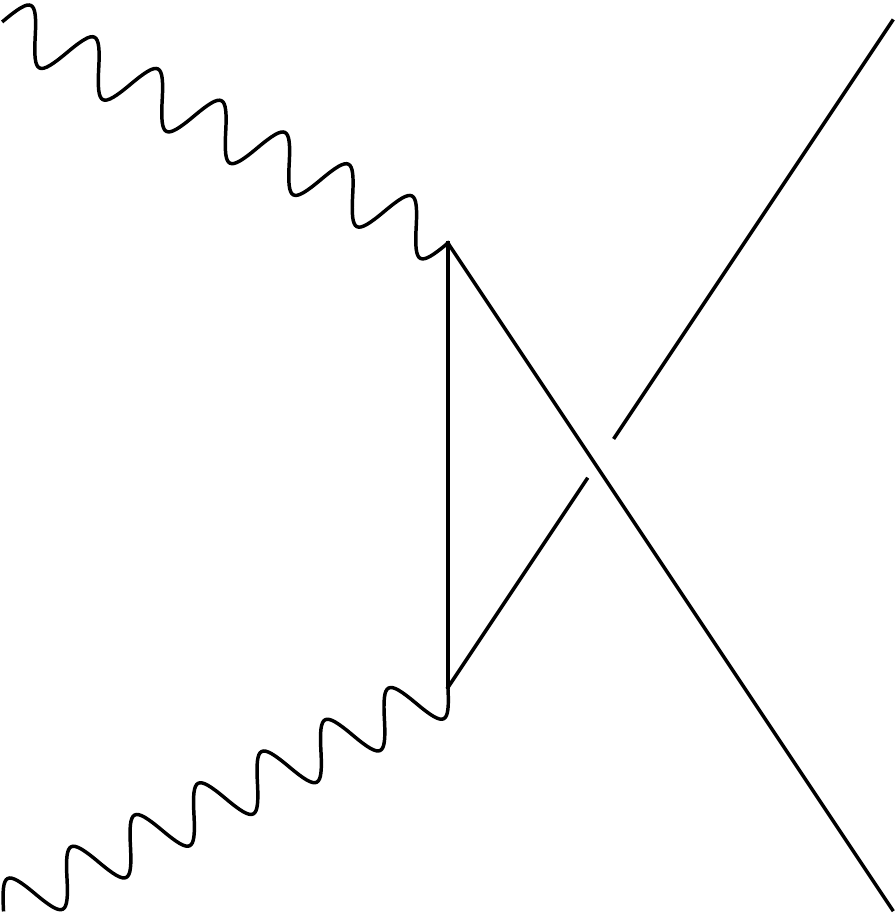}{2cm} + \minidiagSize{HLbL}{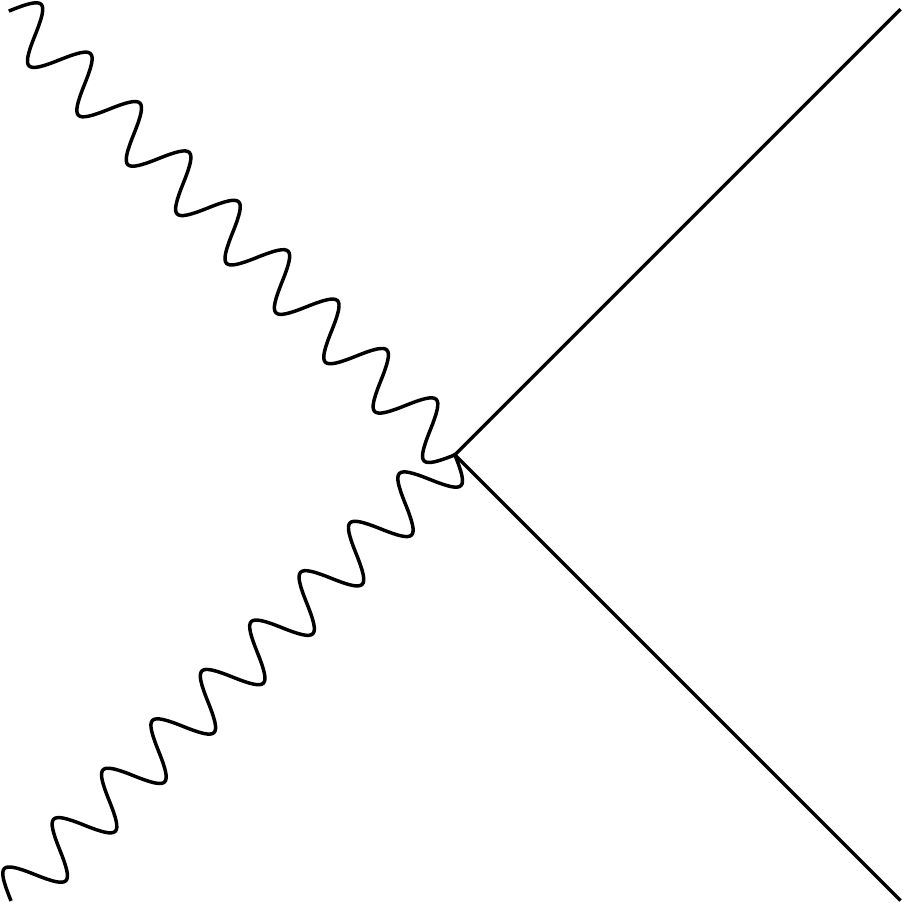}{2cm} \; \right)
	\end{gather*}
	\caption{Correspondence of the pion-pole contribution to $\gamma^*\gamma^*\to\pi^+\pi^-$ in terms of unitarity diagrams and the Born contribution in terms of sQED Feynman diagrams. The dashed lines in the unitarity diagrams indicate a cut line, hence the internal pion is on-shell.}
	\label{img:Poleggpipi}
\end{figure}

The pion-pole contribution to the scalar functions is therefore
\begin{align}
	\begin{split}
		\label{eq:ggpipiPionPoleScalarFunctions}
		A_1^\pi &= - F_\pi^V(q_1^2) F_\pi^V(q_2^2) \left( \frac{1}{t - M_\pi^2} + \frac{1}{u - M_\pi^2} \right) , \\
		A_4^\pi &= - F_\pi^V(q_1^2) F_\pi^V(q_2^2) \frac{2}{s - q_1^2 - q_2^2} \left( \frac{1}{t - M_\pi^2} + \frac{1}{u - M_\pi^2} \right) , \\
		A_2^\pi &= A_3^\pi = A_5^\pi = 0 .
	\end{split}
\end{align}
We compare the pion-pole contribution with the Born contribution in sQED:
\begin{align}
	\begin{split}
		\label{eq:ggpipisQEDBorn}
		i e^2 W^{\mu\nu}_\mathrm{Born} &= \minidiagSize{HLbL}{SubTPoleFeyn}{1.5cm} +  \minidiagSize{HLbL}{SubUPoleFeyn}{1.5cm} + \minidiagSize{HLbL}{SubSeagullFeyn}{1.5cm} \\
			&= i e^2 (2 p_1^\mu - q_1^\mu)(2p_2^\nu - q_2^\nu) \frac{1}{t-M_\pi^2} + i e^2 (2p_2^\mu - q_1^\mu)(2p_1^\nu - q_2^\nu) \frac{1}{u - M_\pi^2} + 2 i e^2 g^{\mu\nu} ,
	\end{split}
\end{align}
and read off the Born values of the scalar functions:
\begin{align}
	\begin{split}
		A_1^\mathrm{Born} &= - \left( \frac{1}{t-M_\pi^2} + \frac{1}{u-M_\pi^2}\right) , \\
		A_4^\mathrm{Born} &= - \frac{2}{s - q_1^2 - q_2^2} \left( \frac{1}{t-M_\pi^2} + \frac{1}{u-M_\pi^2}\right) , \\
		A_2^\mathrm{Born} &= A_3^\mathrm{Born} = A_5^\mathrm{Born} = 0 .
	\end{split}
\end{align}
We find that the pion-pole contribution corresponds exactly to the sQED Born contribution multiplied by electromagnetic pion form factors for the two off-shell photons.\footnote{Therefore, the dispersive definition of the pion pole~(\ref{eq:ggpipiPionPoleScalarFunctions}) coincides with the gauge-invariant pole contribution of the `soft-photon amplitude' in~\cite{Fearing1998}. We thank S.~Scherer for pointing this out.} Note that, if we think in terms of unitarity diagrams, we have now considered the pure pole contribution to the scalar functions. However, in terms of Feynman diagrams in sQED this corresponds to a sum of two pole diagrams and the seagull diagram.\footnote{The equivalence of the pion pole and the Born term is surprising given the fact that~(\ref{eq:ggpipisQEDBorn}) contains a term with $g^{\mu\nu}$, while the imaginary parts~(\ref{eq:ggpipiPionPoleUnitarityTChannel}) and~(\ref{eq:ggpipiPionPoleUnitarityUChannel}) do not. Tracing the above steps backwards, one sees that in the $t$- or $u$-channel imaginary parts the coefficient of $g^{\mu\nu}$ is proportional to $(t-M_\pi^2)\delta(t-M_\pi^2)$ or $(u-M_\pi^2)\delta(u-M_\pi^2)$ and hence vanishes due to the delta function.} It is important to be aware of the different meaning of a topology in the sense of unitarity and a Feynman diagram, see Fig.~\ref{img:Poleggpipi}.
As will be shown in Sect.~\ref{sec:MandelstamRepresentation}, it is exactly this distinction that makes the sQED pion loop in HLbL coincide with box-type unitarity diagrams representing $\pi\pi$ intermediate states with a pion-pole LHC, although, in terms of Feynman diagrams, it is composed of the sum of box, triangle, and bulb topologies.


\section{Lorentz structure of the HLbL tensor}

\label{sec:LorentzStructureHLbLTensor}

\subsection{Definitions}

In order to study the contribution of HLbL scattering to the anomalous magnetic moment of the muon, we need first of all a description of the 
HLbL tensor. The object in question is the hadronic Green's function of four electromagnetic currents, evaluated in pure QCD (i.e.\ with fine-structure constant $\alpha=e^2/(4\pi) = 0$):
\begin{align}
	\label{eq:HLbLTensorDefinition}
	\Pi^{\mu\nu\lambda\sigma}(q_1,q_2,q_3) = -i \int d^4x \, d^4y \, d^4z \, e^{-i(q_1 \cdot x + q_2 \cdot y + q_3 \cdot z)} \< 0 | T \{ j_\mathrm{em}^\mu(x) j_\mathrm{em}^\nu(y) j_\mathrm{em}^\lambda(z) j_\mathrm{em}^\sigma(0) \} | 0 \> .
\end{align}
The electromagnetic current includes only the three lightest quarks:
\begin{align}
	j_\mathrm{em}^\mu := \bar q Q \gamma^\mu q ,
\end{align}
where $q = ( u , d, s )^T$ and $Q = \mathrm{diag}(\frac{2}{3}, -\frac{1}{3}, -\frac{1}{3})$.

\begin{figure}[t]
	\centering
	\includegraphics[width=5.5cm]{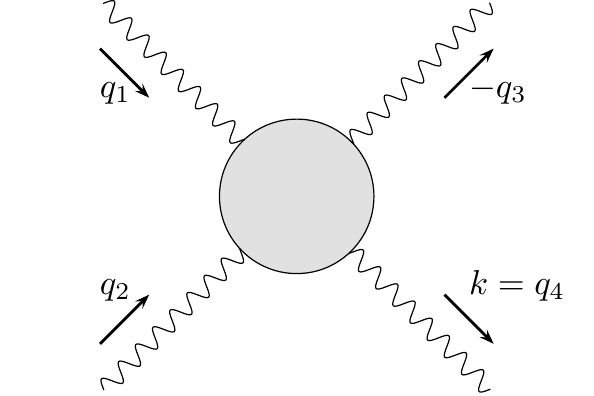}
	\caption{Kinematics of the light-by-light scattering amplitude.}
	\label{img:FullHLbL}
\end{figure}

The contraction of the HLbL tensor with polarization vectors gives the hadronic contribution to the helicity amplitudes for (off-shell) photon--photon scattering:
\begin{align}
	\label{eq:HLbLHelicityAmplitudesDefinition}
	H_{\lambda_1\lambda_2,\lambda_3\lambda_4} = \epsilon_\mu^{\lambda_1}(q_1) \epsilon_\nu^{\lambda_2}(q_2) {\epsilon_\lambda^{\lambda_3}}^*(-q_3) {\epsilon_\sigma^{\lambda_4}}^*(k) \Pi^{\mu\nu\lambda\sigma}(q_1,q_2,q_3) .
\end{align}
For notational convenience, we define
\begin{align}
	q_4 := k = q_1 + q_2 + q_3 .
\end{align}
The kinematics is illustrated in Fig.~\ref{img:FullHLbL}.

We use the following Lorentz scalars as kinematic variables --- these are the usual Mandelstam variables:
\begin{align}
	s := (q_1+q_2)^2, \quad t := (q_1+q_3)^2, \quad u := (q_2 + q_3)^2,
\end{align}
which fulfill (we will take $k^2 = 0$ at some later point)
\begin{align}
	s + t + u = \sum_{i=1}^4 q_i^2 =: \Sigma .
\end{align}

Gauge invariance requires the HLbL tensor to satisfy the Ward--Takahashi identities
\begin{align}
	\label{eq:WardIdentitiesHLbLTensor}
	\{q_1^\mu, q_2^\nu, q_3^\lambda, q_4^\sigma\} \Pi_{\mu\nu\lambda\sigma}(q_1,q_2,q_3) = 0 .
\end{align}

\subsection{Tensor decomposition}

\label{sec:HLbLTensorBTTDecomposition}

In general, the HLbL tensor can be decomposed into 138 Lorentz structures~\cite{Karplus1950, Leo1975, Bijnens1996}:
\begin{align}
	\begin{split}
		\label{eq:HLbLTensor138StructuresLSM}
		\Pi^{\mu\nu\lambda\sigma} &= g^{\mu\nu} g^{\lambda\sigma} \, \Pi^1 + g^{\mu\lambda} g^{\nu\sigma} \, \Pi^2 + g^{\mu\sigma} g^{\nu\lambda} \, \Pi^3 \\
			& + \sum_{\substack{k=1,2,4 \\ l=1,2,3}} g^{\mu\nu} q_k^\lambda q_l^\sigma \, \Pi^4_{kl}
			 + \sum_{\substack{j=1,3,4 \\ l=1,2,3}} g^{\mu\lambda} q_j^\nu q_l^\sigma \, \Pi^5_{jl}
			 + \sum_{\substack{j=1,3,4 \\ k=1,2,4}} g^{\mu\sigma} q_j^\nu q_k^\lambda \, \Pi^6_{jk} \\
			& + \sum_{\substack{i=2,3,4 \\ l=1,2,3}} g^{\nu\lambda} q_i^\mu q_l^\sigma \, \Pi^7_{il}
			 + \sum_{\substack{i=2,3,4 \\ k=1,2,4}} g^{\nu\sigma} q_i^\mu q_k^\lambda \, \Pi^8_{ik}
			 + \sum_{\substack{i=2,3,4 \\ j=1,3,4}} g^{\lambda\sigma} q_i^\mu q_j^\nu \, \Pi^9_{ij} \\
			& + \sum_{\substack{i=2,3,4 \\ j=1,3,4}} \sum_{\substack{k=1,2,4 \\ l=1,2,3}} q_i^\mu q_j^\nu q_k^\lambda q_l^\sigma \, \Pi^{10}_{ijkl} \\
			&=: \sum_{i=1}^{138} L_i^{\mu\nu\lambda\sigma} \, \Xi_i.
	\end{split}
\end{align}
The 138 scalar functions
\begin{align}
	\{\Xi_i\} := \{ \Pi^1, \Pi^2, \Pi^3, \Pi^4_{kl}, \Pi^5_{jl}, \Pi^6_{jk}, \Pi^7_{il}, \Pi^8_{ik}, \Pi^9_{ij}, \Pi^{10}_{ijkl}\}
\end{align}
depend on six independent kinematic variables, e.g.\ on two Mandelstam variables $s$ and $t$ and the virtualities $q_1^2$, $q_2^2$, $q_3^2$, and $q_4^2$. They are free of kinematic singularities but contain kinematic zeros, because they have to fulfill kinematic constraints required by gauge invariance. The Ward identities~\eqref{eq:WardIdentitiesHLbLTensor} impose 95 linearly independent relations on the scalar functions, reducing the set to 43 functions.

As we did in Sect.~\ref{sec:ggtopipiTensorDecomposition} for the case of $\gamma^*\gamma^*\to\pi\pi$, we will now construct a set of Lorentz structures and scalar functions, such that the scalar functions contain neither kinematic singularities nor zeros. Compared to $\gamma^*\gamma^*\to\pi\pi$, the application of the recipe given by Bardeen, Tung~\cite{Bardeen1968}, and Tarrach~\cite{Tarrach1975} is much more involved. Again, the recipe by Bardeen and Tung does not lead to a kinematic-free minimal basis (which would consist here of 43 scalar functions).\footnote{We use `basis' in a loose terminology: as we will discuss in Sect.~\ref{sec:BTTRedundancies}, a basis in the strict mathematical sense consists of 41 elements due to two peculiar redundancies in four space-time dimensions.} Following Tarrach, we will construct a redundant set of 54 structures, which is free of kinematic singularities and zeros.

In a first step, we define the two projectors
\begin{align}
	I_{12}^{\mu\nu} := g^{\mu\nu} - \frac{q_2^\mu q_1^\nu}{q_1 \cdot q_2} , \quad I_{34}^{\lambda\sigma} := g^{\lambda\sigma} - \frac{q_4^\lambda q_3^\sigma}{q_3 \cdot q_4} ,
\end{align}
which have the following properties:
\begin{align*}
	\begin{alignedat}{2}
		q_1^\mu I^{12}_{\mu\nu} &= 0 , \quad &  q_2^\nu I^{12}_{\mu\nu} &= 0 , \\
		q_3^\lambda I^{34}_{\lambda\sigma} &= 0 , \quad &  q_4^\sigma I^{34}_{\lambda\sigma} &= 0 , \\
		I_{12}^{\mu\mu^\prime} \Pi_{\mu^\prime\nu\lambda\sigma} &= \Pi^\mu{}_{\nu\lambda\sigma} , \quad & I_{12}^{\nu^\prime\nu} \Pi_{\mu\nu^\prime\lambda\sigma} &= \Pi_\mu{}^\nu{}_{\lambda\sigma} , \\
		I_{34}^{\lambda\lambda^\prime} \Pi_{\mu\nu\lambda^\prime\sigma} &= \Pi_{\mu\nu}{}^\lambda{}_\sigma , \quad & I_{34}^{\sigma^\prime\sigma} \Pi_{\mu\nu\lambda\sigma^\prime} &= \Pi_{\mu\nu\lambda}{}^\sigma ,
	\end{alignedat} \mytag
\end{align*}
i.e.\ the HLbL tensor is invariant under contraction with the projectors, but the contraction of every Lorentz structure produces a gauge-invariant structure. Hence, we project the tensor 
\begin{align}
	\begin{split}
		\Pi^{\mu\nu\lambda\sigma} &= I_{12}^{\mu\mu^\prime} I_{12}^{\nu^\prime\nu} I_{34}^{\lambda\lambda^\prime} I_{34}^{\sigma^\prime\sigma} \Pi_{\mu^\prime\nu^\prime\lambda^\prime\sigma^\prime} \\
			&= \sum_{i=1}^{138} I_{12}^{\mu\mu^\prime} I_{12}^{\nu^\prime\nu} I_{34}^{\lambda\lambda^\prime} I_{34}^{\sigma^\prime\sigma} L^i_{\mu^\prime\nu^\prime\lambda^\prime\sigma^\prime} \, \Xi_i \\
			&=: \sum_{i=1}^{138} \bar L_i^{\mu\nu\lambda\sigma} \, \Xi_i =  \sum_{j=1}^{43} \bar L_{i_j}^{\mu\nu\lambda\sigma} \, \Xi_{i_j} .
	\end{split}
\end{align}
Only 43 of the 138 projected structures $\bar L_i^{\mu\nu\lambda\sigma}$ are non-zero, i.e.~all constraints imposed by gauge invariance are already manifestly implemented. Since the projected structures are still multiplied by the original scalar functions $\Xi_i$, no kinematic singularities have been introduced into the scalar functions. We now have to remove the kinematic zeros from the scalar functions by removing the single and double poles in $q_1 \cdot q_2$ and $q_3 \cdot q_4$, which are present in the structures $\bar L_i^{\mu\nu\lambda\sigma}$. We adapt the recipe of \cite{Bardeen1968} (cf.\ Sect.~\ref{sec:ggtopipiTensorDecomposition}):
\begin{itemize}
	\item remove as many ($q_1 \cdot q_2$, $q_3 \cdot q_4$) double-double poles as possible by adding to the structures linear combinations of other structures with coefficients containing no poles,
	\item if no more double-double poles can be removed in this way, multiply the structures that still contain double-double poles by either $q_1 \cdot q_2$ or $q_3 \cdot q_4$ (the choice is irrelevant in the end),
	\item proceed in the same way with double-single, single-double poles, etc.\ until no poles at all are left in the structures.
\end{itemize}
As already mentioned, it is again impossible to avoid introducing kinematic singularities into the scalar functions by applying this procedure \cite{Tarrach1975}. However, the only step where kinematic singularities can be introduced is the multiplication of the structures by $q_1\cdot q_2$ or $q_3 \cdot q_4$ (i.e.~the division of the scalar functions by these terms). This means that the only possible singularities are (double or single) poles in $q_1\cdot q_2$ or $q_3 \cdot q_4$. The precise form of these poles can be easily determined: they correspond to degeneracies of the obtained basis of Lorentz structures in the limit $q_1\cdot q_2 \to 0$ and/or $q_3 \cdot q_4 \to 0$. Therefore, the 43-dimensional basis has to be extended by additional structures, which are found by studying the null-space of the present structures in the mentioned limits. 11 such structures can be found. The extended generating set of 54 structures exhibits all possible crossing symmetries in a manifest way.

Explicitly, the resulting representation of the HLbL tensor reads
\begin{align}
	\label{eqn:HLbLTensorKinematicFreeStructures}
	\Pi^{\mu\nu\lambda\sigma} &= \sum_{i=1}^{54} T_i^{\mu\nu\lambda\sigma} \Pi_i , 
\end{align}
where
\begin{align*}
		\label{eq:HLbLBTTStructures}
		T_1^{\mu\nu\lambda\sigma} &= \epsilon^{\mu\nu\alpha\beta} \epsilon^{\lambda\sigma\gamma\delta} {q_1}_\alpha {q_2}_\beta {q_3}_\gamma {q_4}_\delta , \\
		T_4^{\mu\nu\lambda\sigma} &= \Big(q_2^\mu q_1^\nu - q_1 \cdot q_2 g^{\mu \nu} \Big) \Big( q_4^\lambda q_3^\sigma - q_3 \cdot q_4 g^{\lambda \sigma} \Big) , \\
		T_7^{\mu\nu\lambda\sigma} &= \Big(q_2^\mu q_1^\nu - q_1 \cdot q_2 g^{\mu \nu } \Big) \Big( q_1 \cdot q_4 \left(q_1^\lambda q_3^\sigma -q_1 \cdot q_3 g^{\lambda \sigma} \right) + q_4^\lambda q_1^\sigma q_1 \cdot q_3 - q_1^\lambda q_1^\sigma q_3 \cdot q_4 \Big) , \\
		 T_{19}^{\mu\nu\lambda\sigma} &= \Big( q_2^\mu q_1^\nu - q_1 \cdot q_2 g^{\mu \nu } \Big) \Big(q_2 \cdot q_4 \left(q_1^\lambda q_3^\sigma - q_1 \cdot q_3 g^{\lambda\sigma} \right)+q_4^\lambda q_2^\sigma q_1 \cdot q_3 - q_1^\lambda q_2^\sigma q_3 \cdot q_4 \Big) , \\
		T_{31}^{\mu\nu\lambda\sigma} &= \Big(q_2^\mu q_1^\nu - q_1\cdot q_2 g^{\mu\nu}\Big) \Big(q_2^\lambda q_1\cdot q_3 - q_1^\lambda q_2\cdot q_3\Big) \Big(q_2^\sigma q_1\cdot q_4 - q_1^\sigma q_2\cdot q_4\Big) , \\
		T_{37}^{\mu\nu\lambda\sigma} &= \Big( q_3^\mu q_1\cdot q_4 - q_4^\mu q_1\cdot q_3\Big) \begin{aligned}[t]
			& \Big( q_3^\nu q_4^\lambda q_2^\sigma - q_4^\nu q_2^\lambda q_3^\sigma + g^{\lambda\sigma} \left(q_4^\nu q_2\cdot q_3 - q_3^\nu q_2\cdot q_4\right) \\
			& + g^{\nu\sigma} \left( q_2^\lambda q_3\cdot q_4 - q_4^\lambda q_2\cdot q_3 \right) + g^{\lambda\nu} \left( q_3^\sigma q_2\cdot q_4 - q_2^\sigma q_3\cdot q_4 \right) \Big) , \end{aligned} \\
		T_{49}^{\mu\nu\lambda\sigma} &= q_3^\sigma  \begin{aligned}[t]
				& \Big( q_1\cdot q_3 q_2\cdot q_4 q_4^\mu g^{\lambda\nu} - q_2\cdot q_3 q_1\cdot q_4 q_4^\nu g^{\lambda\mu} + q_4^\mu q_4^\nu \left( q_1^\lambda q_2\cdot q_3 - q_2^\lambda q_1\cdot q_3 \right) \\
				& + q_1\cdot q_4 q_3^\mu q_4^\nu q_2^\lambda - q_2\cdot q_4 q_4^\mu q_3^\nu q_1^\lambda + q_1\cdot q_4 q_2\cdot q_4 \left(q_3^\nu g^{\lambda\mu} - q_3^\mu g^{\lambda\nu}\right) \Big) \end{aligned} \\
			& - q_4^\lambda \begin{aligned}[t]
				& \Big( q_1\cdot q_4 q_2\cdot q_3 q_3^\mu g^{\nu\sigma} - q_2\cdot q_4 q_1\cdot q_3 q_3^\nu g^{\mu\sigma} + q_3^\mu q_3^\nu \left(q_1^\sigma q_2\cdot q_4 - q_2^\sigma q_1\cdot q_4\right) \\
				& + q_1\cdot q_3 q_4^\mu q_3^\nu q_2^\sigma - q_2\cdot q_3 q_3^\mu q_4^\nu q_1^\sigma + q_1\cdot q_3 q_2\cdot q_3 \left( q_4^\nu g^{\mu\sigma} - q_4^\mu g^{\nu\sigma} \right) \Big) \end{aligned} \\
			& + q_3\cdot q_4 \Big(\left(q_1^\lambda q_4^\mu - q_1\cdot q_4 g^{\lambda\mu}\right) \left(q_3^\nu q_2^\sigma - q_2\cdot q_3 g^{\nu\sigma}\right) - \left(q_2^\lambda q_4^\nu - q_2\cdot q_4 g^{\lambda\nu}\right) \left(q_3^\mu q_1^\sigma - q_1\cdot q_3 g^{\mu\sigma}\right)\Big) . \quad \mytag
\end{align*}
These structures satisfy the following crossing symmetries:
\begin{align*}
		\label{eq:BTTInternalCrossingSymmetries}
		T_1^{\mu\nu\lambda\sigma} &= \mathcal{C}_{12}[ T_1^{\mu\nu\lambda\sigma} ] = \mathcal{C}_{34}[ T_1^{\mu\nu\lambda\sigma} ] = \mathcal{C}_{34}[ \mathcal{C}_{12}[ T_1^{\mu\nu\lambda\sigma} ] ] = \mathcal{C}_{24}[ \mathcal{C}_{13}[ T_1^{\mu\nu\lambda\sigma} ]] \\
			&= \mathcal{C}_{23}[ \mathcal{C}_{14}[ T_1^{\mu\nu\lambda\sigma} ] ] = \mathcal{C}_{24}[ \mathcal{C}_{13}[ \mathcal{C}_{34}[ T_1^{\mu\nu\lambda\sigma} ] ] ] = \mathcal{C}_{23}[ \mathcal{C}_{14}[ \mathcal{C}_{34}[ T_1^{\mu\nu\lambda\sigma} ] ] ] , \\
		T_4^{\mu\nu\lambda\sigma} &= \mathcal{C}_{12}[ T_4^{\mu\nu\lambda\sigma} ] = \mathcal{C}_{34}[ T_4^{\mu\nu\lambda\sigma} ] = \mathcal{C}_{34}[ \mathcal{C}_{12}[ T_4^{\mu\nu\lambda\sigma} ] ] = \mathcal{C}_{24}[ \mathcal{C}_{13}[ T_4^{\mu\nu\lambda\sigma} ]] \\
			&= \mathcal{C}_{23}[ \mathcal{C}_{14}[ T_4^{\mu\nu\lambda\sigma} ] ] = \mathcal{C}_{24}[ \mathcal{C}_{13}[ \mathcal{C}_{34}[ T_4^{\mu\nu\lambda\sigma} ] ] ] = \mathcal{C}_{23}[ \mathcal{C}_{14}[ \mathcal{C}_{34}[ T_4^{\mu\nu\lambda\sigma} ] ] ] , \\
		T_7^{\mu\nu\lambda\sigma} &= \mathcal{C}_{34}[ T_7^{\mu\nu\lambda\sigma} ] , \\
		T_{19}^{\mu\nu\lambda\sigma} &= \mathcal{C}_{34}[ \mathcal{C}_{12}[ T_{19}^{\mu\nu\lambda\sigma} ] ] , \\
		T_{31}^{\mu\nu\lambda\sigma} &= \mathcal{C}_{12}[ T_{31}^{\mu\nu\lambda\sigma} ] = \mathcal{C}_{34}[ T_{31}^{\mu\nu\lambda\sigma} ] = \mathcal{C}_{34}[ \mathcal{C}_{12}[ T_{31}^{\mu\nu\lambda\sigma} ] ]  , \\
		T_{37}^{\mu\nu\lambda\sigma} &= \mathcal{C}_{34}[ T_{37}^{\mu\nu\lambda\sigma} ] , \\
		T_{49}^{\mu\nu\lambda\sigma} &= - \mathcal{C}_{12}[ T_{49}^{\mu\nu\lambda\sigma} ] = - \mathcal{C}_{34}[ T_{49}^{\mu\nu\lambda\sigma} ] = \mathcal{C}_{34}[ \mathcal{C}_{12}[ T_{49}^{\mu\nu\lambda\sigma} ] ]  , \mytag
\end{align*}
where the crossing operators $\mathcal{C}_{ij}$ exchange momenta and Lorentz indices of the photons $i$ and $j$, e.g.\footnote{The composition of two crossing operators is understood to act e.g.\ in the following way: $\mathcal{C}_{12}[\mathcal{C}_{23}[f(q_1,q_2,q_3,q_4)]] = \mathcal{C}_{12}[f(q_1,q_3,q_2,q_4)] = f(q_2,q_3,q_1,q_4)$.\vspace{0.1cm}}
\begin{align}
	\mathcal{C}_{12}[f] := f( \mu \leftrightarrow \nu, q_1 \leftrightarrow q_2 ) , \quad \mathcal{C}_{14}[f] := f( \mu \leftrightarrow \sigma, q_1 \leftrightarrow -q_4 ) .
\end{align}
All the remaining structures are just crossed versions of the above seven
structures, as shown in App.~\ref{sec:AppendixCrossingLorentzStructures}. 
Since the HLbL tensor $\Pi^{\mu\nu\lambda\sigma}$ is totally crossing
symmetric, the scalar functions $\Pi_i$ have to fulfill exactly the same
crossing properties of the corresponding Lorentz structures as given
in~\eqref{eq:BTTInternalCrossingSymmetries} (note the antisymmetric
crossing relations in $\Pi_{49}$). Therefore, only seven different scalar
functions $\Pi_i$ appear, together with their crossed versions. These
scalar functions are free of kinematic singularities and zeros and hence
fulfill a Mandelstam representation.\footnote{In principle, the crossing
  antisymmetry of $\Pi_{49}$ implies a kinematic zero. We ignore it here
  and show in App.~\ref{sec:KinematicZeroCrossingAntisymmetryHLbL} that
  this zero has no impact on the dispersion relation for the HLbL tensor.}
They are suitable quantities for a dispersive description. 

The subset consisting of the following 43 Lorentz structures forms a basis:
\begin{align}
	\label{eq:HLbLBasisStructures}
	\{\mathcal{B}^{\mu\nu\lambda\sigma}_i\} := \Big\{ T^{\mu\nu\lambda\sigma}_i \big| i \in\{ 1,\ldots,21,23,25,27,29,30,33,\ldots,36,38,\ldots,45,49,\ldots,53 \} \Big\} .
\end{align}
The corresponding scalar coefficient functions $\tilde \Pi_i$, defined by
\begin{align}
	\label{eq:HLbLTensor43Basis}
	\Pi^{\mu\nu\lambda\sigma} &= \sum_{i=1}^{43} \mathcal{B}_i^{\mu\nu\lambda\sigma} \tilde\Pi_i ,
\end{align}
exhibit kinematic singularities in $q_1 \cdot q_2$ and $q_3 \cdot q_4$. The exact form of these kinematic singularities can be determined by projecting~\eqref{eqn:HLbLTensorKinematicFreeStructures} on this basis. The relation between the basis coefficient functions $\tilde \Pi_i$ and the scalar functions $\Pi_i$ is given in App.~\ref{sec:AppendixBTTProjectionAndBasisCrossingRelations}.

While crossing symmetry is manifest if the HLbL tensor is expressed in terms of the redundant set of 54 structures, it is partially obscured in the basis of 43 elements. Still, many basis functions $\tilde\Pi_i$ are related by crossing symmetry, see~\eqref{eq:BTBasisCrossingRelations} in App.~\ref{sec:AppendixBTTProjectionAndBasisCrossingRelations}. The basis coefficient functions are completely specified by crossing symmetry and 9 representative functions, e.g.
\begin{align}
	\label{eq:BTBasisFunctionsRepresentatives}
	\left\{ \tilde\Pi_1, \tilde\Pi_4, \tilde\Pi_7, \tilde\Pi_9, \tilde\Pi_{19}, \tilde\Pi_{21}, \tilde\Pi_{36}, \tilde\Pi_{39}, \tilde\Pi_{40} \right\} .
\end{align} 

We stress again that our
decomposition~\eqref{eqn:HLbLTensorKinematicFreeStructures} is both
manifestly crossing symmetric and gauge invariant. In particular, this
implies that all soft-photon zeros~\cite{Low1958} of the helicity
amplitudes~\eqref{eq:HLbLHelicityAmplitudesDefinition} are already
implemented, in contrast to the helicity basis employed
in~\cite{Colangelo2014a}. In this case, only specific combinations of basis
functions were free of kinematic singularities and zeros and thus admitted
a dispersive representation, while the soft-photon constraints needed to be
imposed by hand. The BTT set~\eqref{eq:HLbLBTTStructures} provides a
Lorentz decomposition that incorporates soft-photon constraints and avoids
kinematics by construction.  On the other hand, the main motivation for
choosing the helicity basis was that the unitarity relations in the form of
helicity partial-wave amplitudes become diagonal.  Of course, this will not
be true in the BTT basis, so that the derivation of an explicit dispersive
representation for the BTT amplitudes requires a basis change. In practice,
this change between BTT and helicity bases proceeds by means
of~\eqref{eq:HLbLTensor138StructuresLSM} as an intermediate step: the
helicity amplitudes can be easily expressed in terms
of~\eqref{eq:HLbLTensor138StructuresLSM}, while the remaining basis change
can be performed along the lines explained in
App.~\ref{sec:AppendixProjection}.

\subsection{Redundancies in the Lorentz decomposition}

\label{sec:BTTRedundancies}

Because the set of 54 scalar functions is redundant, there is obviously an ambiguity in the definition of the coefficient functions. However, crossing symmetry and the absence of kinematic singularities restrict this ambiguity in a very specific way. The shifts
\begin{align}
	\Pi_i \mapsto \Pi_i + \delta\Pi_i
\end{align}
leaving the HLbL tensor $\Pi^{\mu\nu\lambda\sigma}$ invariant are easily found by requiring that the basis elements $\tilde\Pi_i$ remain invariant. We find
\begin{align}
	\begin{alignedat}{3}
		\label{eq:RedundanciesLorentzDecomposition}
		\delta\Pi_1 &= 0 , \quad & 
		\delta\Pi_4 &= 0 , \\
		\delta\Pi_7 &= - q_2 \cdot q_3 q_2 \cdot q_4 \Delta_{19} , \quad & 
		\delta\Pi_{19} &= q_1 \cdot q_4 q_2 \cdot q_3 \Delta_{19} , \quad & 
		\delta\Pi_{31} &= - q_3 \cdot q_4 \Delta_{19} , \\
		\delta\Pi_{37} &= q_1 \cdot q_2 \Delta_{37} , \quad & 
		\delta\Pi_{49} &= q_1 \cdot q_2 \Delta_{49} ,
	\end{alignedat}
\end{align}
where the shifts have to satisfy
\begin{align}
	\begin{split}
		\label{eq:BTTScalarFunctionsAmbiguities}
		\Delta_{19} &= \mathcal{C}_{12}[ \Delta_{19} ] = \mathcal{C}_{34}[ \Delta_{19} ] =  \mathcal{C}_{34}[ \mathcal{C}_{12}[ \Delta_{19} ]] , \\
		\Delta_{37} &= \mathcal{C}_{24}[ \Delta_{37} ] = \mathcal{C}_{23}[ \Delta_{37} ] = \mathcal{C}_{34}[ \Delta_{37} ] = \mathcal{C}_{24}[ \mathcal{C}_{34}[ \Delta_{37} ]] = \mathcal{C}_{24}[ \mathcal{C}_{23}[ \Delta_{37} ]] , \\
		\Delta_{49} &= -\mathcal{C}_{12}[ \Delta_{49} ] = -\mathcal{C}_{13}[ \Delta_{49} ] =  -\mathcal{C}_{14}[ \Delta_{49} ] =  -\mathcal{C}_{23}[ \Delta_{49} ] =  -\mathcal{C}_{24}[ \Delta_{49} ] =  -\mathcal{C}_{34}[ \Delta_{49} ] = \ldots.
	\end{split}
\end{align}
The ellipses in the last line stand for the permutations that are not listed explicitly: $\Delta_{49}$ is symmetric (antisymmetric) under all even (odd) crossing permutations.
The shifts $\delta\Pi_i$ that are not specified follow from the crossing relations~\eqref{eq:BTTCrossingRelations}.

Apart from the redundancy introduced by extending the basis to a set of 54
elements, there exists another ambiguity in the definition of the scalar
function. As it was pointed out in~\cite{Eichmann2014}, the 138 initial
Lorentz structures of the HLbL tensor fulfill two linear relations in $d=4$
space-time dimensions. This is not obvious in a covariant notation, but can
be easily shown in an explicit reference frame. The two linear relations
can be translated into two linear relations between the 43 basis elements
(which means that these structures actually do not form a basis in a strict
sense). Therefore, in four space-time dimensions there are only 41
independent scalar functions, which matches the number of fully off-shell
helicity amplitudes. This is analogous to the case of
$\gamma^*\gamma^*\to\pi\pi$, where the number of helicity amplitudes ($5$)
also matches the number of independent scalar functions. 

The two linear relations can be further translated into relations amongst
the 54 elements of the redundant crossing-symmetric set of Lorentz
structures. In fact, it turns out that expressed in this way, the two
relations become particularly simple. They are given by:
\begin{align}
	\label{eq:136vs138LinearRelations}
	\sum_{i=1}^{54} c_i^{1,2} T_i^{\mu\nu\lambda\sigma} = 0 ,
\end{align}
where
\begin{align}
	\begin{split}
		c^1_{1} &= -s + t + u , \quad
		c^1_{3} = -s - t + u , \quad
		c^1_{4} = -5 s - t + u , \quad
		c^1_{5} = 2 (-s + u) , \quad
		c^1_{6} = -s + t + 5 u , \\
		c^1_{7} &= c^1_{8} = c^1_{17} = c^1_{18} = c^1_{19} = c^1_{20} = c^1_{29} = c^1_{30} = c^1_{50} = c^1_{53} = 4 , \\
		c^1_{11} &= c^1_{12} = c^1_{13} = c^1_{14} = c^1_{23} = c^1_{24} = c^1_{25} = c^1_{26} = -4 , \\
		c^1_{37} &= c^1_{46} = c^1_{47} = c^1_{48} = c^1_{51} = c^1_{52} = 2 , \\
		c^1_{38} &= c^1_{40} = c^1_{43} = c^1_{44} = c^1_{49} = c^1_{54} = -2 ,
	\end{split}
\end{align}
and
\begin{align}
	\begin{split}
		c^2_{2} &= s - t + u, \quad
		c^2_{3} = -s - t + u, \quad
		c^2_{4} = 2 (-t + u), \quad
		c^2_{5} = -s - 5 t + u, \quad
		c^2_{6} = s - t + 5 u, \\
		c^2_{9} &= c^2_{10} = c^2_{15} = c^2_{16} = c^2_{21} = c^2_{22} = c^2_{27} = c^2_{28} = 4, \\
		c^2_{11} &= c^2_{12} = c^2_{13} = c^2_{14} = c^2_{14} = c^2_{23} = c^2_{24} = c^2_{25} = c^2_{26} = c^2_{49} = c^2_{54} = -4 \\
		c^2_{38} &= c^2_{40} = c^2_{43} = c^2_{44} = c^2_{51} = c^2_{52} = -2, \\
		c^2_{39} &= c^2_{41} = c^2_{42} = c^2_{45} = c^2_{50} = c^2_{53} = 2,
	\end{split}
\end{align}
and where all unspecified $c_i^{1,2}$ are
zero. Eq.~\eqref{eq:136vs138LinearRelations} is most easily verified by
contracting the equation with the 138 tensor structures. We have checked
explicitly that the matrix
$\{L_i^{\mu\nu\lambda\sigma}L_{\mu\nu\lambda\sigma}^k\}_{ik}$ is of rank
136. The linear relations between the structures imply the following
ambiguities in the scalar functions: 
\begin{align}
	\Pi_i \mapsto \Pi_i + c_i^1 f_1 + c_i^2 f_2 ,
\end{align}
where $f_{1,2}$ are a priori unspecified functions free of kinematic singularities. The requirement that the scalar functions satisfy crossing relations of the same form as~\eqref{eq:BTTInternalCrossingSymmetries} and \eqref{eq:BTTCrossingRelations} imposes crossing relations on the functions $f_{1,2}$:
\begin{align}
	f_1 &= \mathcal{C}_{12}[ f_2 ] = \mathcal{C}_{34}[ \mathcal{C}_{12}[ f_1 ]] = \mathcal{C}_{24} [\mathcal{C}_{13}[ f_1 ]] = \mathcal{C}_{14}[ f_1 ] = - \mathcal{C}_{12}[ f_1 ] - \mathcal{C}_{13}[ f_1 ] .
\end{align}
Let us write
\begin{align}
	\begin{split}
		\Pi_{49} &= \frac{1}{4} \left( \Pi_{49} + \mathcal{C}_{24}[ \mathcal{C}_{13}[ \Pi_{49} ]] + \mathcal{C}_{14}[ \Pi_{49} ] + \mathcal{C}_{14}[ \mathcal{C}_{24}[ \mathcal{C}_{13}[ \Pi_{49} ]]] \right) \\
			&+ \frac{1}{4} \left( \Pi_{49} + \mathcal{C}_{24}[ \mathcal{C}_{13}[ \Pi_{49} ]] - \mathcal{C}_{14}[ \Pi_{49} ] - \mathcal{C}_{14}[ \mathcal{C}_{24}[ \mathcal{C}_{13}[ \Pi_{49} ]]] \right) \\
			&+ \frac{1}{2} \left( \Pi_{49} - \mathcal{C}_{24}[ \mathcal{C}_{13}[ \Pi_{49} ]] \right) .
	\end{split}
\end{align}
The first bracket in this expression is given by $\Pi_{49} - \Pi_{50} - \Pi_{53} + \Pi_{54}$ and has exactly the crossing properties of $f_1$.
Therefore, the ambiguity can be fixed by the condition
\begin{align}
	\label{eq:AmbiguityFixingCondition}
	\Pi_{49} - \Pi_{50} - \Pi_{53} + \Pi_{54} = 0 .
\end{align}

Very recently, a Lorentz decomposition of the HLbL tensor into 72 redundant
structures has been presented in~\cite{Eichmann2015}. A comparison with our
redundant set of 54 structures, as first given in~\cite{Stoffer2014}, has
been performed, confirming the absence of kinematic singularities. By
studying systematically the irreducible representations of the permutation
group $S_4$, the authors of~\cite{Eichmann2015} aim at constructing a
minimal basis consisting of 41 totally crossing symmetric elements free of
kinematic singularities. Note, however, that \cite{Eichmann2015} neither
provides this basis nor a concrete recipe for how to obtain it. Therefore,
it remains to be seen if a minimal basis free of kinematic singularities
exists at all. For our purpose the question whether such a basis exists is
immaterial, since for a dispersive representation of the HLbL tensor all
that is required is a decomposition into scalar amplitudes free of
kinematic zeros and singularities. 

To conclude this subsection, we stress that the presence of kinematic
singularities in any known basis requires us to work with a redundant set
of $54$ Lorentz structures and scalar functions $\Pi_i$. Although the
redundancy of the Lorentz structures introduces an ambiguity in the
functions $\Pi_i$ (which is already very restricted by crossing symmetry
and will be further restricted by analyticity), this ambiguity will cancel
in the calculation of any physical quantity such as $(g-2)_\mu$. In
particular, this implies that the additional redundancy in $d=4$ dimensions
does not affect our calculation in any way, since the minimal set suited
for a dispersive representation still involves $54$ functions. In the rest
of the paper we will continue to work with the BTT amplitudes defined
in~\eqref{eq:HLbLBTTStructures}.


\section{HLbL contribution to $\boldsymbol{(g-2)_\mu}$}

\label{sec:HLbLContributionToGminus2}

In Sect.~\ref{sec:ProjectorTechniquesg-2}, we review the definition and calculation of $a_\mu$, as well as general techniques for the calculation of $a_\mu^\text{HLbL}$ (see e.g.~\cite{Jegerlehner2008}). The well-known general formula requires still a rather long calculation before a number can be finally obtained. For the pion-pole contribution, these steps have been worked out long ago \cite{Knecht2002}. With our complete set of 54 kinematic-free structures (\ref{eqn:HLbLTensorKinematicFreeStructures}), this procedure can be repeated for the whole HLbL contribution in full generality, as we explain in Sects.~\ref{sec:LoopIntegration} and \ref{sec:MasterFormula}. In Sect.~\ref{sec:WickRotation}, we study the validity of the applied Wick rotation in the presence of anomalous thresholds.

\subsection{Projector techniques}

\label{sec:ProjectorTechniquesg-2}

The $T$-matrix element of the interaction of a muon with the electromagnetic field is defined by
\begin{align}
	\mathcal{T}^\mu(p_1,p_2) := - e \<\mu^-(p_2,s_2) | j_\mathrm{em}^\mu(0) | \mu^-(p_1,s_1) \>,
\end{align}
where $e = |e|$. Diagrammatically
\begin{align}
	\minidiag{HLbL}{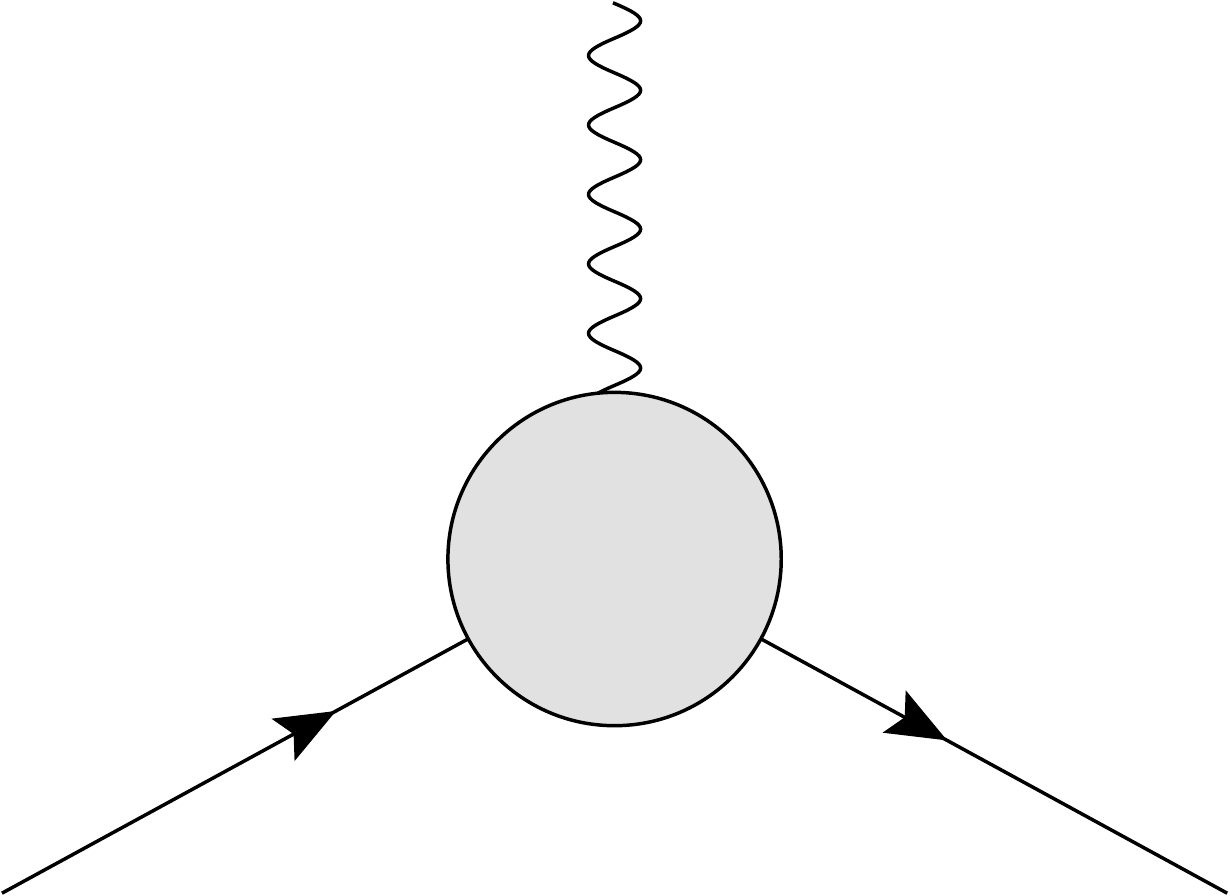} = i \mathcal{T}^\mu(p_1,p_2) = (-ie) \bar u(p_2) \Gamma^\mu(p_1, p_2) u(p_1) .
\end{align}
Assuming parity conservation, the vertex function $\Gamma^\mu$ can be decomposed into form factors as
\begin{align}
	\Gamma^\mu(p_1,p_2) &= \gamma^\mu F_1(k^2) - i \frac{\sigma^{\mu\nu} k_\nu}{2 m_\mu} F_2(k^2) ,
\end{align}
where $k = p_1 - p_2$ and $\sigma^{\mu\nu} := \frac{i}{2} [ \gamma^\mu, \gamma^\nu]$. $F_1$ is the electric charge or Dirac form factor, $F_2$ the magnetic or Pauli form factor.
The anomalous magnetic moment of the muon is given by
\begin{align}
	a_\mu = \frac{1}{2}(g - 2)_\mu = F_2(0) .
\end{align}
We expand the vertex function $\Gamma^\mu$ to first order in powers of $k^\mu$:
\begin{align}
	\begin{split}
		\Gamma^\mu(p_1,p_2) &= \Gamma^\mu(p,p) + k_\nu \frac{\p}{\p k_\nu} \Gamma^\mu(p_1,p_2) \bigg|_{k=0} + \ldots  \\
			&=: V^\mu(p) + k_\nu \Gamma^{\mu\nu}(p) + \ldots ,
	\end{split}
\end{align}
where $p := \frac{1}{2}(p_1 + p_2)$.

Using projector techniques together with angular averages (see \cite{Knecht2002,Jegerlehner2008}), the anomalous magnetic moment of the muon can be calculated as
\begin{align}
	\begin{split}
		a_\mu &= \mathrm{Tr}\left( \left(\frac{1}{12} \gamma^\mu - \frac{1}{3} \left( \frac{p^\mu \slashed p}{m_\mu^2} \right) - \frac{1}{4} \frac{p^\mu}{m_\mu} \right) V_\mu(p) \right) \\
			 &- \frac{1}{48 m_\mu} \mathrm{Tr}\left( (\slashed p + m_\mu) [\gamma^\mu,\gamma^\rho] (\slashed p + m_\mu) \Gamma_{\mu\rho}(p) \right) ,
	\end{split}
\end{align}
where now $p^2 = m_\mu^2$.\footnote{Note that we have defined $k$ as outgoing, resulting in the different sign of the second term with respect to \cite{Jegerlehner2008}.}

We are interested in the contribution of the HLbL tensor to $a_\mu$, diagrammatically
\begin{align}
	\minidiag{HLbL}{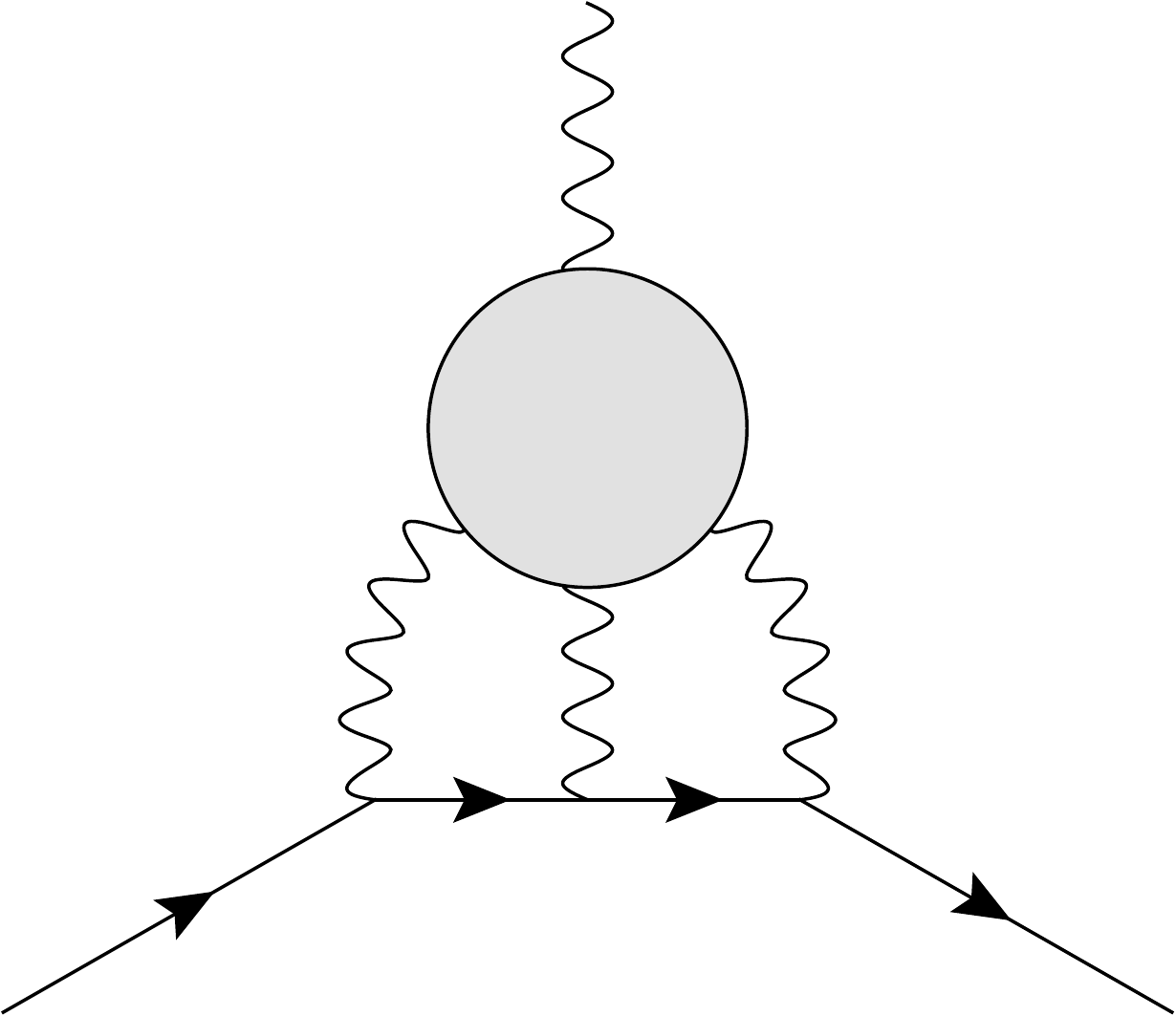} = (-ie) \bar u(p_2) \Gamma^\mu_\mathrm{HLbL}(p_1,p_2) u(p_1) ,
\end{align}
where
\begin{align}
	\begin{split}
		\Gamma^\sigma_\mathrm{HLbL}(p_1,p_2) &= - e^6 \int \frac{d^4q_1}{(2\pi)^4} \frac{d^4q_2}{(2\pi)^4} \gamma_\mu \frac{(\slashed p_2 + \slashed q_1 + m_\mu)}{(p_2+q_1)^2 - m_\mu^2} \gamma_\lambda  \frac{(\slashed p_1 - \slashed q_2 + m_\mu)}{(p_1-q_2)^2 - m_\mu^2} \gamma_\nu \\
			& \quad \times \frac{1}{q_1^2 q_2^2 (p_1-p_2-q_1-q_2)^2} \Pi^{\mu\nu\lambda\sigma}(q_1,q_2,p_1-p_2-q_1-q_2) .
	\end{split}
\end{align}
The HLbL tensor is defined in~\eqref{eq:HLbLTensorDefinition}. Differentiating the fourth Ward identity in~\eqref{eq:WardIdentitiesHLbLTensor} with respect to $k_\rho=(q_1+q_2+q_3)_\rho$ yields
\begin{align}
	\label{eq:DiffWardIdentity}
	\Pi_{\mu\nu\lambda\rho}(q_1,q_2,k-q_1-q_2) = - k^\sigma \frac{\p}{\p k^\rho} \Pi_{\mu\nu\lambda\sigma}(q_1,q_2,k-q_1-q_2) .
\end{align}
It was already argued in~\cite{Aldins1970} that $\Pi_{\mu\nu\lambda\sigma}$ vanishes linearly with $k$ (i.e.\ the derivative contains no singularity), and so must $\Gamma_\sigma^\mathrm{HLbL}$. This is easily verified with our tensor decomposition~\eqref{eqn:HLbLTensorKinematicFreeStructures}. Therefore, the HLbL contribution to the anomalous magnetic moment is given by
\begin{align}
	\label{eq:amuTraceFormula}
	a_\mu^\mathrm{HLbL} = - \frac{1}{48 m_\mu} \mathrm{Tr}\left( (\slashed p + m_\mu) [\gamma^\rho,\gamma^\sigma] (\slashed p + m_\mu) \Gamma_{\rho\sigma}^\mathrm{HLbL}(p) \right) ,
\end{align}
where
\begin{align}
	\Gamma^\mathrm{HLbL}_{\rho\sigma}(p) = \frac{\p}{\p k^\sigma} \Gamma^\mathrm{HLbL}_\rho(p_1,p_2) \bigg|_{k=0} .
\end{align}
We use the Ward identity~\eqref{eq:DiffWardIdentity} to write
\begin{align}
	\begin{split}
		\Gamma_\rho^\mathrm{HLbL}(p_1,p_2) &= e^6 \int \frac{d^4q_1}{(2\pi)^4} \frac{d^4q_2}{(2\pi)^4} \gamma^\mu \frac{(\slashed p_2 + \slashed q_1 + m_\mu)}{(p_2+q_1)^2 - m_\mu^2} \gamma^\lambda  \frac{(\slashed p_1 - \slashed q_2 + m_\mu)}{(p_1-q_2)^2 - m_\mu^2} \gamma^\nu \\
			& \quad \times \frac{1}{q_1^2 q_2^2 (p_1-p_2-q_1-q_2)^2} k^\sigma \frac{\p}{\p k^\rho} \Pi_{\mu\nu\lambda\sigma}(q_1,q_2,k-q_1-q_2) .
	\end{split}
\end{align}
Taking the derivative and limit leads to the well-known expression
\begin{align}
	\begin{split}
		\label{eq:GammaHLbLTwoLoop}
		\Gamma_{\rho\sigma}^\mathrm{HLbL}(p) &= e^6 \int \frac{d^4q_1}{(2\pi)^4} \frac{d^4q_2}{(2\pi)^4} \gamma^\mu \frac{(\slashed p + \slashed q_1 + m_\mu)}{(p+q_1)^2 - m_\mu^2} \gamma^\lambda  \frac{(\slashed p - \slashed q_2 + m_\mu)}{(p-q_2)^2 - m_\mu^2} \gamma^\nu \\
			& \quad \times \frac{1}{q_1^2 q_2^2 (q_1+q_2)^2} \frac{\p}{\p k^\rho} \Pi_{\mu\nu\lambda\sigma}(q_1,q_2,k-q_1-q_2) \bigg|_{k=0}.
	\end{split}
\end{align}

\subsection{Loop integration}

\label{sec:LoopIntegration}

In order to compute the contribution to $(g-2)_\mu$, one has to take the trace in~\eqref{eq:amuTraceFormula} and perform the two-loop integral of equation~\eqref{eq:GammaHLbLTwoLoop}. Five of the eight integrals can be carried out analytically with the help of Gegenbauer polynomial techniques \cite{Rosner1967}. To this end, we employ the representation of the HLbL tensor in terms of the 54 Lorentz structures $T_i^{\mu\nu\lambda\sigma}$:
\begin{align}
	\begin{split}
		a_\mu^\mathrm{HLbL} &= - \frac{e^6}{48 m_\mu}  \int \frac{d^4q_1}{(2\pi)^4} \frac{d^4q_2}{(2\pi)^4} \frac{1}{q_1^2 q_2^2 (q_1+q_2)^2} \frac{1}{(p+q_1)^2 - m_\mu^2} \frac{1}{(p-q_2)^2 - m_\mu^2} \\
			& \quad \times \mathrm{Tr}\left( (\slashed p + m_\mu) [\gamma^\rho,\gamma^\sigma] (\slashed p + m_\mu) \gamma^\mu (\slashed p + \slashed q_1 + m_\mu) \gamma^\lambda (\slashed p - \slashed q_2 + m_\mu) \gamma^\nu \right)  \\
			& \quad \times  \sum_{i=1}^{54} \left( \frac{\p}{\p k^\rho} T^i_{\mu\nu\lambda\sigma}(q_1,q_2,k-q_1-q_2) \right) \bigg|_{k=0} \Pi_i(q_1,q_2,-q_1-q_2) .
	\end{split}
\end{align}
It turns out that there are only 19 independent linear combinations of the structures $T_i^{\mu\nu\lambda\sigma}$ that contribute to $(g-2)_\mu$. It is possible to make a basis change in the 54 structures
\begin{align}
	\Pi^{\mu\nu\lambda\sigma} = \sum_{i=1}^{54} T_i^{\mu\nu\lambda\sigma} \Pi_i =  \sum_{i=1}^{54} \hat T_i^{\mu\nu\lambda\sigma} \hat \Pi_i ,
\end{align}
in such a way that in the limit $k\to0$ the derivative of 35 structures $\hat T_i^{\mu\nu\lambda\sigma}$ vanishes. Although this change of basis does not introduce kinematic singularities into the scalar functions $\hat \Pi_i$, it somewhat obscures crossing symmetry. The 19 structures $\hat T_i^{\mu\nu\lambda\sigma}$ that contribute to $(g-2)_\mu$ can be chosen as follows:
\begin{align}
	\begin{split}
		\left\{ \hat T_i^{\mu\nu\lambda\sigma} \Big| i = 1, \ldots, 19 \right\} &= \left\{ T_i^{\mu\nu\lambda\sigma} \Big| i = 1,\ldots,11,13,14,16,17 \right\} \\
			& \qquad \cup \left\{ T_{39}^{\mu\nu\lambda\sigma} + T_{40}^{\mu\nu\lambda\sigma}, T_{42}^{\mu\nu\lambda\sigma}, T_{43}^{\mu\nu\lambda\sigma}, T_{50}^{\mu\nu\lambda\sigma} - T_{51}^{\mu\nu\lambda\sigma} \right\}.
	\end{split}
\end{align}
The corresponding 19 scalar functions $\hat \Pi_i$ are linear combinations of 33 scalar functions $\Pi_i$, see App.~\ref{sec:AppendixScalarFunctionsContributingTog-2}.
The 21 scalar functions not involved in these relations
\begin{align}
	\left\{ \Pi_i \Big| i = 12, 15, 18, 23, 24, 27, 28, 29, 30, 32, 35, 36, 37, 38, 41, 44, 45, 48, 49, 52, 53 \right\}
\end{align}
are irrelevant for the calculation of $(g-2)_\mu$.

The HLbL contribution to $(g-2)_\mu$ can now be written as
\begin{align}
	\begin{split}
		a_\mu^\mathrm{HLbL} &= - e^6  \int \frac{d^4q_1}{(2\pi)^4} \frac{d^4q_2}{(2\pi)^4} \frac{1}{q_1^2 q_2^2 (q_1+q_2)^2} \frac{1}{(p+q_1)^2 - m_\mu^2} \frac{1}{(p-q_2)^2 - m_\mu^2} \\
			& \quad \times  \sum_{i=1}^{19} \hat T_i(q_1,q_2;p) \hat\Pi_i(q_1,q_2,-q_1-q_2) ,
	\end{split}
\end{align}
where
\begin{align}
	\begin{split}
		\label{eq:DefinitionIntermediateKernels}
		\hat T_i(q_1,q_2;p) := \frac{1}{48 m_\mu} \mathrm{Tr}&\left( (\slashed p + m_\mu) [\gamma^\rho,\gamma^\sigma] (\slashed p + m_\mu) \gamma^\mu (\slashed p + \slashed q_1 + m_\mu) \gamma^\lambda (\slashed p - \slashed q_2 + m_\mu) \gamma^\nu \right)  \\
			& \times \left( \frac{\p}{\p k^\rho} \hat T^i_{\mu\nu\lambda\sigma}(q_1,q_2,k-q_1-q_2) \right) \bigg|_{k=0}
	\end{split}
\end{align}
and the $\hat\Pi_i$ are needed for the reduced kinematics
\begin{align}
	\label{eqn:ReducedKinematics}
	s = (q_1 + q_2)^2 , \quad t = q_2^2 , \quad u = q_1^2 , \quad q_1^2, \quad q_2^2, \quad q_3^2 = (q_1 + q_2)^2 , \quad k^2 = q_4^2 = 0.
\end{align}
The explicit result of the trace calculation and the contraction of the Lorentz indices is given in App.~\ref{sec:AppendixIntermediateKernels}.

We can reduce the number of terms contributing to $(g-2)_\mu$ further by using the symmetry under the exchange of the momenta $q_1 \leftrightarrow -q_2$: the loop integration measure and the product of propagators are invariant under this transformation, while the kernels $\hat T_i$ transform under $q_1 \leftrightarrow -q_2$ as
\begin{align*}
	\begin{alignedat}{4}
		\hat T_1 &\longleftrightarrow \hat T_1 , \quad & \hat T_2 &\longleftrightarrow \hat T_3 , \quad & \hat T_4 &\longleftrightarrow \hat T_4 , \quad & \hat T_5 &\longleftrightarrow \hat T_6 , \\
		\hat T_7 &\longleftrightarrow \hat T_8 , \quad & \hat T_9 &\longleftrightarrow \hat T_{12} , \quad & \hat T_{10} &\longleftrightarrow \hat T_{13} , \quad & \hat T_{11} &\longleftrightarrow \hat T_{14} , \\
		\hat T_{15} &\longleftrightarrow \hat T_{15} , \quad & \hat T_{16} &\longleftrightarrow \hat T_{16} , \quad & \hat T_{17} &\longleftrightarrow \hat T_{18} , \quad & \hat T_{19} &\longleftrightarrow -\hat T_{19} .
	\end{alignedat} \mytag
\end{align*}
For the reduced kinematics~\eqref{eqn:ReducedKinematics} the exchange $q_1 \leftrightarrow -q_2$ is equivalent to the crossing transformation $t\leftrightarrow u$, $q_1^2 \leftrightarrow q_2^2$. With the help of the crossing relations of the scalar functions $\Pi_i$, it is easy to check that the $\vphantom{\Big|}\hat \Pi_i$ transform analogously to the kernels $\hat T_i$, i.e.
\begin{align*}
	\begin{alignedat}{4}
		\hat \Pi_1 &\longleftrightarrow \hat \Pi_1 , \quad & \hat \Pi_2 &\longleftrightarrow \hat \Pi_3 , \quad & \hat \Pi_4 &\longleftrightarrow \hat \Pi_4 , \quad & \hat \Pi_5 &\longleftrightarrow \hat \Pi_6 , \\
		\hat \Pi_7 &\longleftrightarrow \hat \Pi_8 , \quad & \hat \Pi_9 &\longleftrightarrow \hat \Pi_{12} , \quad & \hat \Pi_{10} &\longleftrightarrow \hat \Pi_{13} , \quad & \hat \Pi_{11} &\longleftrightarrow \hat \Pi_{14} , \\
		\hat \Pi_{15} &\longleftrightarrow \hat \Pi_{15} , \quad & \hat \Pi_{16} &\longleftrightarrow \hat \Pi_{16} , \quad & \hat \Pi_{17} &\longleftrightarrow \hat \Pi_{18} , \quad & \hat \Pi_{19} &\longleftrightarrow -\hat \Pi_{19} .
	\end{alignedat} \mytag
\end{align*}
Therefore, it is convenient to write the HLbL contribution to $(g-2)_\mu$ as a sum of 12 terms:
\begin{align}
	\begin{split}
		\label{eq:HLbLMasterFormula8dim}
		a_\mu^\mathrm{HLbL} &= - e^6  \int \frac{d^4q_1}{(2\pi)^4} \frac{d^4q_2}{(2\pi)^4} \frac{1}{q_1^2 q_2^2 (q_1+q_2)^2} \frac{1}{(p+q_1)^2 - m_\mu^2} \frac{1}{(p-q_2)^2 - m_\mu^2} \\
			& \quad \times  \sum_{j=1}^{12} \xi_j \hat T_{i_j}(q_1,q_2;p) \hat\Pi_{i_j}(q_1,q_2,-q_1-q_2) ,
	\end{split}
\end{align}
where
\begin{align}
	\begin{split}
		\{ i_j | j=1,\ldots,12\} &= \{ 1, 2, 4, 5, 7, 9, 10, 14, 15, 16, 17, 19 \} , \\
		\{ \xi_j | j=1, \ldots, 12\} &= \{ 1, 2, 1, 2, 2, 2, 2, 2, 1, 1, 2, 1 \} .
	\end{split}
\end{align}
Note that the first two terms in this sum correspond to the well-known result for the pion-pole contribution~\cite{Knecht2002} (up to conventions: exchange of $\hat T_1$ and $\hat T_2$, the explicit factor $\xi_2=2$, and symmetrization of $\hat T_1$).

In~\eqref{eq:HLbLMasterFormula8dim}, the integrand depends on the five scalar products $q_1^2$, $q_2^2$, $q_1 \cdot q_2$, $p \cdot q_1$, and $p \cdot q_2$, where the dependence on the last two is given explicitly  (the scalar functions only depend on $q_1^2$, $q_2^2$, and $q_1 \cdot q_2$). Therefore, five of the eight integrals can be performed without knowledge of the scalar functions. The same integrals as in the case of the pion-pole contribution occur~\cite{Knecht2002, Jegerlehner2009}, which have been solved with the technique of Gegenbauer polynomials~\cite{Rosner1967}.
This method has been applied before to the full HLbL contribution in the context of vector-meson-dominance and hidden-local-symmetry models~\cite{Bijnens2012,Abyaneh2012}.

We perform a Wick rotation of the momenta $q_1$, $q_2$, and $p$ (see Sect.~\ref{sec:WickRotation}) and denote the Wick-rotated Euclidean momenta by capital letters $Q_1$, $Q_2$, and $P$. Note that $Q_1^2 = -q_1^2$, $Q_2^2 = -q_2^2$, $P^2 = -m_\mu^2$. Since $a_\mu^\mathrm{HLbL}$ is a pure number, it does not depend on the direction of the momentum $P$ of the muon, hence we can take the angular average by integrating over the four-dimensional hypersphere:
\begin{align}
	a_\mu^\mathrm{HLbL} = \int  \frac{d\Omega_4(P)}{2\pi^2} a_\mu^\mathrm{HLbL} .
\end{align}
The kernels $\hat T_i$ are at most quadratic in $p$, therefore we need the following angular integrals~\cite{Jegerlehner2009}:
\begin{align}
	\label{eq:GegenbauerIntegrals}
	\begin{split}
		 \int  \frac{d\Omega_4(P)}{2\pi^2} \frac{1}{(P+Q_1)^2 + m_\mu^2}\frac{1}{(P-Q_2)^2 + m_\mu^2} &= \frac{1}{m_\mu^2 R_{12}} \atan\left(\frac{z x}{1 - z \tau}\right) , \\
		 \int  \frac{d\Omega_4(P)}{2\pi^2} \frac{1}{(P+Q_1)^2 + m_\mu^2} &= - \frac{1 - \sigma^E_1}{2m_\mu^2} , \\
		 \int  \frac{d\Omega_4(P)}{2\pi^2} \frac{1}{(P-Q_2)^2 + m_\mu^2} &= - \frac{1 - \sigma^E_2}{2m_\mu^2} , \\
		 \int  \frac{d\Omega_4(P)}{2\pi^2} \frac{P \cdot Q_2}{(P+Q_1)^2 + m_\mu^2} &= Q_1 \cdot Q_2 \frac{(1 - \sigma^E_1)^2}{8 m_\mu^2} , \\
		 \int  \frac{d\Omega_4(P)}{2\pi^2} \frac{P \cdot Q_1}{(P-Q_2)^2 + m_\mu^2} &= - Q_1 \cdot Q_2 \frac{(1 - \sigma^E_2)^2}{8 m_\mu^2} ,
	\end{split}
\end{align}
where $\tau = \cos\theta_4$, defined by $Q_1 \cdot Q_2 = |Q_1| |Q_2| \tau$, is the cosine of the angle between the Euclidean four-momenta $Q_1$ and $Q_2$, and further
\begin{align}
	\begin{split}
		\sigma^E_i &:= \sqrt{ 1 + \frac{4 m_\mu^2}{Q_i^2} } , \quad R_{12} := |Q_1| |Q_2| x , \quad x := \sqrt{1 - \tau^2} , \\
		z &:= \frac{|Q_1||Q_2|}{4m_\mu^2} (1-\sigma^E_1)(1-\sigma^E_2) .
	\end{split}
\end{align}

\subsection{Master formula}

\label{sec:MasterFormula}

After using the angular integrals~\eqref{eq:GegenbauerIntegrals}, we can immediately perform five of the eight loop integrals by changing to spherical coordinates in four dimensions. This leads us to a master formula for the HLbL contribution to the anomalous magnetic moment of the muon:
\begin{align}
	\label{eq:MasterFormula3Dim}
	a_\mu^\mathrm{HLbL} &= \frac{2 \alpha^3}{3 \pi^2} \int_0^\infty dQ_1 \int_0^\infty dQ_2 \int_{-1}^1 d\tau \sqrt{1-\tau^2} Q_1^3 Q_2^3 \sum_{i=1}^{12} T_i(Q_1,Q_2,\tau) \bar \Pi_i(Q_1,Q_2,\tau) ,
\end{align}
where $Q_1 := |Q_1|$, $Q_2 := |Q_2|$. The hadronic scalar functions $\bar \Pi_i$ are just a subset of the $\hat \Pi_i$ and defined in~\eqref{eq:ScalarFunctionsForMasterFormula}.
They have to be evaluated for the reduced kinematics
\begin{align}
	\begin{split}
		s &= - Q_3^2 = -Q_1^2 - 2 Q_1 Q_2 \tau - Q_2^2 , \quad t = -Q_2^2 , \quad u = -Q_1^2 , \\
		q_1^2 &= -Q_1^2, \quad q_2^2 = -Q_2^2, \quad q_3^2 = - Q_3^2 = - Q_1^2 - 2 Q_1 Q_2 \tau - Q_2^2 , \quad k^2 = q_4^2 = 0.
	\end{split}
\end{align}
The integral kernels $T_i$ listed in App.~\ref{sec:AppendixMasterFormulaKernels} are fully general for any light-by-light process, while the scalar functions $\Pi_i$ parametrize the hadronic content of the master formula.
In particular, \eqref{eq:MasterFormula3Dim} can be considered a generalization of the three-dimensional integral formula for the pion-pole contribution~\cite{Jegerlehner2009}. It is valid for the whole HLbL contribution and completely generic, i.e.\ it can be used to compute the HLbL contribution to $(g-2)_\mu$ for any representation of the HLbL tensor, irrespective of whether the scalar functions are subsequently specified dispersively or taken from a model calculation. For an arbitrary representation of the HLbL tensor, the scalar functions $\Pi_i$ can be obtained by projection, see App.~\ref{sec:AppendixProjection} and App.~\ref{sec:AppendixDRScalarFunctionsForMasterFormula}.

An analogous master formula was derived in~\cite{Colangelo2014a} in the case of a helicity basis for the HLbL tensor. This step in the calculation is completely equivalent, in both cases the calculation of $a_\mu^\text{HLbL}$ proceeds via an evaluation of the trace in~\eqref{eq:amuTraceFormula} and subsequent reduction of the two-loop integral with Gegenbauer techniques. From a technical point of view, the BTT approach offers several simplifications, since the $D$-wave-related amplitudes in~\cite{Colangelo2014a} require another angular average to define the $k\to 0$ limit as well as more complicated Gegenbauer integrals than the standard ones given in~\eqref{eq:GegenbauerIntegrals}.

In close analogy to the pion-pole contribution~\cite{Knecht2002}, the main benefit of the master formula~\eqref{eq:MasterFormula3Dim} is the fact that
it contains only a three-dimensional integral, and thus is well-suited for a direct numerical implementation. In particular, the energy regions generating the bulk of the contribution can be identified by numerically integrating over $\tau$ and plotting the integrand as a function of $Q_1$ and $Q_2$~\cite{Knecht2002,Bijnens2012,Abyaneh2012,Pauk:2014rta}. 

Before turning to the main part of this paper, the foundations for a model-independent calculation of the scalar functions $\Pi_i$ by making use of dispersion relations, we next consider a subtlety in the application of the Gegenbauer integrals in~\eqref{eq:GegenbauerIntegrals}. To obtain these integrals a Wick rotation has been performed, whose validity might be questioned if the HLbL tensor possesses non-trivial analytic properties. For instance, already a simple triangle loop function would give rise to anomalous thresholds due to the integration over all photon virtualities, see~\cite{Mandelstam:1960zz,Lucha:2006vc,Hoferichter:2013ama}. In the following subsection, we prove that the Wick rotation remains valid even in the presence of anomalous thresholds. 

\subsection{Wick rotation and anomalous thresholds}

\label{sec:WickRotation}


Performing a Wick rotation before calculating a loop integral is a standard
procedure. The applicability of this transformation depends on whether the
integrand is free of singularities in the region swept by the integration
path during the rotation. In our case we have to perform the two-loop
integral~\eqref{eq:HLbLMasterFormula8dim}, with an integrand containing
products of propagators multiplying functions satisfying a Mandelstam
representation, so having the analytic properties of one-loop
functions. The case of the propagators is standard. The one-loop functions
occurring in our integrand, however, need some more discussion. The
complication is not due to the presence of cuts instead of poles: cuts of
loop functions are usually on the positive real axis above some threshold,
and the functions are evaluated on the upper rim of the cut. The Wick
rotation can be applied without problem even in these cases. The need for a
more detailed discussion is due to the presence of anomalous cuts --- cuts which
appear first in three-point functions and which, depending on the
value of the momenta squared of the external legs, may intrude into the
first Riemann sheet away from the real axis.  In the present paper we need
to deal with three- and four-point one-loop functions, such as
$C_0(s;q_1^2,q_2^2)$ and $D_0(s,t;q_1^2,q_2^2,q_3^2,q_4^2)$ 
(see App.~\ref{sec:AppendixDispersiveLoops}). For a discussion of the
analytic properties of these loop functions, 
see~\cite{Frederiksen1971,Frederiksen1973b,Frederiksen1973,Frederiksen1973a,Lucha:2006vc}.
We will
illustrate what happens in the presence of such anomalous cuts by discussing
the case of the $C_0$ function: the conclusion is that the Wick rotation
can be applied without problem even in these cases.

The function $C_0(s;q_1^2,q_2^2)$ has an anomalous cut for $q_1^2>0$,
$q_2^2>0$, and $q_1^2+q_2^2> 4 m^2$, where $m$ is the mass of the particle
running in the loop (assuming it is only one mass). The origin of this
anomalous cut is that $C_0$ admits a dispersive representation having a
discontinuity on the positive real axis for $s \geq 4 m^2$. The
discontinuity, however, has itself a cut between the branch points:
\begin{equation}
s_\pm = q_1^2+q_2^2 - \frac{q_1^2 q_2^2}{2 m^2} \mp \frac{1}{2m^2} \sqrt{
  q_1^2 (q_1^2-4 m^2) q_2^2 (q_2^2-4m^2)}.
\end{equation}
While for $q_{1,2}^2< 0$ these branch points lie on the second Riemann
sheet, for $q_1^2>0$, $q_2^2>0$, and $q_1^2+q_2^2> 4 m^2$ one of them moves
into the first Riemann sheet by crossing the real axis above $s=4m^2$. In
this case the dispersive representation of the function $C_0$ needs to be
modified by adding a further integral along the anomalous cut from $4 m^2$
up to $s_+$. The most convenient way to see how the branch points move as a
function of $q_1^2$ and $q_2^2$ is to introduce the variables~\cite{Lucha:2006vc}
\begin{equation}
q_i^2=-m^2 \frac{(1-\xi_i)^2}{\xi_i}.
\end{equation}
The upper half-plane of the complex variables $q_i^2$ are mapped onto the
upper unit semicircle of unit radius: the negative real axis of $q_i^2$ is
mapped onto the segment $0 < \xi_i \leq 1$, the region $0 \leq q_i^2 \leq
4m^2$ onto the semicircle $\xi_i=e^{i \phi_i}$, with $0\leq \phi_i \leq
\pi$, and finally the semiaxis $4 m^2 \leq q_i^2$ onto the segment 
$-1 \leq \xi_i <0$. The branch points $s_\pm$ can be expressed as follows
in terms of these variables
\begin{equation}
\frac{s_\pm}{m^2} = -\frac{(1-\xi_1)^2}{\xi_1}- \frac{(1-\xi_2)^2}{\xi_2}
-\frac{(1-\xi_1)^2(1-\xi_2)^2}{2 \xi_1 \xi_2} \mp 
\frac{(1-\xi_1^2)(1-\xi_2^2)}{2 \xi_1 \xi_2}
\end{equation}
and from this it is easy to check that $s_-$ always stays on the second
sheet, whereas $s_+$ moves into the first one under the conditions given
above. It is important to stress that when it enters the first Riemann
sheet, $s_+$ stays always below the real axis, either on its lower rim (for
$q_1^2<4 m^2$ and $q_2^2<4 m^2$ or for $q_1^2 > 4 m^2$ and $q_2^2> 4 m^2$),
or into the lower complex half-plane (for $q_1^2<4 m^2$ and $q_2^2>4 m^2$
or vice versa), in contrast to what is shown in Fig.~2 of~\cite{Lucha:2006vc}.

When we perform the Wick rotation for $q_1$ and $q_2$ (to avoid weird
trajectories in the complex $s$-plane we rotate both
simultaneously), their squares (in Minkowski space) become transformed into
minus their Euclidean squares ($-Q_i^2$), and can only take negative
values. But also $s=(q_1+q_2)^2$ is mapped onto $s_E\equiv -(Q_1+Q_2)^2$,
which equally can only be negative. If we start from a positive value of
$s$ and perform the Wick rotation, the trajectory of $s$ follows an arc in
the upper complex half plane and lands on the negative axis at $s_E$. The
potential danger is if we start from $s> 4m^2$, just on the upper rim of
the cut, but the Wick rotation moves it away both from the normal as well
as from the anomalous cut, which, as mentioned above, is always below the
real axis.  If we start from negative values of $s$ we also follow an arc
in the upper complex half plane and land at $s_E<s$, without ever coming
even close to the cuts.

\begin{figure}[t]
	\centering
	\begin{subfigure}[b]{0.4\textwidth}
		\includegraphics[width=7cm]{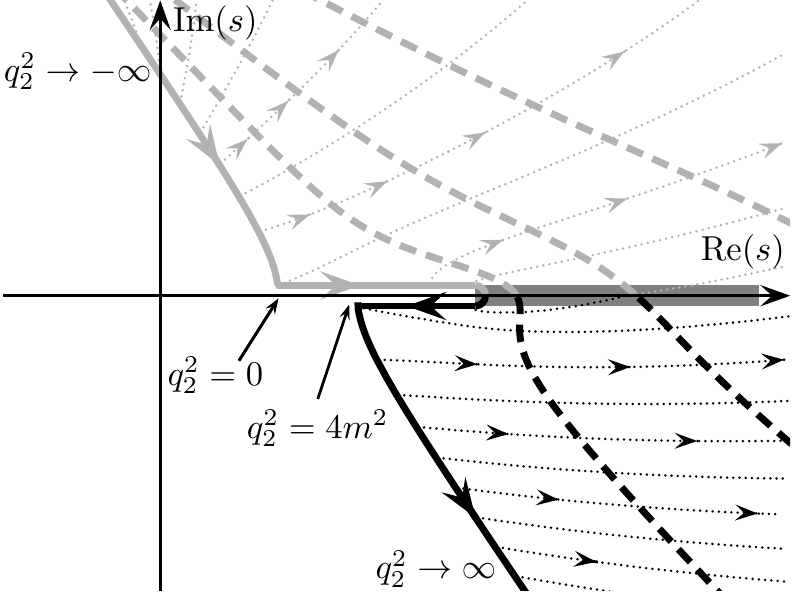}
		\caption{$0<q_1^2<4m^2$}
		\label{img:TrajectoryWickRotationA}
	\end{subfigure}
	\hspace{1.5cm}
	\begin{subfigure}[b]{0.4\textwidth}
		\includegraphics[width=7cm]{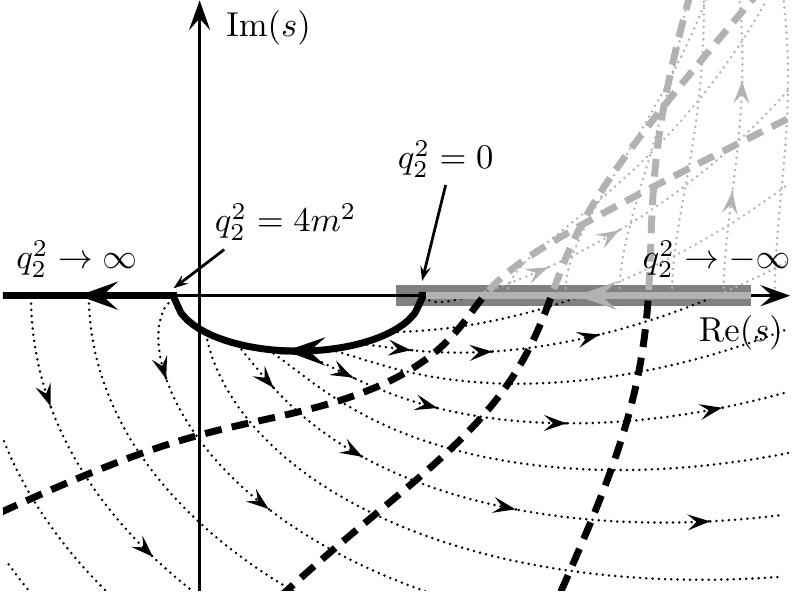}
		\caption{$4m^2<q_1^2$}
		\label{img:TrajectoryWickRotationB}
	\end{subfigure}
	\caption{Movement of the trajectory of the anomalous threshold
          $s_+$ during the Wick rotation. The thick gray line on the real
          axis is the unitarity cut for $s>4m^2$. The solid line denotes
          the trajectory of $s_+$ for a fixed $q_1^2>0$: the anomalous
          threshold enters the physical sheet through the unitarity cut
          when $q_2^2$ increases. The dotted lines show the movement of
          $s_+$ when $q_1$ and $q_2$ are simultaneously Wick rotated: it
          retreats through the unitarity cut into the second Riemann sheet.
          The dashed lines show intermediate trajectories of $s_+$ during
          the Wick rotation. The part of the trajectories that lies on the second sheet is
          shown in light gray, the part on the physical sheet is black.}
	\label{img:TrajectoryWickRotation}
\end{figure}

The discussion above is still incomplete, however, because we have so far
neglected the fact that as we make the Wick rotation not only $s$ moves,
but also the $q_i^2$ and with them $s_+$, which determines where the
anomalous cut is. In principle, $s_+$ could move into the complex upper
half-plane and interfere with the trajectory of $s$ during the Wick
rotation. An explicit calculation shows, however, that this does not
happen: at the end of the Wick rotation both $q_i^2$ have become $-Q_i^2$
and therefore negative, which implies that the positivity conditions
required for the existence of the anomalous cut are not fulfilled
anymore. The trajectory of $s_+$ during the Wick rotation is shown in
Fig.~\ref{img:TrajectoryWickRotation}. There one sees that in all possible
cases $s_+$ moves through the unitarity cut back into the second sheet, so
that indeed, the anomalous cut disappears. The two figures show the two possible
cases, where the anomalous threshold enters the physical sheet. Fig.~\ref{img:TrajectoryWickRotationA}
shows the trajectory of the anomalous threshold for a fixed value of $0 < q_1^2 < 4 m^2$ and $q_2^2$ varying
from $-\infty$ to $\infty$ (solid line, light gray indicates the part of the trajectory on the second sheet, whereas the part on the physical sheet is shown in black). Fig.~\ref{img:TrajectoryWickRotationB}
shows the analogous trajectory for a fixed value of $q_1^2 > 4 m^2$.
The dashed lines are intermediate trajectories, when the two momenta are simultaneously Wick rotated by the same phase.
The dotted lines indicate the movement of selected points on the trajectory of the anomalous threshold during the Wick rotation.
All points of the trajectory that originally lie on the physical sheet move back through the unitarity cut into the second sheet.

As a further check we have calculated numerically a simplified toy example
that exhibits all relevant features discussed above.  
We consider
\begin{equation}
\label{toy_def}
a_\mu^\text{toy}=(4\pi)^4\int\frac{d^4 q_1}{(2\pi)^4}\int\frac{d^4 q_2}{(2\pi)^4}\frac{m_\mu^2}{\big(q_1^2+i\eps\big)\big(q_2^2+i\eps\big) (s+i\eps)\big(s-m_\mu^2+i\eps\big)}\int_0^1d x\int_0^{1-x}d y\, \Delta(x,y)^{-1},
\end{equation}
where 
\begin{equation}
\label{Delta_def}
\Delta(x,y)=M_\pi^2-x y s-x(1-x-y)q_1^2-y(1-x-y)q_2^2-i\eps.
\end{equation}
The function defined by the last two integrals is proportional to the
Feynman-parameter representation of the triangle loop function
$C_0(s;q_1^2,q_2^2)$, while the remainder of $a_\mu^\text{toy}$ is chosen
in as close analogy to the real HLbL integral as possible, including a
sufficient number of propagators to make the integral converge.  
In the standard loop-integral approach the intricacies due to anomalous
thresholds are automatically taken into account by the $i\eps$-prescription
in~\eqref{toy_def} and \eqref{Delta_def}, leading to the following
Feynman-parameter representation 
\begin{align}
\label{toy_Feynman}
 a_\mu^\text{toy}&=-\int_0^1d x \int_0^{1-x}d y \int_0^1d x_1
 \int_0^{1-x_1}d x_2 \int_0^1d
 x_3\frac{\log\Big(1+\frac{x_2}{x_1}\frac{m_\mu^2}{M_\pi^2}\Big)}{\Delta_1(x_1,y)\Delta_2(x_1,x_2,x_3,x,y)^2},\notag\\ 
 \Delta_1(x_1,y)&=1-x_1+x_1 y(1-y),\notag\\
 \Delta_2(x_1,x_2,x_3,x,y)&=\big(1-x_3\big)\Delta_1(x_1,y) +x_3\big(x_2+x_1
 x(1-x)\big)-x_3\frac{(x_2+x_1 x y)^2}{\Delta_1(x_1,y)}. 
\end{align}
In analogy to the HLbL master formula~\eqref{eq:MasterFormula3Dim} the
Wick-rotated representation becomes (the $Q_i$ have been rescaled by
$m_\mu$)  
\begin{align}
\label{toy_Wick}
a_\mu^\text{toy}&=-\frac{8}{\pi}\int_0^\infty d Q_1 \int_0^\infty d
Q_2\int_{-1}^1d \tau\frac{\sqrt{1-\tau^2}Q_1 Q_2}{Q_3^2\big(1+Q_3^2\big)} 
\int_0^1d x\int_0^{1-x}d y\, \tilde \Delta(x,y)^{-1},\notag\\
\tilde \Delta(x,y)&=\frac{M_\pi^2}{m_\mu^2}+x y Q_3^2+x(1-x-y)Q_1^2+y(1-x-y)Q_2^2.
\end{align}
We checked numerically that indeed the representations~\eqref{toy_Feynman}
and~\eqref{toy_Wick} are identical. This confirms that 
the Wick rotation is permitted despite the occurrence of anomalous
thresholds, all signs of which disappear in the Feynman parameterization as
long as the $i\eps$-prescription is applied consistently. Similarly, the
final result for the Wick rotation displays no remnants of the anomalous
thresholds. As shown in Fig.~\ref{img:TrajectoryWickRotation}, in this case
the deeper reason can be traced back to the trajectory of the anomalous
threshold during the Wick rotation towards the second sheet.  

Finally, we note that the above discussion of the anomalous threshold of
the triangle diagram is already sufficient for the full HLbL
calculation. Box diagrams do not lead to further complications, since, in
the master formula~\eqref{eq:HLbLMasterFormula8dim}, the limit $k\to0$ is
taken before the Wick rotation of the loop momenta, hence the analytic
structure of box diagrams is already reduced to the one of triangle
diagrams.


\section{Mandelstam representation}
\label{sec:MandelstamRepresentation}

In the previous section, we have obtained a master formula~\eqref{eq:MasterFormula3Dim} for the HLbL contribution to the anomalous magnetic moment of the muon, where the hadronic dynamics is parametrized in terms of the scalar functions $\Pi_i$. Since these functions are free of kinematic singularities and zeros, they are the quantities that should satisfy a Mandelstam representation~\cite{Mandelstam1958}. We need to determine seven scalar functions that are not related to each other by crossing symmetry.
Due to the complexity of the problem, we cannot obtain an exact solution for the scalar functions but have to rely on approximations. Our strategy is to order the contributions to the scalar functions by the mass of the intermediate states. The lightest intermediate states are expected to give the most important contribution, heavier states are suppressed by the higher threshold and smaller phase space. In this paper, we will consider the two lowest-lying contributions: one- and two-pion intermediate states in all channels, i.e.\ pion-pole and pion-box topologies. These two contributions can be described without any further approximation. Of course, pion-pole and pion-loop contributions have been considered before in various model calculations. The important point of the dispersive approach is that these contributions become unambiguously defined: the pion pole, with on-shell pion transition form factor, corresponds precisely to one-pion intermediate states, and the sQED pion loop dressed with pion vector form factors to two-pion intermediate states with pion-pole LHC. The explicit proof of these identifications will be presented in this section, based on a Mandelstam representation for the BTT scalar functions. 

\subsection{Derivation of the double-spectral representation}

\label{sec:DoubleSpectralRepresentationDerivation}

\begin{figure}[t]
	\centering
	\includegraphics[width=2.5cm]{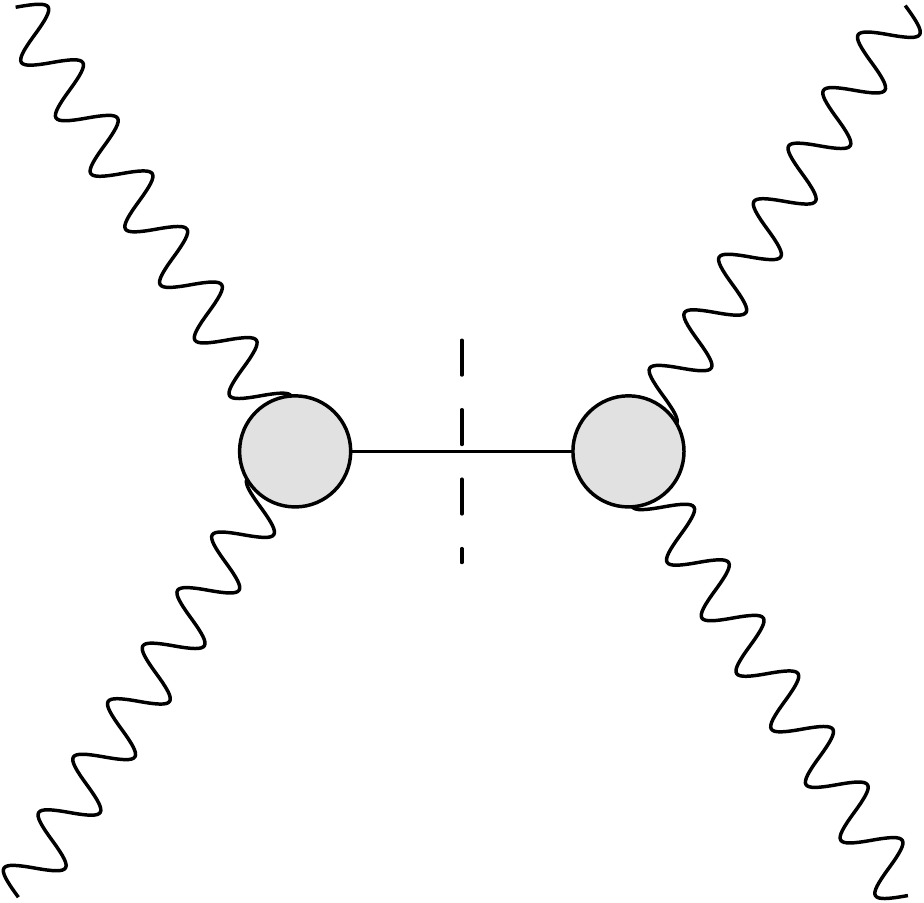}
	\hspace{1cm}
	\includegraphics[width=2.5cm]{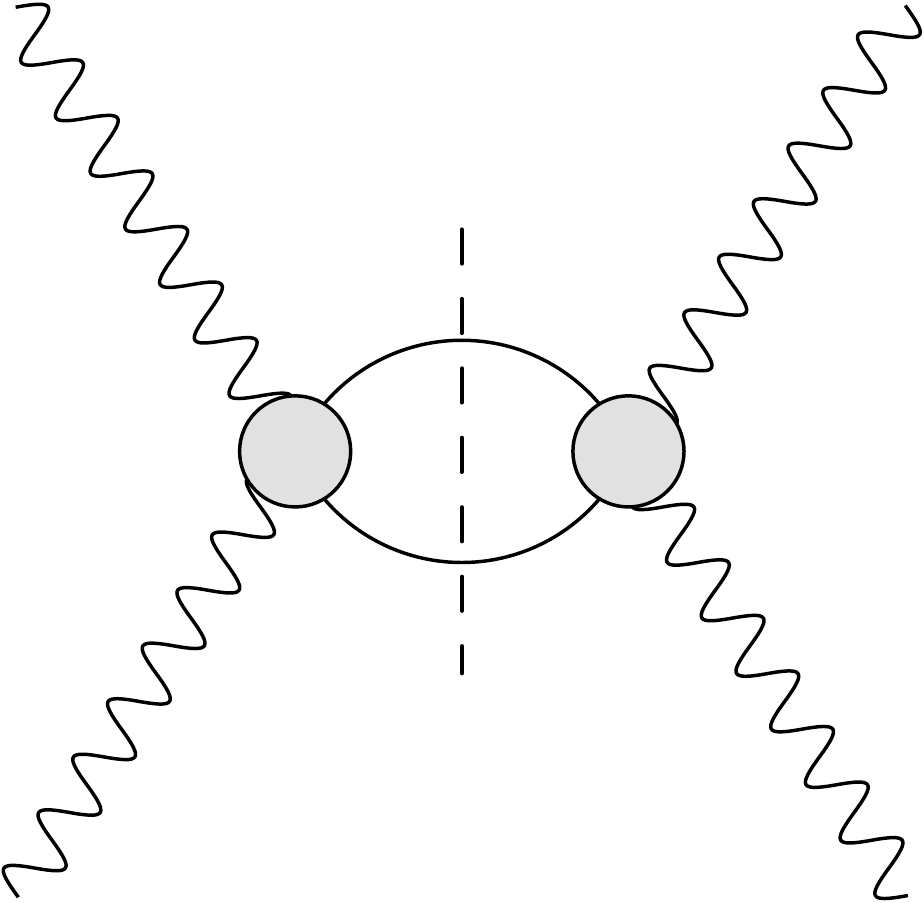}
	\caption{Intermediate states in the direct channel: pion pole and two-pion cut.}
	\label{img:HLbLIntermediateStates}
\end{figure}

For the derivation of a Mandelstam representation of the scalar functions, we follow the discussion in~\cite{Martin1970}. We assume that the photon virtualities $q_i^2$ are fixed and small enough so that no anomalous thresholds are present. A parameter-free description of the HLbL tensor and therefore an unsubtracted dispersion relation is crucial. The behavior of the imaginary parts, which is determined by the asymptotics of the sub-processes, does indeed suggest that no subtractions are needed. Furthermore, even a quark-loop contribution to the HLbL tensor has an asymptotic behavior that requires no subtractions.\footnote{Contrary to possible subtractions in the sub-processes, the presence of subtraction constants in the HLbL scalar functions would imply a contribution to $a_\mu$ that is not determined by unitarity.} Hence, for a generic scalar function $\Pi_i$, we write a fixed-$t$ dispersion relation without any subtractions:
\begin{align}
	\Pi_i^t(s,t,u) &= c_i^t + \frac{\rho_{i;s}^t}{s - M_\pi^2} + \frac{\rho_{i;u}^t}{u - M_\pi^2} + \frac{1}{\pi} \int_{4M_\pi^2}^\infty ds^\prime \frac{ \Im_s \Pi_i^t(s^\prime,t,u^\prime)}{s^\prime - s} + \frac{1}{\pi} \int_{4M_\pi^2}^\infty du^\prime \frac{\Im_u \Pi_i^t(s^\prime,t,u^\prime)}{u^\prime-u} ,
\end{align}
where $c_i^t$ is supposed to behave as $\lim\limits_{t\to0} c_i^t = 0$ and takes into account the $t$-channel pole. The imaginary parts are understood to be evaluated just above the corresponding cut. The primed variables fulfill
\begin{align}
	s^\prime + t + u^\prime = \Sigma := \sum_{i=1}^4 q_i^2 .
\end{align}
If we continue the fixed-$t$ dispersion relation analytically in $t$, we have to replace the imaginary parts by the discontinuities, defined by
\begin{align}
	\begin{split}
		D_{i;s}^t(s^\prime) &:= \frac{1}{2i} \left( \Pi_i^t(s^\prime+i\epsilon,t,u^\prime) - \Pi_i^t(s^\prime-i\epsilon,t,u^\prime) \right) , \\
		D_{i;u}^t(u^\prime) &:= \frac{1}{2i} \left( \Pi_i^t(s^\prime,t,u^\prime+i\epsilon) - \Pi_i^t(s^\prime,t,u^\prime-i\epsilon) \right) ,
	\end{split}
\end{align}
hence
\begin{align}
	\label{eq:HLbLUnsubtractedDRforScalarFunctions}
	\Pi_i^t(s,t,u) &= c_i^t + \frac{\rho_{i;s}^t}{s - M_\pi^2} + \frac{\rho_{i;u}^t}{u - M_\pi^2} + \frac{1}{\pi} \int_{4M_\pi^2}^\infty ds^\prime \frac{ D_{i;s}^t(s^\prime)}{s^\prime - s} + \frac{1}{\pi} \int_{4M_\pi^2}^\infty du^\prime \frac{D_{i;u}^t(u^\prime)}{u^\prime-u} .
\end{align}
Both the discontinuities as well as the pole residues are determined by $s$- or $u$-channel unitarity, which also defines their analytic continuation in $t$. While $\rho^t_{i;s,u}$ are due to a one-pion intermediate state, $D_{i;s,u}^t$ are due to multi-particle intermediate states, see Fig.~\ref{img:HLbLIntermediateStates}. We limit ourselves to two-pion intermediate states and neglect the contribution of heavier intermediate states to the discontinuities.

First, we study the pion-pole contribution by analyzing the unitarity relation:
\begin{align}
	\begin{split}
		\Im_s &\left( e^4 (2\pi)^4 \delta^{(4)}(q_1 + q_2 + q_3 - q_4) H_{\lambda_1\lambda_2,\lambda_3\lambda_4} \right) \\
			&= \sum_n \frac{1}{2S_n} \left( \prod_{i=1}^n \int \widetilde{dp_i} \right) \< \gamma^*(-q_3,\lambda_3) \gamma^*(q_4,\lambda_4) | n; \{p_i\} \>^* \< \gamma^*(q_1,\lambda_1) \gamma^*(q_2,\lambda_2) | n; \{p_i\} \> ,
	\end{split}
\end{align}
where $S_n$ is the symmetry factor of the intermediate state $|n\>$. We consider now only the $\pi^0$ intermediate state in the sum:
\begin{align}
	\begin{split}
		\Im_s^\pi &\left( e^4 (2\pi)^4 \delta^{(4)}(q_1 + q_2 + q_3 - q_4) H_{\lambda_1\lambda_2,\lambda_3\lambda_4} \right) \\
			&= \frac{1}{2} \int \widetilde{dp} \; \< \gamma^*(-q_3,\lambda_3) \gamma^*(q_4,\lambda_4) | \pi^0(p) \>^* \< \gamma^*(q_1,\lambda_1) \gamma^*(q_2,\lambda_2) | \pi^0(p) \> .
	\end{split}
\end{align}
After reducing the matrix elements and using the definition of the pion transition form factor
\begin{align}
	i \int d^4x \; e^{i q \cdot x} \< 0 | T \{ j^\mu_\mathrm{em}(x) j^\nu_\mathrm{em}(0) \} | \pi^0(p) \> = \epsilon^{\mu\nu\alpha\beta} q_\alpha p_\beta \mathcal{F}_{\pi^0\gamma^*\gamma^*}\big(q^2, (q-p)^2\big) ,
\end{align}
we find
\begin{align}
	\begin{split}
		\Im_s^\pi \Pi^{\mu\nu\lambda\sigma} &= - \frac{1}{2} \begin{aligned}[t]
				& \int \widetilde{dp} \; (2\pi)^4 \delta^{(4)}(q_1+q_2-p) \epsilon^{\mu\nu\alpha\beta} \epsilon^{\lambda\sigma\gamma\delta} {q_1}_\alpha {q_2}_\beta {q_3}_\gamma {q_4}_\delta \\
				&\quad \times \mathcal{F}_{\pi^0\gamma^*\gamma^*}\big(q_1^2,q_2^2\big) \mathcal{F}_{\pi^0\gamma^*\gamma^*}\big(q_3^2,q_4^2\big) \end{aligned} \\
			&= - \pi \delta( s - M_\pi^2 ) \epsilon^{\mu\nu\alpha\beta} \epsilon^{\lambda\sigma\gamma\delta} {q_1}_\alpha {q_2}_\beta {q_3}_\gamma {q_4}_\delta \mathcal{F}_{\pi^0\gamma^*\gamma^*}\big(q_1^2,q_2^2\big) \mathcal{F}_{\pi^0\gamma^*\gamma^*}\big(q_3^2,q_4^2\big) .
	\end{split}
\end{align}
By projecting onto the scalar functions $\Pi_i$, this leads to
\begin{align}
	\rho_{i;s}^t &= \left\{ \begin{matrix} \mathcal{F}_{\pi^0\gamma^*\gamma^*}\big(q_1^2,q_2^2\big) \mathcal{F}_{\pi^0\gamma^*\gamma^*}\big(q_3^2,q_4^2\big) & i = 1 , \\
									0 & i \neq 1 , \end{matrix} \right.
\end{align}
and, analogously,
\begin{align}
	\rho_{i;u}^t &= \left\{ \begin{matrix} \mathcal{F}_{\pi^0\gamma^*\gamma^*}\big(q_1^2,q_4^2\big) \mathcal{F}_{\pi^0\gamma^*\gamma^*}\big(q_2^2,q_3^2\big) & i = 3 , \\
									0 & i \neq 3 . \end{matrix} \right.
\end{align}

In order to identify the discontinuities, we project the unitarity relation selecting two-pion intermediate states:
\begin{align}
	\begin{split}
		& \Im_s^{\pi\pi} \left( e^4 (2\pi)^4 \delta^{(4)}(q_1 + q_2 + q_3 - q_4) H_{\lambda_1\lambda_2,\lambda_3\lambda_4} \right) \\
			&= \frac{1}{2} \int \widetilde{dp}_1 \widetilde{dp}_2 \< \pi^+(p_1) \pi^-(p_2) | \gamma^*(-q_3,\lambda_3) \gamma^*(q_4,\lambda_4) \>^* \< \pi^+(p_1) \pi^-(p_2) | \gamma^*(q_1,\lambda_1) \gamma^*(q_2,\lambda_2) \> \\
			&\quad + \frac{1}{4} \int \widetilde{dp}_1 \widetilde{dp}_2 \< \pi^0(p_1) \pi^0(p_2) | \gamma^*(-q_3,\lambda_3) \gamma^*(q_4,\lambda_4) \>^* \< \pi^0(p_1) \pi^0(p_2) | \gamma^*(q_1,\lambda_1) \gamma^*(q_2,\lambda_2) \> ,
	\end{split}
\end{align}
hence
\begin{align*}
	\Im_s^{\pi\pi} \Pi^{\mu\nu\lambda\sigma} &= \frac{1}{32\pi^2} \frac{\sigma_\pi(s)}{2} \int d\Omega_s^\dprime \begin{aligned}[t]
		& \bigg( W_{+-}^{\mu\nu}(p_1,p_2,q_1) {W_{+-}^{\lambda\sigma}}^*(p_1,p_2,-q_3) \\
		& + \frac{1}{2} W_{00}^{\mu\nu}(p_1,p_2,q_1) {W_{00}^{\lambda\sigma}}^*(p_1,p_2,-q_3) \bigg) , \end{aligned} \mytag
\end{align*}
where the subscripts $\{+-, 00\}$ denote the pion charges. The analytic continuation of the unitarity relation can be obtained if the $\gamma^*\gamma^*\to\pi\pi$ matrix element $W^{\mu\nu}$ is expressed in terms of the fixed-$s$ dispersion relation~\eqref{eq:FixedSDispRelggpipi} for its scalar functions:
\begin{align}
	\begin{split}
		W_{+-}^{\mu\nu} &= \sum_{i=1}^5 T^{\mu\nu}_i \begin{aligned}[t]
			& \Bigg( \frac{\hat\rho_{i;t}^{s;+-}(s)}{t-M_\pi^2} + \frac{\hat\rho_{i;u}^{s;+-}(s)}{u-M_\pi^2} + \frac{1}{\pi} \int_{4M_\pi^2}^\infty dt_1 \frac{\hat D_{i;t}^{s;+-}(t_1;s)}{t_1 - t} + \frac{1}{\pi} \int_{4M_\pi^2}^\infty du_1 \frac{\hat D_{i;u}^{s;+-}(u_1;s)}{u_1 - u} \Bigg) , \end{aligned} \\
		W_{00}^{\mu\nu} &= \sum_{i=1}^5 T^{\mu\nu}_i \begin{aligned}[t]
			& \Bigg( \frac{1}{\pi} \int_{4M_\pi^2}^\infty dt_1 \frac{\hat D_{i;t}^{s;00}(t_1;s)}{t_1 - t} + \frac{1}{\pi} \int_{4M_\pi^2}^\infty du_1 \frac{\hat D_{i;u}^{s;00}(u_1;s)}{u_1 - u} \Bigg) . \end{aligned}
	\end{split}
\end{align}
Note that $W_{00}^{\mu\nu}$ does not contain any pole terms because the photon does not couple to two neutral pions due to angular momentum conservation and Bose symmetry.

If we pick the contribution of the pole terms on both sides of the cut, we single out box topologies:
\begin{align}
	\Im_s^{\pi\pi} \Pi^{\mu\nu\lambda\sigma} \Big|_\mathrm{box} &= \frac{1}{32\pi^2} \frac{\sigma_\pi(s)}{2} \int d\Omega_s^\dprime  \sum_{i,j=1,4} T^{\mu\nu}_i T^{\lambda\sigma}_j 
		\Bigg( \frac{\hat\rho_{i;t}^{s;+-}(s)}{t^\prime-M_\pi^2} + \frac{\hat\rho_{i;u}^{s;+-}(s)}{u^\prime-M_\pi^2} \Bigg) \Bigg( \frac{\hat\rho_{j;t}^{s;+-}(s)}{t^\dprime-M_\pi^2} + \frac{\hat\rho_{j;u}^{s;+-}(s)}{u^\dprime-M_\pi^2} \Bigg)^* ,
\end{align}
where the primed variables belong to the sub-process on the left-hand side and the double-primed variables to the sub-process on the right-hand side of the cut.

We could now apply a tensor reduction to obtain phase-space integrals for the reduced scalar quantities. The projection on the scalar functions $\Pi_i$ would allow us to identify the discontinuities $D_{i;s}^t$ due to box structures. The reduced scalar integrals could then be transformed into another dispersive integral. Together with the dispersion integral $ds^\prime$ of the primary cut, this produces the double-spectral representation. The case of the simplest scalar phase-space integral is explained in~\cite{Stoffer2014}.

Unfortunately, the tensor reduction is enormously complex because it involves the inversion of a $138\times138$ square matrix. We will therefore follow a different strategy for the box topologies. We note that the non-zero pole residues $\hat\rho_{i;t,u}^{s;+-}$ contain two electromagnetic pion form factors for the off-shell photons. These form factors can be factored out and multiply then the discontinuity that would be obtained by applying Cutkosky's rules~\cite{Cutkosky1960} to the sQED pion loop calculation. This becomes clear from the relation between the pole terms and the sQED Born terms as discussed Sect.~\ref{sec:PionPoleggpipi}. Therefore, the box contribution is nothing else but the sQED contribution multiplied by a vector form factor $F_\pi^V(q_i^2)$ for each of the off-shell photons. As in the case of the sub-process, the difference between unitarity diagrams and Feynman diagrams is absolutely crucial: the sQED contribution consists of boxes, triangles, and bulb Feynman diagrams, but corresponds to the pure box topology in terms of unitarity. In Sect.~\ref{sec:UniquenessBoxFsQED}, we will prove that this identification is correct and unique.

Finally, there are the contributions with discontinuities either in one or in both of the sub-processes. They contain for example resonance contributions in the sub-process, but also two-pion rescattering effects in the direct channel.

\subsection{Symmetrization and classification into topologies}

In the previous subsection, we have explained how the double-spectral
representation can be derived from a fixed-$t$ dispersion relation by
taking the analytic continuation in $t$, which is defined by the unitarity
relation. In the $s$-channel unitarity relation, a fixed-$s$ dispersion
relation of the sub-process is inserted (in the unitarity relation for the
$u$-channel contribution, the variable $u$ is kept fixed, which, however,
plays again the role of $s$ in the sub-process). Of course, one could have
started with a fixed-$u$ or fixed-$s$ dispersion relation in the first
place. The requirement that this lead to the same result allows us to
identify a symmetric representation, which treats the Mandelstam variables
on an equal footing and therefore implements crossing symmetry. In this
symmetric representation, we classify the different contributions in terms
of topologies. Note that in the case of HLbL, we get the two other
possibilities (i.e.\ taking fixed-$u$ and fixed-$s$ dispersion relations as
the starting point) for free, because we consider a totally crossing
symmetric process.

If we compare the three different representations of unsubtracted
dispersion relations~\eqref{eq:HLbLUnsubtractedDRforScalarFunctions}, we
immediately see that the contributions in one representation are either
contained explicitly in the other representations, can be understood as
part of the respective constant $c_i^t$, or correspond to a contribution of
neglected higher intermediate states, as we are now going to discuss.

\subsubsection{Pion-pole contribution}

\begin{figure}[t]
	\centering
	\includegraphics[width=3cm]{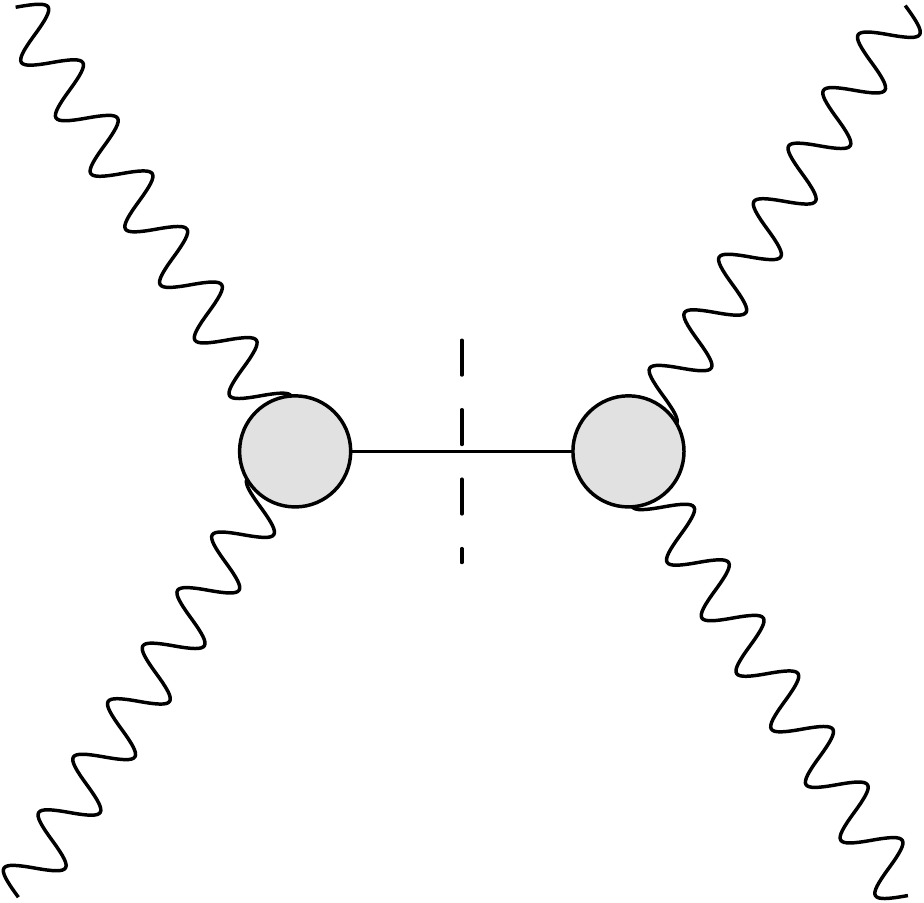}
	\caption{Unitarity diagram representing the pion-pole contribution in one channel.}
	\label{img:HLbLPionPole}
\end{figure}

First, we consider the pion-pole topology (see Fig.~\ref{img:HLbLPionPole}). The fixed-$t$ dispersion relation contains the poles in the $s$- and $u$-channel explicitly. Analogously, the fixed-$s$ (fixed-$u$) dispersion relation contains the poles in the $t$- and $u$-channel ($s$- and $t$-channel). The $t$-channel pole contribution, which is not explicit in the fixed-$t$ representation, can be identified with $c_i^t$ as it vanishes in the limit $t\to\infty$. Hence, the total pion-pole contribution is just given by
\begin{align}
	\label{eq:PionPoleScalarFunctions}
	\Pi_i^{\pi^0\text{-pole}}(s,t,u) &= \frac{\rho_{i;s}}{s-M_\pi^2} + \frac{\rho_{i;t}}{t-M_\pi^2} + \frac{\rho_{i;u}}{u-M_\pi^2} ,
\end{align}
where the pole residues are products of pion transition form factors:
\begin{align}
	\begin{split}
		\rho_{i,s} &= \delta_{i1} \; \mathcal{F}_{\pi^0\gamma^*\gamma^*}\big(q_1^2,q_2^2\big) \mathcal{F}_{\pi^0\gamma^*\gamma^*}\big(q_3^2,q_4^2\big) , \\
		\rho_{i,t} &= \delta_{i2} \; \mathcal{F}_{\pi^0\gamma^*\gamma^*}\big(q_1^2,q_3^2\big) \mathcal{F}_{\pi^0\gamma^*\gamma^*}\big(q_2^2,q_4^2\big) , \\
		\rho_{i,u} &= \delta_{i3} \; \mathcal{F}_{\pi^0\gamma^*\gamma^*}\big(q_1^2,q_4^2\big) \mathcal{F}_{\pi^0\gamma^*\gamma^*}\big(q_2^2,q_3^2\big) , \\
	\end{split}
\end{align}
where $\delta_{ij}$ is the Kronecker delta.

Since the Lorentz structures of the usual pion-pole contribution, see e.g.\ \cite{Knecht2002}, coincide with the first three BTT structures, this proves that in a dispersive approach the pion pole is unambiguously defined as given in~\cite{Colangelo2014a} (and already in~\cite{Knecht2002}), in particular with the transition form factor defined for an on-shell pion. 

\subsubsection{Box contribution}

\label{sec:BoxContribution}

\begin{figure}[t]
	\centering
	\begin{subfigure}[b]{0.3\textwidth}
		\centering
		\includegraphics[width=3cm]{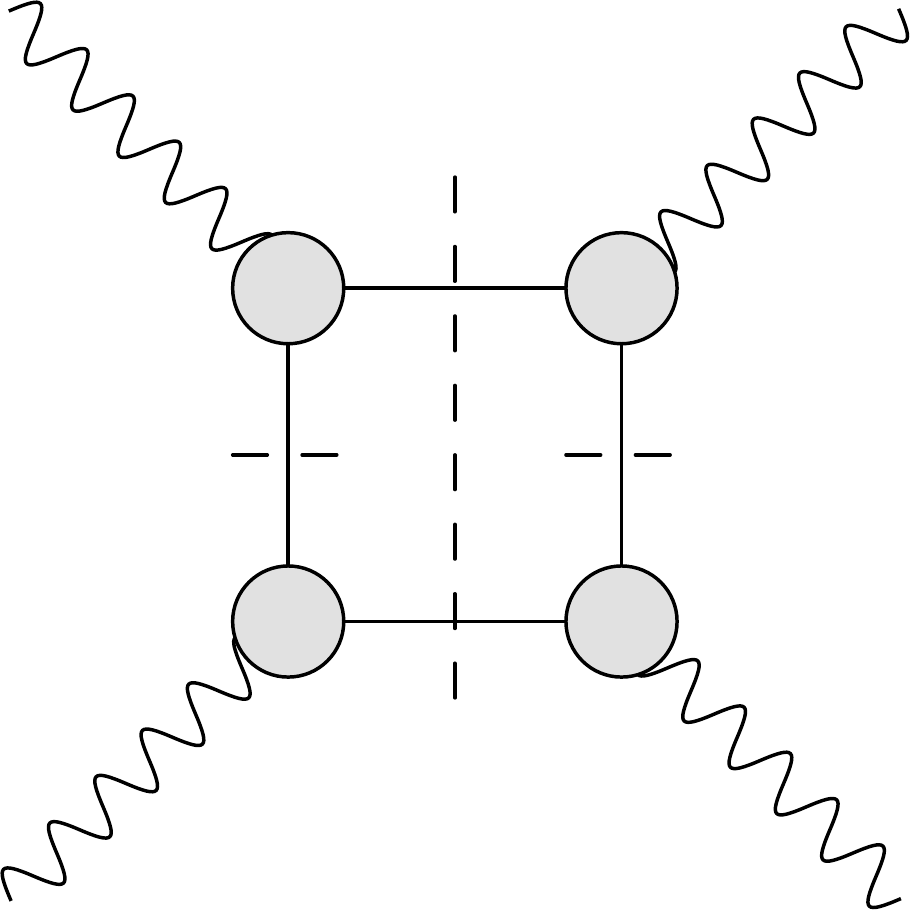}
		\caption{}
		\label{img:HLbLBoxA}
	\end{subfigure}
	\begin{subfigure}[b]{0.3\textwidth}
		\centering
		\includegraphics[width=3cm]{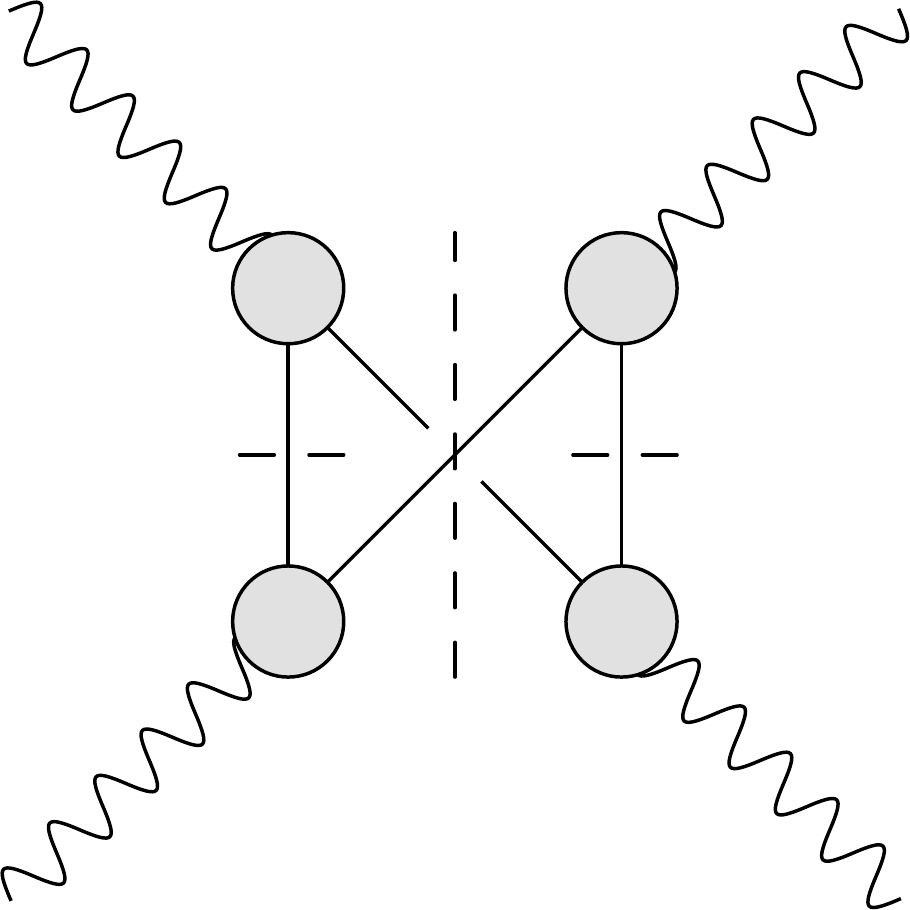}
		\caption{}
		\label{img:HLbLBoxB}
	\end{subfigure}
	\begin{subfigure}[b]{0.3\textwidth}
		\centering
		\includegraphics[width=3cm]{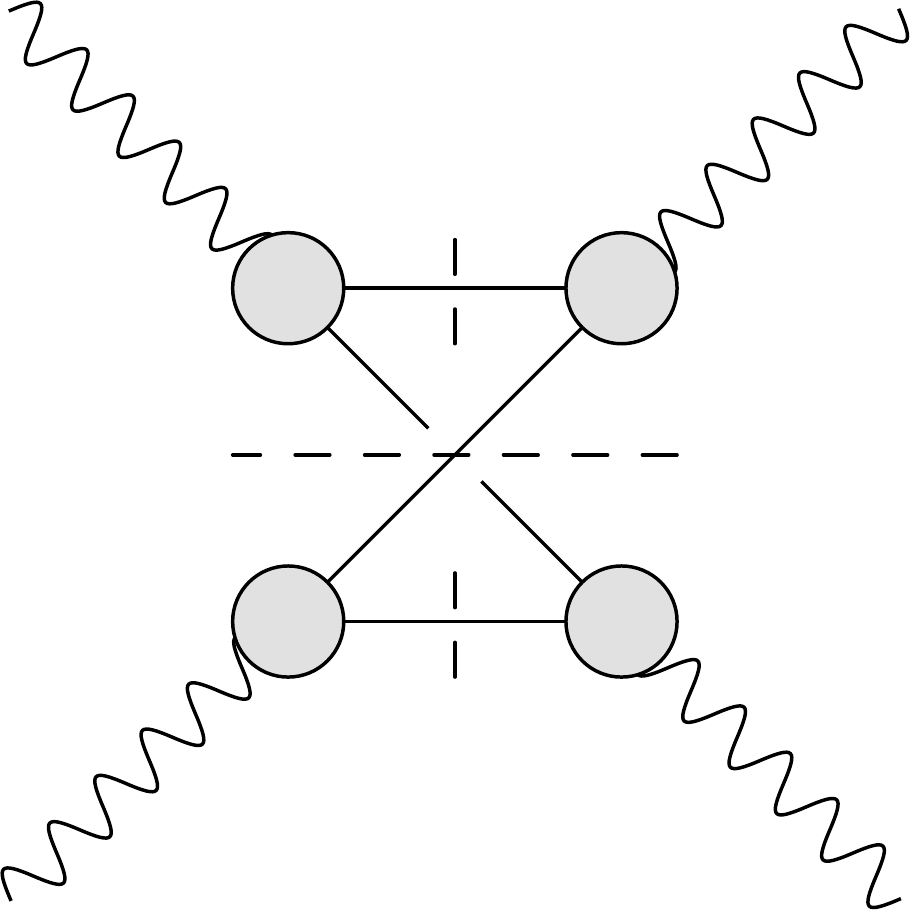}
		\caption{}
		\label{img:HLbLBoxC}
	\end{subfigure}
	\caption{Unitarity diagrams representing the box contributions.}
	\label{img:HLbLBox}
\end{figure}

\begin{figure}[t]
	\centering
	\includegraphics[width=12cm]{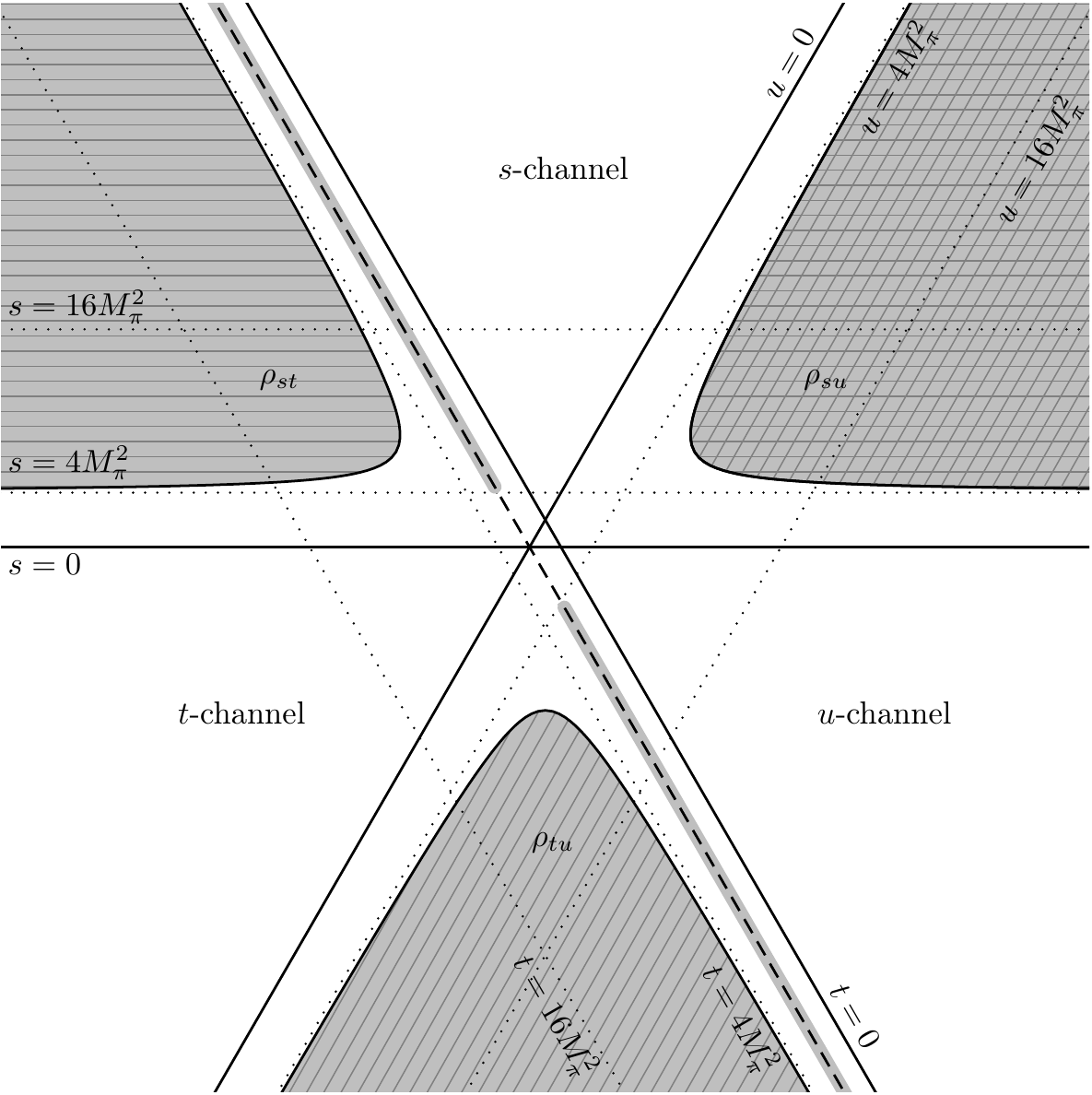}
	\caption{Mandelstam diagram for HLbL scattering for the case $q_i^2=0.5M_\pi^2$ with double-spectral regions for box topologies. The dashed line marks a line of fixed $t$ with its $s$- and $u$-channel cuts highlighted in gray.}
	\label{img:HLbLMandelstamDiagramBox}
\end{figure}

For the contribution of box topologies (see Fig.~\ref{img:HLbLBox}), Mandelstam diagrams prove very useful for the discussion of double-spectral regions. In Fig.~\ref{img:HLbLMandelstamDiagramBox}, such a diagram is shown for the case $q_i^2 = 0.5 M_\pi^2$. A dashed line indicates a line of fixed $t$, used for writing the fixed-$t$ dispersion relation. The two cuts are highlighted in gray. The discontinuities along these cuts can be written again as a dispersive integral over double-spectral functions. The three regions of non-vanishing double-spectral functions are labeled in Fig.~\ref{img:HLbLMandelstamDiagramBox} by $\rho_{st}$, $\rho_{su}$, and $\rho_{tu}$.

The $s$-channel cut receives contributions from the double-spectral regions $\rho_{st}$ and $\rho_{su}$, according to the unitarity diagrams~\ref{img:HLbLBoxA} and \ref{img:HLbLBoxB}, where first the vertical cut, then the horizontal cut is applied.\footnote{In fact, each of the shown diagrams corresponds to two topologies because the pion is charged and its line has a direction.} The $u$-channel cut receives contributions again from $\rho_{su}$ and from $\rho_{tu}$, according to the unitarity diagrams~\ref{img:HLbLBoxB} and \ref{img:HLbLBoxC}. In diagram~\ref{img:HLbLBoxB}, the horizontal cut is now applied first.

Hence, the fixed-$t$ dispersion relation leads to a priori four double-spectral integrals: one for each of the regions $\rho_{st}$ and $\rho_{tu}$ and two for the region $\rho_{su}$. However, it turns out that the sum of the two double-spectral integrals for the region $\rho_{su}$ equals the crossed version of one of the other double-spectral integrals. This is illustrated for the example of a simple scalar box diagram in App.~\ref{sec:AppendixScalar4PointFunction}. Therefore, the box contributions constructed from a fixed-$t$ dispersion relation are already crossing symmetric and identical to the box contributions that are obtained from a fixed-$s$ or fixed-$u$ dispersion relation.

This discussion already anticipates the idea of the uniqueness proof in Sect.~\ref{sec:UniquenessBoxFsQED}: isolating the pion-pole contribution to the LHC in $\pi\pi$ intermediate states, the double-spectral functions become identical to the ones of the box diagrams in the sQED loop calculation.
What needs to be shown is that the Lorentz structures in the BTT set not only produce the right Lorentz structures for the sQED box diagrams, but also that the triangle- and bulb-type sQED diagrams are correctly reproduced. Moreover, this identification has to be proven for the full set of BTT functions, including those which cannot immediately be identified with basis functions due to the occurrence of the Tarrach amplitudes.

\subsubsection{Higher intermediate states}

Topologies where we replace either one or both of the pion poles in the sub-process by a multi-particle cut are the last contribution with a two-pion intermediate state in the direct channel. 
Such contributions could be captured in several ways, either relying on a partial-wave picture or approximating the multi-pion states by an effective resonance description.
While the case of $S$-waves was already discussed in~\cite{Colangelo2014a}, the BTT formulation facilitates the generalization to $D$-waves, since the technical complications related to angular averages and kinematic singularities, the latter requiring the introduction of off-diagonal kernels, are avoided. The application to rescattering effects will be presented in a forthcoming publication.

\subsection{Uniqueness of the box topologies}

\label{sec:UniquenessBoxFsQED}

In this subsection, we will prove that the box topologies are equal to the FsQED contribution (sQED multiplied by pion vector form factors) and that this contribution is unique. The proof is based on the Mandelstam representation and consists of two arguments:
\begin{itemize}
	\item the double-spectral densities of box topologies and FsQED agree,
	\item both representations fulfill an unsubtracted Mandelstam representation.
\end{itemize}
The argument for the first point is the following: in the case of the box topologies, the discontinuities are defined by the unitarity relation as phase-space integrals of products of two pure pole contributions for $\gamma^*\gamma^*\to\pi\pi$. In the case of FsQED, the discontinuities are most easily found by applying Cutkosky's rules~\cite{Cutkosky1960}, i.e.\ by cutting the loop diagrams. We see immediately that the discontinuities in FsQED are phase-space integrals of products of two Born terms for $\gamma^*\gamma^*\to\pi\pi$. In Sect.~\ref{sec:PionPoleggpipi}, we have already shown that the Born contribution is the same as the pure pole, hence the discontinuities (and therefore also the double-spectral densities) of the box topologies and FsQED are equal.

To complete the proof it remains to be shown that the scalar functions in FsQED fulfill a Mandelstam representation (the scalar functions of the box topologies are defined by the Mandelstam representation). Apart from the first six scalar functions $\Pi_1, \ldots, \Pi_6$, which are unaffected by the Tarrach amplitudes, the scalar BTT functions  $\Pi_i$ are defined only up to the ambiguity~\eqref{eq:BTTScalarFunctionsAmbiguities}, so that we cannot compare the representations for these functions immediately. Therefore, we will derive the double-spectral representation of the basis functions $\tilde \Pi_i$ that is implied by an unsubtracted Mandelstam representation for the scalar functions $\Pi_i$ and show that the FsQED basis functions fulfill a double-spectral representation of this form. This will then complete the proof of the uniqueness and the equality of the box topologies and the FsQED contribution.

First, we calculate the fully off-shell sQED loop contribution. It consists of six box diagrams, twelve triangles, and three bulb diagrams:
\begin{align}
	i e^4 \Pi_\mathrm{sQED}^{\mu\nu\lambda\sigma} = 3 \times \hspace{-0.25cm} \minidiagSize{HLbL}{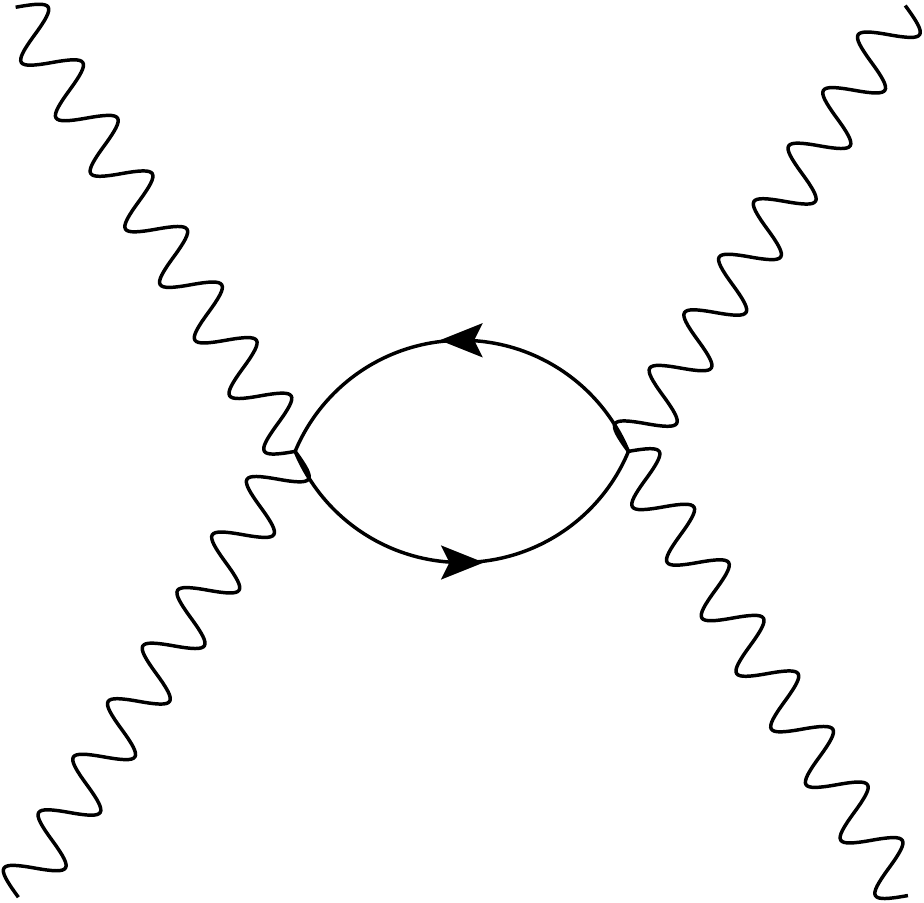}{1.5cm} + 12 \times \hspace{-0.25cm} \minidiagSize{HLbL}{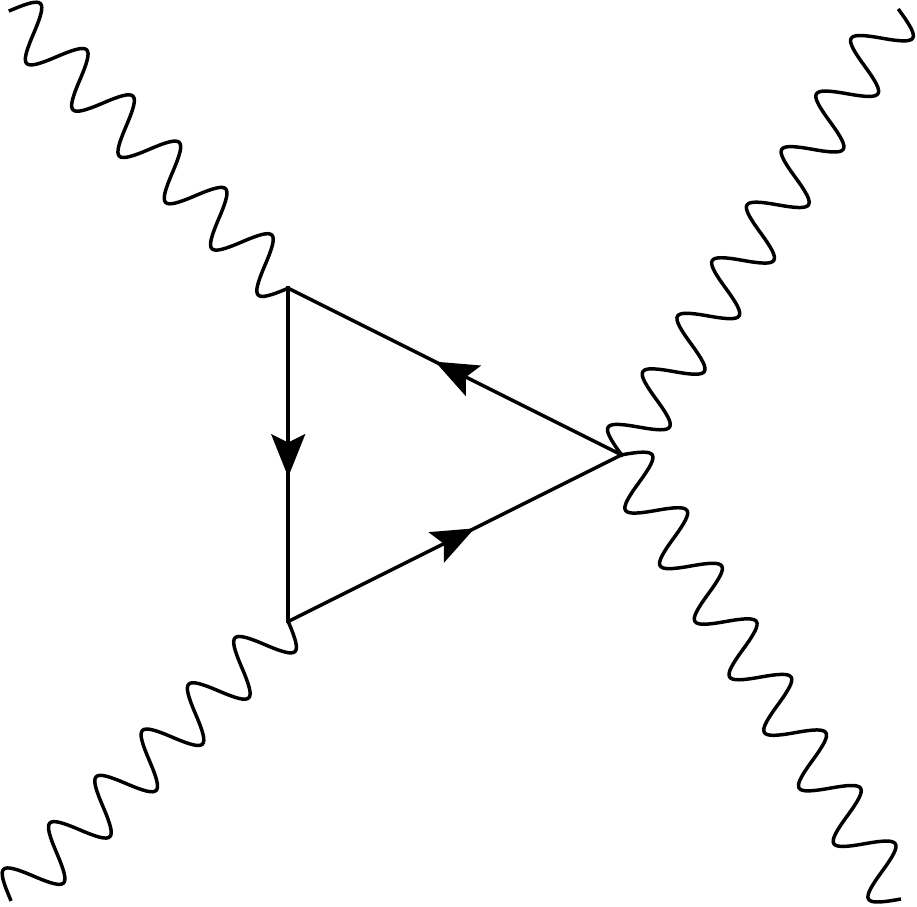}{1.5cm} + 6 \times \hspace{-0.25cm} \minidiagSize{HLbL}{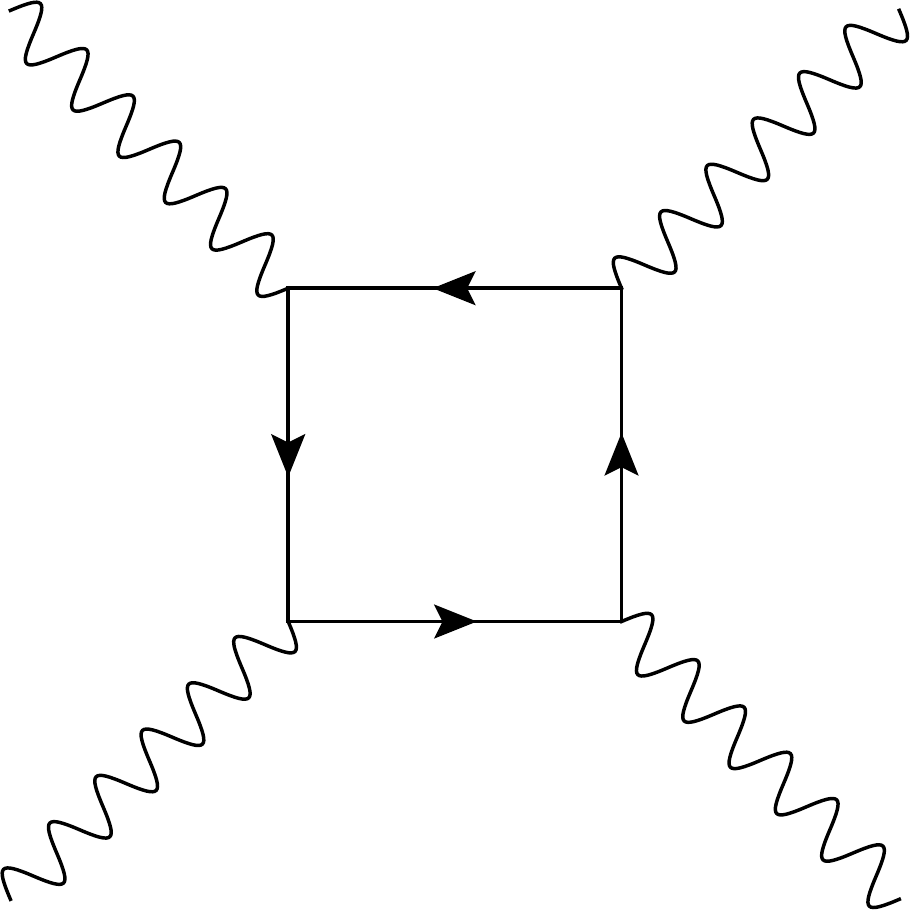}{1.5cm} .
\end{align}
The three bulb diagrams are given by
\begin{align}
	3 \times \hspace{-0.25cm} \minidiagSize{HLbL}{sQED_Bulb}{1.5cm} = 4 i e^4 \Big( g^{\mu\nu} g^{\lambda\sigma} B_0(s,M_\pi^2,M_\pi^2) + g^{\mu\lambda} g^{\nu\sigma} B_0(t,M_\pi^2,M_\pi^2) + g^{\mu\sigma} g^{\nu\lambda} B_0(u,M_\pi^2,M_\pi^2) \Big) ,
\end{align}
where the scalar loop function $B_0$ is defined in~\eqref{eq:DefinitionScalarTwoPointFunction}. One of the twelve triangle diagrams can be represented as
\begin{align*}
		\minidiagSize{HLbL}{sQED_Triangle}{1.5cm} &= 4 i e^4 g^{\lambda \sigma} \begin{aligned}[t]
			& \bigg( \frac{1}{2} q_1^{\mu } q_2^{\nu } C_0(q_1^2,s,q_2^2,M_\pi^2,M_\pi^2,M_\pi^2) - q_1^{\mu } \left(q_1^{\nu}-q_2^{\nu }\right) C_1(q_1^2,s,q_2^2,M_\pi^2,M_\pi^2,M_\pi^2) \\
			& + q_2^{\nu } \left(q_1^{\mu}-q_2^{\mu }\right) C_2(q_1^2,s,q_2^2,M_\pi^2,M_\pi^2,M_\pi^2) \\
			& - 2 g^{\mu\nu} C_{00}(q_1^2,s,q_2^2,M_\pi^2,M_\pi^2,M_\pi^2) + 2 \left(q_1^{\nu } q_2^{\mu }+q_1^{\mu } q_2^{\nu }\right) C_{12}(q_1^2,s,q_2^2,M_\pi^2,M_\pi^2,M_\pi^2) \\
			& - 2 q_1^{\mu } q_1^{\nu } C_{11}(q_1^2,s,q_2^2,M_\pi^2,M_\pi^2,M_\pi^2) - 2 q_2^{\mu } q_2^{\nu } C_{22}(q_1^2,s,q_2^2,M_\pi^2,M_\pi^2,M_\pi^2) \bigg) , \end{aligned} \mytag
\end{align*}
where the scalar loop function $C_0$ is defined in~\eqref{eq:DefinitionScalarThreePointFunction}. For the tensor coefficient functions $C_{i}$, $C_{ij}$, we use the convention of~\cite{Denner1993} up to a normalization factor: $16\pi^2 C_0 = C_0^\text{\cite{Denner1993}}$ etc. The remaining eleven triangles follow by crossing the external momenta.
Similarly, the fully off-shell formula for the box diagrams can also be expressed in terms of the scalar four-point function $D_0$ and the tensor coefficient functions $D_i$, $D_{ij}$, $D_{ijk}$, $D_{ijkl}$, but the result is rather long and need not be reproduced here. 

In this way, we obtain a representation in terms of scalar loop functions $B_0$, $C_0$, $D_0$, and the pertinent tensor coefficient functions. Next, we project this expression onto the HLbL tensor basis~\eqref{eq:HLbLTensor43Basis}, see App.~\ref{sec:AppendixProjection}, in order to identify the sQED contribution to the basis functions $\tilde\Pi_i$. Using FeynCalc~\cite{Mertig1991}, we perform a Passarino--Veltman reduction~\cite{Hooft1979, Passarino1979} of the tensor coefficient functions, so that the basis functions are given by linear combinations of the scalar loop functions:
\begin{align}
	\begin{split}
		\label{eq:sQEDBasisScalarLoopFunctions}
		\tilde\Pi_i^\mathrm{sQED} &= p_i + a_i A_0(M_\pi^2) \\
			&\quad + b_i^1 B_0(q_1^2, M_\pi^2, M_\pi^2) + b_i^2 B_0(q_2^2, M_\pi^2, M_\pi^2) + b_i^3 B_0(q_3^2, M_\pi^2, M_\pi^2) + b_i^4 B_0(q_4^2, M_\pi^2, M_\pi^2) \\
			&\quad + b_i^s B_0(s, M_\pi^2, M_\pi^2) + b_i^t B_0(t, M_\pi^2, M_\pi^2) + b_i^u B_0(u, M_\pi^2, M_\pi^2) \\
			&\quad + c_i^{12} C_0(q_1^2, q_2^2, s, M_\pi^2, M_\pi^2, M_\pi^2) + c_i^{13} C_0(q_1^2, q_3^2, t, M_\pi^2, M_\pi^2, M_\pi^2) + c_i^{14} C_0(q_1^2, q_4^2, u, M_\pi^2, M_\pi^2, M_\pi^2) \\
			&\quad + c_i^{34} C_0(q_3^2, q_4^2, s, M_\pi^2, M_\pi^2, M_\pi^2) + c_i^{24} C_0(q_2^2, q_4^2, t, M_\pi^2, M_\pi^2, M_\pi^2) + c_i^{23} C_0(q_2^2, q_3^2, u, M_\pi^2, M_\pi^2, M_\pi^2) \\
			&\quad + d_i^{st} D_0(q_1^2, q_2^2, q_4^2, q_3^2, s, t, M_\pi^2, M_\pi^2, M_\pi^2, M_\pi^2) \\
			&\quad + d_i^{su} D_0(q_1^2, q_2^2, q_3^2, q_4^2, s, u, M_\pi^2, M_\pi^2, M_\pi^2, M_\pi^2) \\
			&\quad + d_i^{tu} D_0(q_1^2, q_3^2, q_2^2, q_4^2, t, u, M_\pi^2, M_\pi^2, M_\pi^2, M_\pi^2) ,
	\end{split}
\end{align}
where the coefficients $p_i$, $a_i$, $b_i^j$, $c_i^j$, $d_i^j$ are meromorphic functions of the virtualities $q_i^2$ and the Mandelstam variables $s$, $t$, $u$. The Passarino--Veltman reduction results unfortunately in extremely long expressions for the coefficients, which, in addition, contain many kinematic singularities. However, at the kinematic points of these singularities the scalar loop functions fulfill linear relations which cancel most of the singularities. By evaluating the expressions for the functions $\tilde\Pi_i^\mathrm{sQED}$ numerically, we have checked that the only surviving kinematic singularities are exactly the ones required by the projection of the 54 scalar functions $\Pi_i$ onto the basis~\eqref{eq:HLbLBTTProjectedOnBasis}.

As already mentioned, the verification of the Mandelstam representation is simplest for the functions $\Pi_1, \ldots, \Pi_6$, because they are identical to the basis functions $\tilde\Pi_1, \ldots, \tilde\Pi_6$, i.e.\ their projection does not involve kinematic singularities.
We have to show that these functions fulfill a Mandelstam representation of the form (suppressing the $i\epsilon$)
\begin{align}
	\begin{split}
		\label{eq:sQEDScalarFunctionsDoubleSpectralRepresentation}
		\Pi_i^\mathrm{sQED}(s,t,u) &= \frac{1}{\pi^2} \int_{4M_\pi^2}^\infty ds^\prime  \int_{t^+(s^\prime)}^\infty dt^\prime \frac{\rho_{i;st}^\mathrm{sQED}(s^\prime,t^\prime)}{(s^\prime - s)(t^\prime-t)} \\
			& +  \frac{1}{\pi^2} \int_{4M_\pi^2}^\infty ds^\prime  \int_{u^+(s^\prime)}^\infty du^\prime \frac{\rho_{i;su}^\mathrm{sQED}(s^\prime,u^\prime)}{(s^\prime - s)(u^\prime-u)} \\
			& +  \frac{1}{\pi^2} \int_{4M_\pi^2}^\infty dt^\prime  \int_{u^+(t^\prime)}^\infty du^\prime \frac{\rho_{i;tu}^\mathrm{sQED}(t^\prime,u^\prime)}{(t^\prime - t)(u^\prime-u)} ,
	\end{split}
\end{align}
where the borders of the double-spectral regions $t^+,u^+$ are defined in~\eqref{eq:BorderDoubleSpectralRegion}.

As the scalar four-point functions $D_0$ are the only ones in~\eqref{eq:sQEDBasisScalarLoopFunctions} with a double-spectral region, we can immediately identify the double-spectral densities:
\begin{align*}
\label{eq:rho_sQED}
	\left. \begin{alignedat}{1}
		\rho_{i;st}^\mathrm{sQED}(s,t) &= d_i^{st}(s,t) \rho_0(s,t;q_1^2,q_2^2,q_4^2,q_3^2) , \\
		\rho_{i;su}^\mathrm{sQED}(s,u) &= d_i^{su}(s,u) \rho_0(s,u;q_1^2,q_2^2,q_3^2,q_4^2) , \\
		\rho_{i;tu}^\mathrm{sQED}(t,u) &= d_i^{tu}(t,u) \rho_0(t,u;q_1^2,q_3^2,q_2^2,q_4^2)
		\end{alignedat} \quad \right\} \quad i = \{ 1, \ldots, 6 \} , \mytag
\end{align*}
where $\rho_0$ is the double-spectral density of the $D_0$ scalar loop function, defined in~\eqref{eq:DoubleSpectralDensityD0}. By evaluating numerically the double-spectral integrals~\eqref{eq:sQEDScalarFunctionsDoubleSpectralRepresentation} at some random kinematic points below the appearance of anomalous thresholds, we have checked that the Mandelstam representation for the functions $\Pi_1^\mathrm{sQED}, \ldots, \Pi_6^\mathrm{sQED}$ indeed agrees with the expression in terms of scalar loop functions~\eqref{eq:sQEDBasisScalarLoopFunctions}. Even though showing the identity of the two expressions algebraically would have been desirable, already the sheer size of these expressions has prevented an analytic comparison. However, the numerical check does not leave any room open for differences.

This result implies that the double-spectral densities of the sQED box diagrams~\eqref{eq:rho_sQED} have the correct form that, when inserted into the Mandelstam representation, the non-residue pieces can be separated in such a way that all triangle and bulb loop functions in~\eqref{eq:sQEDBasisScalarLoopFunctions} are reproduced with the correct coefficients. The explicit verification of this property is the crucial point in the argument. Since the double-spectral densities of box topologies and FsQED agree, this proves the claim for $i=1,\ldots,6$.

The remaining basis functions exhibit kinematic singularities. In order to work out their double-dispersive representation, we can limit ourselves to the seven remaining representatives of~\eqref{eq:BTBasisFunctionsRepresentatives}. All other basis functions are related by crossing symmetry~\eqref{eq:BTBasisCrossingRelations}.

We write the basis functions of interest as
\begin{align*}
	\label{eq:InterestingBasisFunctionsInTermsOfBTT}
	-4 q_3 \cdot q_4 \tilde\Pi_7 &= 2 (s - s_b) \Pi_7 - (u - u_a) (t - t_b) \Pi_{31} , \\
	-2 q_3 \cdot q_4 \tilde\Pi_9 &= (s - s_b) \Pi_9 + (u - u_b) \Pi_{22} , \\
	-4 q_3 \cdot q_4 \tilde\Pi_{19} &= 2 (s - s_b) \Pi_{19} + (u - u_a) (u - u_b) \Pi_{31} , \\
	-4 q_1\cdot q_2 q_3\cdot q_4 \tilde\Pi_{21} &= (s - s_a) (s - s_b) \Pi_{21} - (u - u_a) (u - u_b) \Pi_{22} , \\
	2 q_1\cdot q_2 \tilde\Pi_{36} &= (s - s_a) \Pi_{43} - (u - u_b) \Pi_{37} , \\
	-2 q_3 \cdot q_4 \tilde\Pi_{39} &= (s - s_b) \Pi_{49} - (s - s_a) \Pi_{54} , \\
	-2 q_3 \cdot q_4 \tilde\Pi_{40} &= (s - s_b) \Pi_{50} - (t - t_b) \Pi_{54} , \mytag
\end{align*}
where $s_a := q_1^2 + q_2^2$, $s_b := q_3^2 + q_4^2$, $u_a := q_2^2 + q_3^2$, $u_b := q_1^2 + q_4^2$, $t_b := q_2^2 + q_4^2$. Next, we introduce subtractions for the scalar functions $\Pi_i$: each function is subtracted once or twice at the subtraction point that appears in the coefficient of the other function $\Pi_i$ in the above linear combinations. For instance, in the equation for $\tilde\Pi_9$, we subtract $\Pi_9$ at $u=u_b$ and $\Pi_{22}$ at $s=s_b$. This leads to representations of the basis functions $\tilde\Pi_i$ where both double-spectral contributions are multiplied by a common prefactor (e.g.\ $(s-s_b)(u-u_b)$ in the case of $\tilde\Pi_9$), while only the subtraction terms, which are single-dispersion integrals, have different prefactors. The results for these double-spectral representations of the basis functions $\tilde\Pi_i$ are given in~\eqref{eq:DoubleSpectralBTBasisFunctionsSingleSubtr} and~\eqref{eq:DoubleSpectralBTBasisFunctionsDoubleSubtr} in App.~\ref{sec:AppendixDoubleSpectralBasisFunctions}.

The important point is that in these double-spectral representations both the discontinuities of the single-dispersion integrals and the combined double-spectral densities $\tilde\rho_i$ are unambiguously defined. We do not attempt to split the double-spectral densities into two contributions from the $\Pi_i$: this splitting involves exactly the ambiguity~\eqref{eq:BTTScalarFunctionsAmbiguities}.
For the sQED contribution, the discontinuities and double-spectral densities can be extracted from the loop representation of the basis functions~\eqref{eq:sQEDBasisScalarLoopFunctions} as follows:
\begin{itemize}
	\item The discontinuities are extracted by taking the appropriate limit of the basis functions $\tilde\Pi_i$. For instance, the discontinuity of the subtraction term of $\Pi_9$ is obtained by considering the limit $\lim\limits_{u\to u_b} (-2 q_3\cdot q_4 \tilde\Pi_9 )$.
	\item The combined double-spectral densities $\tilde\rho_i$ are (up to a kinematic factor) just the double-spectral densities of the basis functions $\tilde\Pi_i$. Therefore, they can be obtained again from the coefficients of the $D_0$ functions in the loop representation, in analogy to the case of $\tilde\Pi_1,\ldots,\tilde\Pi_6$.
\end{itemize}	
After having identified all the discontinuities and double-spectral densities in~\eqref{eq:DoubleSpectralBTBasisFunctionsSingleSubtr} and~\eqref{eq:DoubleSpectralBTBasisFunctionsDoubleSubtr}, we have checked numerically for random kinematic points (below the appearance of anomalous thresholds) that the dispersive representations of the functions $\tilde\Pi_i$ agrees with the loop representation. It turns out that $\tilde\Pi_{39}^\mathrm{sQED} = \tilde\Pi_{40}^\mathrm{sQED} = 0$, hence we can set
\begin{align}
	\Pi_{49}^\mathrm{sQED} = \ldots = \Pi_{54}^\mathrm{sQED} = 0 ,
\end{align}
which also fixes the redundancy~\eqref{eq:136vs138LinearRelations}.

This completes our proof of the uniqueness of the pion-box contribution. Cutkosky's rules tell us that the discontinuities of the FsQED contribution are the same as the ones of the pion-box topologies in the sense of unitarity. The FsQED contribution fulfills the same double-spectral representation as the pure pion-box topologies. Therefore, the two representations are the same. Unitarity and Mandelstam analyticity define the pion-box contribution, i.e.\ $\pi\pi$ intermediate states with a pion-pole LHC, in a unique way.

Finally, we stress that these calculations also provide a strong test of the Lorentz decomposition~\eqref{eqn:HLbLTensorKinematicFreeStructures}. Apart from the function $\Pi_{49}$, which does not receive a contribution from the pion loop, all scalar functions have been shown to be free of kinematics,
so that the sQED pion-loop amplitude behaves exactly as expected from the general BTT formalism.

\subsection[Contribution to $(g-2)_\mu$]{Contribution to $\boldsymbol{(g-2)_\mu}$}

In this subsection, we insert our dispersive representation of the scalar functions into the master formula~\eqref{eq:MasterFormula3Dim} to obtain the contribution to $a_\mu$.

\subsubsection{Pion-pole contribution}

With~\eqref{eq:PionPoleScalarFunctions} and using the master formula~\eqref{eq:MasterFormula3Dim}, we recover the well-known result for the pion-pole contribution to $a_\mu$~\cite{Knecht2002}:
\begin{align}
	\begin{split}
		a_\mu^{\pi^0\text{-pole}} &= \frac{2 \alpha^3}{3 \pi^2} \int_0^\infty dQ_1 \int_0^\infty dQ_2 \int_{-1}^1 d\tau \sqrt{1-\tau^2} Q_1^3 Q_2^3 \\
			&\quad \times \left( T_1(Q_1,Q_2,\tau) \bar \Pi_1^{\pi^0\text{-pole}}(Q_1,Q_2,\tau) + T_2(Q_1,Q_2,\tau) \bar \Pi_2^{\pi^0\text{-pole}}(Q_1,Q_2,\tau) \right) ,
	\end{split}
\end{align}
with
\begin{align}
	\begin{split}
		\bar \Pi_1^{\pi^0\text{-pole}} &= - \frac{\mathcal{F}_{\pi^0\gamma^*\gamma^*}\big({-Q_1^2},-Q_2^2\big) \mathcal{F}_{\pi^0\gamma^*\gamma^*}\big({-Q_3^2},0\big)}{Q_3^2+M_\pi^2} , \\
		\bar \Pi_2^{\pi^0\text{-pole}} &= - \frac{\mathcal{F}_{\pi^0\gamma^*\gamma^*}\big( {-Q_1^2},-Q_3^2\big) \mathcal{F}_{\pi^0\gamma^*\gamma^*}\big({-Q_2^2},0\big)}{Q_2^2+M_\pi^2} ,
	\end{split}
\end{align}
where $Q_3^2 = Q_1^2 + 2 Q_1 Q_2 \tau + Q_2^2$ and the integral kernels $T_i$ are given in App.~\ref{sec:AppendixMasterFormulaKernels}.

\subsubsection{Pion-box contribution}

The single-integral discontinuities and the double-spectral densities in the dispersive representations of the basis functions~\eqref{eq:DoubleSpectralBTBasisFunctionsSingleSubtr} and \eqref{eq:DoubleSpectralBTBasisFunctionsDoubleSubtr} are quantities that can be extracted directly from the projected basis functions $\tilde\Pi_i$. To the contrary, the separation of the double-spectral densities $\tilde\rho_i$ into the two contributions from the different scalar functions $\Pi_i$ is not unambiguously possible, which reflects just the redundancy~\eqref{eq:BTTScalarFunctionsAmbiguities}. However, such a separation is not necessary: for the calculation of $a_\mu$, we need the scalar functions $\Pi_i$ only in the limit $k\to0$. In this limit, all the scalar functions $\Pi_i$ appearing in the master formula~\eqref{eq:ScalarFunctionsForMasterFormula} can be expressed in terms of single-dispersion integrals, where the discontinuities are directly related to the basis functions $\tilde\Pi_i$. All the subtracted double-spectral integrals, which are not unambiguously defined, drop out in the limit $k\to0$.

We stress that the ambiguities in the definition of the BTT functions $\Pi_i$ are related to the fact that these functions are not observables. In the calculation of any physical quantity, such as helicity amplitudes or the HLbL contribution to $a_\mu$, any ambiguity due to the redundancy in the BTT set has to drop out.

The pion-box contribution to $a_\mu$ is therefore given by
\begin{align*}
		a_\mu^{\pi\text{-box}}=a_\mu^\text{FsQED} &= \frac{2 \alpha^3}{3 \pi^2} \int_0^\infty dQ_1 \int_0^\infty dQ_2 \int_{-1}^1 d\tau \sqrt{1-\tau^2} Q_1^3 Q_2^3 \begin{aligned}[t]
			& F_\pi^V(-Q_1^2) F_\pi^V(-Q_2^2) F_\pi^V(-Q_3^2) \\
			& \times \sum_{i=1}^{12} T_i(Q_1,Q_2,\tau) \bar \Pi_i^\mathrm{sQED}(Q_1,Q_2,\tau) , \end{aligned} \mytag
\end{align*}
where the functions $\bar\Pi_i$ are defined in~\eqref{eq:ScalarFunctionsForMasterFormula}. They are linear combinations of the scalar functions $\Pi_i$ in the limit $k\to0$. The required functions $\Pi_i$ can be obtained in this limit from the basis functions $\tilde\Pi_i$, see~\eqref{eq:RelationBTTvsBasisLimitZeroK} in App.~\ref{sec:AppendixDRScalarFunctionsForMasterFormula}. Their explicit dispersive representation is given in~\eqref{eq:DispersiveRepresentationBTTforMasterFormula}. In fact, this representation is applicable not only to the pion-box contribution, but whenever a dispersive representation for the $\Pi_i$ can be constructed. In the special case of the pion box, one has in addition 
$\Pi_{50}^\mathrm{sQED} = \Pi_{51}^\mathrm{sQED} = \Pi_{54}^\mathrm{sQED} = 0$, hence $\bar\Pi_{12}^\mathrm{sQED} = 0$.


\section{Conclusion and outlook}
\label{sec:HLbLDiscussionConclusion}

We have derived a decomposition of the HLbL tensor into scalar
functions~\eqref{eqn:HLbLTensorKinematicFreeStructures}, following the
general recipe by Bardeen, Tung, and Tarrach, which allowed us to derive a
master formula~\eqref{eq:MasterFormula3Dim} for the HLbL contribution to
the anomalous magnetic moment of the muon $(g-2)_\mu$ in terms of scalar
functions that by construction are free of kinematic singularities and
zeros.  In particular, these scalar functions fulfill a Mandelstam
double-spectral representation. In the present work, we have considered
only the lowest-lying intermediate states, pion-pole and pion-box
topologies. These two contributions can be studied without any
approximation. The central result is that the pion pole corresponds exactly
to the contribution of one-pion intermediate states if interpreted with an
on-shell pion transition form factor, and the pion-box, i.e.\ $\pi\pi$
intermediate states with a pion-pole LHC, to the sQED pion loop dressed
with pion vector form factors.  These results confirm both the definition
of the pion pole and prove explicitly the validity of the separation of
Born and non-Born $\pi\pi$ contributions given in~\cite{Colangelo2014a}.

Moreover, while the formulation in terms of helicity amplitudes adopted
in~\cite{Colangelo2014a} required the introduction of off-diagonal kernel
functions to account for kinematic singularities, the BTT version avoids
these complications and thus facilitates the generalization to partial
waves beyond $S$-waves. In order to exploit unitarity in a partial-wave
picture, the partial-wave unitarity relations analyzed in the helicity
basis need to be transformed to the BTT basis, the formalism for which we
provided in the present paper.

This treatment is based on fundamental principles of particle physics:
gauge invariance and crossing symmetry are already implemented in the
decomposition of the HLbL tensor into scalar functions. The dispersive
description uses analyticity and unitarity to establish a relation between
the HLbL contribution and different on-shell quantities. These on-shell
quantities are in principle either experimentally accessible or can be
reconstructed from data with dispersive methods. For the two cases covered
in this article, the corresponding input is parametrized by the pion
transition form factor $\mathcal{F}_{\pi^0\gamma^*\gamma^*}$ in case of the
pion pole, and by the pion vector form factor $F_\pi^V$ for the pion box,
both form factors being needed for negative virtualities of the off-shell
photons. For general $\pi\pi$ intermediate states also information on the
partial waves for $\gamma^*\gamma^*\to\pi\pi$ is required.  An overview
over which processes can help in the dispersive reconstruction of the pion
transition form
factor~\cite{Niecknig:2012sj,Schneider:2012ez,Hoferichter:2012pm,Hoferichter2014}
or
$\gamma^*\gamma^*\to\pi\pi$~\cite{Garcia-Martin2010,Hoferichter2011,Moussallam2013,Hoferichter:2013ama,Colangelo2014a}
is provided in~\cite{Colangelo2014b}.

An extension of the presented dispersive treatment is possible within
certain limits. It is straightforward to include higher pseudoscalar poles,
i.e.\ the $\eta$ and $\eta^\prime$ mesons by just adding their contribution
in complete analogy to the $\pi^0$ pole. The input quantities will be the
transition form factors of these heavier pseudoscalars,
see~\cite{Stollenwerk:2011zz,Hanhart:2013vba,Kubis:2015sga} for work
towards their dispersive calculation. Although these mesons are unstable in
QCD, their decay width is certainly small enough to justify the treatment
as a pure pole. A bit more difficult is the inclusion of higher
two-particle intermediate states: an extension to $K\bar K$ intermediate
states is still straightforward, but the experimental input will be less
accurate. In general, the approach applies to the low and intermediate
energies $\lesssim 1.5\,\text{GeV}$, where a few channels dominate. All
model calculations done so far indeed support the assumption that the
lowest-lying singularities govern the HLbL tensor.

A careful numerical analysis of the formalism presented in this paper is in
progress. It will reveal the relative importance of the different
contributions and should allow us to identify which input quantities will
have the largest impact concerning the reduction of the hadronic
uncertainty in $a_\mu^\mathrm{HLbL}$. In this way, we are confident that
the presented treatment of HLbL scattering shows a path towards a
data-driven and thus less model-dependent evaluation of $(g-2)_\mu$.

\section*{Acknowledgements}
\addcontentsline{toc}{section}{Acknowledgements}

We thank J.~Gasser, B.~Kubis, H.~Leutwyler, and S.~Scherer for interesting and useful discussions and B.~Kubis for comments on the manuscript.
Financial support by the Swiss National Science Foundation, BMBF ARCHES, the Helmholtz Alliance HA216/EMMI, the DFG (CRC 16, ``Subnuclear Structure of Matter''), and the DOE (Grant No.\ DE-FG02-00ER41132) is gratefully acknowledged.

\begin{appendices}


\section{Partial-wave representation for $\boldsymbol{\gamma^*\gamma^*\to\pi\pi}$}

\label{sec:AppendixPartialWavesggpipi}

Here, we give the expressions for the scalar basis functions of the sub-process $\gamma^*\gamma^*\to\pi\pi$, defined in~\eqref{eq:ggpipiFinalBasisFunctions}, in terms of the helicity partial waves including $D$-waves:
\begin{align*}
	\label{eq:DWavesSubProcessScalarFunctions}
	A_1 &= \frac{2}{\lambda_{12}(s)} \begin{aligned}[t]
		& \Bigg\{ 2 q_1^2 q_2^2 \left( h_{0,5}(s) + \frac{5}{2} (3 z^2-1) h_{2,5}(s) \right) - (s-q_1^2-q_2^2) \left( h_{0,1}(s) + \frac{5}{2} (3 z^2-1) h_{2,1}(s) \right) \\
		& + \frac{5 \sqrt{6} \left( \left((q_1^2 - q_2^2)^2-s (q_1^2+q_2^2)\right) z^2 - s (s - q_1^2 - q_2^2)\right)}{4 s} h_{2,2}(s) \\
		& - \frac{5 \sqrt{3}}{2 \sqrt{s}} \left(s (q_1^2 + q_2^2) - (q_1^2 - q_2^2)^2\right) z^2 h_{2,3}(s) 
		 + \frac{5 \sqrt{3}}{2 \sqrt{s}} (q_1^2-q_2^2) (s-q_1^2-q_2^2) z^2 h_{2,4}(s)  \Bigg\} , \end{aligned} \\
	A_2 &= \frac{2}{\lambda_{12}(s)} \begin{aligned}[t]
		& \Bigg\{ 2 \left( h_{0,1}(s) + \frac{5}{2} (3 z^2-1) h_{2,1}(s) \right) - (s-q_1^2-q_2^2) \left( h_{0,5}(s) + \frac{5}{2} (3 z^2-1) h_{2,5}(s) \right) \\
		& + 5 \sqrt{\frac{3}{2}} (1+z^2) h_{2,2}(s) + 5 \sqrt{3s} z^2 h_{2,3}(s) 
		 - \frac{5 \sqrt{3} \left(2(q_1^2-q_2^2)^2 - \lambda_{12}(s)\right) z^2}{2 \sqrt{s}} \frac{h_{2,4}(s)}{q_1^2 - q_2^2} \Bigg\} , \end{aligned} \\
	A_3 &= - \frac{5 \sqrt{6}}{(s-4M_\pi^2)\lambda_{12}(s)} \bigg( h_{2,2}(s) + \sqrt{\frac{s}{2}} h_{2,3}(s) - \sqrt{\frac{s}{2}} (q_1^2 + q_2^2) \frac{h_{2,4}(s)}{q_1^2-q_2^2} \bigg) , \\
	A_4 &= - \frac{5 \sqrt{6}}{(s-4M_\pi^2)\lambda_{12}(s)} \bigg( (s - q_1^2 - q_2^2) h_{2,2}(s) + 2\sqrt{2s} q_1^2 q_2^2 \frac{h_{2,4}(s)}{q_1^2-q_2^2} \bigg) , \\
	A_5 &= \frac{10 \sqrt{6}}{(s-4M_\pi^2)\lambda_{12}(s)} \bigg( h_{2,2}(s) + \sqrt{\frac{s}{2}}(s-q_1^2-q_2^2) \frac{h_{2,4}(s)}{q_1^2 - q_2^2} \bigg) . \mytag
\end{align*}
We note that under $q_1^2 \leftrightarrow q_2^2$, $h_{2,4}$ changes sign, while all the other $S$- and $D$-waves are invariant. Therefore, the apparent singularities in $q_1^2-q_2^2$ in~\eqref{eq:DWavesSubProcessScalarFunctions} are canceled by the corresponding kinematic zero in $h_{2,4}$.


\section{Tensor basis and crossing relations}

\subsection{Crossing relations between Lorentz structures}

\label{sec:AppendixCrossingLorentzStructures}

The redundant set of generating Lorentz structures of the HLbL tensor consists of 54 elements. While seven structures are defined in~\eqref{eq:HLbLBTTStructures}, all the remaining structures are given by crossed versions thereof:
\begin{align*}
		\label{eq:BTTCrossingRelations}
		T_2^{\mu\nu\lambda\sigma} &= \mathcal{C}_{14}[ T_1^{\mu\nu\lambda\sigma} ] , \quad & T_3^{\mu\nu\lambda\sigma} &= \mathcal{C}_{13}[ T_1^{\mu\nu\lambda\sigma} ] , \\
		T_5^{\mu\nu\lambda\sigma} &= \mathcal{C}_{14}[ T_4^{\mu\nu\lambda\sigma} ] , \quad & T_6^{\mu\nu\lambda\sigma} &= \mathcal{C}_{13}[ T_4^{\mu\nu\lambda\sigma} ] , \\
		T_8^{\mu\nu\lambda\sigma} &= \mathcal{C}_{12}[ T_7^{\mu\nu\lambda\sigma} ] , \quad & T_9^{\mu\nu\lambda\sigma} &= \mathcal{C}_{13}[ \mathcal{C}_{23}[ T_7^{\mu\nu\lambda\sigma} ] ] , \quad & T_{10}^{\mu\nu\lambda\sigma} &= \mathcal{C}_{23}[ T_7^{\mu\nu\lambda\sigma} ] , \\
		T_{11}^{\mu\nu\lambda\sigma} &= \mathcal{C}_{24}[ T_7^{\mu\nu\lambda\sigma} ] , \quad & T_{12}^{\mu\nu\lambda\sigma} &= \mathcal{C}_{14}[ \mathcal{C}_{24}[ T_7^{\mu\nu\lambda\sigma} ] ]  , \quad & T_{13}^{\mu\nu\lambda\sigma} &= \mathcal{C}_{13}[ T_7^{\mu\nu\lambda\sigma} ] , \\
		T_{14}^{\mu\nu\lambda\sigma} &= \mathcal{C}_{23}[  \mathcal{C}_{13}[ T_7^{\mu\nu\lambda\sigma} ] ] , \quad & T_{15}^{\mu\nu\lambda\sigma} &= \mathcal{C}_{14}[ T_7^{\mu\nu\lambda\sigma} ] , \quad & T_{16}^{\mu\nu\lambda\sigma} &= \mathcal{C}_{24}[ \mathcal{C}_{14}[ T_7^{\mu\nu\lambda\sigma} ] ] , \\
		T_{17}^{\mu\nu\lambda\sigma} &= \mathcal{C}_{24}[  \mathcal{C}_{13}[ T_7^{\mu\nu\lambda\sigma} ] ] , \quad & T_{18}^{\mu\nu\lambda\sigma} &= \mathcal{C}_{23}[  \mathcal{C}_{14}[ T_7^{\mu\nu\lambda\sigma} ] ] , \\
		T_{20}^{\mu\nu\lambda\sigma} &= \mathcal{C}_{34}[ T_{19}^{\mu\nu\lambda\sigma} ] , \quad & T_{21}^{\mu\nu\lambda\sigma} &= \mathcal{C}_{23}[ T_{19}^{\mu\nu\lambda\sigma} ] , \quad & T_{22}^{\mu\nu\lambda\sigma} &= \mathcal{C}_{24}[ \mathcal{C}_{23}[ T_{19}^{\mu\nu\lambda\sigma} ] ] , \\
		T_{23}^{\mu\nu\lambda\sigma} &= \mathcal{C}_{23}[ \mathcal{C}_{24}[ T_{19}^{\mu\nu\lambda\sigma} ] ] , \quad & T_{24}^{\mu\nu\lambda\sigma} &= \mathcal{C}_{24}[ T_{19}^{\mu\nu\lambda\sigma} ] , \quad & T_{25}^{\mu\nu\lambda\sigma} &= \mathcal{C}_{23}[ \mathcal{C}_{13}[ T_{19}^{\mu\nu\lambda\sigma} ] ] , \\
		T_{26}^{\mu\nu\lambda\sigma} &= \mathcal{C}_{13}[ T_{19}^{\mu\nu\lambda\sigma} ] , \quad & T_{27}^{\mu\nu\lambda\sigma} &= \mathcal{C}_{14}[ T_{19}^{\mu\nu\lambda\sigma} ] , \quad & T_{28}^{\mu\nu\lambda\sigma} &= \mathcal{C}_{24}[ \mathcal{C}_{14}[ T_{19}^{\mu\nu\lambda\sigma} ] ] , \\
		T_{29}^{\mu\nu\lambda\sigma} &= \mathcal{C}_{24}[ \mathcal{C}_{13}[ T_{19}^{\mu\nu\lambda\sigma} ] ] , \quad & T_{30}^{\mu\nu\lambda\sigma} &= \mathcal{C}_{34}[ \mathcal{C}_{24}[ \mathcal{C}_{13}[ T_{19}^{\mu\nu\lambda\sigma} ]]] , \\
		T_{32}^{\mu\nu\lambda\sigma} &= \mathcal{C}_{24}[ \mathcal{C}_{13}[ T_{31}^{\mu\nu\lambda\sigma} ] ] , \quad & T_{33}^{\mu\nu\lambda\sigma} &= \mathcal{C}_{23}[ T_{31}^{\mu\nu\lambda\sigma} ] , \quad & T_{34}^{\mu\nu\lambda\sigma} &= \mathcal{C}_{13}[ T_{31}^{\mu\nu\lambda\sigma} ] , \\
		T_{35}^{\mu\nu\lambda\sigma} &= \mathcal{C}_{24}[ T_{31}^{\mu\nu\lambda\sigma} ] , \quad & T_{36}^{\mu\nu\lambda\sigma} &= \mathcal{C}_{14}[ T_{31}^{\mu\nu\lambda\sigma} ] , \\
		T_{38}^{\mu\nu\lambda\sigma} &= \mathcal{C}_{34}[ \mathcal{C}_{14}[ T_{37}^{\mu\nu\lambda\sigma} ] ] , \quad & T_{39}^{\mu\nu\lambda\sigma} &= \mathcal{C}_{14}[ T_{37}^{\mu\nu\lambda\sigma} ] , \quad & T_{40}^{\mu\nu\lambda\sigma} &= \mathcal{C}_{12}[ \mathcal{C}_{14}[ T_{37}^{\mu\nu\lambda\sigma} ] ] , \\
		T_{41}^{\mu\nu\lambda\sigma} &= \mathcal{C}_{23}[ \mathcal{C}_{12}[ T_{37}^{\mu\nu\lambda\sigma} ] ] , \quad & T_{42}^{\mu\nu\lambda\sigma} &= \mathcal{C}_{12}[ \mathcal{C}_{24}[ T_{37}^{\mu\nu\lambda\sigma} ] ] , \quad & T_{43}^{\mu\nu\lambda\sigma} &= \mathcal{C}_{24}[ T_{37}^{\mu\nu\lambda\sigma} ] , \\
		T_{44}^{\mu\nu\lambda\sigma} &= \mathcal{C}_{12}[ \mathcal{C}_{23}[ T_{37}^{\mu\nu\lambda\sigma} ] ] , \quad & T_{45}^{\mu\nu\lambda\sigma} &= \mathcal{C}_{23}[ T_{37}^{\mu\nu\lambda\sigma} ] , \quad & T_{46}^{\mu\nu\lambda\sigma} &= \mathcal{C}_{14}[ \mathcal{C}_{23}[ T_{37}^{\mu\nu\lambda\sigma} ] ] , \\
		T_{47}^{\mu\nu\lambda\sigma} &= \mathcal{C}_{24}[ \mathcal{C}_{13}[ T_{37}^{\mu\nu\lambda\sigma} ] ] , \quad & T_{48}^{\mu\nu\lambda\sigma} &= \mathcal{C}_{12}[ T_{37}^{\mu\nu\lambda\sigma} ] , \\
		T_{50}^{\mu\nu\lambda\sigma} &= \mathcal{C}_{12}[ \mathcal{C}_{24}[ T_{49}^{\mu\nu\lambda\sigma} ] ] , \quad & T_{51}^{\mu\nu\lambda\sigma} &= \mathcal{C}_{24}[ T_{49}^{\mu\nu\lambda\sigma} ] , \quad & T_{52}^{\mu\nu\lambda\sigma} &= \mathcal{C}_{13}[ T_{49}^{\mu\nu\lambda\sigma} ] , \\
		T_{53}^{\mu\nu\lambda\sigma} &= \mathcal{C}_{12}[ \mathcal{C}_{13}[ T_{49}^{\mu\nu\lambda\sigma} ] ] , \quad & T_{54}^{\mu\nu\lambda\sigma} &= \mathcal{C}_{23}[ \mathcal{C}_{14}[ T_{49}^{\mu\nu\lambda\sigma} ] ] .
	\mytag
\end{align*}

\subsection{Basis coefficient functions}

\label{sec:AppendixBTTProjectionAndBasisCrossingRelations}

The relation between the 43 basis coefficient functions $\tilde \Pi_i$ and the 54 scalar functions of the redundant set can be obtained by projecting~\eqref{eqn:HLbLTensorKinematicFreeStructures} on the basis. The result defines the exact form of the kinematic singularities in $q_1\cdot q_2$ and $q_3 \cdot q_4$ in the functions $\tilde \Pi_i$:
\begin{align*}
		\label{eq:HLbLBTTProjectedOnBasis}
		\tilde\Pi_1 &= \Pi_1 , & \quad
		\tilde\Pi_2 &= \Pi_2 , & \quad
		\tilde\Pi_3 &= \Pi_3 , \\
		\tilde\Pi_4 &= \Pi_4 , & \quad
		\tilde\Pi_5 &= \Pi_5 , & \quad
		\tilde\Pi_6 &= \Pi_6 , \\
		\tilde\Pi_7 &= \Pi_7 - \frac{ q_2\cdot q_3 q_2\cdot q_4}{q_3\cdot q_4} \Pi_{31} , & \quad
		\tilde\Pi_8 &= \Pi_8-\frac{q_1\cdot q_3 q_1\cdot q_4}{q_3\cdot q_4} \Pi_{31} , \\
		\tilde\Pi_9 &= \Pi_9+\frac{q_1\cdot q_4}{q_3\cdot q_4} \Pi_{22} , & \quad
		\tilde\Pi_{10} &= \Pi_{10}+\frac{q_2\cdot q_3}{q_1\cdot q_2} \Pi_{22} , & \\
		\tilde\Pi_{11} &= \Pi_{11}-\frac{q_2\cdot q_4}{q_1\cdot q_2} \Pi_{24} , & \quad
		\tilde\Pi_{12} &= \Pi_{12}-\frac{q_1\cdot q_3}{q_3\cdot q_4} \Pi_{24} , \\
		\tilde\Pi_{13} &= \Pi_{13}+\frac{q_2\cdot q_4}{q_3\cdot q_4} \Pi_{26} , & \quad
		\tilde\Pi_{14} &= \Pi_{14}+\frac{q_1\cdot q_3}{q_1\cdot q_2} \Pi_{26} , \\
		\tilde\Pi_{15} &= \Pi_{15}-\frac{q_2\cdot q_3}{q_3\cdot q_4} \Pi_{28} , & \quad
		\tilde\Pi_{16} &= \Pi_{16}-\frac{q_1\cdot q_4}{q_1\cdot q_2} \Pi_{28} , \\
		\tilde\Pi_{17} &= \Pi_{17}-\frac{q_1\cdot q_4 q_2\cdot q_4}{q_1\cdot q_2} \Pi_{32} , & \quad
		\tilde\Pi_{18} &= \Pi_{18}-\frac{q_1\cdot q_3 q_2\cdot q_3}{q_1\cdot q_2} \Pi_{32} , \\
		\tilde\Pi_{19} &= \Pi_{19}+\frac{q_1\cdot q_4 q_2\cdot q_3}{q_3\cdot q_4} \Pi_{31} , & \quad
		\tilde\Pi_{20} &= \Pi_{20}+\frac{q_1\cdot q_3 q_2\cdot q_4}{q_3\cdot q_4} \Pi_{31} , \\
		\tilde\Pi_{21} &= \Pi_{21}-\frac{q_1\cdot q_4 q_2\cdot q_3}{q_1\cdot q_2 q_3\cdot q_4} \Pi_{22} , & \quad
		\tilde\Pi_{22} &= \Pi_{23}-\frac{q_1\cdot q_3 q_2\cdot q_4}{q_1\cdot q_2 q_3\cdot q_4} \Pi_{24} , \\
		\tilde\Pi_{23} &= \Pi_{25}-\frac{q_1\cdot q_3 q_2\cdot q_4}{q_1\cdot q_2 q_3\cdot q_4} \Pi_{26} , & \quad
		\tilde\Pi_{24} &= \Pi_{27}-\frac{q_1\cdot q_4 q_2\cdot q_3}{q_1\cdot q_2 q_3\cdot q_4} \Pi_{28} , \\
		\tilde\Pi_{25} &= \Pi_{29}-\frac{q_1\cdot q_4 q_2\cdot q_3}{q_1\cdot q_2} \Pi_{32} , & \quad
		\tilde\Pi_{26} &= \Pi_{30}-\frac{q_1\cdot q_3 q_2\cdot q_4}{q_1\cdot q_2} \Pi_{32} , \\
		\tilde\Pi_{27} &= \Pi_{33} + \frac{q_2\cdot q_4}{q_1\cdot q_2 q_3\cdot q_4} \Pi_{22} , & \quad
		\tilde\Pi_{28} &= \Pi_{34} + \frac{q_1\cdot q_4}{q_1\cdot q_2 q_3\cdot q_4} \Pi_{26} , \\
		\tilde\Pi_{29} &= \Pi_{35}-\frac{q_2\cdot q_3}{q_1\cdot q_2 q_3\cdot q_4} \Pi_{24} , & \quad
		\tilde\Pi_{30} &= \Pi_{36}-\frac{q_1\cdot q_3}{q_1\cdot q_2 q_3\cdot q_4} \Pi_{28} , \\
		\tilde\Pi_{31} &= \Pi_{38}+\frac{q_2\cdot q_3}{q_3\cdot q_4} \Pi_{47} , & \quad
		\tilde\Pi_{32} &= \Pi_{39}-\frac{q_2\cdot q_4}{q_3\cdot q_4} \Pi_{46} , & \quad
		\tilde\Pi_{33} &= \Pi_{40}-\frac{q_1\cdot q_4}{q_3\cdot q_4} \Pi_{46} , \\
		\tilde\Pi_{34} &= \Pi_{41}+\frac{q_1\cdot q_3}{q_3\cdot q_4} \Pi_{47} , & \quad
		\tilde\Pi_{35} &= \Pi_{42}+\frac{q_2\cdot q_4}{q_1\cdot q_2} \Pi_{48} , & \quad
		\tilde\Pi_{36} &= \Pi_{43} + \frac{q_1\cdot q_4}{q_1\cdot q_2} \Pi_{37} , \\
		\tilde\Pi_{37} &= \Pi_{44}-\frac{q_2\cdot q_3}{q_1\cdot q_2} \Pi_{48} , & \quad
		\tilde\Pi_{38} &= \Pi_{45}-\frac{q_1\cdot q_3}{q_1\cdot q_2} \Pi_{37} , \\
		\tilde\Pi_{39} &= \Pi_{49}+\frac{q_1\cdot q_2}{q_3\cdot q_4} \Pi_{54} , & \quad
		\tilde\Pi_{40} &= \Pi_{50}-\frac{q_2\cdot q_4}{q_3\cdot q_4} \Pi_{54} , & \quad
		\tilde\Pi_{41} &= \Pi_{51}+\frac{q_1\cdot q_4}{q_3\cdot q_4} \Pi_{54} , \\
		\tilde\Pi_{42} &= \Pi_{52}-\frac{q_2\cdot q_3}{q_3\cdot q_4} \Pi_{54} , & \quad
		\tilde\Pi_{43} &= \Pi_{53}+\frac{q_1\cdot q_3}{q_3\cdot q_4} \Pi_{54} .
	\mytag
\end{align*}
Crossing symmetry for the functions $\Pi_i$ implies the following relations between the basis functions $\tilde\Pi_i$:
\begin{align*}
		\label{eq:BTBasisCrossingRelations}
		\tilde\Pi_2 &= \mathcal{C}_{14}[ \tilde\Pi_1 ] , \quad &
		\tilde\Pi_3 &= \mathcal{C}_{13}[ \tilde\Pi_1 ] , \\
		\tilde\Pi_5 &= \mathcal{C}_{14}[ \tilde\Pi_4 ] , \quad &
		\tilde\Pi_6 &= \mathcal{C}_{13}[ \tilde\Pi_4 ] , \\
		\tilde\Pi_8 &= \mathcal{C}_{12}[ \tilde\Pi_7 ] , \quad &
		\tilde\Pi_{17} &= \mathcal{C}_{24}[\mathcal{C}_{13}[ \tilde\Pi_7 ]] , \quad &
		\tilde\Pi_{18} &= \mathcal{C}_{23}[\mathcal{C}_{14}[ \tilde\Pi_7 ]] , \\
		\tilde\Pi_{10} &= \mathcal{C}_{24}[ \mathcal{C}_{13}[ \tilde\Pi_9 ]] , \quad &
		\tilde\Pi_{11} &= \mathcal{C}_{34}[ \mathcal{C}_{24}[ \mathcal{C}_{13}[ \tilde\Pi_9 ]]] , \quad &
		\tilde\Pi_{12} &= \mathcal{C}_{34}[ \tilde\Pi_9 ] , \\
		\tilde\Pi_{13} &= \mathcal{C}_{12}[ \tilde\Pi_9 ] , \quad &
		\tilde\Pi_{14} &= \mathcal{C}_{12}[ \mathcal{C}_{24}[ \mathcal{C}_{13}[ \tilde\Pi_9 ]]] , \quad &
		\tilde\Pi_{15} &= \mathcal{C}_{34}[ \mathcal{C}_{12}[ \tilde\Pi_9 ]] , \\
		\tilde\Pi_{16} &= \mathcal{C}_{23}[ \mathcal{C}_{14}[ \tilde\Pi_9 ]] , \\
		\tilde\Pi_{22} &= \mathcal{C}_{34}[ \tilde\Pi_{21} ] , \quad &
		\tilde\Pi_{23} &= \mathcal{C}_{12}[ \tilde\Pi_{21} ] , \quad &
		\tilde\Pi_{24} &= \mathcal{C}_{34}[ \mathcal{C}_{12}[ \tilde\Pi_{21} ]] , \\
		\tilde\Pi_{20} &= \mathcal{C}_{34}[ \tilde\Pi_{19} ] , \quad &
		\tilde\Pi_{25} &= \mathcal{C}_{23}[ \mathcal{C}_{14}[ \tilde\Pi_{19} ]] , \quad &
		\tilde\Pi_{26} &= \mathcal{C}_{34}[ \mathcal{C}_{23}[ \mathcal{C}_{14}[ \tilde\Pi_{19} ]]] , \\
		\tilde\Pi_{27} &= \frac{q_2\cdot q_4}{q_1 \cdot q_2 q_3\cdot q_4} \mathcal{C}_{23}[ \mathcal{C}_{34}[ \tilde\Pi_{19} ]] , \quad &
		\tilde\Pi_{28} &= \frac{q_1\cdot q_4}{q_1 \cdot q_2 q_3\cdot q_4} \mathcal{C}_{13}[ \tilde\Pi_{19} ] , \quad &
		\tilde\Pi_{29} &= - \frac{q_2\cdot q_3}{q_1 \cdot q_2 q_3\cdot q_4} \mathcal{C}_{24}[ \tilde\Pi_{19} ] , \\
		\tilde\Pi_{30} &= - \frac{q_1\cdot q_3}{q_1 \cdot q_2 q_3\cdot q_4} \mathcal{C}_{24}[\mathcal{C}_{14}[ \tilde\Pi_{19} ]] , \\
		\tilde\Pi_{31} &= \mathcal{C}_{24}[ \mathcal{C}_{13}[ \tilde\Pi_{36} ]] , \quad &
		\tilde\Pi_{32} &= \mathcal{C}_{34}[ \mathcal{C}_{24}[ \mathcal{C}_{13}[ \tilde\Pi_{36} ]]] , \quad &
		\tilde\Pi_{33} &= \mathcal{C}_{23}[ \mathcal{C}_{14}[ \tilde\Pi_{36} ]] , \\
		\tilde\Pi_{34} &= \mathcal{C}_{12}[ \mathcal{C}_{24}[ \mathcal{C}_{13}[ \tilde\Pi_{36} ]]] , \quad &
		\tilde\Pi_{35} &= \mathcal{C}_{12}[ \tilde\Pi_{36} ] , \quad & 
		\tilde\Pi_{37} &= \mathcal{C}_{34}[ \mathcal{C}_{12}[ \tilde\Pi_{36} ]] , \\
		\tilde\Pi_{38} &= \mathcal{C}_{34}[ \tilde\Pi_{36} ] , \\
		\tilde\Pi_{41} &= \mathcal{C}_{12}[ \tilde\Pi_{40} ] , \quad &
		\tilde\Pi_{42} &= \mathcal{C}_{34}[ \tilde\Pi_{40} ] , \quad &
		\tilde\Pi_{43} &= \mathcal{C}_{34}[ \mathcal{C}_{12}[ \tilde\Pi_{40} ]] .
	\mytag
\end{align*}


\section{Projection of the scalar functions}

\label{sec:AppendixProjection}

Given any representation of the HLbL tensor $\Pi^{\mu\nu\lambda\sigma}$, the following procedure allows the identification of the basis coefficients $\tilde \Pi_i$ in~\eqref{eq:HLbLTensor43Basis}:
\begin{itemize}
	\item write the tensor in terms of the 138 elementary structures~\eqref{eq:HLbLTensor138StructuresLSM} and identify the scalar coefficients $\Xi_i$,
	\item take the subset consisting of the following 43 scalar coefficients:
	\begin{align}
		\begin{split}
			\{ \tilde \Xi_i \} = \big\{ \Xi_i | i ={} & 1,2,3,4,5,7,8,16,17,19,20,25,26,28,29,\\
				& 34,35,37,38,43,44,46,47,53,54,56,57,\\
				& 94,95,97,98,103,104,106,107,\\
				& 121,122,124,125,130,131,133,134 \big\}
		\end{split},
	\end{align}
	\item perform a basis change according to
	\begin{align}
		\tilde \Pi_j = \sum_{i=1}^{43} \tilde \Xi_i P_{i,j},
	\end{align}
	where $P$ is an invertible $43\times43$ matrix, defined below.
\end{itemize}

The matrix describing the change of basis is sparse and has the following non-zero entries:
\begin{align*}
		P_{1,4} &= \frac{1}{q_{12} q_{34}}, \\
		P_{2,3} &= \frac{1}{q_{12} q_{34}}, \quad
		P_{2,4} = \frac{q_{13} q_{24}}{q_{12}^2 q_{34}^2}+\frac{1}{q_{12} q_{34}}, \quad
		P_{2,5} = \frac{1}{q_{12} q_{34}}, \quad
		P_{2,19} = -\frac{1}{q_{12}^2 q_{34}}, \quad
		P_{2,25} = \frac{1}{q_{12} q_{34}^2}, \\
		P_{3,2} &= \frac{1}{q_{12} q_{34}}, \quad
		P_{3,4} = \frac{q_{14} q_{23}}{q_{12}^2 q_{34}^2}+\frac{1}{q_{12} q_{34}}, \quad
		P_{3,6} = \frac{1}{q_{12} q_{34}}, \quad
		P_{3,20} = -\frac{1}{q_{12}^2 q_{34}}, \quad
		P_{3,26} = \frac{1}{q_{12} q_{34}^2}, \\
		P_{4,4} &= -\frac{q_{13} q_{14}}{q_{12} q_{34}^2}, \quad
		P_{4,7} = \frac{1}{q_{12} q_{34}}, \quad
		P_{5,4} = -\frac{q_{13} q_{24}}{q_{12} q_{34}^2}, \quad
		P_{5,19} = \frac{1}{q_{12} q_{34}}, \\
		P_{6,4} &= -\frac{q_{14} q_{23}}{q_{12} q_{34}^2}, \quad
		P_{6,20} = \frac{1}{q_{12} q_{34}}, \quad
		P_{7,4} = -\frac{q_{23} q_{24}}{q_{12} q_{34}^2}, \quad
		P_{7,8} = \frac{1}{q_{12} q_{34}}, \\
		P_{8,2} &= \frac{q_{13}}{2 q_{12} q_{34}}, \quad
		P_{8,4} = \frac{q_{13}}{2 q_{12} q_{34}}+\frac{q_{14} q_{23} q_{13}}{q_{12}^2 q_{34}^2}, \quad
		P_{8,6} = \frac{q_{13}}{2 q_{12} q_{34}}, \quad
		P_{8,7} = -\frac{q_{23}}{q_{12}^2 q_{34}}, \\
		P_{8,17} &= -\frac{q_{14}}{q_{12} q_{34}^2}, \quad
		P_{8,32} = -\frac{1}{2 q_{12} q_{34}}, \quad
		P_{8,35} = -\frac{1}{2 q_{12} q_{34}}, \quad
		P_{8,40} = \frac{1}{2 q_{12} q_{34}}, \\
		P_{9,3} &= \frac{q_{23}}{2 q_{12} q_{34}}, \quad
		P_{9,4} = \frac{q_{23}}{2 q_{12} q_{34}}+\frac{q_{13} q_{24} q_{23}}{q_{12}^2 q_{34}^2}, \quad
		P_{9,5} = \frac{q_{23}}{2 q_{12} q_{34}}, \quad
		P_{9,17} = -\frac{q_{24}}{q_{12} q_{34}^2}, \\
		P_{9,19} &= -\frac{q_{23}}{q_{12}^2 q_{34}}, \quad
		P_{9,33} = \frac{1}{2 q_{12} q_{34}}, \quad
		P_{9,36} = -\frac{1}{2 q_{12} q_{34}}, \quad
		P_{9,41} = \frac{1}{2 q_{12} q_{34}}, \\
		P_{10,3} &= \frac{q_{14}}{2 q_{12} q_{34}}, \quad
		P_{10,4} = \frac{q_{14}}{2 q_{12} q_{34}}+\frac{q_{13} q_{24} q_{14}}{q_{12}^2 q_{34}^2}, \quad
		P_{10,5} = \frac{q_{14}}{2 q_{12} q_{34}}, \quad
		P_{10,7} = -\frac{q_{24}}{q_{12}^2 q_{34}}, \\
		P_{10,25} &= \frac{q_{14}}{q_{12} q_{34}^2}, \quad
		P_{10,31} = \frac{1}{2 q_{12} q_{34}}, \quad
		P_{10,37} = -\frac{1}{2 q_{12} q_{34}}, \quad
		P_{10,42} = -\frac{1}{2 q_{12} q_{34}}, \\
		P_{11,3} &= \frac{q_{24}}{q_{12} q_{34}}, \quad
		P_{11,4} = \frac{q_{13} q_{24}^2}{q_{12}^2 q_{34}^2}+\frac{q_{24}}{q_{12} q_{34}}, \quad
		P_{11,5} = \frac{q_{24}}{q_{12} q_{34}}, \quad
		P_{11,19} = -\frac{q_{24}}{q_{12}^2 q_{34}}, \\
		P_{11,24} &= \frac{1}{q_{12} q_{34}}, \quad
		P_{11,25} = \frac{q_{24}}{q_{12} q_{34}^2}, \\
		P_{12,2} &= \frac{q_{13}}{2 q_{12} q_{34}}, \quad
		P_{12,4} = \frac{q_{13}}{2 q_{12} q_{34}}+\frac{q_{14} q_{23} q_{13}}{q_{12}^2 q_{34}^2}, \quad
		P_{12,6} = \frac{q_{13}}{2 q_{12} q_{34}}, \quad
		P_{12,7} = -\frac{q_{23}}{q_{12}^2 q_{34}}, \\
		P_{12,26} &= \frac{q_{13}}{q_{12} q_{34}^2}, \quad
		P_{12,32} = -\frac{1}{2 q_{12} q_{34}}, \quad
		P_{12,35} = \frac{1}{2 q_{12} q_{34}}, \quad
		P_{12,40} = \frac{1}{2 q_{12} q_{34}}, \\
		P_{13,2} &= \frac{q_{23}}{q_{12} q_{34}}, \quad
		P_{13,4} = \frac{q_{14} q_{23}^2}{q_{12}^2 q_{34}^2}+\frac{q_{23}}{q_{12} q_{34}}, \quad
		P_{13,6} = \frac{q_{23}}{q_{12} q_{34}}, \quad
		P_{13,20} = -\frac{q_{23}}{q_{12}^2 q_{34}}, \\
		P_{13,23} &= -\frac{1}{q_{12} q_{34}}, \quad
		P_{13,26} = \frac{q_{23}}{q_{12} q_{34}^2}, \\
		P_{14,3} &= \frac{q_{14}}{2 q_{12} q_{34}}, \quad
		P_{14,4} = \frac{q_{14}}{2 q_{12} q_{34}}+\frac{q_{13} q_{24} q_{14}}{q_{12}^2 q_{34}^2}, \quad
		P_{14,5} = \frac{q_{14}}{2 q_{12} q_{34}}, \quad
		P_{14,7} = -\frac{q_{24}}{q_{12}^2 q_{34}}, \\
		P_{14,18} &= -\frac{q_{13}}{q_{12} q_{34}^2}, \quad
		P_{14,31} = \frac{1}{2 q_{12} q_{34}}, \quad
		P_{14,37} = \frac{1}{2 q_{12} q_{34}}, \quad
		P_{14,42} = -\frac{1}{2 q_{12} q_{34}}, \\
		P_{15,2} &= \frac{q_{24}}{2 q_{12} q_{34}}, \quad
		P_{15,4} = \frac{q_{24}}{2 q_{12} q_{34}}+\frac{q_{14} q_{23} q_{24}}{q_{12}^2 q_{34}^2}, \quad
		P_{15,6} = \frac{q_{24}}{2 q_{12} q_{34}}, \quad
		P_{15,18} = -\frac{q_{23}}{q_{12} q_{34}^2}, \\
		P_{15,20} &= -\frac{q_{24}}{q_{12}^2 q_{34}}, \quad
		P_{15,34} = -\frac{1}{2 q_{12} q_{34}}, \quad
		P_{15,38} = \frac{1}{2 q_{12} q_{34}}, \quad
		P_{15,43} = -\frac{1}{2 q_{12} q_{34}}, \\
		P_{16,2} &= \frac{q_{13}}{2 q_{12} q_{34}}, \quad
		P_{16,4} = \frac{q_{13}}{2 q_{12} q_{34}}+\frac{q_{14} q_{23} q_{13}}{q_{12}^2 q_{34}^2}, \quad
		P_{16,6} = \frac{q_{13}}{2 q_{12} q_{34}}, \quad
		P_{16,17} = -\frac{q_{14}}{q_{12} q_{34}^2}, \\
		P_{16,20} &= -\frac{q_{13}}{q_{12}^2 q_{34}}, \quad
		P_{16,32} = \frac{1}{2 q_{12} q_{34}}, \quad
		P_{16,35} = -\frac{1}{2 q_{12} q_{34}}, \quad
		P_{16,40} = \frac{1}{2 q_{12} q_{34}}, \\
		P_{17,3} &= \frac{q_{23}}{2 q_{12} q_{34}}, \quad
		P_{17,4} = \frac{q_{23}}{2 q_{12} q_{34}}+\frac{q_{13} q_{24} q_{23}}{q_{12}^2 q_{34}^2}, \quad
		P_{17,5} = \frac{q_{23}}{2 q_{12} q_{34}}, \quad
		P_{17,8} = -\frac{q_{13}}{q_{12}^2 q_{34}}, \\
		P_{17,17} &= -\frac{q_{24}}{q_{12} q_{34}^2}, \quad
		P_{17,33} = -\frac{1}{2 q_{12} q_{34}}, \quad
		P_{17,36} = -\frac{1}{2 q_{12} q_{34}}, \quad
		P_{17,41} = \frac{1}{2 q_{12} q_{34}}, \\
		P_{18,2} &= \frac{q_{14}}{q_{12} q_{34}}, \quad
		P_{18,4} = \frac{q_{23} q_{14}^2}{q_{12}^2 q_{34}^2}+\frac{q_{14}}{q_{12} q_{34}}, \quad
		P_{18,6} = \frac{q_{14}}{q_{12} q_{34}}, \quad
		P_{18,20} = -\frac{q_{14}}{q_{12}^2 q_{34}}, \\
		P_{18,22} &= \frac{1}{q_{12} q_{34}}, \quad
		P_{18,26} = \frac{q_{14}}{q_{12} q_{34}^2}, \\
		P_{19,2} &= \frac{q_{24}}{2 q_{12} q_{34}}, \quad
		P_{19,4} = \frac{q_{24}}{2 q_{12} q_{34}}+\frac{q_{14} q_{23} q_{24}}{q_{12}^2 q_{34}^2}, \quad
		P_{19,6} = \frac{q_{24}}{2 q_{12} q_{34}}, \quad
		P_{19,8} = -\frac{q_{14}}{q_{12}^2 q_{34}}, \\
		P_{19,26} &= \frac{q_{24}}{q_{12} q_{34}^2}, \quad
		P_{19,34} = \frac{1}{2 q_{12} q_{34}}, \quad
		P_{19,38} = -\frac{1}{2 q_{12} q_{34}}, \quad
		P_{19,43} = -\frac{1}{2 q_{12} q_{34}}, \\
		P_{20,3} &= \frac{q_{13}}{q_{12} q_{34}}, \quad
		P_{20,4} = \frac{q_{24} q_{13}^2}{q_{12}^2 q_{34}^2}+\frac{q_{13}}{q_{12} q_{34}}, \quad
		P_{20,5} = \frac{q_{13}}{q_{12} q_{34}}, \quad
		P_{20,19} = -\frac{q_{13}}{q_{12}^2 q_{34}}, \\
		P_{20,21} &= -\frac{1}{q_{12} q_{34}}, \quad
		P_{20,25} = \frac{q_{13}}{q_{12} q_{34}^2}, \\
		P_{21,3} &= \frac{q_{23}}{2 q_{12} q_{34}}, \quad
		P_{21,4} = \frac{q_{23}}{2 q_{12} q_{34}}+\frac{q_{13} q_{24} q_{23}}{q_{12}^2 q_{34}^2}, \quad
		P_{21,5} = \frac{q_{23}}{2 q_{12} q_{34}}, \quad
		P_{21,8} = -\frac{q_{13}}{q_{12}^2 q_{34}}, \\
		P_{21,25} &= \frac{q_{23}}{q_{12} q_{34}^2}, \quad
		P_{21,33} = -\frac{1}{2 q_{12} q_{34}}, \quad
		P_{21,36} = \frac{1}{2 q_{12} q_{34}}, \quad
		P_{21,41} = \frac{1}{2 q_{12} q_{34}}, \\
		P_{22,3} &= \frac{q_{14}}{2 q_{12} q_{34}}, \quad
		P_{22,4} = \frac{q_{14}}{2 q_{12} q_{34}}+\frac{q_{13} q_{24} q_{14}}{q_{12}^2 q_{34}^2}, \quad
		P_{22,5} = \frac{q_{14}}{2 q_{12} q_{34}}, \quad
		P_{22,18} = -\frac{q_{13}}{q_{12} q_{34}^2}, \\
		P_{22,19} &= -\frac{q_{14}}{q_{12}^2 q_{34}}, \quad
		P_{22,31} = -\frac{1}{2 q_{12} q_{34}}, \quad
		P_{22,37} = \frac{1}{2 q_{12} q_{34}}, \quad
		P_{22,42} = -\frac{1}{2 q_{12} q_{34}}, \\
		P_{23,2} &= \frac{q_{24}}{2 q_{12} q_{34}}, \quad
		P_{23,4} = \frac{q_{24}}{2 q_{12} q_{34}}+\frac{q_{14} q_{23} q_{24}}{q_{12}^2 q_{34}^2}, \quad
		P_{23,6} = \frac{q_{24}}{2 q_{12} q_{34}}, \quad
		P_{23,8} = -\frac{q_{14}}{q_{12}^2 q_{34}}, \\
		P_{23,18} &= -\frac{q_{23}}{q_{12} q_{34}^2}, \quad
		P_{23,34} = \frac{1}{2 q_{12} q_{34}}, \quad
		P_{23,38} = \frac{1}{2 q_{12} q_{34}}, \quad
		P_{23,43} = -\frac{1}{2 q_{12} q_{34}}, \\
		P_{24,4} &= -\frac{q_{13} q_{23}}{q_{12}^2 q_{34}}, \quad
		P_{24,17} = \frac{1}{q_{12} q_{34}}, \quad
		P_{25,4} = -\frac{q_{13} q_{24}}{q_{12}^2 q_{34}}, \quad
		P_{25,25} = -\frac{1}{q_{12} q_{34}}, \\
		P_{26,4} &= -\frac{q_{14} q_{23}}{q_{12}^2 q_{34}}, \quad
		P_{26,26} = -\frac{1}{q_{12} q_{34}}, \quad
		P_{27,4} = -\frac{q_{14} q_{24}}{q_{12}^2 q_{34}}, \quad
		P_{27,18} = \frac{1}{q_{12} q_{34}}, \\
		P_{28,2} &= \frac{q_{13}^2}{2 q_{12} q_{34}}, \quad
		P_{28,4} = \frac{q_{13}^2}{2 q_{12} q_{34}}+\frac{q_{14} q_{23} q_{13}^2}{q_{12}^2 q_{34}^2}, \quad
		P_{28,6} = \frac{q_{13}^2}{2 q_{12} q_{34}}, \quad
		P_{28,7} = -\frac{q_{13} q_{23}}{q_{12}^2 q_{34}}, \\
		P_{28,17} &= -\frac{q_{13} q_{14}}{q_{12} q_{34}^2}, \quad
		P_{28,27} = -\frac{1}{q_{12} q_{34}}, \quad
		P_{28,32} = -\frac{q_{13}}{2 q_{12} q_{34}}, \quad
		P_{28,35} = -\frac{q_{13}}{2 q_{12} q_{34}}, \quad
		P_{28,40} = \frac{q_{13}}{2 q_{12} q_{34}}, \\
		P_{29,3} &= \frac{q_{13} q_{23}}{2 q_{12} q_{34}}, \quad
		P_{29,4} = \frac{q_{23} q_{24} q_{13}^2}{q_{12}^2 q_{34}^2}+\frac{q_{23} q_{13}}{2 q_{12} q_{34}}, \quad
		P_{29,5} = \frac{q_{13} q_{23}}{2 q_{12} q_{34}}, \quad
		P_{29,9} = \frac{1}{q_{34}}, \\
		P_{29,17} &= -\frac{q_{13} q_{24}}{q_{12} q_{34}^2}, \quad
		P_{29,19} = -\frac{q_{13} q_{23}}{q_{12}^2 q_{34}}, \quad
		P_{29,21} = -\frac{q_{23}}{q_{12} q_{34}}, \quad
		P_{29,33} = \frac{q_{13}}{2 q_{12} q_{34}}, \\
		P_{29,36} &= -\frac{q_{13}}{2 q_{12} q_{34}}, \quad
		P_{29,41} = \frac{q_{13}}{2 q_{12} q_{34}}, \\
		P_{30,2} &= \frac{q_{13} q_{23}}{2 q_{12} q_{34}}, \quad
		P_{30,4} = \frac{q_{13} q_{14} q_{23}^2}{q_{12}^2 q_{34}^2}+\frac{q_{13} q_{23}}{2 q_{12} q_{34}}, \quad
		P_{30,6} = \frac{q_{13} q_{23}}{2 q_{12} q_{34}}, \quad
		P_{30,13} = \frac{1}{q_{34}}, \\
		P_{30,17} &= -\frac{q_{14} q_{23}}{q_{12} q_{34}^2}, \quad
		P_{30,20} = -\frac{q_{13} q_{23}}{q_{12}^2 q_{34}}, \quad
		P_{30,23} = -\frac{q_{13}}{q_{12} q_{34}}, \quad
		P_{30,32} = \frac{q_{23}}{2 q_{12} q_{34}}, \\
		P_{30,35} &= -\frac{q_{23}}{2 q_{12} q_{34}}, \quad
		P_{30,40} = \frac{q_{23}}{2 q_{12} q_{34}}, \\
		P_{31,3} &= \frac{q_{23}^2}{2 q_{12} q_{34}}, \quad
		P_{31,4} = \frac{q_{23}^2}{2 q_{12} q_{34}}+\frac{q_{13} q_{24} q_{23}^2}{q_{12}^2 q_{34}^2}, \quad
		P_{31,5} = \frac{q_{23}^2}{2 q_{12} q_{34}}, \quad
		P_{31,8} = -\frac{q_{13} q_{23}}{q_{12}^2 q_{34}}, \\
		P_{31,17} &= -\frac{q_{23} q_{24}}{q_{12} q_{34}^2}, \quad
		P_{31,28} = -\frac{1}{q_{12} q_{34}}, \quad
		P_{31,33} = -\frac{q_{23}}{2 q_{12} q_{34}}, \quad
		P_{31,36} = -\frac{q_{23}}{2 q_{12} q_{34}}, \quad
		P_{31,41} = \frac{q_{23}}{2 q_{12} q_{34}}, \\
		P_{32,3} &= \frac{q_{13} q_{14}}{2 q_{12} q_{34}}, \quad
		P_{32,4} = \frac{q_{14} q_{24} q_{13}^2}{q_{12}^2 q_{34}^2}+\frac{q_{14} q_{13}}{2 q_{12} q_{34}}, \quad
		P_{32,5} = \frac{q_{13} q_{14}}{2 q_{12} q_{34}}, \quad
		P_{32,7} = -\frac{q_{13} q_{24}}{q_{12}^2 q_{34}}, \\
		P_{32,10} &= \frac{1}{q_{12}}, \quad
		P_{32,21} = -\frac{q_{14}}{q_{12} q_{34}}, \quad
		P_{32,25} = \frac{q_{13} q_{14}}{q_{12} q_{34}^2}, \quad
		P_{32,31} = \frac{q_{13}}{2 q_{12} q_{34}}, \\
		P_{32,37} &= -\frac{q_{13}}{2 q_{12} q_{34}}, \quad
		P_{32,42} = -\frac{q_{13}}{2 q_{12} q_{34}}, \\
		P_{33,3} &= \frac{q_{13} q_{24}}{q_{12} q_{34}}, \quad
		P_{33,4} = \frac{q_{13}^2 q_{24}^2}{q_{12}^2 q_{34}^2}+\frac{q_{13} q_{24}}{q_{12} q_{34}}, \quad
		P_{33,5} = \frac{q_{13} q_{24}}{q_{12} q_{34}}+1, \quad
		P_{33,19} = -\frac{q_{13} q_{24}}{q_{12}^2 q_{34}}, \\
		P_{33,21} &= -\frac{q_{24}}{q_{12} q_{34}}, \quad
		P_{33,24} = \frac{q_{13}}{q_{12} q_{34}}, \quad
		P_{33,25} = \frac{q_{13} q_{24}}{q_{12} q_{34}^2}, \\
		P_{34,1} &= \frac{1}{2}, \quad
		P_{34,4} = \frac{q_{13} q_{14} q_{23} q_{24}}{q_{12}^2 q_{34}^2}, \quad
		P_{34,5} = \frac{1}{2}, \quad
		P_{34,6} = \frac{1}{2}, \quad
		P_{34,20} = -\frac{q_{13} q_{24}}{q_{12}^2 q_{34}}, \\
		P_{34,25} &= \frac{q_{14} q_{23}}{q_{12} q_{34}^2}, \quad
		P_{34,32} = \frac{q_{24}}{2 q_{12} q_{34}}, \quad
		P_{34,34} = -\frac{q_{13}}{2 q_{12} q_{34}}, \quad
		P_{34,36} = \frac{q_{14}}{2 q_{12} q_{34}}, \\
		P_{34,37} &= -\frac{q_{23}}{2 q_{12} q_{34}}, \quad
		P_{34,39} = -\frac{1}{2 q_{34}}, \quad
		P_{34,40} = \frac{q_{24}}{2 q_{12} q_{34}}, \quad
		P_{34,43} = -\frac{q_{13}}{2 q_{12} q_{34}}, \\
		P_{35,3} &= \frac{q_{23} q_{24}}{2 q_{12} q_{34}}, \quad
		P_{35,4} = \frac{q_{13} q_{23} q_{24}^2}{q_{12}^2 q_{34}^2}+\frac{q_{23} q_{24}}{2 q_{12} q_{34}}, \quad
		P_{35,5} = \frac{q_{23} q_{24}}{2 q_{12} q_{34}}, \quad
		P_{35,8} = -\frac{q_{13} q_{24}}{q_{12}^2 q_{34}}, \\
		P_{35,16} &= \frac{1}{q_{12}}, \quad
		P_{35,24} = \frac{q_{23}}{q_{12} q_{34}}, \quad
		P_{35,25} = \frac{q_{23} q_{24}}{q_{12} q_{34}^2}, \quad
		P_{35,33} = -\frac{q_{24}}{2 q_{12} q_{34}}, \\
		P_{35,36} &= \frac{q_{24}}{2 q_{12} q_{34}}, \quad
		P_{35,41} = \frac{q_{24}}{2 q_{12} q_{34}}, \\
		P_{36,2} &= \frac{q_{13} q_{14}}{2 q_{12} q_{34}}, \quad
		P_{36,4} = \frac{q_{13} q_{23} q_{14}^2}{q_{12}^2 q_{34}^2}+\frac{q_{13} q_{14}}{2 q_{12} q_{34}}, \quad
		P_{36,6} = \frac{q_{13} q_{14}}{2 q_{12} q_{34}}, \quad
		P_{36,7} = -\frac{q_{14} q_{23}}{q_{12}^2 q_{34}}, \\
		P_{36,11} &= \frac{1}{q_{12}}, \quad
		P_{36,22} = \frac{q_{13}}{q_{12} q_{34}}, \quad
		P_{36,26} = \frac{q_{13} q_{14}}{q_{12} q_{34}^2}, \quad
		P_{36,32} = -\frac{q_{14}}{2 q_{12} q_{34}}, \\
		P_{36,35} &= \frac{q_{14}}{2 q_{12} q_{34}}, \quad
		P_{36,40} = \frac{q_{14}}{2 q_{12} q_{34}}, \\
		P_{37,1} &= \frac{1}{2}, \quad
		P_{37,4} = \frac{q_{13} q_{14} q_{23} q_{24}}{q_{12}^2 q_{34}^2}, \quad
		P_{37,5} = \frac{1}{2}, \quad
		P_{37,6} = \frac{1}{2}, \quad
		P_{37,19} = -\frac{q_{14} q_{23}}{q_{12}^2 q_{34}}, \\
		P_{37,26} &= \frac{q_{13} q_{24}}{q_{12} q_{34}^2}, \quad
		P_{37,31} = -\frac{q_{23}}{2 q_{12} q_{34}}, \quad
		P_{37,33} = \frac{q_{14}}{2 q_{12} q_{34}}, \quad
		P_{37,35} = \frac{q_{24}}{2 q_{12} q_{34}}, \\
		P_{37,38} &= -\frac{q_{13}}{2 q_{12} q_{34}}, \quad
		P_{37,39} = \frac{1}{2 q_{34}}, \quad
		P_{37,41} = \frac{q_{14}}{2 q_{12} q_{34}}, \quad
		P_{37,42} = -\frac{q_{23}}{2 q_{12} q_{34}}, \\
		P_{38,2} &= \frac{q_{14} q_{23}}{q_{12} q_{34}}, \quad
		P_{38,4} = \frac{q_{14}^2 q_{23}^2}{q_{12}^2 q_{34}^2}+\frac{q_{14} q_{23}}{q_{12} q_{34}}, \quad
		P_{38,6} = \frac{q_{14} q_{23}}{q_{12} q_{34}}+1, \quad
		P_{38,20} = -\frac{q_{14} q_{23}}{q_{12}^2 q_{34}}, \\
		P_{38,22} &= \frac{q_{23}}{q_{12} q_{34}}, \quad
		P_{38,23} = -\frac{q_{14}}{q_{12} q_{34}}, \quad
		P_{38,26} = \frac{q_{14} q_{23}}{q_{12} q_{34}^2}, \\
		P_{39,2} &= \frac{q_{23} q_{24}}{2 q_{12} q_{34}}, \quad
		P_{39,4} = \frac{q_{14} q_{24} q_{23}^2}{q_{12}^2 q_{34}^2}+\frac{q_{24} q_{23}}{2 q_{12} q_{34}}, \quad
		P_{39,6} = \frac{q_{23} q_{24}}{2 q_{12} q_{34}}, \quad
		P_{39,8} = -\frac{q_{14} q_{23}}{q_{12}^2 q_{34}}, \\
		P_{39,14} &= \frac{1}{q_{12}}, \quad
		P_{39,23} = -\frac{q_{24}}{q_{12} q_{34}}, \quad
		P_{39,26} = \frac{q_{23} q_{24}}{q_{12} q_{34}^2}, \quad
		P_{39,34} = \frac{q_{23}}{2 q_{12} q_{34}}, \\
		P_{39,38} &= -\frac{q_{23}}{2 q_{12} q_{34}}, \quad
		P_{39,43} = -\frac{q_{23}}{2 q_{12} q_{34}}, \\
		P_{40,3} &= \frac{q_{14}^2}{2 q_{12} q_{34}}, \quad
		P_{40,4} = \frac{q_{14}^2}{2 q_{12} q_{34}}+\frac{q_{13} q_{24} q_{14}^2}{q_{12}^2 q_{34}^2}, \quad
		P_{40,5} = \frac{q_{14}^2}{2 q_{12} q_{34}}, \quad
		P_{40,7} = -\frac{q_{14} q_{24}}{q_{12}^2 q_{34}}, \\
		P_{40,18} &= -\frac{q_{13} q_{14}}{q_{12} q_{34}^2}, \quad
		P_{40,29} = -\frac{1}{q_{12} q_{34}}, \quad
		P_{40,31} = \frac{q_{14}}{2 q_{12} q_{34}}, \quad
		P_{40,37} = \frac{q_{14}}{2 q_{12} q_{34}}, \\
		P_{40,42} &= -\frac{q_{14}}{2 q_{12} q_{34}}, \\
		P_{41,3} &= \frac{q_{14} q_{24}}{2 q_{12} q_{34}}, \quad
		P_{41,4} = \frac{q_{13} q_{14} q_{24}^2}{q_{12}^2 q_{34}^2}+\frac{q_{14} q_{24}}{2 q_{12} q_{34}}, \quad
		P_{41,5} = \frac{q_{14} q_{24}}{2 q_{12} q_{34}}, \quad
		P_{41,15} = \frac{1}{q_{34}}, \\
		P_{41,18} &= -\frac{q_{13} q_{24}}{q_{12} q_{34}^2}, \quad
		P_{41,19} = -\frac{q_{14} q_{24}}{q_{12}^2 q_{34}}, \quad
		P_{41,24} = \frac{q_{14}}{q_{12} q_{34}}, \quad
		P_{41,31} = -\frac{q_{24}}{2 q_{12} q_{34}}, \\
		P_{41,37} &= \frac{q_{24}}{2 q_{12} q_{34}}, \quad
		P_{41,42} = -\frac{q_{24}}{2 q_{12} q_{34}}, \\
		P_{42,2} &= \frac{q_{14} q_{24}}{2 q_{12} q_{34}}, \quad
		P_{42,4} = \frac{q_{23} q_{24} q_{14}^2}{q_{12}^2 q_{34}^2}+\frac{q_{24} q_{14}}{2 q_{12} q_{34}}, \quad
		P_{42,6} = \frac{q_{14} q_{24}}{2 q_{12} q_{34}}, \quad
		P_{42,12} = \frac{1}{q_{34}}, \\
		P_{42,18} &= -\frac{q_{14} q_{23}}{q_{12} q_{34}^2}, \quad
		P_{42,20} = -\frac{q_{14} q_{24}}{q_{12}^2 q_{34}}, \quad
		P_{42,22} = \frac{q_{24}}{q_{12} q_{34}}, \quad
		P_{42,34} = -\frac{q_{14}}{2 q_{12} q_{34}}, \\
		P_{42,38} &= \frac{q_{14}}{2 q_{12} q_{34}}, \quad
		P_{42,43} = -\frac{q_{14}}{2 q_{12} q_{34}}, \\
		P_{43,2} &= \frac{q_{24}^2}{2 q_{12} q_{34}}, \quad
		P_{43,4} = \frac{q_{24}^2}{2 q_{12} q_{34}}+\frac{q_{14} q_{23} q_{24}^2}{q_{12}^2 q_{34}^2}, \quad
		P_{43,6} = \frac{q_{24}^2}{2 q_{12} q_{34}}, \quad
		P_{43,8} = -\frac{q_{14} q_{24}}{q_{12}^2 q_{34}}, \\
		P_{43,18} &= -\frac{q_{23} q_{24}}{q_{12} q_{34}^2}, \quad
		P_{43,30} = -\frac{1}{q_{12} q_{34}}, \quad
		P_{43,34} = \frac{q_{24}}{2 q_{12} q_{34}}, \\
		P_{43,38} &= \frac{q_{24}}{2 q_{12} q_{34}}, \quad
		P_{43,43} = -\frac{q_{24}}{2 q_{12} q_{34}} , \mytag
\end{align*}
where $q_{ij} := q_i \cdot q_j$.


\section{Scalar functions contributing to $\boldsymbol{(g-2)_\mu}$}

\label{sec:AppendixScalarFunctionsContributingTog-2}

The following 19 linear combinations of 33 scalar functions $\Pi_i$ contribute to $(g-2)_\mu$:
\begin{align*}
	\label{eq:ScalarFunctionsContributingToAmu}
	\hat \Pi_1 &= \Pi_1 + q_1 \cdot q_2 \Pi_{47} , \\
	\hat \Pi_2 &= \Pi_2 - \frac{1}{2} \left(q_1^2 + q_1 \cdot q_2 \right) \left( 2 \Pi_{47} - \Pi_{50} - \Pi_{51} - \Pi_{54} \right) , \\
	\hat \Pi_3 &= \Pi_3 - \frac{1}{2} \left(q_1 \cdot q_2 + q_2^2\right) \left( 2 \Pi_{47} - \Pi_{50} - \Pi_{51} + \Pi_{54} \right) , \\
	\hat \Pi_4 &= \Pi_4 + \left(q_1^2 + q_1 \cdot q_2 \right) \Pi_{19} + \left(q_1 \cdot q_2 + q_2^2 \right) \Pi_{20} \\
		& \quad + \left(q_1^2 + q_1 \cdot q_2 \right) \left(q_1 \cdot q_2+q_2^2\right) \Pi_{31} - \frac{s}{2} \left( 2 \Pi_{47} - \Pi_{50} - \Pi_{51} \right) + \frac{1}{2} \left( q_1^2 - q_2^2\right) \Pi_{54} , \\
	\hat \Pi_5 &= \Pi_5 - q_1 \cdot q_2 \Pi_{21} + \frac{1}{2} \left(q_1 \cdot q_2+q_2^2\right) \left( 2 \Pi_{22} - 2 q_1 \cdot q_2 \Pi_{33}  + \Pi_{50}  + \Pi_{51} - \Pi_{54}\right) - q_2^2 \Pi_{47} , \\
	\hat \Pi_6 &= \Pi_6 - q_1 \cdot q_2 \Pi_{25} + \frac{1}{2} \left(q_1^2+q_1 \cdot q_2\right) \left( 2 \Pi_{26} - 2 q_1 \cdot q_2 \Pi_{34} + \Pi_{50} + \Pi_{51} + \Pi_{54} \right) - q_1^2 \Pi_{47} , \\
	\hat \Pi_7 &= \Pi_7 - \Pi_{19} - \left( q_1 \cdot q_2 + q_2^2\right) \Pi_{31} , \\
	\hat \Pi_8 &= \Pi_8 - \Pi_{20} - \left(q_1^2 + q_1 \cdot q_2\right) \Pi_{31} , \\
	\hat \Pi_9 &= \Pi_9 - \Pi_{22} + q_1 \cdot q_2 \Pi_{33} , \\
	\hat \Pi_{10} &= \Pi_{10} - \Pi_{21} - \left( q_1 \cdot q_2 + q_2^2 \right) \Pi_{33} , \\
	\hat \Pi_{11} &= \Pi_{11} + \Pi_{47} - \Pi_{54} , \\
	\hat \Pi_{12} &= \Pi_{13} - \Pi_{26} + q_1 \cdot q_2 \Pi_{34} , \\
	\hat \Pi_{13} &= \Pi_{14} - \Pi_{25} - \left( q_1^2 + q_1 \cdot q_2\right) \Pi_{34} , \\
	\hat \Pi_{14} &= \Pi_{16} + \Pi_{47} + \Pi_{54} , \\
	\hat \Pi_{15} &= \Pi_{17} + \Pi_{47} - \Pi_{50} - \Pi_{51} , \\
	\hat \Pi_{16} &= \frac{1}{2} \left( \Pi_{39} + \Pi_{40} + \Pi_{46} \right) , \\
	\hat \Pi_{17} &= \Pi_{42} - \Pi_{47} + \frac{1}{2} \left( \Pi_{50} + \Pi_{51} + \Pi_{54} \right) , \\
	\hat \Pi_{18} &= \Pi_{43} - \Pi_{47} + \frac{1}{2} \left( \Pi_{50} + \Pi_{51} - \Pi_{54} \right) , \\
	\hat \Pi_{19} &= \frac{1}{2} \left( \Pi_{50} - \Pi_{51} + \Pi_{54} \right) . \mytag
\end{align*}
Because of the symmetries under the exchange of the momenta $q_1\leftrightarrow -q_2$, only a subset of 12 of these functions appears in the master formula~\eqref{eq:MasterFormula3Dim} for $(g-2)_\mu$:
\begin{align*}
		\label{eq:ScalarFunctionsForMasterFormula}
		\bar \Pi_1 &= \Pi_1 + q_1 \cdot q_2 \Pi_{47} , \\
		\bar \Pi_2 &= \Pi_2 - \frac{1}{2} \left(q_1^2 + q_1 \cdot q_2 \right) \left( 2 \Pi_{47} - \Pi_{50} - \Pi_{51} - \Pi_{54} \right) , \\
		\bar \Pi_3 &= \Pi_4 + \left(q_1^2 + q_1 \cdot q_2 \right) \Pi_{19} + \left(q_1 \cdot q_2 + q_2^2 \right) \Pi_{20} \\
			& \quad + \left(q_1^2 + q_1 \cdot q_2 \right) \left(q_1 \cdot q_2+q_2^2\right) \Pi_{31} - \frac{s}{2} \left( 2 \Pi_{47} - \Pi_{50} - \Pi_{51} \right) + \frac{1}{2} \left( q_1^2 - q_2^2\right) \Pi_{54} , \\
		\bar \Pi_4 &= \Pi_5 - q_1 \cdot q_2 \Pi_{21} + \frac{1}{2} \left(q_1 \cdot q_2+q_2^2\right) \left( 2 \Pi_{22} - 2 q_1 \cdot q_2 \Pi_{33}  + \Pi_{50}  + \Pi_{51} - \Pi_{54}\right) - q_2^2 \Pi_{47} , \\
		\bar \Pi_5 &= \Pi_7 - \Pi_{19} - \left( q_1 \cdot q_2 + q_2^2\right) \Pi_{31} , \\
		\bar \Pi_6 &= \Pi_9 - \Pi_{22} + q_1 \cdot q_2 \Pi_{33} , \\
		\bar \Pi_7 &= \Pi_{10} - \Pi_{21} - \left( q_1 \cdot q_2 + q_2^2 \right) \Pi_{33} , \\
		\bar \Pi_8 &= \Pi_{16} + \Pi_{47} + \Pi_{54} , \\
		\bar \Pi_9 &= \Pi_{17} + \Pi_{47} - \Pi_{50} - \Pi_{51} , \\
		\bar \Pi_{10} &= \frac{1}{2} \left( \Pi_{39} + \Pi_{40} + \Pi_{46} \right) , \\
		\bar \Pi_{11} &= \Pi_{42} - \Pi_{47} + \frac{1}{2} \left( \Pi_{50} + \Pi_{51} + \Pi_{54} \right) , \\
		\bar \Pi_{12} &= \frac{1}{2} \left( \Pi_{50} - \Pi_{51} + \Pi_{54} \right) . \mytag
\end{align*}


\section{Integral kernels}

\subsection{Intermediate kernels}

\label{sec:AppendixIntermediateKernels}

After calculating the trace and performing the contraction of the Lorentz indices in~\eqref{eq:DefinitionIntermediateKernels}, one finds the following integral kernels:
\begin{align*}
		\hat T_1(q_1,q_2;p) &= -\frac{8}{3} \left((q_1 \cdot q_2)^2-q_1^2 q_2^2\right) m_\mu^2-\frac{8}{3} q_2^2 (p \cdot q_1)^2-\frac{8}{3} q_1^2 (p \cdot q_2)^2 \\
			&\quad -\frac{4}{3} q_1^2 p \cdot q_2 \left(q_2^2+q_1 \cdot q_2\right)+p \cdot q_1 \left(\frac{4}{3} \left(q_1^2+q_1 \cdot q_2\right) q_2^2+\frac{16}{3} p \cdot q_2 q_1 \cdot q_2\right) , \\
   		\hat T_2(q_1,q_2;p) &= -\frac{8}{3} \left((q_1 \cdot q_2)^2-q_1^2 q_2^2\right) m_\mu^2-\frac{8}{3} q_1^2 (p \cdot q_2)^2+p \cdot q_1\left(\frac{8}{3} p \cdot q_2 q_1 \cdot q_2-\frac{4}{3} q_2^2 q_1 \cdot q_2\right) \\
			&\quad +p \cdot q_2 \left(4 q_1^2 q_2^2-\frac{8}{3} (q_1 \cdot q_2)^2\right) , \\
   		\hat T_3(q_1,q_2;p) &= -\frac{8}{3} \left((q_1 \cdot q_2)^2-q_1^2 q_2^2\right) m_\mu^2-\frac{8}{3} q_2^2 (p \cdot q_1)^2+\frac{4}{3} q_1^2 p \cdot q_2 q_1 \cdot q_2 \\
			&\quad +p \cdot q_1 \left(-4 q_1^2 q_2^2+\frac{8}{3} (q_1 \cdot q_2)^2+\frac{8}{3} p \cdot q_2 q_1 \cdot q_2\right) , \\
		\hat T_4(q_1,q_2;p) &= \frac{8}{3} q_1 \cdot q_2 \left(q_1^2+q_2^2+2 q_1 \cdot q_2\right) m_\mu^2+\frac{8}{3} q_2^2 (p \cdot q_1)^2+\frac{8}{3} q_1^2 (p \cdot q_2)^2 \\*
			&\quad -\frac{8}{3} p \cdot q_1 p \cdot q_2 \left(q_1^2+q_2^2+4 q_1 \cdot q_2\right) , \\
		\hat T_5(q_1,q_2;p) &= \frac{8}{3} (p \cdot q_1)^2 q_2^2-\frac{8}{3} m_\mu^2 \left(q_1^2+q_1 \cdot q_2\right) q_2^2+\frac{8}{3} q_1^2 (p \cdot q_2)^2-\frac{4}{3} p \cdot q_2 q_1 \cdot q_2 \left(3 q_1^2+2 q_1 \cdot q_2\right) \\
			&\quad +p \cdot q_1 \left(\frac{4}{3} q_2^2 \left(3 q_1^2+2 q_1 \cdot q_2\right)-\frac{8}{3} p \cdot q_2 \left(q_1 \cdot q_2-q_2^2\right)\right) , \\
		\hat T_6(q_1,q_2;p) &= \frac{8}{3} (p \cdot q_2)^2 q_1^2-\frac{8}{3} m_\mu^2 \left(q_2^2+q_1 \cdot q_2\right) q_1^2-\frac{4}{3} p \cdot q_2 \left(3 q_2^2+2 q_1 \cdot q_2\right) q_1^2 \\
			&\quad +\frac{8}{3} q_2^2 (p \cdot q_1)^2+p \cdot q_1 \left(\frac{8}{3} p \cdot q_2 \left(q_1^2-q_1 \cdot q_2\right)+\frac{4}{3} q_1 \cdot q_2 \left(3 q_2^2+2 q_1 \cdot q_2\right)\right) , \\
		\hat T_7(q_1,q_2;p) &= \frac{4}{3} q_1 \cdot q_2 \left(2 q_1^4+\left(q_2^2+4 q_1 \cdot q_2\right) q_1^2+(q_1 \cdot q_2)^2\right) m_\mu^2-\frac{4}{3} q_2^2 (p \cdot q_1)^2 q_1 \cdot q_2 \\
			&\quad +\frac{4}{3} q_1^2 (p \cdot q_2)^2 \left(2 q_1^2+q_1 \cdot q_2\right)-\frac{8}{3} p \cdot q_1 p \cdot q_2 \left(q_1^4+3 q_1 \cdot q_2 q_1^2+(q_1 \cdot q_2)^2\right) , \\
		\hat T_8(q_1,q_2;p) &= \frac{4}{3} q_1 \cdot q_2 \left(2 q_2^4+\left(q_1^2+4 q_1 \cdot q_2\right) q_2^2+(q_1 \cdot q_2)^2\right) m_\mu^2-\frac{4}{3} q_1^2 (p \cdot q_2)^2 q_1 \cdot q_2 \\
			&\quad +\frac{4}{3} q_2^2 (p \cdot q_1)^2 \left(2 q_2^2+q_1 \cdot q_2\right)-\frac{8}{3} p \cdot q_1 p \cdot q_2 \left(q_2^4+3 q_1 \cdot q_2 q_2^2+(q_1 \cdot q_2)^2\right) , \\
		\hat T_9(q_1,q_2;p) &= -\frac{4}{3} \left(q_1^2+q_1 \cdot q_2\right) \left(2 q_2^4+\left(q_1^2+4 q_1 \cdot q_2\right) q_2^2+(q_1 \cdot q_2)^2\right) m_\mu^2 \\
			&\quad +\frac{4}{3} q_1^2 (p \cdot q_2)^2 \left(q_1^2+q_1 \cdot q_2\right) -\frac{4}{3} p \cdot q_2 \left(q_1^2+q_1 \cdot q_2\right) \left(q_2^2+q_1 \cdot q_2\right) \left(3 q_1^2+2 q_1 \cdot q_2\right) \\
			&\quad +\frac{4}{3} (p \cdot q_1)^2 \left(2 q_2^4+\left(q_1^2+5 q_1 \cdot q_2\right) q_2^2+2 (q_1 \cdot q_2)^2\right) \\
			&\quad +p \cdot q_1 \left(\frac{4}{3} \left(3 q_1^2+2 q_1 \cdot q_2\right) \left(q_2^2+q_1 \cdot q_2\right)^2+\frac{8}{3} p \cdot q_2 \left(q_2^2 \left(q_2^2+2 q_1 \cdot q_2\right)-q_1^2 q_1 \cdot q_2\right)\right) , \\
		\hat T_{10}(q_1,q_2;p) &= -\frac{4}{3} \left(q_1^2+q_1 \cdot q_2\right) \left(q_1^2 q_2^2+(q_1 \cdot q_2)^2\right) m_\mu^2+\frac{4}{3} q_1^2 (p \cdot q_2)^2 \left(q_1^2-q_1 \cdot q_2\right) \\
			&\quad -\frac{4}{3} q_1^2 p \cdot q_2 q_1 \cdot q_2 \left(3 q_1^2+2 q_1 \cdot q_2\right)+\frac{4}{3} (p \cdot q_1)^2 \left(\left(q_1^2+q_1 \cdot q_2\right) q_2^2+2 (q_1 \cdot q_2)^2\right) \\
			&\quad +p \cdot q_1 \left(\frac{4}{3} \left(3 q_1^2+2 q_1 \cdot q_2\right) (q_1 \cdot q_2)^2+\frac{8}{3} p \cdot q_2 \left(q_1 \cdot q_2-q_1^2\right) q_1 \cdot q_2\right) , \\
		\hat T_{11}(q_1,q_2;p) &= -\frac{8}{3} m_\mu^2 q_1 \cdot q_2 \left(q_1^2+q_1 \cdot q_2\right) q_1^2-\frac{4}{3} p \cdot q_2 \left(\left(q_2^2+2 q_1 \cdot q_2\right) q_1^2+2 (q_1 \cdot q_2)^2\right) q_1^2 \\*
			&\quad +p \cdot q_1 \left(\frac{8}{3} p \cdot q_2 \left(q_1^2+q_1 \cdot q_2\right) q_1^2+\frac{4}{3} q_1 \cdot q_2 \left(\left(q_2^2+2 q_1 \cdot q_2\right) q_1^2+2 (q_1 \cdot q_2)^2\right)\right) , \\
		\hat T_{12}(q_1,q_2;p) &= -\frac{4}{3} \left(q_2^2+q_1 \cdot q_2\right) \left(2 q_1^4+\left(q_2^2+4 q_1 \cdot q_2\right) q_1^2+(q_1 \cdot q_2)^2\right) m_\mu^2 \\
			&\quad +\frac{4}{3} q_2^2 (p \cdot q_1)^2 \left(q_2^2+q_1 \cdot q_2\right)-\frac{4}{3} p \cdot q_2 \left(q_1^2+q_1 \cdot q_2\right)^2 \left(3 q_2^2+2 q_1 \cdot q_2\right) \\
			&\quad +\frac{4}{3} (p \cdot q_2)^2 \left(2 q_1^4+\left(q_2^2+5 q_1 \cdot q_2\right) q_1^2+2 (q_1 \cdot q_2)^2\right) \\
			&\quad +p \cdot q_1 \begin{aligned}[t]
				& \bigg(\frac{4}{3} \left(q_1^2+q_1 \cdot q_2\right) \left(q_2^2+q_1 \cdot q_2\right) \left(3 q_2^2+2 q_1 \cdot q_2\right) \\
				& +\frac{8}{3} p \cdot q_2 \left(q_1^4+2 q_1 \cdot q_2 q_1^2-q_2^2 q_1 \cdot q_2\right)\bigg) , \end{aligned} \\
		\hat T_{13}(q_1,q_2;p) &= -\frac{4}{3} \left(q_2^2+q_1 \cdot q_2\right) \left(q_1^2 q_2^2+(q_1 \cdot q_2)^2\right) m_\mu^2+\frac{4}{3} q_2^2 (p \cdot q_1)^2 \left(q_2^2-q_1 \cdot q_2\right) \\
			&\quad -\frac{4}{3} p \cdot q_2 (q_1 \cdot q_2)^2 \left(3 q_2^2+2 q_1 \cdot q_2\right)+\frac{4}{3} (p \cdot q_2)^2 \left(\left(q_2^2+q_1 \cdot q_2\right) q_1^2+2 (q_1 \cdot q_2)^2\right) \\
			&\quad +p \cdot q_1 \left(\frac{4}{3} q_1 \cdot q_2 \left(3 q_2^2+2 q_1 \cdot q_2\right) q_2^2+\frac{8}{3} p \cdot q_2 q_1 \cdot q_2 \left(q_1 \cdot q_2-q_2^2\right)\right) , \\
		\hat T_{14}(q_1,q_2;p) &= -\frac{8}{3} m_\mu^2 q_1 \cdot q_2 \left(q_2^2+q_1 \cdot q_2\right) q_2^2+p \cdot q_2 \left(-\frac{8}{3} (q_1 \cdot q_2)^3-\frac{4}{3} q_2^2 \left(q_1^2+2 q_1 \cdot q_2\right) q_1 \cdot q_2\right) \\
			&\quad +p \cdot q_1 \left(\frac{8}{3} p \cdot q_2 \left(q_2^2+q_1 \cdot q_2\right) q_2^2+\frac{4}{3} \left(\left(q_1^2+2 q_1 \cdot q_2\right) q_2^2+2 (q_1 \cdot q_2)^2\right) q_2^2\right) , \\
		\hat T_{15}(q_1,q_2;p) &= \frac{8}{3} m_\mu^2 \left(q_1^2+q_1 \cdot q_2\right) \left(q_2^2+q_1 \cdot q_2\right) \left(q_1^2+q_2^2+2 q_1 \cdot q_2\right) -\frac{8}{3} p \cdot q_1 p \cdot q_2 \left(q_1^2+q_2^2+2 q_1 \cdot q_2\right)^2 , \\
		\hat T_{16}(q_1,q_2;p) &= \frac{8}{3} \left(q_1^2+q_2^2+q_1 \cdot q_2\right) \left(q_1^2 q_2^2-(q_1 \cdot q_2)^2\right) m_\mu^2 -\frac{8}{3} q_2^2 (p \cdot q_1)^2 \left(q_1^2+q_2^2+q_1 \cdot q_2\right) \\
			&\quad -\frac{8}{3} q_1^2 (p \cdot q_2)^2 \left(q_1^2+q_2^2+q_1 \cdot q_2\right)+\frac{16}{3} p \cdot q_1 p \cdot q_2 q_1 \cdot q_2 \left(q_1^2+q_2^2+q_1 \cdot q_2\right) , \\
		\hat T_{17}(q_1,q_2;p) &= -\frac{4}{3} q_1^2 p \cdot q_2 \left(q_1^2+q_1 \cdot q_2\right) q_2^2+\frac{4}{3} m_\mu^2 \left(q_1^2 q_2^2-(q_1 \cdot q_2)^2\right) q_2^2 \\
			&\quad -\frac{4}{3} (p \cdot q_1)^2 \left(q_2^2+6 \left(q_1^2+q_1 \cdot q_2\right)\right) q_2^2+\frac{4}{3} q_1^2 (p \cdot q_2)^2 \left(q_2^2+2 q_1 \cdot q_2\right) \\
			&\quad +p \cdot q_1 \left(\frac{4}{3} q_1^2 \left(q_2^2+q_1 \cdot q_2\right) q_2^2+\frac{8}{3} p \cdot q_2 q_1 \cdot q_2 \left(3 q_1^2+2 q_1 \cdot q_2\right)\right) , \\
		\hat T_{18}(q_1,q_2;p) &= -\frac{4}{3} q_2^2 p \cdot q_2 \left(q_1^2+q_1 \cdot q_2\right) q_1^2+\frac{4}{3} m_\mu^2 \left(q_1^2 q_2^2-(q_1 \cdot q_2)^2\right) q_1^2 \\
			&\quad -\frac{4}{3} (p \cdot q_2)^2 \left(q_1^2+6 \left(q_2^2+q_1 \cdot q_2\right)\right) q_1^2+\frac{4}{3} q_2^2 (p \cdot q_1)^2 \left(q_1^2+2 q_1 \cdot q_2\right) \\
			&\quad +p \cdot q_1 \left(\frac{4}{3} q_1^2 \left(q_2^2+q_1 \cdot q_2\right) q_2^2+\frac{8}{3} p \cdot q_2 q_1 \cdot q_2 \left(3 q_2^2+2 q_1 \cdot q_2\right)\right) , \\
		\hat T_{19}(q_1,q_2;p) &= -\frac{4}{3} m_\mu^2 \left(q_1^2-q_2^2\right) \left(q_1^2 \left(4 q_1 \cdot q_2+q_2^2\right)+q_1 \cdot q_2 \left(7 q_1 \cdot q_2+4 q_2^2\right)\right) \\
			&\quad -\frac{8}{3} p \cdot q_2 \left(q_1^2-q_2^2\right) q_1 \cdot q_2 \left(q_1^2+q_1 \cdot q_2\right)+\frac{4}{3} q_2^2 (p \cdot q_1)^2 \left(q_1^2+2 q_1 \cdot q_2+q_2^2\right) \\
			&\quad -\frac{4}{3} q_1^2 (p \cdot q_2)^2 \left(q_1^2+2 q_1 \cdot q_2+q_2^2\right) \\
			&\quad +p \cdot q_1 \left(\frac{16}{3} p \cdot q_2 \left(q_1^2-q_2^2\right) \left(q_1^2+2 q_1 \cdot q_2+q_2^2\right)+\frac{8}{3} \left(q_1^2-q_2^2\right) q_1 \cdot q_2 \left(q_1 \cdot q_2+q_2^2\right)\right) . \mytag
\end{align*}

\subsection{Kernels for the master formula}

\label{sec:AppendixMasterFormulaKernels}

In the master formula~\eqref{eq:MasterFormula3Dim}, the following integral kernels appear:
\begin{align*}
		T_1 &= \frac{Q_1^2 \tau  \left(\sigma_1^E-1\right) \left(\sigma_1^E+5\right)+Q_2^2 \tau  \left(\sigma_2^E-1\right) \left(\sigma_2^E+5\right)+4 Q_1 Q_2 \left(\sigma_1^E+\sigma_2^E-2\right)-8 \tau  m_{\mu }^2}{2 Q_1 Q_2 Q_3^2 m_{\mu }^2} \\
			&\quad + X \left(\frac{8 \left(\tau ^2-1\right)}{Q_3^2}-\frac{4}{m_{\mu }^2}\right) , \\
		T_2 &= \frac{Q_1 \left(\sigma _1^E-1\right) \left(Q_1 \tau  \left(\sigma _1^E+1\right)+4 Q_2 \left(\tau ^2-1\right)\right)-4 \tau  m_{\mu}^2}{Q_1 Q_2 Q_3^2 m_{\mu }^2} +  X \frac{8 \left(\tau ^2-1\right) \left(2 m_{\mu }^2-Q_2^2\right)}{Q_3^2 m_{\mu }^2} , \\
		T_3 &= \frac{1}{Q_3^2} \begin{aligned}[t]
			& \bigg( -\frac{2 \left(\sigma _1^E+\sigma _2^E-2\right)}{m_{\mu }^2}-\frac{Q_1 \tau  \left(\sigma _1^E-1\right) \left(\sigma_1^E+7\right)}{2 Q_2 m_{\mu }^2} + \frac{8 \tau }{Q_1Q_2} \\
			& -\frac{Q_2 \tau  \left(\sigma _2^E-1\right) \left(\sigma _2^E+7\right)}{2 Q_1 m_{\mu }^2} + \frac{Q_1^2 \left(1-\sigma _1^E\right)}{Q_2^2 m_{\mu }^2} + \frac{Q_2^2 \left(1-\sigma _2^E\right)}{Q_1^2 m_{\mu}^2}+\frac{2}{Q_1^2}+\frac{2}{Q_2^2}\bigg) \end{aligned} \\
			&\quad + X \left(\frac{4}{m_{\mu }^2}-\frac{8 \tau }{Q_1 Q_2}\right) , \\
		T_4 &= \frac{1}{Q_3^2} \begin{aligned}[t]
			& \Bigg(\frac{4 \left(\tau ^2 \left(\sigma _1^E-1\right)+\sigma _2^E-1\right)}{m_{\mu }^2}-\frac{Q_1 \tau  \left(\sigma _1^E-5\right) \left(\sigma _1^E-1\right)}{Q_2 m_{\mu }^2} + \frac{4 \tau }{Q_1Q_2} \\
			& - \frac{Q_2 \tau  \left(\sigma _2^E-3\right) \left(\sigma_2^E-1\right)}{Q_1m_{\mu }^2} +\frac{2 Q_2^2 \left(\sigma _2^E-1\right)}{Q_1^2m_{\mu }^2}-\frac{4}{Q_1^2} \\
   			& +X \left(-\frac{8 Q_2^2 \tau^2}{m_{\mu }^2}-\frac{16 Q_2 Q_1 \tau }{m_{\mu }^2}-\frac{8 Q_1^2}{m_{\mu }^2}+\frac{16 Q_2 \tau }{Q_1}+16\right)\Bigg) , \end{aligned} \\
		T_5 &= \frac{1}{Q_3^2} \begin{aligned}[t]
			& \Bigg( Q_1^2 \left(\frac{\tau ^2 \left(\sigma _1^E-1\right) \left(\sigma _1^E+3\right)+4 \left(\sigma _1^E+\sigma _2^E-2\right)}{2 m_{\mu }^2} -\frac{4}{Q_2^2}\right) -\frac{Q_2^2 \tau ^2 \left(\sigma _2^E-5\right) \left(\sigma _2^E-1\right)}{2 m_{\mu}^2} \\
			& +\frac{Q_1^3 \tau  \left(\sigma _1^E-1\right) \left(\sigma _1^E+5\right)}{Q_2 m_{\mu }^2} + Q_1 \left(\frac{Q_2 \tau  \left(\sigma_1^E+5 \sigma _2^E-6\right)}{m_{\mu }^2}-\frac{12 \tau }{Q_2}\right)+\frac{2 Q_1^4 \left(\sigma _1^E-1\right)}{Q_2^2 m_{\mu }^2} \\
			&  - 4 \tau ^2 +X \begin{aligned}[t]
				& \Bigg(Q_1 \left(8 Q_2 \left(\tau ^3+\tau \right)-\frac{2 Q_2^3 \tau }{m_{\mu }^2}\right)+Q_1^2 \bigg(32 \tau ^2-\frac{4 Q_2^2 \left(\tau ^2+1\right)}{m_{\mu }^2}\bigg) \\
				& +Q_1^3 \left(\frac{16 \tau }{Q_2}-\frac{10 Q_2 \tau }{m_{\mu }^2}\right)-\frac{4 Q_1^4}{m_{\mu }^2}\Bigg) \Bigg) , \end{aligned} \end{aligned} \\
   		T_6 &= \frac{1}{Q_3^2} \begin{aligned}[t]
			& \Bigg(\frac{Q_1^2 \left(\tau ^2 \left(\left(\sigma _1^E-22\right) \sigma _1^E-8 \sigma _2^E+29\right)+2 \left(-5 \sigma _1^E+\sigma_2^E+4\right)\right)}{2 m_{\mu }^2} \\
			& +Q_1 \left(\frac{Q_2 \tau  \left(2 \tau ^2 \left(\left(\sigma _2^E-3\right)^2-4 \sigma_1^E\right)-26 \sigma _1^E+\sigma _2^E \left(\sigma _2^E-12\right)+37\right)}{2 m_{\mu }^2}-\frac{4 \tau }{Q_2}\right) \\
			& +\frac{Q_2^2 \left(\tau ^2 \left(-8 \sigma _1^E+\sigma _2^E \left(5 \sigma _2^E-26\right)+29\right)-4 \left(\sigma _1^E+2 \sigma_2^E-3\right)\right)}{2 m_{\mu }^2}+\frac{Q_1^3 \tau  \left(\sigma _1^E-9\right) \left(\sigma _1^E-1\right)}{2 Q_2 m_{\mu}^2} \\
			& +\frac{Q_2^3 \tau  \left(\sigma _2^E-9\right) \left(\sigma _2^E-1\right)}{Q_1m_{\mu }^2}+\frac{8 Q_2 \tau }{Q_1}+\frac{2 Q_2^4 \left(1- \sigma _2^E\right)}{Q_1^2 m_{\mu }^2}+\frac{4 Q_2^2}{Q_1^2} \\
			& +X \begin{aligned}[t]
				& \Bigg(\frac{Q_2 Q_1^3 \left(8 \tau ^3+22 \tau \right)}{m_{\mu}^2}+\frac{Q_1^4 \left(8 \tau ^2-2\right)}{m_{\mu }^2}+Q_1^2 \left(\frac{Q_2^2 \left(36 \tau ^2+18\right)}{m_{\mu }^2}-8 \left(\tau^2+1\right)\right) \\
				& +\frac{Q_2^4 \left(8 \tau ^2+4\right)}{m_{\mu }^2}+Q_1 \left(\frac{Q_2^3 \left(8 \tau ^3+34 \tau \right)}{m_{\mu}^2}-8 Q_2 \tau  \left(\tau ^2+5\right)\right) \\
				& -16 Q_2^2 \left(2 \tau ^2+1\right)-\frac{16 Q_2^3 \tau }{Q_1}\Bigg)\Bigg) , \end{aligned} \end{aligned} \\
   		T_7 &= \frac{1}{Q_3^2} \begin{aligned}[t]
			& \Bigg( \frac{Q_1^2 \left(2 \left(\sigma _1^E+\sigma _2^E-2\right)-\tau ^2 \left(\left(\sigma _1^E+10\right) \sigma _1^E+8 \sigma_2^E-19\right)\right)}{2 m_{\mu }^2} \\
			& +Q_1 \left(\frac{Q_2 \tau  \left(2 \tau ^2 \left(\sigma _2^E-5\right) \left(\sigma_2^E-1\right)-2 \sigma _1^E+\sigma _2^E \left(\sigma _2^E+4\right)-3\right)}{2 m_{\mu }^2}-\frac{4 \tau }{Q_2}\right) \\
			& +\frac{Q_2^2 \tau ^2 \left(\sigma _2^E-5\right) \left(\sigma _2^E-1\right)}{2 m_{\mu }^2}+\frac{Q_1^3 \tau  \left(\sigma _1^E-9\right) \left(\sigma _1^E-1\right)}{2 Q_2 m_{\mu }^2} + 4 \tau ^2 \\
			& +X \begin{aligned}[t]
				& \Bigg(\frac{Q_2 Q_1^3 \left(8 \tau ^3+6 \tau \right)}{m_{\mu }^2}+Q_1 \left(\frac{2 Q_2^3 \tau }{m_{\mu }^2}-8 Q_2 \left(\tau ^3+\tau \right)\right) \\
				& +\frac{Q_1^4 \left(8 \tau ^2-2\right)}{m_{\mu }^2}+Q_1^2 \left(\frac{2 Q_2^2 \left(6 \tau ^2-1\right)}{m_{\mu }^2}-8 \left(\tau ^2+1\right)\right)\Bigg) \Bigg) , \end{aligned} \end{aligned} \\
   		T_8 &= \frac{1}{Q_3^2} \begin{aligned}[t]
			& \Bigg( Q_1^2 \left(\frac{4}{Q_2^2}-\frac{2 \left(2 \tau ^2+1\right) \left(\sigma _1^E+\sigma _2^E-2\right)}{m_{\mu }^2}\right)+Q_1 \left(\frac{4 \tau }{Q_2}-\frac{4 Q_2 \tau  \left(\tau ^2+1\right) \left(\sigma _2^E-1\right)}{m_{\mu }^2}\right) \\
			& -\frac{6 Q_1^3 \tau \left(\sigma _1^E-1\right)}{Q_2 m_{\mu }^2}+\frac{Q_1^4 \left(2-2 \sigma _1^E\right)}{Q_2^2 m_{\mu }^2} \\
			& +X \left(\frac{Q_1^4 \left(8 \tau ^2+4\right)}{m_{\mu }^2}+Q_1^3 \left(\frac{8 Q_2 \tau  \left(\tau ^2+2\right)}{m_{\mu }^2}-\frac{16 \tau }{Q_2}\right)+Q_1^2 \left(\frac{Q_2^2 \left(8 \tau ^2+4\right)}{m_{\mu }^2}-16 \tau ^2\right)\right)\Bigg) , \end{aligned} \\ 
   		T_9 &=  Q_3^2 \left(\frac{\sigma _1^E-1}{Q_2^2 m_{\mu }^2}+\frac{\sigma _2^E-1}{Q_1^2 m_{\mu }^2}-\frac{2}{Q_1^2Q_2^2} \right)+X \left(-\frac{2 Q_3^2}{m_{\mu }^2}+\frac{8 Q_2 \tau }{Q_1}+\frac{8 Q_1 \tau }{Q_2}+8 \left(\tau ^2+1\right)\right) , \\
   		T_{10} &= \frac{1}{Q_3^2} \begin{aligned}[t]
			& \Bigg( -\frac{Q_1^2 \left(\tau ^2 \left(\sigma _1^E-1\right) \left(\sigma _1^E+3\right)+2 \left(\sigma _1^E+\sigma_2^E-2\right)\right)}{m_{\mu }^2} -\frac{Q_2^3 \tau  \left(\sigma _2^E-1\right) \left(\sigma_2^E+3\right)}{Q_1m_{\mu }^2} \\
			& -\frac{Q_2^2 \left(\tau ^2 \left(\sigma _2^E-1\right) \left(\sigma _2^E+3\right)+2 \left(\sigma_1^E+\sigma _2^E-2\right)\right)}{m_{\mu }^2}-\frac{Q_1^3 \tau  \left(\sigma _1^E-1\right) \left(\sigma _1^E+3\right)}{Q_2 m_{\mu}^2} \\
			& +Q_1 \left(\frac{8 \tau }{Q_2}-\frac{Q_2 \tau  \left(\left(\sigma _1^E+4\right) \sigma _1^E+\sigma _2^E \left(\sigma_2^E+4\right)-10\right)}{m_{\mu }^2}\right)+\frac{8 Q_2 \tau}{Q_1} \\
			& +8 \tau ^2 +X \left(-16 Q_1^2 \left(\tau ^2-1\right)-16 Q_2 Q_1 \tau  \left(\tau ^2-1\right)-16 Q_2^2 \left(\tau ^2-1\right)\right) \Bigg) \end{aligned} \\
   			&\quad +X \left(\frac{4 Q_2 Q_1 \tau }{m_{\mu }^2}+\frac{4 Q_1^2}{m_{\mu }^2}+\frac{4 Q_2^2}{m_{\mu }^2}\right) , \\
   		T_{11} &= \frac{1}{Q_3^2} \begin{aligned}[t]
			& \Bigg( \frac{Q_1^2 \left(\tau ^2 \left(\sigma _1^E-5\right) \left(\sigma _1^E-1\right)-2 \left(\sigma _1^E+3 \sigma_2^E-4\right)\right)}{m_{\mu }^2} \\
			& +\frac{Q_2^2 \left(\tau ^2 \left(\left(2-3 \sigma _2^E\right) \sigma _2^E+1\right)+\sigma _1^E-3 \sigma _2^E+2\right)}{m_{\mu }^2}-\frac{6 Q_1^3 \tau  \left(\sigma _1^E-1\right)}{Q_2 m_{\mu }^2} \\
			& +Q_1 \left(\frac{Q_2 \tau \left(\left(\sigma _1^E-2\right) \sigma _1^E-2 \sigma _2^E \left(3 \sigma _2^E+8\right)+23\right)}{2 m_{\mu }^2}+\frac{12 \tau}{Q_2}\right)-\frac{Q_2^3 \tau  \left(\sigma _2^E-1\right)^2}{2 Q_1 m_{\mu }^2} \\
			& +X \left(\frac{Q_2^2 Q_1^2 \left(8 \tau^2+10\right)}{m_{\mu }^2}+\frac{28 Q_2 Q_1^3 \tau }{m_{\mu }^2}+\frac{12 Q_1^4}{m_{\mu }^2}-\frac{2 Q_2^4}{m_{\mu }^2}+Q_2^2 \left(8-8 \tau ^2\right)\right)+8 \tau ^2 \Bigg) , \end{aligned} \\
   		T_{12} &= -\frac{Q_1 \tau  \left({\sigma _1^E}^2-7\right)}{4 Q_2 m_{\mu }^2}+\frac{Q_2 \tau  \left({\sigma_2^E}^2 -7\right)}{4 Q_1 m_{\mu}^2}  +\frac{2 Q_1^2}{Q_2^2 m_{\mu }^2} - \frac{2 Q_2^2}{Q_1^2m_{\mu}^2} - \frac{4}{Q_1^2}+\frac{4}{Q_2^2} \\*
			&\quad +\frac{1}{Q_3^2} \begin{aligned}[t]
				& \Bigg(\frac{Q_2^2 \left(4 \tau ^2 \sigma _1^E+2 \tau ^2 \sigma _2^E+3 \sigma _1^E+\sigma _2^E\right)}{2 m_{\mu }^2}-\frac{Q_1^2 \left(2 \tau ^2 \sigma _1^E+4 \tau ^2 \sigma _2^E+\sigma _1^E+3 \sigma _2^E\right)}{2 m_{\mu }^2} \\
				& -\frac{11 Q_1^3 \tau  \sigma_1^E}{2 Q_2 m_{\mu }^2}+\frac{11 Q_2^3 \tau  \sigma _2^E}{2 Q_1 m_{\mu }^2}+\frac{11 Q_1 Q_2 \tau  \left(\sigma _1^E-\sigma_2^E\right)}{2 m_{\mu }^2}-\frac{2 Q_1^4 \sigma _1^E}{Q_2^2 m_{\mu }^2}+\frac{2 Q_2^4 \sigma _2^E}{Q_1^2 m_{\mu }^2} \\
				& +X \begin{aligned}[t]
					& \Bigg(\frac{Q_1^4 \left(4 \tau ^2+3\right)}{m_{\mu }^2}-\frac{Q_2^4 \left(4 \tau ^2+3\right)}{m_{\mu }^2}+Q_1^3 \left(\frac{14 Q_2 \tau }{m_{\mu }^2}-\frac{16 \tau }{Q_2}\right)-\frac{14 Q_2^3 Q_1 \tau }{m_{\mu }^2} \\
					& -4 Q_1^2 \left(7 \tau ^2+1\right)+Q_2^2 \left(28 \tau ^2+4\right)+\frac{16 Q_2^3 \tau }{Q_1}\Bigg)\Bigg) , \end{aligned}  \end{aligned} \mytag
\end{align*}
where
\begin{align}
	\begin{split}
		X &= \frac{1}{Q_1 Q_2 x} \atan\left( \frac{z x}{1 - z \tau} \right) , \quad x = \sqrt{1 - \tau^2} , \\
		z &= \frac{Q_1 Q_2}{4m_\mu^2} (1-\sigma^E_1)(1-\sigma^E_2) , \quad \sigma^E_i = \sqrt{ 1 + \frac{4 m_\mu^2}{Q_i^2} } , \\
		Q_3^2 &= Q_1^2 + 2 Q_1 Q_2 \tau + Q_2^2 .
	\end{split}
\end{align}


\section{Dispersive representations of scalar functions}

\subsection{Double-spectral representation of basis functions}

\label{sec:AppendixDoubleSpectralBasisFunctions}

In order to derive the double-spectral representation of the basis functions $\tilde\Pi_i$ that is implied by the Mandelstam representation of the BTT functions $\Pi_i$, we have to apply subtractions to the double-spectral integrals. For the subtraction of double-spectral integrals at $u=u_0$, we use the following relations:
\begin{align*}
	\label{eq:SubtractedDoubleSpectralIntegrals}
		\frac{1}{\pi^2} \int ds^\prime du^\prime \frac{1}{s^\prime-s} \frac{1}{u^\prime - u} \rho_{su}(s^\prime,u^\prime) &= \frac{1}{\pi^2} \int ds^\prime du^\prime \frac{1}{s^\prime - s} \frac{\rho_{su}(s^\prime,u^\prime)}{u^\prime - u_0} \\
			&\quad + \frac{u-u_0}{\pi^2} \int ds^\prime du^\prime \frac{1}{s^\prime - s} \frac{1}{u^\prime - u} \frac{\rho_{su}(s^\prime, u^\prime)}{u^\prime - u_0} , \\
		\frac{1}{\pi^2} \int dt^\prime du^\prime \frac{1}{t^\prime - t} \frac{1}{u^\prime - u} \rho_{tu}(t^\prime,u^\prime) &= \frac{1}{\pi^2} \int dt^\prime du^\prime \frac{1}{t^\prime - t} \frac{\rho_{tu}(t^\prime,u^\prime)}{u^\prime - u_0} \\
			&\quad + \frac{u-u_0}{\pi^2} \int dt^\prime du^\prime \frac{1}{t^\prime - t} \frac{1}{u^\prime - u} \frac{\rho_{tu}(t^\prime, u^\prime)}{u^\prime - u_0} , \\
		\frac{1}{\pi^2} \int ds^\prime dt^\prime \frac{1}{s^\prime - s} \frac{1}{t^\prime - t} \rho_{st}(s^\prime,t^\prime) &= - \frac{1}{\pi^2} \int ds^\prime dt^\prime \frac{1}{s^\prime - s} \frac{\rho_{st}(s^\prime, t^\prime)}{u^\prime - u_0} - \frac{1}{\pi^2} \int ds^\prime dt^\prime \frac{1}{t^\prime-t} \frac{\rho_{st}(s^\prime,t^\prime)}{u^\prime - u_0} \\*
			&\quad + \frac{u - u_0}{\pi^2} \int ds^\prime dt^\prime \frac{1}{s^\prime - s} \frac{1}{t^\prime - t} \frac{\rho_{st}(s^\prime, t^\prime)}{u^\prime - u_0} . \mytag
\end{align*}
The first two relations appear like subtractions of a single-dispersion integral, while the $(st)$-double spectral integral involves two subtraction terms. Analogous relations are used for subtractions at $s=s_0$ or $t=t_0$.

We insert in~\eqref{eq:InterestingBasisFunctionsInTermsOfBTT} once- or twice-subtracted double-dispersion relations for the scalar functions $\Pi_i$ to obtain double-spectral representations for the basis functions $\tilde\Pi_i$. The functions requiring one subtraction are given by
\begin{align*}
		\label{eq:DoubleSpectralBTBasisFunctionsSingleSubtr}
		-2 q_3 \cdot q_4 \tilde\Pi_9 &= (s-s_b) \begin{aligned}[t]
				& \bigg\{ \frac{1}{\pi} \int_{4M_\pi^2}^\infty ds^\prime \frac{D^{s;u}_9(s^\prime;u_b)}{s^\prime - s} + \frac{1}{\pi} \int_{4M_\pi^2}^\infty dt^\prime \frac{D^{t;u}_9(t^\prime;u_b)}{t^\prime - t} \bigg\} \end{aligned} \\
				& + (u-u_b) \begin{aligned}[t]
					& \bigg\{ \frac{1}{\pi} \int_{4M_\pi^2}^\infty dt^\prime \frac{D^{t;s}_{22}(t^\prime;s_b)}{t^\prime-t} + \frac{1}{\pi} \int_{4M_\pi^2}^\infty du^\prime \frac{D^{u;s}_{22}(u^\prime;s_b)}{u^\prime - u} \bigg\} \end{aligned} \\
				& + (s-s_b)(u-u_b) \begin{aligned}[t]
					&\bigg\{ \frac{1}{\pi^2} \int ds^\prime dt^\prime \frac{\tilde\rho_{9;st}(s^\prime,t^\prime)}{(s^\prime - s)(t^\prime - t)} + \frac{1}{\pi^2} \int ds^\prime du^\prime \frac{\tilde\rho_{9;su}(s^\prime,u^\prime)}{(s^\prime - s)(u^\prime - u)} \\
					& + \frac{1}{\pi^2} \int dt^\prime du^\prime \frac{\tilde\rho_{9;tu}(t^\prime,u^\prime)}{(t^\prime - t)(u^\prime - u)} \bigg\} ,  \end{aligned} \\
		2 q_1 \cdot q_2 \tilde\Pi_{36} &= (s-s_a) \begin{aligned}[t]
				& \bigg\{ \frac{1}{\pi} \int_{4M_\pi^2}^\infty ds^\prime \frac{D^{s;u}_{43}(s^\prime;u_b)}{s^\prime - s} + \frac{1}{\pi} \int_{4M_\pi^2}^\infty dt^\prime \frac{D^{t;u}_{43}(t^\prime;u_b)}{t^\prime - t} \bigg\} \end{aligned} \\
				& - (u-u_b) \begin{aligned}[t]
					& \bigg\{ \frac{1}{\pi} \int_{4M_\pi^2}^\infty dt^\prime \frac{D^{t;s}_{37}(t^\prime;s_a)}{t^\prime-t} + \frac{1}{\pi} \int_{4M_\pi^2}^\infty du^\prime \frac{D^{u;s}_{37}(u^\prime;s_a)}{u^\prime - u} \bigg\} \end{aligned} \\
				& + (s-s_a)(u-u_b) \begin{aligned}[t]
					&\bigg\{ \frac{1}{\pi^2} \int ds^\prime dt^\prime \frac{\tilde\rho_{36;st}(s^\prime,t^\prime)}{(s^\prime - s)(t^\prime - t)} + \frac{1}{\pi^2} \int ds^\prime du^\prime \frac{\tilde\rho_{36;su}(s^\prime,u^\prime)}{(s^\prime - s)(u^\prime - u)} \\
					& + \frac{1}{\pi^2} \int dt^\prime du^\prime \frac{\tilde\rho_{36;tu}(t^\prime,u^\prime)}{(t^\prime - t)(u^\prime - u)} \bigg\} ,  \end{aligned} \\
		-2 q_3 \cdot q_4 \tilde\Pi_{39} &= (s-s_b) \begin{aligned}[t]
				& \bigg\{ \frac{1}{\pi} \int_{4M_\pi^2}^\infty dt^\prime \frac{D^{t;s}_{49}(t^\prime;s_a)}{t^\prime - t} + \frac{1}{\pi} \int_{4M_\pi^2}^\infty du^\prime \frac{D^{u;s}_{49}(u^\prime;s_a)}{u^\prime - u} \bigg\} \end{aligned} \\
				& - (s-s_a) \begin{aligned}[t]
					& \bigg\{ \frac{1}{\pi} \int_{4M_\pi^2}^\infty dt^\prime \frac{D^{t;s}_{54}(t^\prime;s_b)}{t^\prime-t} + \frac{1}{\pi} \int_{4M_\pi^2}^\infty du^\prime \frac{D^{u;s}_{54}(u^\prime;s_b)}{u^\prime - u} \bigg\} \end{aligned} \\
				& + (s-s_a)(s-s_b) \begin{aligned}[t]
					&\bigg\{ \frac{1}{\pi^2} \int ds^\prime dt^\prime \frac{\tilde\rho_{39;st}(s^\prime,t^\prime)}{(s^\prime - s)(t^\prime - t)} + \frac{1}{\pi^2} \int ds^\prime du^\prime \frac{\tilde\rho_{39;su}(s^\prime,u^\prime)}{(s^\prime - s)(u^\prime - u)} \\
					& + \frac{1}{\pi^2} \int dt^\prime du^\prime \frac{\tilde\rho_{39;tu}(t^\prime,u^\prime)}{(t^\prime - t)(u^\prime - u)} \bigg\} ,  \end{aligned} \\
		-2 q_3 \cdot q_4 \tilde\Pi_{40} &= (s-s_b) \begin{aligned}[t]
				& \bigg\{ \frac{1}{\pi} \int_{4M_\pi^2}^\infty ds^\prime \frac{D^{s;t}_{50}(s^\prime;t_b)}{s^\prime - s} + \frac{1}{\pi} \int_{4M_\pi^2}^\infty du^\prime \frac{D^{u;t}_{50}(u^\prime;t_b)}{u^\prime - u} \bigg\} \end{aligned} \\
				& - (t-t_b) \begin{aligned}[t]
					& \bigg\{ \frac{1}{\pi} \int_{4M_\pi^2}^\infty dt^\prime \frac{D^{t;s}_{54}(t^\prime;s_b)}{t^\prime-t} + \frac{1}{\pi} \int_{4M_\pi^2}^\infty du^\prime \frac{D^{u;s}_{54}(u^\prime;s_b)}{u^\prime - u} \bigg\} \end{aligned} \\
				& + (s-s_b)(t-t_b) \begin{aligned}[t]
					&\bigg\{ \frac{1}{\pi^2} \int ds^\prime dt^\prime \frac{\tilde\rho_{40;st}(s^\prime,t^\prime)}{(s^\prime - s)(t^\prime - t)} + \frac{1}{\pi^2} \int ds^\prime du^\prime \frac{\tilde\rho_{40;su}(s^\prime,u^\prime)}{(s^\prime - s)(u^\prime - u)} \\
					& + \frac{1}{\pi^2} \int dt^\prime du^\prime \frac{\tilde\rho_{40;tu}(t^\prime,u^\prime)}{(t^\prime - t)(u^\prime - u)} \bigg\} ,  \end{aligned} \mytag
\end{align*}
where the discontinuities are defined as (note that in all second
integrals, the variable in the denominator should be traded for the
integration variable by using $s'+t'+u'=\Sigma$)
\begin{align*}
	\label{eq:DiscontinuitiesDefThroughDoubleSpecDens}
	D^{s;u}_i(s^\prime;u_x) &:= \frac{1}{\pi} \int du^\prime \frac{\rho_{i;su}(s^\prime,u^\prime)}{u^\prime - u_x} - \frac{1}{\pi} \int dt^\prime \frac{\rho_{i;st}(s^\prime, t^\prime)}{u^\prime - u_x} , \\
	D^{t;u}_i(t^\prime;u_x) &:= \frac{1}{\pi} \int du^\prime \frac{\rho_{i;tu}(t^\prime,u^\prime)}{u^\prime - u_x} - \frac{1}{\pi} \int ds^\prime \frac{\rho_{i;st}(s^\prime,t^\prime)}{u^\prime - u_x} , \\
	D^{s;t}_i(s^\prime;t_x) &:= \frac{1}{\pi} \int dt^\prime \frac{\rho_{i;st}(s^\prime, t^\prime)}{t^\prime - t_x} - \frac{1}{\pi} \int du^\prime \frac{\rho_{i;su}(s^\prime,u^\prime)}{t^\prime - t_x} , \\
	D^{u;t}_i(u^\prime;t_x) &:= \frac{1}{\pi} \int dt^\prime \frac{\rho_{i;tu}(t^\prime,u^\prime)}{t^\prime - t_x} - \frac{1}{\pi} \int ds^\prime \frac{\rho_{i;su}(s^\prime,u^\prime)}{t^\prime - t_x} , \\
	D^{t;s}_i(t^\prime;s_x) &:= \frac{1}{\pi} \int ds^\prime \frac{\rho_{i;st}(s^\prime,t^\prime)}{s^\prime - s_x}  - \frac{1}{\pi} \int du^\prime \frac{\rho_{i;tu}(t^\prime,u^\prime)}{s^\prime-s_x} , \\
	D^{u;s}_i(u^\prime;s_x) &:= \frac{1}{\pi} \int ds^\prime \frac{\rho_{i;su}(s^\prime,u^\prime)}{s^\prime - s_x} - \frac{1}{\pi} \int dt^\prime \frac{\rho_{i;tu}(t^\prime,u^\prime)}{s^\prime - s_x} , \mytag
\end{align*}
and the double-spectral densities are given by
\begin{align*}
		\tilde\rho_{9;st}(s^\prime,t^\prime) &:= \frac{\rho_{9;st}(s^\prime, t^\prime)}{u^\prime - u_b} + \frac{\rho_{22;st}(s^\prime,t^\prime)}{s^\prime-s_b} , \\
		\tilde\rho_{36;st}(s^\prime,t^\prime) &:= \frac{\rho_{43;st}(s^\prime, t^\prime)}{u^\prime - u_b} - \frac{\rho_{37;st}(s^\prime,t^\prime)}{s^\prime-s_a} , \\
		\tilde\rho_{39;st}(s^\prime,t^\prime) &:= \frac{\rho_{49;st}(s^\prime, t^\prime)}{s^\prime - s_a} - \frac{\rho_{54;st}(s^\prime,t^\prime)}{s^\prime-s_b} , \\
		\tilde\rho_{40;st}(s^\prime,t^\prime) &:= \frac{\rho_{50;st}(s^\prime, t^\prime)}{t^\prime - t_b} - \frac{\rho_{54;st}(s^\prime,t^\prime)}{s^\prime-s_b} . \mytag
\end{align*}
The densities of the $(su)$- and $(tu)$-double spectral regions are defined analogously.

The more complicated functions requiring two subtractions are
\begin{align*}
		\label{eq:DoubleSpectralBTBasisFunctionsDoubleSubtr}
		-4 q_3 \cdot q_4 \tilde\Pi_7 &= 2 (s - s_b) \bigg\{ \frac{1}{\pi} \int_{4M_\pi^2}^\infty ds^\prime \frac{D_{7}^{s;t}(s^\prime;t_b)}{s^\prime - s} + \frac{1}{\pi} \int_{4M_\pi^2}^\infty du^\prime \frac{D_{7}^{u;t}(u^\prime;t_b)}{u^\prime - u} \bigg\} \\
			& + 2 (s - s_b)(t-t_b) \bigg\{  \frac{1}{\pi} \int_{4M_\pi^2}^\infty ds^\prime \frac{D_{7}^{s;t,u}(s^\prime;t_b,u_a)}{s^\prime - s} + \frac{1}{\pi} \int_{4M_\pi^2}^\infty dt^\prime \frac{D_{7}^{t;t,u}(t^\prime;t_b,u_a)}{t^\prime - t} \bigg\} \\
			& - (u - u_a) (t - t_b) \bigg\{ \frac{1}{\pi} \int_{4M_\pi^2}^\infty dt^\prime \frac{D_{31}^{t;s}(t^\prime;s_b)}{t^\prime-t} + \frac{1}{\pi} \int_{4M_\pi^2}^\infty du^\prime \frac{D_{31}^{u;s}(u^\prime;s_b)}{u^\prime-u} \bigg\} \\
			& + (s - s_b)(u - u_a)(t-t_b) \bigg\{ \frac{1}{\pi^2} \int ds^\prime dt^\prime \frac{\tilde\rho_{7;st}(s^\prime,t^\prime)}{(s^\prime - s)(t^\prime - t)} \\
			&\qquad\qquad + \frac{1}{\pi^2} \int ds^\prime du^\prime \frac{\tilde\rho_{7;su}(s^\prime,u^\prime)}{(s^\prime - s)(u^\prime - u)} + \frac{1}{\pi^2} \int dt^\prime du^\prime \frac{\tilde\rho_{7;tu}(t^\prime,u^\prime)}{(t^\prime - t)(u^\prime - u)} \bigg\} , \\
		-4 q_3 \cdot q_4 \tilde\Pi_{19} &= 2 (s - s_b) \bigg\{ \frac{1}{\pi} \int_{4M_\pi^2}^\infty ds^\prime \frac{D_{19}^{s;u}(s^\prime;u_b)}{s^\prime - s} + \frac{1}{\pi} \int_{4M_\pi^2}^\infty dt^\prime \frac{D_{19}^{t;u}(t^\prime;u_b)}{t^\prime - t} \bigg\} \\
			& + 2 (s - s_b)(u-u_b) \bigg\{  \frac{1}{\pi} \int_{4M_\pi^2}^\infty ds^\prime \frac{D_{19}^{s;u,u}(s^\prime;u_a,u_b)}{s^\prime - s} + \frac{1}{\pi} \int_{4M_\pi^2}^\infty dt^\prime \frac{D_{19}^{t;u,u}(t^\prime;u_a,u_b)}{t^\prime - t} \bigg\} \\
			& + (u - u_a) (u - u_b) \bigg\{ \frac{1}{\pi} \int_{4M_\pi^2}^\infty dt^\prime \frac{D_{31}^{t;s}(t^\prime;s_b)}{t^\prime-t} + \frac{1}{\pi} \int_{4M_\pi^2}^\infty du^\prime \frac{D_{31}^{u;s}(u^\prime;s_b)}{u^\prime-u} \bigg\} \\
			& + (s - s_b)(u - u_a)(u-u_b) \bigg\{ \frac{1}{\pi^2} \int ds^\prime dt^\prime \frac{\tilde\rho_{19;st}(s^\prime,t^\prime)}{(s^\prime - s)(t^\prime - t)} \\
			&\qquad\qquad + \frac{1}{\pi^2} \int ds^\prime du^\prime \frac{\tilde\rho_{19;su}(s^\prime,u^\prime)}{(s^\prime - s)(u^\prime - u)} + \frac{1}{\pi^2} \int dt^\prime du^\prime \frac{\tilde\rho_{19;tu}(t^\prime,u^\prime)}{(t^\prime - t)(u^\prime - u)} \bigg\} , \\
		-4 q_1\cdot q_2 q_3\cdot q_4 \tilde\Pi_{21} &= (s - s_a) (s - s_b) \bigg\{ \frac{1}{\pi} \int_{4M_\pi^2}^\infty ds^\prime \frac{D_{21}^{s;u}(s^\prime;u_b)}{s^\prime - s} + \frac{1}{\pi} \int_{4M_\pi^2}^\infty dt^\prime \frac{D_{21}^{t;u}(t^\prime;u_b)}{t^\prime - t} \bigg\} \\
			& + (s - s_a) (s - s_b)(u-u_b) \bigg\{  \frac{1}{\pi} \int_{4M_\pi^2}^\infty ds^\prime \frac{D_{21}^{s;u,u}(s^\prime;u_a,u_b)}{s^\prime - s} + \frac{1}{\pi} \int_{4M_\pi^2}^\infty dt^\prime \frac{D_{21}^{t;u,u}(t^\prime;u_a,u_b)}{t^\prime - t} \bigg\} \\
			& - (u - u_a) (u - u_b) \bigg\{ \frac{1}{\pi} \int_{4M_\pi^2}^\infty dt^\prime \frac{D_{22}^{t;s}(t^\prime;s_b)}{t^\prime-t} + \frac{1}{\pi} \int_{4M_\pi^2}^\infty du^\prime \frac{D_{22}^{u;s}(u^\prime;s_b)}{u^\prime-u} \bigg\} \\
			& - (s-s_b)(u - u_a) (u - u_b) \bigg\{ \frac{1}{\pi} \int_{4M_\pi^2}^\infty dt^\prime \frac{D_{22}^{t;s,s}(t^\prime;s_a,s_b)}{t^\prime-t} +  \frac{1}{\pi} \int_{4M_\pi^2}^\infty du^\prime \frac{D_{22}^{u;s,s}(u^\prime;s_a,s_b)}{u^\prime-u} \bigg\} \\
			& + (s - s_a) (s - s_b)(u-u_a)(u - u_b) \bigg\{ \frac{1}{\pi^2} \int ds^\prime dt^\prime \frac{\tilde\rho_{21;st}(s^\prime,t^\prime)}{(s^\prime - s)(t^\prime - t)} \\*
			&\qquad\qquad + \frac{1}{\pi^2} \int ds^\prime du^\prime \frac{\tilde\rho_{21;su}(s^\prime,u^\prime)}{(s^\prime - s)(u^\prime - u)} + \frac{1}{\pi^2} \int dt^\prime du^\prime \frac{\tilde\rho_{21;tu}(t^\prime,u^\prime)}{(t^\prime - t)(u^\prime - u)} \bigg\} , \mytag
\end{align*}
where
\begin{align*}
	\tilde\rho_{7;st}(s^\prime,t^\prime) &:= \frac{2 \rho_{7;st}(s^\prime, t^\prime)}{(t^\prime-t_b)(u^\prime-u_a)} - \frac{\rho_{31;st}(s^\prime,t^\prime)}{s^\prime-s_b} , \\
	\tilde\rho_{19;st}(s^\prime,t^\prime) &:= \frac{2 \rho_{19;st}(s^\prime, t^\prime)}{(u^\prime-u_a)(u^\prime-u_b)} + \frac{\rho_{31;st}(s^\prime,t^\prime)}{s^\prime-s_b} , \\
	\tilde\rho_{21;st}(s^\prime,t^\prime) &:= \frac{\rho_{21;st}(s^\prime, t^\prime)}{(u^\prime-u_a)(u^\prime-u_b)} - \frac{\rho_{22;st}(s^\prime,t^\prime)}{(s^\prime-s_a)(s^\prime-s_b)} . \mytag
\end{align*}
Again, the densities of the $(su)$- and $(tu)$-double spectral regions are defined analogously. The discontinuities of the twice-subtracted integrals are:
\begin{align*}
		D^{s;t,u}_i(s^\prime;t_x,u_y) &:= \frac{1}{\pi} \int du^\prime \frac{\rho_{i;su}(s^\prime,u^\prime)}{(t^\prime - t_x)(u^\prime - u_y)} - \frac{1}{\pi} \int dt^\prime \frac{\rho_{i;st}(s^\prime, t^\prime)}{(t^\prime - t_x)(u^\prime - u_y)} , \\
		D^{t;t,u}_i(t^\prime;t_x,u_y) &:= \frac{1}{\pi} \int du^\prime \frac{\rho_{i;tu}(t^\prime,u^\prime)}{(t^\prime - t_x)(u^\prime - u_y)} - \frac{1}{\pi} \int ds^\prime \frac{\rho_{i;st}(s^\prime, t^\prime)}{(t^\prime - t_x)(u^\prime - u_y)}  , \\
		D^{s;u,u}_i(s^\prime;u_x,u_y) &:= \frac{1}{\pi} \int du^\prime \frac{\rho_{i;su}(s^\prime,u^\prime)}{(u^\prime - u_x)(u^\prime - u_y)} - \frac{1}{\pi} \int dt^\prime \frac{\rho_{i;st}(s^\prime, t^\prime)}{(u^\prime - u_x)(u^\prime - u_y)}  , \\
		D^{t;u,u}_i(t^\prime;u_x,u_y) &:= \frac{1}{\pi} \int du^\prime \frac{\rho_{i;tu}(t^\prime,u^\prime)}{(u^\prime - u_x)(u^\prime - u_y)} - \frac{1}{\pi} \int ds^\prime \frac{\rho_{i;st}(s^\prime, t^\prime)}{(u^\prime - u_x)(u^\prime - u_y)}  , \\
		D^{t;s,s}_i(t^\prime;s_x,s_y) &:= \frac{1}{\pi} \int ds^\prime \frac{\rho_{i;st}(s^\prime,t^\prime)}{(s^\prime - s_x)(s^\prime - s_y)} - \frac{1}{\pi} \int du^\prime \frac{\rho_{i;tu}(t^\prime, u^\prime)}{(s^\prime - s_x)(s^\prime - s_y)}  , \\
		D^{u;s,s}_i(u^\prime;s_x,s_y) &:= \frac{1}{\pi} \int ds^\prime \frac{\rho_{i;su}(s^\prime,u^\prime)}{(s^\prime - s_x)(s^\prime - s_y)} - \frac{1}{\pi} \int dt^\prime \frac{\rho_{i;tu}(t^\prime, u^\prime)}{(s^\prime - s_x)(s^\prime - s_y)} . \mytag
\end{align*}
The signs are determined by the second subtraction.

\subsection[Dispersion relation for the pion-box input for $(g-2)_\mu$]{Dispersion relation for the pion-box input for $\boldsymbol{(g-2)_\mu}$}

\label{sec:AppendixDRScalarFunctionsForMasterFormula}

In the limit $k\to0$, the scalar functions $\Pi_i(Q_1,Q_2,\tau)$ that are required as an input in the master formula~\eqref{eq:MasterFormula3Dim} can be obtained from the basis functions $\tilde\Pi_i$. Due to the presence of kinematic singularities in the basis functions, it is important to reach the limit $k\to0$ in a very specific way in order to correctly identify the BTT functions $\Pi_i$. One possibility is the following:
\begin{align*}
	\label{eq:RelationBTTvsBasisLimitZeroK}
	\Pi_{1}(Q_1,Q_2,\tau) &= \lim_{k\to0} \tilde\Pi_{1} , \\
	\Pi_{2}(Q_1,Q_2,\tau) &= \lim_{k\to0} \mathcal{C}_{14}\big[ \tilde\Pi_{1} \big] , \\
	\Pi_{4}(Q_1,Q_2,\tau) &= \lim_{k\to0} \tilde\Pi_{4} , \\
	\Pi_{5}(Q_1,Q_2,\tau) &= \lim_{k\to0} \mathcal{C}_{14}\big[ \tilde\Pi_{4} \big] , \\
	\Pi_{7}(Q_1,Q_2,\tau) &= \lim_{k\to0} \Big( \lim_{t\to t_b} \tilde\Pi_7 \Big) , \\
	\Pi_{9}(Q_1,Q_2,\tau) &= \lim_{k\to0} \Big( \mathcal{C}_{13}\big[ \mathcal{C}_{23}\big[ \lim_{t\to t_b} \tilde\Pi_7 \big]\big] \Big) , \\
	\Pi_{10}(Q_1,Q_2,\tau) &= \lim_{k\to0} \Big( \mathcal{C}_{23}\big[ \lim_{t\to t_b} \tilde\Pi_7 \big] \Big) , \\
	\Pi_{16}(Q_1,Q_2,\tau) &= \lim_{k\to0} \Big( \mathcal{C}_{24}\big[ \mathcal{C}_{14}\big[ \lim_{t\to t_b} \tilde\Pi_7 \big]\big] \Big) , \\
	\Pi_{17}(Q_1,Q_2,\tau) &= \lim_{k\to0} \Big( \mathcal{C}_{24}\big[ \mathcal{C}_{13}\big[ \lim_{t\to t_b} \tilde\Pi_7 \big]\big] \Big) , \\
	\Pi_{19}(Q_1,Q_2,\tau) &= \lim_{k\to0} \Big( \lim_{u\to u_b} \tilde\Pi_{19} \Big) , \\
	\Pi_{20}(Q_1,Q_2,\tau) &= \lim_{k\to0} \Big( \mathcal{C}_{23}\Big[ \lim_{s\to s_b} \Big( -\frac{q_1 \cdot q_2 q_3\cdot q_4}{q_1\cdot q_4 q_2\cdot q_3} \tilde\Pi_{21} \Big) \Big] \Big) , \\
	\Pi_{21}(Q_1,Q_2,\tau) &= \lim_{k\to0} \Big( \mathcal{C}_{23}\big[ \lim_{u\to u_b} \tilde\Pi_{19} \big] \Big) , \\
	\Pi_{22}(Q_1,Q_2,\tau) &= \lim_{k\to0} \Big( \lim_{s\to s_b} \Big( -\frac{q_1 \cdot q_2 q_3\cdot q_4}{q_1\cdot q_4 q_2\cdot q_3} \tilde\Pi_{21} \Big) \Big) , \\
	\Pi_{31}(Q_1,Q_2,\tau) &= \lim_{k\to0} \Big( \lim_{s\to s_b} \Big( \frac{q_3 \cdot q_4}{q_1 \cdot q_4 q_2 \cdot q_3} \tilde\Pi_{19} \Big) \Big) , \\
	\Pi_{33}(Q_1,Q_2,\tau) &= \lim_{k\to0} \Big( \mathcal{C}_{23}\Big[ \lim_{s\to s_b} \Big( \frac{q_3 \cdot q_4}{q_1 \cdot q_4 q_2 \cdot q_3} \tilde\Pi_{19} \Big) \Big] \Big) , \\
	\Pi_{39}(Q_1,Q_2,\tau) &= \lim_{k\to0} \Big( \mathcal{C}_{14}\Big[ \lim_{s\to s_a} \Big( \frac{q_1 \cdot q_2}{q_1 \cdot q_4} \tilde\Pi_{36} \Big) \Big] \Big) , \\
	\Pi_{40}(Q_1,Q_2,\tau) &= \lim_{k\to0} \Big( \mathcal{C}_{12}\Big[ \mathcal{C}_{14}\Big[ \lim_{s\to s_a} \Big( \frac{q_1 \cdot q_2}{q_1 \cdot q_4} \tilde\Pi_{36} \Big) \Big] \Big] \Big) , \\
	\Pi_{42}(Q_1,Q_2,\tau) &= \lim_{k\to0} \Big( \mathcal{C}_{12}\Big[ \mathcal{C}_{24}\Big[ \lim_{s\to s_a} \Big( \frac{q_1 \cdot q_2}{q_1 \cdot q_4} \tilde\Pi_{36} \Big) \Big] \Big] \Big) , \\
	\Pi_{46}(Q_1,Q_2,\tau) &= \lim_{k\to0} \Big( \mathcal{C}_{14}\Big[ \mathcal{C}_{23}\Big[ \lim_{s\to s_a} \Big( \frac{q_1 \cdot q_2}{q_1 \cdot q_4} \tilde\Pi_{36} \Big) \Big] \Big] \Big) , \\
	\Pi_{47}(Q_1,Q_2,\tau) &= \lim_{k\to0} \Big( \mathcal{C}_{24}\Big[ \mathcal{C}_{13}\Big[ \lim_{s\to s_a} \Big( \frac{q_1 \cdot q_2}{q_1 \cdot q_4} \tilde\Pi_{36} \Big) \Big] \Big] \Big) , \\
	\Pi_{50}(Q_1,Q_2,\tau) &= \lim_{k\to0} \Big( \lim_{t\to t_b} \tilde\Pi_{40} \Big) , \\
	\Pi_{51}(Q_1,Q_2,\tau) &= \lim_{k\to0} \Big( \mathcal{C}_{12}\big[ \lim_{t\to t_b} \tilde\Pi_{40} \big] \Big) , \\
	\Pi_{54}(Q_1,Q_2,\tau) &= \lim_{k\to0} \Big( \mathcal{C}_{23}\big[ \mathcal{C}_{24}\big[ \lim_{t\to t_b} \tilde\Pi_{40} \big] \big] \Big) . \mytag
\end{align*}
The scalar products in these equations follow from the general relations between the BTT functions and the basis functions~\eqref{eq:HLbLBTTProjectedOnBasis}.

Due to the crossing relations, many other equivalent prescriptions exist. The crossing operator $\mathcal{C}_{ij}$ is understood to act on the Mandelstam variables $s$, $t$, and $u$ as well as on the $q_i^2$. In particular, the limits $s_a$, $s_b$, $\ldots$ are affected as well, which ensures that the limits taken in the argument of the crossing operators lead indeed to a limit compatible with $k\to0$.

We insert the double-spectral representations for the basis functions~\eqref{eq:DoubleSpectralBTBasisFunctionsSingleSubtr} and \eqref{eq:DoubleSpectralBTBasisFunctionsDoubleSubtr} into these relations. The crossing operator acts on the external Mandelstam variable in the Cauchy kernels, but the primed integration variables are understood to be unaffected. We perform the action of the crossing operators on the Cauchy kernels and rename the integration variables where it is convenient. Finally, we obtain the following dispersion relations:
\begin{align*}
	\label{eq:DispersiveRepresentationBTTforMasterFormula}
	\Pi_{1}(Q_1,Q_2,\tau) &= \lim_{q_4^2\to0} \left( \frac{1}{\pi} \int_{4M_\pi^2}^\infty ds^\prime \frac{D_{1}^{s;u}(s^\prime;u_b)}{s^\prime + Q_3^2} + \frac{1}{\pi} \int_{4M_\pi^2}^\infty dt^\prime \frac{D_{1}^{t;u}(t^\prime;u_b)}{t^\prime + Q_2^2} \right) , \\
	\Pi_{2}(Q_1,Q_2,\tau) &= \lim_{q_4^2\to0} \left( \frac{1}{\pi} \int_{4M_\pi^2}^\infty ds^\prime \frac{\mathcal{C}_{14}[ D_{1}^{t;u}(s^\prime;u_b) ]}{s^\prime + Q_3^2} + \frac{1}{\pi} \int_{4M_\pi^2}^\infty dt^\prime \frac{\mathcal{C}_{14}[ D_{1}^{s;u}(t^\prime;u_b) ]}{t^\prime + Q_2^2} \right) , \\
	\Pi_{4}(Q_1,Q_2,\tau) &= \lim_{q_4^2\to0} \left( \frac{1}{\pi} \int_{4M_\pi^2}^\infty ds^\prime \frac{D_{4}^{s;u}(s^\prime;u_b)}{s^\prime + Q_3^2} + \frac{1}{\pi} \int_{4M_\pi^2}^\infty dt^\prime \frac{D_{4}^{t;u}(t^\prime;u_b)}{t^\prime + Q_2^2} \right) , \\
	\Pi_{5}(Q_1,Q_2,\tau) &= \lim_{q_4^2\to0} \left( \frac{1}{\pi} \int_{4M_\pi^2}^\infty ds^\prime \frac{\mathcal{C}_{14}[ D_{4}^{t;u}(s^\prime;u_b) ]}{s^\prime + Q_3^2} + \frac{1}{\pi} \int_{4M_\pi^2}^\infty dt^\prime \frac{\mathcal{C}_{14}[ D_{4}^{s;u}(t^\prime;u_b) ]}{t^\prime + Q_2^2} \right) , \\
	\Pi_{7}(Q_1,Q_2,\tau) &= \lim_{q_4^2\to0} \left( \frac{1}{\pi} \int_{4M_\pi^2}^\infty ds^\prime \frac{D_{7}^{s;t}(s^\prime;t_b)}{s^\prime + Q_3^2} + \frac{1}{\pi} \int_{4M_\pi^2}^\infty du^\prime \frac{D_{7}^{u;t}(u^\prime;t_b)}{u^\prime + Q_1^2} \right) , \\
	\Pi_{9}(Q_1,Q_2,\tau) &= \lim_{q_4^2\to0} \left( \frac{1}{\pi} \int_{4M_\pi^2}^\infty ds^\prime \frac{\mathcal{C}_{13}[\mathcal{C}_{23}[ D_{7}^{u;t}(s^\prime;t_b) ]]}{s^\prime + Q_3^2} + \frac{1}{\pi} \int_{4M_\pi^2}^\infty dt^\prime \frac{\mathcal{C}_{13}[\mathcal{C}_{23}[ D_{7}^{s;t}(t^\prime;t_b)]]}{t^\prime + Q_2^2} \right) , \\
	\Pi_{10}(Q_1,Q_2,\tau) &= \lim_{q_4^2\to0} \left( \frac{1}{\pi} \int_{4M_\pi^2}^\infty dt^\prime \frac{\mathcal{C}_{23}[ D_{7}^{s;t}(t^\prime;t_b) ]}{t^\prime + Q_2^2} + \frac{1}{\pi} \int_{4M_\pi^2}^\infty du^\prime \frac{\mathcal{C}_{23}[ D_{7}^{u;t}(u^\prime;t_b) ]}{u^\prime + Q_1^2} \right) , \\
	\Pi_{16}(Q_1,Q_2,\tau) &= \lim_{q_4^2\to0} \left( \frac{1}{\pi} \int_{4M_\pi^2}^\infty ds^\prime \frac{\mathcal{C}_{24}[\mathcal{C}_{14}[ D_{7}^{u;t}(s^\prime;t_b) ]]}{s^\prime + Q_3^2} + \frac{1}{\pi} \int_{4M_\pi^2}^\infty dt^\prime \frac{\mathcal{C}_{24}[\mathcal{C}_{14}[ D_{7}^{s;t}(t^\prime;t_b) ]]}{t^\prime + Q_2^2} \right) , \\
	\Pi_{17}(Q_1,Q_2,\tau) &= \lim_{q_4^2\to0} \left( \frac{1}{\pi} \int_{4M_\pi^2}^\infty ds^\prime \frac{\mathcal{C}_{24}[\mathcal{C}_{13}[ D_{7}^{s;t}(s^\prime;t_b) ]]}{s^\prime + Q_3^2} + \frac{1}{\pi} \int_{4M_\pi^2}^\infty du^\prime \frac{\mathcal{C}_{24}[\mathcal{C}_{13}[ D_{7}^{u;t}(u^\prime;t_b) ]]}{u^\prime + Q_1^2} \right) , \\
	\Pi_{19}(Q_1,Q_2,\tau) &= \lim_{q_4^2\to0} \left( \frac{1}{\pi} \int_{4M_\pi^2}^\infty ds^\prime \frac{D_{19}^{s;u}(s^\prime;u_b)}{s^\prime + Q_3^2} + \frac{1}{\pi} \int_{4M_\pi^2}^\infty dt^\prime \frac{D_{19}^{t;u}(t^\prime;u_b)}{t^\prime + Q_2^2} \right) , \\
	\Pi_{20}(Q_1,Q_2,\tau) &= \lim_{q_4^2\to0} \left( \frac{1}{\pi} \int_{4M_\pi^2}^\infty ds^\prime \frac{\mathcal{C}_{23}[ D_{22}^{t;s}(s^\prime;s_b) ]}{s^\prime + Q_3^2} + \frac{1}{\pi} \int_{4M_\pi^2}^\infty du^\prime \frac{\mathcal{C}_{23}[ D_{22}^{u;s}(u^\prime;s_b) ]}{u^\prime + Q_1^2} \right) , \\
	\Pi_{21}(Q_1,Q_2,\tau) &= \lim_{q_4^2\to0} \left( \frac{1}{\pi} \int_{4M_\pi^2}^\infty ds^\prime \frac{\mathcal{C}_{23}[ D_{19}^{t;u}(s^\prime;u_b) ]}{s^\prime + Q_3^2} + \frac{1}{\pi} \int_{4M_\pi^2}^\infty dt^\prime \frac{\mathcal{C}_{23}[ D_{19}^{s;u}(t^\prime;u_b) ]}{t^\prime + Q_2^2} \right) , \\
	\Pi_{22}(Q_1,Q_2,\tau) &= \lim_{q_4^2\to0} \left( \frac{1}{\pi} \int_{4M_\pi^2}^\infty dt^\prime \frac{D_{22}^{t;s}(t^\prime;s_b)}{t^\prime + Q_2^2} + \frac{1}{\pi} \int_{4M_\pi^2}^\infty du^\prime \frac{D_{22}^{u;s}(u^\prime;s_b)}{u^\prime + Q_1^2} \right) , \\
	\Pi_{31}(Q_1,Q_2,\tau) &= \lim_{q_4^2\to0} \left( \frac{1}{\pi} \int_{4M_\pi^2}^\infty dt^\prime \frac{D_{31}^{t;s}(t^\prime;s_b)}{t^\prime + Q_2^2} + \frac{1}{\pi} \int_{4M_\pi^2}^\infty du^\prime \frac{D_{31}^{u;s}(u^\prime;s_b)}{u^\prime + Q_1^2} \right) , \\
	\Pi_{33}(Q_1,Q_2,\tau) &= \lim_{q_4^2\to0} \left( \frac{1}{\pi} \int_{4M_\pi^2}^\infty ds^\prime \frac{\mathcal{C}_{23}[ D_{31}^{t;s}(s^\prime;s_b) ]}{s^\prime + Q_3^2} + \frac{1}{\pi} \int_{4M_\pi^2}^\infty du^\prime \frac{\mathcal{C}_{23}[ D_{31}^{u;s}(u^\prime;s_b) ]}{u^\prime + Q_1^2} \right) , \\
	\Pi_{39}(Q_1,Q_2,\tau) &= \lim_{q_4^2\to0} \left( \frac{1}{\pi} \int_{4M_\pi^2}^\infty ds^\prime \frac{\mathcal{C}_{14}[ D^{t;s}_{37}(s^\prime;s_a) ]}{s^\prime + Q_3^2} + \frac{1}{\pi} \int_{4M_\pi^2}^\infty du^\prime \frac{\mathcal{C}_{14}[ D^{u;s}_{37}(u^\prime;s_a) ]}{u^\prime + Q_1^2} \right) , \\
	\Pi_{40}(Q_1,Q_2,\tau) &= \lim_{q_4^2\to0} \left( \frac{1}{\pi} \int_{4M_\pi^2}^\infty ds^\prime \frac{\mathcal{C}_{12}[ \mathcal{C}_{14}[ D^{t;s}_{37}(s^\prime;s_a) ]]}{s^\prime + Q_3^2} + \frac{1}{\pi} \int_{4M_\pi^2}^\infty dt^\prime \frac{\mathcal{C}_{12}[ \mathcal{C}_{14}[ D^{u;s}_{37}(t^\prime;s_a) ]]}{t^\prime + Q_2^2} \right) , \\
	\Pi_{42}(Q_1,Q_2,\tau) &= \lim_{q_4^2\to0} \left( \frac{1}{\pi} \int_{4M_\pi^2}^\infty ds^\prime \frac{\mathcal{C}_{12}[ \mathcal{C}_{24}[ D^{u;s}_{37}(s^\prime;s_a) ]]}{s^\prime + Q_3^2} + \frac{1}{\pi} \int_{4M_\pi^2}^\infty du^\prime \frac{\mathcal{C}_{12}[ \mathcal{C}_{24}[  D^{t;s}_{37}(u^\prime;s_a) ]]}{u^\prime + Q_1^2} \right) , \\
	\Pi_{46}(Q_1,Q_2,\tau) &= \lim_{q_4^2\to0} \left( \frac{1}{\pi} \int_{4M_\pi^2}^\infty dt^\prime \frac{\mathcal{C}_{14}[ \mathcal{C}_{23}[ D^{t;s}_{37}(t^\prime;s_a) ]]}{t^\prime + Q_2^2} + \frac{1}{\pi} \int_{4M_\pi^2}^\infty du^\prime \frac{\mathcal{C}_{14}[ \mathcal{C}_{23}[ D^{u;s}_{37}(u^\prime;s_a) ]]}{u^\prime + Q_1^2} \right) , \\
	\Pi_{47}(Q_1,Q_2,\tau) &= \lim_{q_4^2\to0} \left( \frac{1}{\pi} \int_{4M_\pi^2}^\infty dt^\prime \frac{\mathcal{C}_{24}[ \mathcal{C}_{13}[ D^{t;s}_{37}(t^\prime;s_a) ]]}{t^\prime + Q_2^2} + \frac{1}{\pi} \int_{4M_\pi^2}^\infty du^\prime \frac{\mathcal{C}_{24}[ \mathcal{C}_{13}[ D^{u;s}_{37}(u^\prime;s_a) ]]}{u^\prime + Q_1^2} \right) , \\
	\Pi_{50}(Q_1,Q_2,\tau) &= \lim_{q_4^2\to0} \left( \frac{1}{\pi} \int_{4M_\pi^2}^\infty ds^\prime \frac{D^{s;t}_{50}(s^\prime;t_b)}{s^\prime + Q_3^2} + \frac{1}{\pi} \int_{4M_\pi^2}^\infty du^\prime \frac{D^{u;t}_{50}(u^\prime;t_b)}{u^\prime + Q_1^2} \right), \\
	\Pi_{51}(Q_1,Q_2,\tau) &= \lim_{q_4^2\to0} \left( \frac{1}{\pi} \int_{4M_\pi^2}^\infty ds^\prime \frac{\mathcal{C}_{12}[ D^{s;t}_{50}(s^\prime;t_b) ]}{s^\prime + Q_3^2} + \frac{1}{\pi} \int_{4M_\pi^2}^\infty dt^\prime \frac{\mathcal{C}_{12}[ D^{u;t}_{50}(t^\prime;t_b) ]}{t^\prime + Q_2^2} \right), \\
	\Pi_{54}(Q_1,Q_2,\tau) &= \lim_{q_4^2\to0} \left( \frac{1}{\pi} \int_{4M_\pi^2}^\infty dt^\prime \frac{\mathcal{C}_{23}[ \mathcal{C}_{24}[ D^{u;t}_{50}(t^\prime;t_b) ]]}{t^\prime + Q_2^2} + \frac{1}{\pi} \int_{4M_\pi^2}^\infty du^\prime \frac{\mathcal{C}_{23}[ \mathcal{C}_{24}[ D^{s;t}_{50}(u^\prime;t_b) ]]}{u^\prime + Q_1^2} \right) . \mytag
\end{align*}
The limit $k\to0$ is already taken in the Cauchy kernels. As the numerators only depend on the integration variables and the $q_i^2$, the remaining limit to be taken is $q_4^2\to0$.

For $\Pi_1$, $\Pi_2$, $\Pi_4$, and $\Pi_5$ we have randomly chosen a fixed-$u$ dispersion relation. Because the first six basis functions are not affected by kinematic singularities, we could equally well choose a fixed-$s$ or fixed-$t$ dispersion relation or use directly the Mandelstam representation.

\subsection{Kinematic zeros due to crossing antisymmetry in HLbL}

\label{sec:KinematicZeroCrossingAntisymmetryHLbL}

Here, we discuss in more detail the crossing property of the BTT function $\Pi_{49}$. According to~\eqref{eq:BTTInternalCrossingSymmetries}, $\Pi_{49}$ and its crossed versions are the only scalar functions with odd intrinsic crossing properties. Crossing antisymmetry implies an additional kinematic zero, but, so far, we have ignored this and written down dispersion relations directly for $\Pi_{49}$. As neither the pion-pole nor the pion-box topologies contribute to $\Pi_{49}$, the treatment of the kinematic zero in $\Pi_{49}$ can be of relevance only for contributions of higher intermediate states than discussed here. We show here in general that the dispersive representation of the basis functions is not affected by the presence of the kinematic zero. In particular, the kinematic zero has no influence on the contribution to $(g-2)_\mu$.

Let us denote
\begin{align}
	\Pi_{49}^{1234}(s,t,u) := \Pi_{49}(s,t,u;q_1^2,q_2^2,q_3^2,q_4^2) .
\end{align}
The crossing properties~\eqref{eq:BTTInternalCrossingSymmetries} of $\Pi_{49}$ are then
\begin{align}
	\Pi_{49}^{1234}(s,t,u) = - \Pi_{49}^{2134}(s,u,t) = - \Pi_{49}^{1243}(s,u,t) = \Pi_{49}^{2143}(s,t,u) .
\end{align}
We build symmetric and antisymmetric combinations under exchange of the Mandelstam variables $t$ and $u$, but fixed virtualities:
\begin{align}
	\begin{split}
		\label{eq:SymmAntisymmPartPi49}
		S_{49}^{1234}(s,t,u) &:= \frac{1}{2} \left( \Pi_{49}^{1234}(s,t,u) + \Pi_{49}^{1234}(s,u,t) \right) , \\
		A_{49}^{1234}(s,t,u) &:= \frac{1}{2} \left( \Pi_{49}^{1234}(s,t,u) - \Pi_{49}^{1234}(s,u,t) \right) .
	\end{split}
\end{align}
They fulfill the crossing relations
\begin{align}
	\begin{split}
		S_{49}^{1234}(s,t,u) &= S_{49}^{1234}(s,u,t) = -S_{49}^{2134}(s,t,u) = -S_{49}^{1243}(s,t,u) , \\
		A_{49}^{1234}(s,t,u) &= -A_{49}^{1234}(s,u,t) = A_{49}^{2134}(s,t,u) = A_{49}^{1243}(s,t,u) 
	\end{split}
\end{align}
and hence exhibit the following kinematic zeros:
\begin{align}
	\begin{split}
		S_{49}^{1234}(s,t,u) &= (q_1^2-q_2^2)(q_3^2-q_4^2) \hat S_{49}^{1234}(s,t,u) , \\
		A_{49}^{1234}(s,t,u) &= (t-u) \hat A_{49}(s,t,u) .
	\end{split}
\end{align}
The functions $\hat S_{49}$ and $\hat A_{49}$ are symmetric under the individual exchanges of $t\leftrightarrow u$, $q_1^2\leftrightarrow q_2^2$, and $q_3^2\leftrightarrow q_4^2$. The question arises now if it makes a difference if we write down dispersion relations for $\Pi_{49}$ or rather for $\hat S_{49}$ and $\hat A_{49}$. We show in the following that the Mandelstam representations are not equal but that the basis functions $\tilde\Pi_i$ remain unaffected. Therefore, the kinematic zero has no influence on any physical quantity.

Let us start with the Mandelstam representation of $\Pi_{49}$:
\begin{align}
	\begin{split}
		\Pi_{49}^{1234}(s,t,u) &= \frac{1}{\pi^2} \int ds^\prime  \int dt^\prime \frac{\rho_{49;st}^{1234}(s^\prime,t^\prime)}{(s^\prime - s)(t^\prime-t)}  +  \frac{1}{\pi^2} \int ds^\prime  \int du^\prime \frac{\rho_{49;su}^{1234}(s^\prime,u^\prime)}{(s^\prime - s)(u^\prime-u)} \\
			& +  \frac{1}{\pi^2} \int dt^\prime  \int du^\prime \frac{\rho_{49;tu}^{1234}(t^\prime,u^\prime)}{(t^\prime - t)(u^\prime-u)} .
	\end{split}
\end{align}
The implicit integration limits are the borders of the double-spectral regions. The crossing properties of $\Pi_{49}$ imply the following relations for the double-spectral densities:
\begin{align}
	\begin{split}
		\rho_{49;st}^{1234}(s,t) &= - \rho_{49;su}^{2134}(s,t) = - \rho_{49;su}^{1243}(s,t) = \rho_{49;st}^{2143}(s,t) , \\
		\rho_{49;tu}^{1234}(t,u) &= - \rho_{49;tu}^{2134}(u,t) = - \rho_{49;tu}^{1243}(u,t) = \rho_{49;tu}^{2143}(t,u) .
	\end{split}
\end{align}
With the Mandelstam representation for $\Pi_{49}$, we obtain for the functions defined in~\eqref{eq:SymmAntisymmPartPi49}:
\begin{align*}
		S_{49}^{1234}(s,t,u) &:= \frac{1}{2} \begin{aligned}[t]
			& \Bigg( \frac{1}{\pi^2} \int ds^\prime  \int dt^\prime \frac{\rho_{49;st}^{1234}(s^\prime,t^\prime)+\rho_{49;su}^{1234}(s^\prime,t^\prime)}{(s^\prime - s)(t^\prime-t)}  +  \frac{1}{\pi^2} \int ds^\prime  \int du^\prime \frac{\rho_{49;su}^{1234}(s^\prime,u^\prime)+\rho_{49;st}^{1234}(s^\prime,u^\prime)}{(s^\prime - s)(u^\prime-u)} \\
			& +  \frac{1}{\pi^2} \int dt^\prime  \int du^\prime \frac{\rho_{49;tu}^{1234}(t^\prime,u^\prime)+\rho_{49;tu}^{1234}(u^\prime,t^\prime)}{(t^\prime - t)(u^\prime-u)}  \Bigg) ,\\ \end{aligned} \\
		A_{49}^{1234}(s,t,u) &:= \frac{1}{2} \begin{aligned}[t]
			& \Bigg( \frac{1}{\pi^2} \int ds^\prime  \int dt^\prime \frac{\rho_{49;st}^{1234}(s^\prime,t^\prime)-\rho_{49;su}^{1234}(s^\prime,t^\prime)}{(s^\prime - s)(t^\prime-t)}  +  \frac{1}{\pi^2} \int ds^\prime  \int du^\prime \frac{\rho_{49;su}^{1234}(s^\prime,u^\prime)-\rho_{49;st}^{1234}(s^\prime,u^\prime)}{(s^\prime - s)(u^\prime-u)} \\
			& +  \frac{1}{\pi^2} \int dt^\prime  \int du^\prime \frac{\rho_{49;tu}^{1234}(t^\prime,u^\prime)-\rho_{49;tu}^{1234}(u^\prime,t^\prime)}{(t^\prime - t)(u^\prime-u)}  \Bigg) . \end{aligned}\mytag
\end{align*}
The crossing properties of the double-spectral densities allow us to factor out kinematic zeros. In the case of $S_{49}$, the combinations of double-spectral densities have kinematic zeros of the form $(q_1^2-q_2^2)(q_3^2-q_4^2)$, which can be taken immediately out of the dispersive integrals, hence we find indeed
\begin{align*}
		S_{49}^{1234}(s,t,u) = (q_1^2-q_2^2)(q_3^2-q_4^2) \begin{aligned}[t]
			& \Bigg( \frac{1}{\pi^2} \int ds^\prime  \int dt^\prime \frac{\sigma_{49;st}^{1234}(s^\prime,t^\prime)}{(s^\prime - s)(t^\prime-t)}  +  \frac{1}{\pi^2} \int ds^\prime  \int du^\prime \frac{\sigma_{49;su}^{1234}(s^\prime,u^\prime)}{(s^\prime - s)(u^\prime-u)} \\
			& +  \frac{1}{\pi^2} \int dt^\prime  \int du^\prime \frac{\sigma_{49;tu}^{1234}(t^\prime,u^\prime)}{(t^\prime - t)(u^\prime-u)} \Bigg) , \end{aligned} \mytag
\end{align*}
where the $\sigma_{49}$ denote the double-spectral densities of $\hat S_{49}$.

In the case of $A_{49}$, some algebra is necessary. We subtract in each term the dispersion integrals so that factors of $(t-u)$ multiply the double-spectral contribution:
\begin{align*}
		\frac{1}{\pi^2} \int ds^\prime \int dt^\prime & \frac{\rho_{49;st}^{1234}(s^\prime,t^\prime)-\rho_{49;su}^{1234}(s^\prime,t^\prime)}{(s^\prime-s)(t^\prime-t)} \\
			&= \frac{1}{\pi} \int ds^\prime \frac{1}{s^\prime-s} \frac{1}{\pi} \int dt^\prime \frac{\rho_{49;st}^{1234}(s^\prime,t^\prime)-\rho_{49;su}^{1234}(s^\prime,t^\prime)}{t^\prime-\frac{\Sigma-s^\prime}{2}} \\
			& + \frac{1}{\pi} \int dt^\prime \frac{1}{t^\prime-t} \frac{1}{\pi} \int ds^\prime \frac{\rho_{49;st}^{1234}(s^\prime,t^\prime)-\rho_{49;su}^{1234}(s^\prime,t^\prime)}{2\left(t^\prime-\frac{\Sigma-s^\prime}{2}\right)} \\
			& + (t-u) \frac{1}{\pi^2} \int ds^\prime \int dt^\prime \frac{1}{(s^\prime-s)(t^\prime-t)} \frac{\rho_{49;st}^{1234}(s^\prime,t^\prime)-\rho_{49;su}^{1234}(s^\prime,t^\prime)}{2\left(t^\prime-\frac{\Sigma-s^\prime}{2}\right)}, \\
		\frac{1}{\pi^2} \int ds^\prime \int du^\prime & \frac{\rho_{49;su}^{1234}(s^\prime,u^\prime)-\rho_{49;st}^{1234}(s^\prime,u^\prime)}{(s^\prime-s)(u^\prime-u)} \\
			&= \frac{1}{\pi} \int ds^\prime \frac{1}{s^\prime-s} \frac{1}{\pi} \int du^\prime \frac{\rho_{49;su}^{1234}(s^\prime,u^\prime)-\rho_{49;st}^{1234}(s^\prime,u^\prime)}{u^\prime-\frac{\Sigma-s^\prime}{2}} \\
			& + \frac{1}{\pi} \int du^\prime \frac{1}{u^\prime-u} \frac{1}{\pi} \int ds^\prime \frac{\rho_{49;su}^{1234}(s^\prime,u^\prime)-\rho_{49;st}^{1234}(s^\prime,u^\prime)}{2\left(u^\prime-\frac{\Sigma-s^\prime}{2}\right)}  \\
			& + (t-u) \frac{1}{\pi^2} \int ds^\prime \int du^\prime \frac{1}{(s^\prime-s)(u^\prime-u)} \frac{\rho_{49;st}^{1234}(s^\prime,u^\prime)-\rho_{49;su}^{1234}(s^\prime,u^\prime)}{2\left(u^\prime-\frac{\Sigma-s^\prime}{2}\right)} , \\
		\frac{1}{\pi^2} \int dt^\prime \frac{1}{\pi} \int du^\prime & \frac{\rho_{49;tu}^{1234}(t^\prime,u^\prime)-\rho_{49;tu}^{1234}(u^\prime,t^\prime)}{(t^\prime-t)(u^\prime-u)} \\
			&= \frac{1}{\pi} \int dt^\prime \frac{1}{t^\prime-t} \frac{1}{\pi} \int du^\prime \frac{\rho_{49;tu}^{1234}(t^\prime,u^\prime)-\rho_{49;tu}^{1234}(u^\prime,t^\prime)}{u^\prime-t^\prime} \\
			& + \frac{1}{\pi} \int du^\prime \frac{1}{u^\prime-u} \frac{1}{\pi} \int dt^\prime \frac{\rho_{49;tu}^{1234}(t^\prime,u^\prime)-\rho_{49;tu}^{1234}(u^\prime,t^\prime)}{t^\prime-u^\prime} \\
			& + (t-u) \frac{1}{\pi} \int dt^\prime \frac{1}{t^\prime-t} \frac{1}{\pi} \int du^\prime \frac{\rho_{49;tu}^{1234}(t^\prime,u^\prime)-\rho_{49;tu}^{1234}(u^\prime,t^\prime)}{(u^\prime-u)(t^\prime-u^\prime)} .\mytag
\end{align*}
We see that in the sum of these three terms, the one-dimensional $s$-channel integrals cancel but that the other subtraction terms do not. Hence we find:
\begin{align}
	\begin{split}
		\label{eq:A49DifferentMandelstamReps}
		A_{49}^{1234}(s,t,u) &= (t-u) \begin{aligned}[t]
			& \Bigg( \frac{1}{\pi^2} \int ds^\prime  \int dt^\prime \frac{\alpha_{49;st}^{1234}(s^\prime,t^\prime)}{(s^\prime - s)(t^\prime-t)}  +  \frac{1}{\pi^2} \int ds^\prime  \int du^\prime \frac{\alpha_{49;su}^{1234}(s^\prime,u^\prime)}{(s^\prime - s)(u^\prime-u)} \\
			& +  \frac{1}{\pi^2} \int dt^\prime  \int du^\prime \frac{\alpha_{49;tu}^{1234}(t^\prime,u^\prime)}{(t^\prime - t)(u^\prime-u)} \Bigg) \end{aligned} \\
			& + \frac{1}{\pi} \int dt^\prime \frac{\Delta_{49}^{1234}(t^\prime)}{t^\prime - t} - \frac{1}{\pi} \int du^\prime \frac{\Delta_{49}^{1234}(u^\prime)}{u^\prime - u} ,
	\end{split}
\end{align}
where the $\alpha_{49}$ denote the double-spectral densities of $\hat A_{49}$ and where
\begin{align}
	\Delta_{49}^{1234}(t^\prime) := \frac{1}{\pi} \int ds^\prime
        \frac{\rho_{49;st}^{1234}(s^\prime,t^\prime)-\rho_{49;su}^{1234}(s^\prime,t^\prime)}{2 t^\prime+s^\prime-\Sigma} + \frac{1}{\pi} \int du^\prime \frac{\rho_{49;tu}^{1234}(t^\prime,u^\prime)-\rho_{49;tu}^{1234}(u^\prime,t^\prime)}{u^\prime-t^\prime} .
\end{align}
We learn that writing a Mandelstam representation for $\Pi_{49}$ is not equivalent to taking explicitly into account the kinematic zeros and writing Mandelstam representations for $\hat S_{49}$ and $\hat A_{49}$. The difference consists of a contribution with $t$- and $u$-channel cuts but no double-spectral regions.

However, due to the redundancy in the set of BTT functions, $\Pi_{49}$ is
not a physical quantity: the HLbL tensor is defined by the basis functions
$\tilde\Pi_i$, and the dispersive representation for these does not change
whether one starts from an unsubtracted Mandelstam representation of
$\Pi_{49}$ or $\hat S_{49}$/$\hat A_{49}$. To prove this, we need to study
the impact of the difference terms in~\eqref{eq:A49DifferentMandelstamReps}
on the basis functions $\tilde\Pi_i$. The double-spectral representations
for the basis functions involving $\Pi_{49}$ are given 
in~\eqref{eq:DoubleSpectralBTBasisFunctionsSingleSubtr}. The difference
terms in~\eqref{eq:A49DifferentMandelstamReps} obviously result in a shift
of the single-variable dispersion integrals, while the double-spectral
contributions are unchanged. Note that the single-variable dispersion
integrals involve only fixed-$s$ representations of $\Pi_{49}$ and crossed
versions thereof. They have exactly the form of the difference terms
in~\eqref{eq:A49DifferentMandelstamReps}, hence the difference between the
choices of $\Pi_{49}$ or $\hat S_{49}$/$\hat A_{49}$ as the functions to
fulfill an unsubtracted Mandelstam representation can be reabsorbed in a
shift of $D^{t;s}_{49}$ and $D^{u;s}_{49}$. This means that the basis
functions $\tilde \Pi_i$ and the HLbL tensor remain unaffected by the
kinematic zero in $\Pi_{49}$. In particular, the form of the dispersion
relation for $(g-2)_\mu$
in~\eqref{eq:DispersiveRepresentationBTTforMasterFormula} is unchanged: the
functions $\Pi_{50}$, $\Pi_{51}$, and $\Pi_{54}$ are defined by crossed
versions of a fixed-$s$ dispersion relation for $\Pi_{49}$, where the
effect of the kinematic zero is reabsorbed in the discontinuities of the
$t$- and $u$-channel integrals. Since in the end the discontinuities will
be calculated directly from the unitarity relation, the observation that
the kinematic zero in $\Pi_{49}$ does not alter the functional form of the
dispersion relation completes the proof that there are no practical
consequences, neither in the construction of the dispersion relation nor in
$(g-2)_\mu$.


\section{Dispersive representation of loop functions}

\label{sec:AppendixDispersiveLoops}

\subsection{Scalar two-point function}

\begin{figure}[ht]
	\centering
	\includegraphics[width=5.26cm]{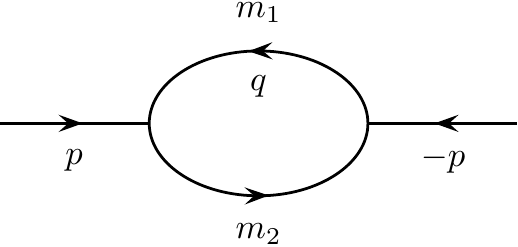}
	\caption{Bulb diagram}
	\label{img:BulbDiagram}
\end{figure}

The scalar two-point function~\cite{Denner1993} corresponding to the loop diagram in Fig.~\ref{img:BulbDiagram} is given by
\begin{align}
	\label{eq:DefinitionScalarTwoPointFunction}
	B_0(p^2,m_1^2,m_2^2) = \frac{1}{i} \int \frac{d^4q}{(2\pi)^4} \frac{1}{[q^2-m_1^2] [(q+p)^2-m_2^2] } .
\end{align}
We define $s:=p^2$. The two-point function has a normal threshold at $s=(m_1+m_2)^2$. According to Cutkosky's rules~\cite{Cutkosky1960}, the discontinuity is given by:
\begin{align}
	2i \Delta_s B_0(s) := B_0(s+i\epsilon) - B_0(s-i\epsilon) = \frac{i}{8\pi} \frac{\lambda^{1/2}(s,m_1^2,m_2^2)}{s} .
\end{align}
$B_0$ is divergent and satisfies a once-subtracted dispersion relation:
\begin{align}
	B_0(s) - B_0(0) = \frac{s}{\pi} \int_{(m_1+m_2)^2}^\infty ds^\prime \frac{\Delta_s B_0(s^\prime)}{(s^\prime - s - i \epsilon)s^\prime} .
\end{align}

\subsection{Scalar three-point function}

\begin{figure}[ht]
	\centering
	\includegraphics[width=4.1cm]{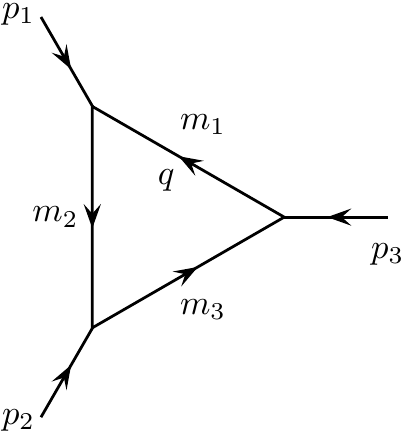}
	\caption{Triangle diagram}
	\label{img:TriangleDiagram}
\end{figure}

Consider the scalar three-point function~\cite{Denner1993} corresponding to the loop diagram in Fig.~\ref{img:TriangleDiagram}:
\begin{align}
	\label{eq:DefinitionScalarThreePointFunction}
	C_0(p_1^2,p_2^2,p_3^2,m_1^2,m_2^2,m_3^2) = \frac{1}{i} \int \frac{d^4q}{(2\pi)^4} \frac{1}{[q^2-m_1^2] [(q+p_1)^2-m_2^2] [(q+p_1+p_2)^2-m_3^2]} .
\end{align}
We define $s := (p_1+p_2)^2 = p_3^2$. The scalar three-point function has a normal threshold at $s=(m_1+m_3)^2$. The discontinuity along the corresponding branch cut can again be found with Cutkosky's rules:
\begin{align}
	2i \Delta_s C_0(s) := C_0(s+i\epsilon) - C_0(s-i\epsilon) = \frac{i}{32\pi^2} \frac{\lambda^{1/2}(s,m_1^2,m_3^2)}{s} \int d\Omega_s^\dprime \frac{1}{(q+p_1)^2-m_2^2} .
\end{align}
Performing the phase space integral leads to:
\begin{align}
	\Delta_s C_0(s) = \frac{1}{16\pi} \frac{1}{\lambda_{12}^{1/2}(s)} \log\left( \frac{2s(\tilde\Delta_{12}+p_1^2) - (s+\tilde\Delta_{13})(s+\Delta_{12})+\tilde\lambda_{13}^{1/2}(s)\lambda_{12}^{1/2}(s)}{2s(\tilde\Delta_{12}+p_1^2) - (s+\tilde\Delta_{13})(s+\Delta_{12})-\tilde\lambda_{13}^{1/2}(s)\lambda_{12}^{1/2}(s)} \right) ,
\end{align}
where $\Delta_{ik} := p_i^2 - p_k^2$, $\tilde\Delta_{ik} := m_i^2-m_k^2$, $\lambda_{ik}(s) := \lambda(s,p_i^2,p_k^2)$, and $\tilde\lambda_{ik}(s) := \lambda(s,m_i^2,m_k^2)$.

$C_0$ satisfies an unsubtracted dispersion relation:
\begin{align}
	C_0(s) = \frac{1}{\pi} \int_{(m_1+m_3)^2}^\infty ds^\prime \frac{\Delta_s C_0(s^\prime)}{s^\prime - s - i \epsilon} .
\end{align}
As the anomalous threshold appears this dispersive representation needs to
be amended by adding a dispersive integral over the anomalous cut
\cite{Frederiksen1971,Lucha:2006vc,Hoferichter:2013ama}.

\subsection{Scalar four-point function}

\label{sec:AppendixScalar4PointFunction}

\begin{figure}[ht]
	\centering
	\includegraphics[width=4.49cm]{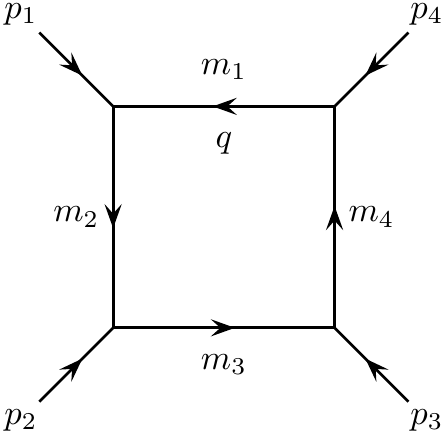}
	\caption{Box diagram}
	\label{img:BoxDiagram}
\end{figure}

Consider the scalar four-point function~\cite{Denner1993} corresponding to the diagram in Fig.~\ref{img:BoxDiagram}:
\begin{align}
	\begin{split}
		D_0&(p_1^2,p_2^2,p_3^2,p_4^2,(p_1+p_2)^2,(p_2+p_3)^2,m_1^2,m_2^2,m_3^2,m_4^2) = \\
		& \frac{1}{i} \int \frac{d^4q}{(2\pi)^4} \frac{1}{[q^2-m_1^2] [(q+p_1)^2-m_2^2] [(q+p_1+p_2)^2-m_3^2] [(q-p_4)^2-m_4^2]} .
	\end{split}
\end{align}
We define
\begin{align}
	\begin{split}
		s &:= (p_1 + p_2)^2 , \\
		t &:= (p_2 + p_3)^2 , \\
		u &:= (p_1+p_3)^2 , \\
		P &:= p_1 + p_2 = - p_3 - p_4 .
	\end{split}
\end{align}
In the $s$-channel, the diagram has a normal threshold at $s=(m_1+m_3)^2$, in the $t$-channel at $t=(m_2+m_4)^2$. For a fixed value of $t$, we find the discontinuity along the $s$-channel cut:
\begin{align}
	\begin{split}
		2i \Delta_s^t D_0(s,t) :={}& D_0(s+i\epsilon,t) - D_0(s-i\epsilon,t) \\
			={}& \frac{i}{32\pi^2} \frac{\lambda^{1/2}(s,m_1^2,m_3^2)}{s} \int d\Omega_s^\dprime \frac{1}{(q+p_1)^2-m_2^2} \frac{1}{(q-p_4)^2-m_4^2} .
	\end{split}
\end{align}
The phase-space integral can be transformed into a dispersive integral, see~\cite{Stoffer2014,Barut1967}:
\begin{align}
	\Delta_s^t D_0(s,t) &= \frac{1}{8\pi} \frac{1}{\tilde\lambda_{13}^{1/2}(s)} \int_{t^+(s)}^\infty dt^\prime \frac{1}{(t^\prime-t - i \epsilon)\sqrt{\big(t^\prime - t^+(s)\big)\big(t^\prime-t^-(s)\big)}} ,
\end{align}
where
\begin{align}
	\begin{split}
		\label{eq:BorderDoubleSpectralRegion}
		t^\pm(s) &:= \frac{\Sigma - s}{2} + \frac{\Delta_{12}\Delta_{34}}{2s} + \frac{\lambda_{12}^{1/2}(s)\lambda_{34}^{1/2}(s)}{2s} z_s^\pm , \\
		z_s^\pm &:= \alpha_1 \alpha_2 \pm \sqrt{\left(\alpha_1^2 - 1\right)\left(\alpha_2^2 - 1\right)} , \\
		\alpha_1 &:= \frac{s( 2 m_2^2 - \Sigma^\prime) + s^2 + \Delta_{12}\tilde\Delta_{13}}{\lambda_{12}^{1/2}(s)\tilde\lambda_{13}^{1/2}(s)} , \\
		\alpha_2 &:= \frac{s( 2 m_4^2 - \Sigma^\dprime) + s^2 - \Delta_{34}\tilde\Delta_{13}}{\lambda_{34}^{1/2}(s)\tilde\lambda_{13}^{1/2}(s)} ,
	\end{split}
\end{align}
and
\begin{align}
	\begin{split}
		\Sigma &:= p_1^2 + p_2^2 + p_3^2 + p_4^2 , \\
		\Sigma^\prime &:= p_1^2 + p_2^2 + m_1^2 + m_3^2 , \\
		\Sigma^\dprime &:= p_3^2+p_4^2+m_1^2+m_3^2 , \\
		\Delta_{ik} &:= p_i^2 - p_k^2 , \\
		\tilde\Delta_{ik} &:= m_i^2-m_k^2 , \\
		\lambda_{ik}(s) &:= \lambda(s,p_i^2,p_k^2) , \\
		 \tilde\lambda_{ik}(s) &:= \lambda(s,m_i^2,m_k^2) .
	\end{split}
\end{align}
Since $D_0$ satisfies an unsubtracted dispersion relation, it can be written as
\begin{align}
	D_0(s,t) &= \frac{1}{\pi^2} \int_{(m_1+m_3)^2}^\infty ds^\prime  \int_{t^+(s^\prime)}^\infty dt^\prime \frac{\rho_0(s^\prime,t^\prime)}{(s^\prime - s - i \epsilon)(t^\prime-t - i \epsilon)} ,
\end{align}
where
\begin{align}
	\label{eq:DoubleSpectralDensityD0}
	\rho_0(s,t) = \frac{1}{8 \tilde\lambda_{13}^{1/2}(s)} \frac{1}{\sqrt{(t - t^+(s))(t-t^-(s))}} .
\end{align}
The validity of the double-spectral representation of the box diagram
breaks down with the appearance of anomalous thresholds. The possibility to
amend it by adding the contribution of a dispersive integral over
anomalous cuts has not been investigated here, as it is irrelevant for the
$(g-2)_\mu$ calculation.

If we do not take a fixed-$t$ dispersion relation as the starting point but a fixed-$u$ dispersion relation, we have to take into account that a line of fixed $u$ in the Mandelstam plane will encounter a right- as well as a left-hand cut:
\begin{align}
	D_0(s,t) &= \frac{1}{\pi} \int_{-\infty}^\infty ds^\prime \frac{\Delta_s^u D_0(s^\prime,\Sigma-u-s^\prime)}{s^\prime - s - i\epsilon} ,
\end{align}
where
\begin{align}
	\begin{split}
		\Delta_s^u &D_0(s^\prime, \Sigma-u-s^\prime) \\
			&= \theta(s^\prime - (m_1+m_3)^2) \frac{D_0(s^\prime + i\epsilon, \Sigma-u-s^\prime)-D_0(s^\prime - i\epsilon, \Sigma-u-s^\prime)}{2i} \\
			&+ \theta(\Sigma-u-s^\prime - (m_2+m_4)^2) \frac{D_0(s^\prime, \Sigma-u-s^\prime - i\epsilon)-D_0(s^\prime, \Sigma-u-s^\prime + i\epsilon)}{2i},
	\end{split}
\end{align}
hence
\begin{align}
	D_0(s,t) &= \frac{1}{\pi} \int_{(m_1+m_3)^2}^\infty ds^\prime \frac{\Delta_s^u D_0(s^\prime,\Sigma-u-s^\prime)}{s^\prime - s - i\epsilon} +  \frac{1}{\pi} \int_{(m_2+m_4)^2}^\infty dt^\prime \frac{\Delta_t^u D_0(\Sigma - u - t^\prime,t^\prime)}{t^\prime - t - i\epsilon} .
\end{align}
Cutkosky's rules lead to the following discontinuities:
\begin{align}
	\begin{split}
		\Delta_s^u D_0(s,\Sigma - u - s) &= \frac{1}{8\pi} \frac{1}{\tilde\lambda_{13}^{1/2}(s)} \int_{t^+(s)}^\infty dt^\prime \frac{1}{(t^\prime - \Sigma + u  + s - i \epsilon)\sqrt{(t^\prime - t^+)(t^\prime-t^-)}} , \\
		\Delta_t^u D_0(\Sigma - u - t,t) &= \frac{1}{8\pi} \frac{1}{\tilde\lambda_{24}^{1/2}(t)} \int_{s^+(t)}^\infty ds^\prime \frac{1}{(s^\prime - \Sigma + u  + t - i \epsilon)\sqrt{(s^\prime - s^+)(s^\prime-s^-)}} ,
	\end{split}
\end{align}
where $s^\pm(t)$ are defined in analogy to $t^\pm(s)$ with the proper permutations of the momenta and masses.

Now, we note that
\begin{align}
	\int_{(m_2+m_4)^2}^\infty dt^\prime \int_{s^+(t^\prime)}^\infty ds^\prime = \int_{(m_1+m_3)^2}^\infty ds^\prime \int_{t^+(s^\prime)}^\infty dt^\prime
\end{align}
and find
\begin{align}
	D_0(s,t) &= \frac{1}{\pi^2} \int_{(m_1+m_3)^2}^\infty ds^\prime  \int_{t^+(s^\prime)}^\infty dt^\prime X(s,t,s^\prime,t^\prime) ,
\end{align}
where
\begin{align}
	\begin{split}
		X(s,t,s^\prime, t^\prime) &= \frac{1}{8 \tilde \lambda_{13}^{1/2}(s^\prime)} \frac{1}{s^\prime - s - i\epsilon} \frac{1}{t^\prime - t - i \epsilon} \frac{1}{\sqrt{(t^\prime - t^+)(t^\prime - t^-)}} \frac{1}{s^\prime - s + t^\prime - t - i\epsilon} \\
			&\quad  \times \left( t^\prime - t + (s^\prime - s) \frac{\tilde \lambda_{13}^{1/2}(s^\prime)}{\tilde \lambda_{24}^{1/2}(t^\prime)} \sqrt{\frac{(t^\prime - t^+(s^\prime))(t^\prime - t^-(s^\prime))}{(s^\prime - s^+(t^\prime))(s^\prime - s^-(t^\prime))}} \right) .
	\end{split}
\end{align}
With some (computer) algebra, we can check that
\begin{align}
	\frac{\tilde \lambda_{13}^{1/2}(s^\prime)}{\tilde \lambda_{24}^{1/2}(t^\prime)} \sqrt{\frac{(t^\prime - t^+(s^\prime))(t^\prime - t^-(s^\prime))}{(s^\prime - s^+(t^\prime))(s^\prime - s^-(t^\prime))}} = 1 ,
\end{align}
so that we find again
\begin{align}
	X(s,t,s^\prime,t^\prime) =  \frac{\rho_0(s^\prime,t^\prime)}{(s^\prime - s - i \epsilon)(t^\prime-t - i \epsilon)} .
\end{align}

\end{appendices}

\renewcommand\bibname{References}
\renewcommand{\bibfont}{\raggedright}
\bibliographystyle{my-physrev}
\phantomsection
\addcontentsline{toc}{section}{References}
\bibliography{Literature}

\end{document}